\documentclass[3p,sort&compress]{elsarticle} 
\usepackage{amsfonts}
\usepackage{amsmath}
\usepackage{amssymb}
\usepackage{bm}
\usepackage{dcolumn}
\usepackage{multirow}

\usepackage{epsfig}
\usepackage{graphicx}
\usepackage{graphics}
\usepackage[latin1]{inputenc}
\usepackage{latexsym}
\usepackage{rotating}
\usepackage[colorlinks=true]{hyperref}

\usepackage{xspace} % Sensible space treatment at end of simple macros
\usepackage[usenames]{color}
\usepackage{mathrsfs}

\usepackage{fancyhdr}
 \usepackage{eucal}

\usepackage{lineno}

\journal{Physics Reports}

\begin{document}

%%%%%%%%%%%%%%%%%%%%%%%%%%%%%%%%%%%%%%%%%%%%%%%%%%%%%%%
%%%%%%%%%%%%%%%%%%%%%%%%%%%%%%%%%%%%%%%%%%%%%%%%%%%%%%%
\newcommand\be{\begin{equation}}
\newcommand\ba{\begin{eqnarray}}
\newcommand\bal{\begin{align}}
\newcommand\eal{\end{align}}
\newcommand\ee{\end{equation}}
\newcommand\ea{\end{eqnarray}}
\newcommand{\ny}[1]{\textcolor{blue}{\it{\textbf{ny: #1}}} }
\newcommand{\ky}[1]{\textcolor{magenta}{\it{\textbf{ky: #1}}} }
\newcommand{\kent}[1]{\textcolor{cyan}{\textbf{ #1}} }

\newcommand{\comment}{\bf}
\newcommand{\pont}{{\,^\ast\!}\mathcal R\,\mathcal R}

%%%
\newcommand{\bb}{\mbox{\boldmath$b$}}
\newcommand{\qq}{\mbox{\boldmath$q$}}
\newcommand{\mm}{\mbox{\boldmath$g$}}
\newcommand{\vm}{{\vec{m}}}
\newcommand{\vn}{{\vec{n}}}
\newcommand{\vl}{{\vec{l}}}
\newcommand{\temp}{{\mbox{\tiny temp}}}
\newcommand{\true}{{\mbox{\tiny true}}}
\newcommand{\inj}{{\mbox{\tiny i}}}
\newcommand{\sys}{{\mbox{\tiny sys}}}
\newcommand{\AB}{{\mbox{\tiny AB}}}

\newcommand{\et}{\textit{et al.}}

\newcommand{\mrm}{\mathrm}

\def\hatn{{\bf \hat n}}
\def\hatnprime{{\bf \hat n'}}
\def\hatnone{{\bf \hat n}_1}
\def\hatntwo{{\bf \hat n}_2}
\def\hatni{{\bf \hat n}_i}
\def\hatnj{{\bf \hat n}_j}
\def\vecx{{\bf x}}
\def\veck{{\bf k}}
\def\hatx{{\bf \hat x}}
\def\hatk{{\bf \hat k}}
\def\hatz{{\bf \hat z}}
\def\VEV#1{{\left\langle #1 \right\rangle}}
\def\cP{{\cal P}}
\def\noise{{\rm noise}}
\def\pix{{\rm pix}}
\def\map{{\rm map}}

\def\o{\over}
\def\a{\alpha}
\def\lmat{\left( \matrix{ }
\def\rmat{ }\right) }
\def\Pl{{\rm Pl}}
\def\hp{h_+}
\def\hc{h_{\times}}
\def\half{\frac{1}{2}}
\def\a{\alpha}
\def\lmat{\left( \matrix{ }
\def\rmat{ }\right) }
\def\yboxit#1#2{\vbox{\hrule height #1 \hbox{\vrule width #1
    \vbox{#2}\vrule width #1 }\hrule height #1 }}
\def\fillbox#1{\hbox to #1{\vbox to #1{\vfil}\hfil}}
\def\ybox{\yboxit{0.4pt}{\fillbox{8pt}}\hskip-0.4pt}
\def\VEV#1{\langle{ #1} \rangle}
\newcommand{\fixme}[1]{\textbf{FIXME: }$\langle$\textit{#1}$\rangle$}
%%%
\def\sss{\scriptscriptstyle}

\def\hp{h_+}
\def\hc{h_{\times}}
\def\yboxit#1#2{\vbox{\hrule height #1 \hbox{\vrule width #1
    \vbox{#2}\vrule width #1 }\hrule height #1 }}
\def\fillbox#1{\hbox to #1{\vbox to #1{\vfil}\hfil}}
\def\ybox{\yboxit{0.4pt}{\fillbox{8pt}}\hskip-0.4pt}

\newcommand{\bomega}{\mbox{\boldmath$\omega$}}
\newcommand{\RRdual}{\mbox{\boldmath$ R \tilde R$}}
\newcommand{\Rdual}{\mbox{$\tilde R$}} 

\newcommand{\mb}[1]{\mbox{\boldmath $#1$}}
\newcommand{\ZM}{{\mbox{\tiny ZM}}}
\newcommand{\CPM}{{\mbox{\tiny CPM}}}
\newcommand{\CS}{{\mbox{\tiny CS}}}
\newcommand{\GR}{{\mbox{\tiny GR}}}
\newcommand{\AXIAL}{{\mbox{\tiny Axial}}}
\newcommand{\POLAR}{{\mbox{\tiny Polar}}}
\newcommand{\HH}{{\mbox{\tiny H}}}
\newcommand{\BL}{{\mbox{\tiny BL}}}
\newcommand{\BH}{{\mbox{\tiny BH}}}
\newcommand{\LE}{{\mbox{\tiny LE}}}
\newcommand{\EH}{{\mbox{\tiny EH}}}
\newcommand{\E}{{\mbox{\tiny E}}}
\newcommand{\mat}{{\mbox{\tiny mat}}}
\newcommand{\Pont}{{\mbox{\tiny pont}}}
\newcommand{\EDGB}{{\mbox{\tiny EDGB}}}
\newcommand{\EiBI}{{\mbox{\tiny EiBI}}}
\newcommand{\Edd}{{\mbox{\tiny Edd}}}
\newcommand{\BD}{{\mbox{\tiny BD}}}
\newcommand{\orb}{{\mbox{\tiny orb}}}
\newcommand{\N}{{\mbox{\tiny N}}}
\newcommand{\IS}{{\mbox{\tiny IS}}}
\newcommand{\SBS}{{\mbox{\tiny SBS}}}

%%% 
\newcommand{\bse}{\begin{subequations}}
\newcommand{\ese}{\end{subequations}}
\newcommand{\barr}{\begin{array}}
\newcommand{\earr}{\end{array}}

\newcommand{\mP}{M_{\textrm{Pl}}}
\newcommand{\ie}{\emph{ie}}
\newcommand{\lsim}{\lesssim}
\newcommand{\gsim}{\gtrsim}
\newcommand{\nn}{\nonumber}

%%%%%%%%%%%%%%%%%%%%%%%%%%%%%%%%%%%%%%%%%%%%%%%%%%%%%%%
%%%%%%%%%%%%%%%%%%%%%%%%%%%%%%%%%%%%%%%%%%%%%%%%%%%%%%%
\begin{frontmatter}

%%%%%%%%%%%%%%%%%%%%%%%%%%%%%%%%%%%%%%%%%%%%%%%%%%%%%%%
%%%%%%%%%%%%%%%%%%%%%%%%%%%%%%%%%%%%%%%%%%%%%%%%%%%%%%%
\title{Approximate Universal Relations \\ for Neutron Stars and Quark Stars}

\author[ad1,ad2]{Kent Yagi}
\cortext[corr]{Corresponding author}
\ead{kyagi@princeton.edu}
\author[ad2]{Nicol\'as Yunes\corref{corr}}
%\cortext[corr]{Corresponding author}
%\ead{nyunes@physics.montana.edu}
\address[ad1]{Department of Physics, Princeton University, Princeton, NJ 08544, USA.}
\address[ad2]{eXtreme Gravity Institute, Department of Physics, Montana State University, Bozeman, MT 59717, USA.}

%%%%%%%%%%%%%%%%%%%%%%%%%%%%%%%%%%%%%%%%%%%%%%%%%%%%%%%
%%%%%%%%%%%%%%%%%%%%%%%%%%%%%%%%%%%%%%%%%%%%%%%%%%%%%%%
\begin{abstract}

Neutron stars and quark stars are ideal laboratories to study fundamental physics at supra nuclear densities and strong gravitational fields. Astrophysical observables, however, depend strongly on the star's internal structure, which is currently unknown due to uncertainties in the equation of state. Universal relations, however, exist among certain stellar observables that do not depend sensitively on the star's internal structure. One such set of relations is between the star's moment of inertia ($I$), its tidal Love number (Love) and its quadrupole moment ($Q$), the so-called \emph{I-Love-Q relations}. Similar relations hold among the star's multipole moments, which resemble the well-known black hole no-hair theorems. Universal relations break degeneracies among astrophysical observables, leading to a variety of applications: (i) X-ray measurements  of the nuclear matter equation of state, (ii) gravitational wave measurements of the intrinsic spin of inspiraling compact objects, and (iii) gravitational and astrophysical tests of General Relativity that are independent of the equation of state. We here review how the universal relations come about and all the applications that have been devised to date. 

\end{abstract}
%%%%%%%%%%%%%%%%%%%%%%%%%%%%%%%%%%%%%%%%%%%%%%%%%%%%%%%
%%%%%%%%%%%%%%%%%%%%%%%%%%%%%%%%%%%%%%%%%%%%%%%%%%%%%%%


\end{frontmatter}

\tableofcontents

\newpage

%%%%%%%%%%%%%%%%%%%%%%%%%%%%%%%%%%%%%%%%%%%%%%%%%%%%%%%%
\section{The Ubiquity of Universality}

Universal behavior, a set of properties of a physical system that arises irrespective of the internal details of the system, is everywhere in Nature. From the probability of an avalanche happening to how cracks form and propagate in materials, from the electrical breakdown of dielectrics to fluid flow in disordered media, from molecular diffusion in fluids to how rocks distribute according to their sizes, from the emergence of opalescence in fluids to the size and spin of black holes, universal behavior is unavoidable. The concept of universality, however, differs slightly from discipline to discipline. Perhaps the most well-studied version of universality stems from the field of statistical mechanics where universal behavior is associated with a large class of systems whose macroscopic properties are independent of their dynamics. On a microscopic level, different parts of the system may seem to act independently, but in a certain scaling limit, the system shows global properties that are independent of the microscopic details of each part. 

Historically, universality in statistical mechanics arose from the study of phase transitions, colloquially speaking, the abrupt and dramatic change of the properties of a system. The latter are typically encoded in a certain order parameter, such as the density $\rho$, which is a function of another parameter of the system, such as the temperature $T$. A phase transition is defined by the existence of a critical parameter or \emph{critical point}, at which the system's properties change abruptly. Universal behavior in phase transitions occurs when the order parameter of the system becomes insensitive to the details of the system as the critical point is approached. For the example considered above, universality occurs if $\rho = \rho_\mrm{cr} [(T - T_\mrm{cr})/T_\mrm{cr}]^{\delta}$ and if the critical exponent $\delta$ is a constant that is independent of the details of the system, where $\rho_\mrm{cr}$ and $T_\mrm{cr}$ are the critical density and temperature. Therefore, systems that differ widely in composition or overall properties can present the same type of universal behavior, the same critical exponent, independent of their internal composition. A well-known example of such a phase transition is ferromagnetism.

Today, the concept of universality has gone beyond the precise statistical mechanics definition to simply refer to properties of a system or an object that can be deduced from a small, finite set of \emph{global} parameters, without requiring local knowledge of the system. A global parameter here means a quantity that depends on the behavior of the entire system, and not just on the behavior of the system in a small subdomain; typically, a global quantity is defined as an integral over the entire manifold. For example, the total mass of a system is a global quantity, while the density at the system's surface is a local quantity.  From Newtonian mechanics and Birkhoff's theorem in General Relativity, we know that the gravitational field outside a spherical mass distribution is completely determined by its total mass, and not by the precise material composition of the distribution. If the distribution is not spherical, however, the exterior gravitational field depends on an infinite set of multipole moments, and thus, the exterior gravitational field of spinning planets and stars is not universal.

%%%%%%%%%%%%%%%%%%%%%%%%%%%%%%%%%%%%%%%%%%%%%%%%%%%%%%%%
\subsection{Universality in Black Holes}

One of the most fascinating and far-reaching results of General Relativity involves the universality of black holes: the exterior gravitational field, and in fact the entire exterior metric tensor field, of stationary, isolated black holes in General Relativity can be entirely described in terms of only three global parameters: the mass, the electric charge and the spin angular momentum\footnote{See~\cite{Herdeiro:2014goa,Herdeiro:2015gia,Herdeiro:2015waa,Herdeiro:2015tia} for recent work on black holes in General Relativity that can acquire scalar hair.}. This statement sometimes goes by the name of the \emph{no-hair theorem} or the \emph{two-hair theorem}~\cite{1996bhut.book.....H,Chrusciel:2012jk}. The latter neglects the electric charge by assuming that black holes quickly and efficiently neutralize due to quantum Schwinger pair-production effects~\cite{Gibbons:1975kk,1982PhRvD..25.2509H}, a vacuum breakdown mechanism~\cite{Goldreich:1969sb,Ruderman:1975ju,Blandford:1977ds} and accretion of intergalactic or disk plasma. According to this universality principle, it does not matter what fell into the black hole to create it in the first place; all that information is hidden inside the black hole's event horizon and it is thus causally disconnected and unobservable by any exterior agent. Because of this principle, astrophysical observations need only care about the mass (and if accurate enough, also the spin) of black holes to, for example, monitor the orbits of stars at the center of the Milky Way~\cite{Alexander:2005jz,Will:2007pp,Gillessen:2008qv,Merritt:2009ex,Liu:2011ae} and search for gravitational waves emitted in binary black hole coalescences~\cite{Abbott:2016blz,Abbott:2016nmj}. 

The history of the no-hair theorem is long and twisted, and summarized for example in Misner, Thorne and Wheeler~\cite{Misner:1973cw}. Hawking proved that all stationary black holes must (i) have a horizon with a spherical topology and (ii) be either static or axially symmetric~\cite{Hawking:1971tu,Hawking:1972vc}. Israel had earlier proved that any static black hole with an event horizon of spherical topology must (i) have external fields that are determined uniquely by its mass and charge and (ii) be either the Schwarzschild or Reissner-Nordstr\"om solution~\cite{Israel:1967wq,Israel:1967za}. Carter extended Israel's proof to axially-symmetric but uncharged black holes with event horizons of spherical topology, whose external fields are uniquely determined by its mass and spin angular momentum~\cite{Carter:1971zc}. This, in turn, was extended by Robinson, who proved the uniqueness of the Kerr solution~\cite{Robinson:1975bv}. Combining all of these results, one arrives at the statement of the no-hair theorem mentioned above. These theorems, however, make several assumptions, such as the non-degeneracy of event horizons and the real analyticity of spacetime, that have not yet been relaxed. 

Real black holes, of course, do have hair. The theorems described above are only valid for \emph{isolated and stationary} black holes, but no system in Nature is truly isolated or stationary. Astrophysical black holes are typically surrounded by other stars or compact objects, even if the latter are a large distance away. Moreover, astrophysical black holes are constantly accreting matter, even if at an infinitesimal rate due to the inextricable presence of dust or radiation. Therefore, the total multipole moments of the spacetime at spatial infinity are different from those of an isolated black hole, but this difference is entirely due to the multipole moments of these additional external fields~\cite{Gurlebeck:2015xpa} (see also~\cite{Damour:2009vw,Binnington:2009bb,Kol:2011vg,Chakrabarti:2013lua,Pani:2015hfa,Landry:2015zfa} for related works). The latter have a truly negligible effect in the exterior gravitational field of astrophysical black holes, which is still effectively dominated just by the black hole's mass and spin angular momentum. The spirit of the no-hair theorems, however, can be strongly violated in theories other than General Relativity, if one considers black holes in spacetimes with dimensions higher than four~\cite{Emparan:2001wn,Emparan:2008eg}, or black holes in the presence of exotica, such as non-Abelian Yang-Mills fields, Proca fields, coupled scalar fields or skyrmions.

Perhaps one of the nicest incarnations of the no-hair theorem is the relation between the multipole moments of the exterior metric of a black hole. In flat space, multipole moments are typically understood as the angle-independent coefficients of the far-field expansion of a solution to the Laplace equation in terms of a Legendre decomposition. Geroch extended this definition to static, curved spacetimes in a coordinate-independent way through the use of the conformal group~\cite{Geroch:1970cc,Geroch:1970cd}. Hansen further extended~\cite{Geroch:1970cc,Geroch:1970cd} to stationary spacetimes and found that the extended Geroch moments of the Kerr solution satisfy the following identity~\cite{Hansen:1974zz}:
\be
\label{eq:BH-no-hair}
M_{\ell} + i S_{\ell} = M (i a)^{\ell}\,,
\ee
where $M_{\ell}$ are mass moments, $S_{\ell}$ are current moments, $M = M_{0}$ is the black hole mass, and $a = S_{1}/M = |\vec{S}|/M$ is the black hole spin angular momentum per unit mass. Notice that only two parameters, $M$ and $a$, are required to completely determine all of the Geroch-Hansen moments of the Kerr solution.  Since the metric tensor of a stationary spacetime can be constructed from knowledge of all the Geroch-Hansen moments of the spacetime~\cite{Backdahl:2005uz,Backdahl:2006ed}, the exterior gravitational field described by the Kerr solution is completely determined by the black hole's mass and spin.  

%%%%%%%%%%%%%%%%%%%%%%%%%%%%%%%%%%%%%%%%%%%%%%%%%%%%%%%%
\subsection{Universality in Non-Vacuum Spacetimes}

Astrophysical objects other than black holes are not expected to share the same type of universality. Black holes are very special solutions to Einstein's theory, with singularities and event horizons; it may stand to reason to expect that information about material that fell into a black hole would be hidden from its exterior by its event horizon. Stars, however, do not possess an event horizon, but rather a stellar surface. The internal composition of a star or of a planet should therefore affect the exterior gravitational field it produces. This is indeed the case, for example, for Earth, whose gravitational field can be represented as a sum over its infinitely many multipole moments. In fact, astrophysical missions, such as GRACE~\cite{2004GeoRL..31.9607T} and GAIA~\cite{Perryman:2001sp,deBruijne:2012xj}, are designed to measure the multipole moments of Earth~\cite{2004GeoRL..31.9607T}, and thus, be able to model and predict the motion of satellites in orbit to high accuracy. 

The same lack of universality may also be expected in compact objects, such as white dwarfs and neutron stars. White dwarfs are mostly composed of degenerate electrons, supported against gravitational collapse by electron-degeneracy pressure. With masses comparable to the Sun and radii comparable to Earth, white dwarfs are very weakly relativistic, with gravitational compactnesses $C = (G M)/(R c^{2}) \sim 10^{-3}$, and their internal composition can thus be well-described as a cold Fermi gas. Neutron stars~\cite{Lattimer:2000nx,Lattimer:2006xb,Lattimer:2012nd}, on the other hand, are supported by neutron-degeneracy pressure, and thus, are much more compact; although their mass is typically comparable to the Sun, their radii are of order ten kilometers, so that $C \in (0.05,0.25)$. The density inside neutron stars can easily exceed the nuclear saturation limit, rendering the cold Fermi gas model inapplicable. The inner core of sufficiently massive neutron stars may contain hyperons and kaon-condensates, as well as quark-gluon plasmas that are color-superconducting~\cite{Alford:2004pf}. In fact, neutron-star-like compact objects may be made out of purely quarks (quark stars)~\cite{SQM}. Therefore, the interior structure of such compact stars\footnote{In this review, the phrase ``compact stars'' refers to both neutron stars and quark stars.} can vary with its mass, as more massive stars allow for higher central densities, and thus, for the possibility of exotica in their inner cores. Variability in the internal structure then suggests variability in their exterior gravitational field, since one may expect stars with exotica in their inner cores to produce a gravitational field different from that of stars without such exotica. 

In spite of this reasoning and of the lack of applicability of the no-hair theorems, the exterior gravitational field of neutron stars and quark stars has recently been found to present certain universality. For example, the current dipole moment (i.e.~the spin angular momentum) and the mass quadrupole moment (i.e.~the quadrupole moment) have been shown to obey \emph{approximately universal relations}, i.e.~relations that are approximately insensitive to the stellar internal composition to percent level. The \emph{I-Love-Q} relations~\cite{Yagi:2013bca,Yagi:2013awa} (between the moment of inertia, the Love number and the quadrupole moment) are an example of this universality. These relations have been extended to higher multipole order~\cite{Pappas:2013naa,Stein:2014wpa,Yagi:2014bxa} and to a large class of neutron stars, including weakly-magnetized~\cite{Haskell:2013vha} and rapidly rotating~\cite{Doneva:2013rha,Pappas:2013naa,Chakrabarti:2013tca,Yagi:2014bxa} ones. Taken together, these results imply the existence of \emph{approximate no-hair relations for neutron stars and quark stars} through which one can approximately determine all of the star's multipole moments with knowledge only of the first three: the mass, the spin and the quadrupole moment~\cite{Stein:2014wpa,Yagi:2014bxa}. The approximate relations take on a form analogous to Eq.~\eqref{eq:BH-no-hair}, which, in turn, allows one to construct the gravitational field.  In fact, one can approximately represent the full metric tensor outside a stationary compact star entirely in terms of only these three quantities~\cite{Pappas:2012nv,Pappas:2015mba,Pappas:2016sye}. In particular, Pappas~\cite{Pappas:2012nv,Pappas:2015mba} use the two-soliton solution found by Manko \et~\cite{1995JMP....36.3063M} (see also Manko and Ruiz~\cite{Manko:2016avv} for a simpler representation of the solution). 

The approximately universal I-Love-Q and no-hair relations that compact stars satisfy are the main topic of this review paper. Many other universal relations have been discovered in neutron stars~\cite{Lattimer:1989zz,1994ApJ...424..846R,Prakash:1996xs,Andersson:1997rn,Lattimer:2000nx,Bejger:2002ty,Carriere:2002bx,Benhar:2004xg,Tsui:2004qd,Lattimer:2004pg,Lattimer:2004nj,Morsink:2007tv,Haensel:2009wa,Lau:2009bu,Kiuchi:2010ze,Kyutoku:2010zd,Bauswein:2011tp,Urbanec:2013fs,Baubock:2013gna,Takami:2014zpa,Bernuzzi:2014kca,AlGendy:2014eua,Sotani:2014goa,Takami:2014tva,Bernuzzi:2015rla,Chirenti:2015dda,Pannarale:2015jia,Blazquez-Salcedo:2015ets,Bauswein:2015yca,Steiner:2015aea,Breu:2016ufb,Silva:2016myw,Rezzolla:2016nxn}, such as the relation between the complex oscillation frequencies of neutron stars and certain functions of mass and radius~\cite{Andersson:1996pn,Andersson:1997rn,Tsui:2004qd}. These relations are different from the I-Love-Q and no-hair ones in that they do not involve the multipole moments of the exterior gravitational field of compact stars. We will here touch on some of these other universal relations only through a description of how they connect to the I-Love-Q and no-hair relations for compact stars.   

%%%%%%%%%%%%%%%%%%%%%%%%%%%%%%%%%%%%%%%%%%%%%%%%%%%%%%%%
\subsection{Why Universality Matters}

The approximate universal relations of neutron stars and quark stars are important for multiple reasons. On the astrophysics front, a measurement of the compact star mass, spin period and moment of inertia can be used to automatically obtain the star's quadrupole moment through the I-Q relations. In turn, this would provide a very precise description of the exterior gravitational field of compact stars~\cite{Berti:2004ny}, allowing an extraordinarily precise prediction of the orbit of objects around such stars. Of course, such precision is currently unnecessary, since not even the effect of the moment of inertia, and thus the dipole moment, on the orbital motion is currently within the level of observational precision~\cite{Lattimer:2004nj,Kramer:2009zza}. 

Perhaps of greater applicability is the use of the approximate universal relations to break degeneracies in compact star observations. For example, the NICER~\cite{2012SPIE.8443E..13G,2014SPIE.9144E..20A} and LOFT~\cite{Feroci:2011jc,2012AAS...21924906R,2012SPIE.8443E..2DF} missions aim to detect X-rays emitted from the surface of hot spots on millisecond pulsars. The waveform model for the X-ray light-curve depends, in principle, on a large set of parameters that include the mass, the spin and the star's higher multipole moments~\cite{Psaltis:2013fha,Psaltis:2013zja}. Through the use of compact star universality one can eliminate some of these parameters from the waveform model, thus analytically breaking degeneracies that allows the remaining parameters to be measured more accurately~\cite{Morsink:2007tv,Baubock:2013gna,Psaltis:2013fha}.  Similarly, gravitational wave astronomy has just recently begun with the direct detection of gravitational waves emitted in black hole binary coalescences~\cite{Abbott:2016blz,Abbott:2016nmj}. The LIGO~\cite{Abramovici:1992ah,Abbott:2007kv,Harry:2010zz}, Virgo~\cite{Giazotto:1988gw,TheVirgo:2014hva} GEO~\cite{Willke:2002bs}, KAGRA~\cite{Somiya:2011np,Aso:2013eba} and LIGO-India~\cite{Unnikrishnan:2013qwa} detectors aim to detect many more gravitational wave signals, such as those produced in the late inspiral and merger of compact star binaries. The waveform model for the gravitational wave signal depends on parameters that include the compact star masses, spins, quadrupole moments and Love numbers. Through the use of universal relations one can analytically break degeneracies in the waveform model, and thus, measure the remaining parameters more accurately~\cite{Yagi:2013bca,Yagi:2013awa,Yagi:2015pkc}. 

An improvement in the estimation of parameters from astrophysical observations can be very valuable. Not only can one extract more physical information from the signal that has been detected, but one can also combine this information across multiple observations to obtain new physical information that would be otherwise inaccessible. For example, a measurement of the moment of inertia of neutron stars from the orbital motion of binary pulsars together with the measurement of the Love number of neutron stars from gravitational wave observations would provide a unique, model-independent and internal-structure independent test of General Relativity~\cite{Yagi:2013bca,Yagi:2013awa}. This is because both measurements (and their error ellipsoids) must lie on the approximately universal I-Love curve of General Relativity if neutron stars are described by Einstein's theory. One can further use universal relations to probe cosmology with gravitational wave observations of neutron star binaries~\cite{Yagi:2015pkc} via measurements of tidal effects in the late inspiral~\cite{Messenger:2011gi,Li:2013via,DelPozzo:2015bna}.

%%%%%%%%%%%%%%%%%%%%%%%%%%%%%%%%%%%%%%%%%%%%%%%%%%%%%%%%
\subsection{Layout and Conventions}

This review paper will focus on approximate universal relations between different properties of neutron stars and quark stars, in particular on the I-Love-Q and the no-hair relations. Section~\ref{sec2:NS-BHrelations} summarizes these universal relations in detail within General Relativity. Section~\ref{sec3:Extensions} discusses extensions of the universality to rapidly and differentially rotating stars, magnetars, binary neutron stars, and stars with anisotropic or non-barotropic fluids. Section~\ref{sec4:Connections} describes connections between other forms of universality in neutron stars and the I-Love-Q and the no-hair relations. Section~\ref{sec5:Why} attempts to explain why the universality in the I-Love-Q relations is present in compact stars, describing in detail one explanation that has not yet been disproven, as well as a few early explanations that have now been shown to be insufficient. Section~\ref{sec6:ILQ-Mod-Grav}  summarizes how the I-Love-Q relations are changed in a few modified theories of gravity and for exotic compact objects such as gravastars. Section~\ref{sec7:Applications} focuses on the different applications of the universal relations, including those in nuclear physics, gravitational wave physics, experimental relativity and cosmology. Section~\ref{sec8:OpenQuestions} concludes and points to future research.

The remainder of the review paper utilizes the following conventions. In general, we follow mostly the conventions of Misner, Thorne and Wheeler~\cite{Misner:1973cw}, where Greek indices $(\alpha, \beta, \ldots)$ in sublists are spacetime indices. We, moreover, use geometric units $G = 1 = c$, which can be converted to physical units using the fact that $M_{\odot} \sim 1.476 \; {\rm{km}} \sim 5 \times 10^{-6} \; {\rm{s}}$. When discussing data analysis topics, Latin indices $(a, b, \ldots)$ in sublists are parameter indices. For ease of reading, we present below a reference list with the definitions of many symbols that appear commonly throughout the review:
\begin{itemize}
\item $M$, $M^\mrm{(max)}$: the mass and the maximum of a compact star,
\item $e \equiv \sqrt{1-a_3^2/a_1^2}$: the eccentricity of an ellipsoidal star, where $a_1$  and $a_3$ are the semi-major and semi-minor axis, and $e_0$ is the eccentricity at the surface,
\item $f_e \equiv 1 - a_3/a_1$: the stellar flattening parameter similar to the stellar eccentricity,
\item $R$, $R(\theta)$, $\bar R \equiv a_1 (1-e^2)^{1/6}$, $R_{\rm e} \equiv R(\pi/2)$: the stellar radius for a non-rotating configuration, the stellar surficial radius in terms of the polar angle, the geometrical mean radius and the equatorial radius of a star respectively,
\item $C \equiv M/R$, $\bar C \equiv M / \bar R$, $C_\mrm{e} \equiv M / R_\mrm{e}$: the stellar compactness defined with respect to $R$, $\bar R$ and $R_\mrm{e}$ respectively,
\item $|\vec S|$, $I$, $\Omega$, $f_s \equiv \Omega/(2 \pi)$, $P \equiv 1/f_s$: the magnitude of the spin angular momentum, the moment of inertia, the spin angular velocity, the spin frequency and the spin period of a compact star, with $\Omega_c$, $\Omega_s$ and $\Omega_\mrm{bk}$ the spin angular velocity at the center, at the surface and at breakup respectively,
\item $Q$, $\lambda_{2}$, $k_2$, $\lambda_2^{(\mrm{rot})}$: the quadrupole moment and the (electric-type) quadrupolar tidal deformability, the tidal Love number and the rotational deformability of a compact star,  
\item $p$, $q_t$, $\rho$, $T$: the internal radial and tangential pressure, the energy density and the temperature, where $p_{c}$ and $\rho_{c}$ are the central pressure and density, while $\rho_0$ is the nuclear saturation density and $\bar \rho$ is the mean density, 
\item $\Gamma$, $n$: the adiabatic and polytropic indices, 
\item $a \equiv |\vec S|/M$, $\chi \equiv |\vec S|/M^{2}$, $\tilde q \equiv - i \sqrt{M_2/M}$: the magnitude of the spin angular momentum per unit mass, the dimensionless spin parameter and the reduced quadrupole moment,
\item $\bar I \equiv {I}/{M^3}$, $\bar \lambda_2 \equiv {\lambda_2}/{M^5}$, $\bar Q \equiv -{Q}/({M^3 \chi^2})$: the dimensionless moment of inertia, tidal deformability and quadrupole moment, 
\item $\tilde I \equiv I/\left(M \bar R^2 \right)$ and $J_2 \equiv - M_2/(M \bar R^2)$: another dimensionless version of the moment of inertia and the quadrupole moment,
\item $M_{\ell}$, $S_{\ell}$: mass and current multipole moments, 
\item $\bar{M}_{\ell} \equiv (-1)^{{\ell}/{2}} {M_{\ell}}/({M^{\ell + 1} \chi^{\ell}})$, $\bar{S}_{\ell} = (-1)^{({\ell-1})/{2}} {S_{\ell}}/({M^{\ell + 1} \chi^{\ell}})$: the dimensionless mass and current multipole moments,
\item $\lambda_\ell$, $\sigma_\ell$: the $\ell$th electric-type and magnetic type tidal deformabilities,
\item $\bar \lambda_\ell \equiv \lambda_\ell/M^{2\ell +1}$, $\bar \sigma_\ell \equiv \sigma_\ell/M^{2\ell +1}$: the dimensionless version of $\lambda_\ell$ and $\sigma_\ell$,
\item $h_\ell$, $\bar \eta_\ell \equiv \{2/[(2\ell-1)!!]\} (h_\ell/C^{2\ell+1})$: the $\ell$th shape Love number and the $\ell$th dimensionless shape tidal deformability,
\item $\vartheta_\LE$: the Lane-Emden function, 
\item $E$, $\Omega_\mrm{orb}$, $f_\mrm{orb} \equiv \Omega_\mrm{orb}/(2\pi)$, $v \equiv (M \Omega_\mrm{orb})^{1/3}$: the energy, orbital angular velocity, orbital frequency and velocity of a test-particle around a compact star, 
\item $B_p$, $\langle B \rangle$: the magnetic field strength at the poles and the averaged field strength, 
\item $\omega_\ell$: the $\ell$th f-mode oscillation (angular) frequency, with $\bar \omega_\ell \equiv M \omega_\ell$ its dimensionless version,
\item $\mathcal{R}$: the Ricci curvature scalar,
\item $m \equiv m_1 + m_2$, $\eta \equiv m_1 m_2/m^2$, $\mathcal M \equiv m \eta^{3/5}$, $q \equiv m_1/m_2$: the total mass, the symmetric mass ratio, the chirp mass and the mass ratio of a binary with masses $m_1$ and $m_2$,
\item $b$: the orbital separation of a binary system,
\item $f$: the gravitational wave frequency,
\item $\vec \chi_s \equiv (\vec \chi_1 + \vec \chi_2)/2$, $\vec \chi_a \equiv (\vec \chi_1 - \vec \chi_2)/2$: the symmetric and antisymmetric combination of the individual spin angular momentum in a binary with $\vec \chi_A 
\equiv \vec S_A / m_A^2$, $\chi_s \equiv |\vec \chi_s|$ and $\chi_a \equiv |\vec \chi_a|$,
\item $\bar Q_s \equiv (\bar Q_1 + \bar Q_2)/2$, $\bar Q_a \equiv (\bar Q_1 - \bar Q_2)/2$: the symmetric and antisymmetric combination of the individual dimensionless quadrupole moments,
\item $\bar \lambda_{\ell,s} \equiv (\bar \lambda_{\ell,1} + \bar \lambda_{\ell,2})/2$, $\bar \lambda_{\ell,a} \equiv (\bar \lambda_{\ell,1} - \bar \lambda_{\ell,2})/2$: the symmetric and antisymmetric combination of the individual $\ell$th-order, dimensionless tidal deformability,
\item $D_L$, $z$: the luminosity distance and the redshift,
\item $m_{z,A} \equiv m_A (1+z)$: the redshifted mass of the $A$th body.
\end{itemize}

%%%%%%%%%%%%%%%%%%%%%%%%%%%%%%%%%%%%%%%%%%%%%%%%%%%%%%%%
\section{Universal Relations for Neutron Stars and Quark Stars in General Relativity}
\label{sec2:NS-BHrelations}

This section reviews the universal I-Love-Q and no-hair relations for neutron stars and quarks stars. We begin by introducing a description of neutron stars and quark stars and by defining a couple of useful approximations that will be used heavily in many calculations. We then proceed to describe the universal relations in General Relativity, both in the so-called Newtonian limit and in the relativistic regime. 

%%%%%%%%%%%%%%%%%%%%%%%%%%%%%%%%%%%%%%%%%%%%%%%%%%%%%%
\subsection{Neutron Stars, Quark Stars and Approximation Schemes}
\label{sec:Preliminaries}

%--------------------
\subsubsection{The Structure of Compact Stars}

Most calculations that aim to model old and cold neutron stars typically use a perfect fluid stress-energy tensor to describe the matter sector. Such a tensor can be written as
\be
\label{eq:Tmunu}
T_{\mu \nu} = \left(\rho + p\right) u_{\mu} u_{\nu} + p \; g_{\mu \nu}\,,
\ee
where $\rho$ is the energy density, $p$ is the (isotropic) pressure, $u^{\mu}$ is the (timelike) four-velocity of the fluid and $g_{\mu \nu}$ is the metric tensor. When modeling real fluids, one typically must also account for shear stresses, anisotropic pressure, viscosity, magnetic fields and heat conduction. These quantities, however, play a subdominant role and can be effectively neglected (we consider some of these effects in Sec.~\ref{sec3:Extensions}). 

Above we referred to old and cold neutrons stars to differentiate them from newly-born and hot proto-neutron stars. By ``old'' we mean stars that have lived long enough to cool down to surface temperatures of roughly $10^5$--$10^6$ K, such that they are cold relative to their Fermi temperature. Such stars are also typically so far from any other body that they can be treated as isolated. These isolated stars rotate rigidly because vorticity is unsourced when the pressure only depends on the stellar density, i.e.~in the barotropic limit. On the other hand, proto-neutron stars formed after a supernova explosion or hypermassive neutron stars formed after the merger of binary neutron stars, are typically hot, highly magnetized and deformed, and rotate differentially. Such stars cannot be modeled with a perfect fluid stress-energy tensor. In this review paper, we mostly focus on old and cold neutron stars, which we shall refer to as simply ``neutron stars'', unless otherwise stated.

The internal structure of stars is fully determined by their equation of state, i.e.~a thermodynamic relation between the matter degrees of freedom in their interior. For neutron stars and quark stars described by a perfect fluid stress-energy tensor, the equation of state is \emph{barotropic}, meaning that it relates the star's internal pressure to only its density. The equation of state of compact stars can be probed experimentally on Earth at around nuclear saturation density $\rho_{0} \approx 2.5 \times 10^{14} \; {\rm{g}}/{\rm{cm}}^{3}$, appropriate to model their crust. Neutron stars and quark stars, however, are extremely compact, with densities that easily exceed the nuclear saturation limit in their interior. The equation of state is thus quite uncertain at densities appropriate to the outer core [$\rho \in (0.5,2) \rho_{0}$], or even worse at densities appropriate to the inner core [$\rho > 2 \rho_{0}$].  

\begin{figure}[htb]
\begin{center}
\begin{tabular}{l}
\includegraphics[width=9.cm,clip=true]{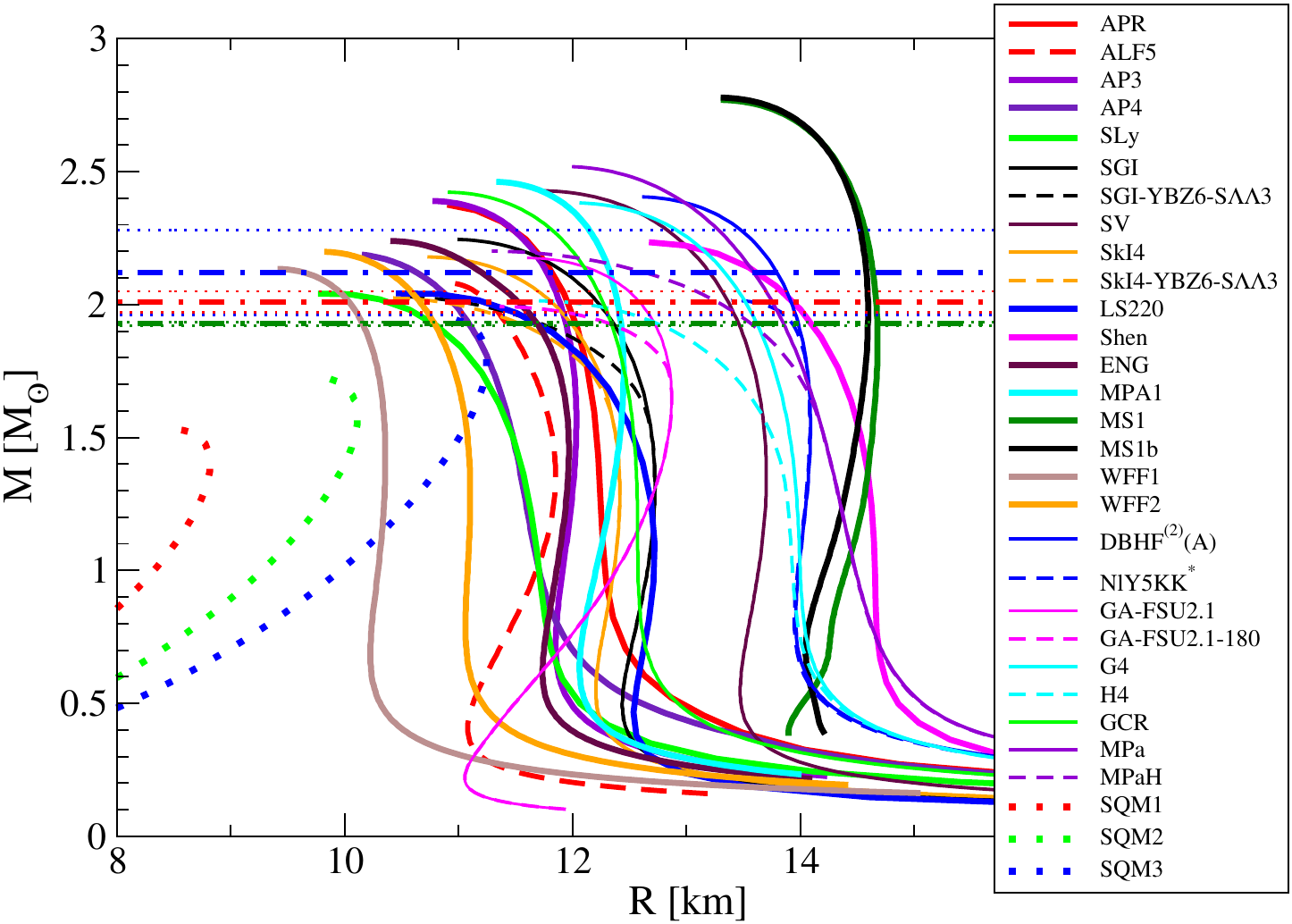} 
\end{tabular}
\caption{\label{fig:M-R} Mass-radius curve for a sequence of isolated, non-rotating compact star solutions labeled by central density. The solid lines correspond to regular neutron star solutions with different equations of state with normal matter composition (i.e.~neutrons, protons electrons and muons)~\cite{APR,Douchin:2001sv,Lim:2013tqa,1995NuPhA.584..467R,LS,Shen1,Shen2,1996ApJ...469..794E,1987PhLB..199..469M,Mueller:1996pm,Wiringa:1988tp,2014arXiv1410.7166K,2013JPhG...40b5203G,Lackey:2005tk,Gandolfi:2011xu,2014PhRvC..90d5805Y}. The dashed lines correspond to solutions for equations of state that also contain exotic matter, namely hyperons, kaon condensates and quarks~\cite{Chatziioannou:2015uea,2014arXiv1412.5722L,2014arXiv1410.7166K,2013JPhG...40b5203G,Lackey:2005tk,2014PhRvC..90d5805Y}. The dotted lines correspond to strange quark star solutions~\cite{SQM}. Observe how the mass-radius curve can be drastically different as one varies the equation of state, with most still passing the constraints imposed by the massive pulsars J1614-2230~\cite{Arzoumanian:2015gjs,1.97NS} (green dot-dashed horizontal line), J0348+0432~\cite{Antoniadis:2013pzd} (red dot-dashed horizontal line) and Vela X-1~\cite{Rawls:2011jw,Ozel:2012ax,Falanga:2015mra} (blue dot-dashed horizontal line), where the measurement error in the latter is shown with dotted horizontal lines.
}
\end{center}
\end{figure}

These uncertainties have led to a large number of proposed, so-called ``realistic'', equations of state for neutron stars. These differ not only in the different approximations used to solve many-body nuclear physics equations (e.g.~the variational method~\cite{Pandharipande:1971up}, the Hartree-Fock approximation~\cite{1956PMag....1.1043S,1996ApJ...469..794E}, the Relativistic Mean Field approximation~\cite{Shen1,Shen2}), but also in the internal matter degrees of freedom that are allowed to be present (e.g.~normal matter in the form of neutrons, protons, electrons, and muons, or kaon condensates~\cite{Kaplan:1986yq,2013JPhG...40b5203G}, hyperons~\cite{2014arXiv1412.5722L,2014arXiv1410.7166K,Lackey:2005tk}, and quark-gluon plasmas~\cite{Alford:2004pf}) or even pure quark matter~\cite{SQM}. Figure~\ref{fig:M-R} shows the mass-radius curve for a sequence of compact stars, constructed by varying the central density (a higher central density leads to a more massive star with a smaller radius). Observe that different equations of state can lead to widely different mass-radius curves, with most of them allowing for stars with masses larger than the recently discovered massive pulsar J1614-2230~\cite{1.97NS}, J0348+0432~\cite{Antoniadis:2013pzd} and Vela X-1~\cite{Rawls:2011jw,Ozel:2012ax,Falanga:2015mra}.

In this review, we will choose a subset of equations of state to study universal behavior in neutron stars and quark stars. When modeling neutron stars composed only of normal matter, we will present results using a group of the following equations of state: APR, AP3 and AP4~\cite{APR}, SLy~\cite{Douchin:2001sv}, SGI and SV~\cite{Lim:2013tqa}, SkI4~\cite{1995NuPhA.584..467R}, LS220~\cite{LS}, Shen~\cite{Shen1,Shen2}, ENG~\cite{1996ApJ...469..794E}, MPA1~\cite{1987PhLB..199..469M}, MS1 and MS1b~\cite{Mueller:1996pm}, WFF1 and WFF2~\cite{Wiringa:1988tp}, DBHF$^{(2)}$(A)~\cite{2014arXiv1410.7166K}, GA-FSU2.1~\cite{2013JPhG...40b5203G}, G4~\cite{Lackey:2005tk}, GCR~\cite{Gandolfi:2011xu}, MPa~\cite{2014PhRvC..90d5805Y}. When modeling neutron stars that also contain exotic matter, we will employ the following equations of state: ALF5~\cite{Chatziioannou:2015uea}, SGI-YBZ6-S$\mathrm{\Lambda\Lambda}$3 and SkI4-YBZ6-S$\mathrm{\Lambda\Lambda}$3~\cite{2014arXiv1412.5722L}, NlY5KK$^*$~\cite{2014arXiv1410.7166K}, GA-FSU2.1-180~\cite{2013JPhG...40b5203G}, H4~\cite{Lackey:2005tk}, MPaH~\cite{2014PhRvC..90d5805Y}. When modeling quark stars, we will employ the SQM1, SQM2 and SQM3~\cite{SQM} equations of state. This set of equations of state include some that present kaon condensates, hyperons and quark-gluon plasmas; see Chatziioannou \et~\cite{Chatziioannou:2015uea} for a more detailed but concise description of these equations of state. 

All of these different equations of state can be effectively approximated as piecewise polytropes~\cite{Lattimer:2000nx,Read:2008iy}. A polytropic equation of state is one defined by the equation
\be
\label{eq:polytropic-EoS}
p = K \rho^{\Gamma} = K \rho^{1 + 1/n}\,,
\ee
where the quantities $K$, $\Gamma$ and $n$ are constants, with $\Gamma$ the adiabatic index and $n$ the polytropic index. The $n=0$ case corresponds to the equation of state of an incompressible (i.e.~constant density) fluid, which is also a good model for strange quark stars~\cite{SQM} in the low pressure regime. The realistic equations of state described above can be modeled through piecewise polytropic equations of state with indices $n \in [0.5,1]$~\cite{Lattimer:2000nx,Read:2008iy}.

%--------------------
\subsubsection{Useful Approximations}
\label{sec:Approxs}

The calculation of the gravitational field generated by a compact star requires the solution to the field equations. In General Relativity, this is a set of ten, non-linear and coupled partial differential equations for ten metric functions of four spacetime coordinates. Needless to say, a solution to these equations has not yet been found analytically in closed form for generic equations of state, even when restricting attention to highly symmetric spacetimes. Fortunately, however, compact stars are rather simple objects that possess enough symmetries to simplify the mathematics significantly. To start with, since compact stars are typically isolated, rigidly rotating objects, one can assume their spacetime is stationary and axially symmetric. Mathematically, these symmetries imply the existence of a temporal and an azimuthal Killing vector (with the azimuthal direction identified with rotation), which in turn implies that the metric functions only depend on the radial and polar angle coordinates. Moreover, these symmetries plus the Einstein equations imply that only four of the ten metric functions need to be non-vanishing, which allows us to write the line element in the standard Weyl-Papapetrou form~\cite{1967ApJ...147..317H,Hartle:1967he} in Boyer-Lindquist-type coordinates
\be
\label{eq:metric-ansatz}
ds^{2} = -e^{\nu(r,\theta)} dt^{2} + e^{\lambda(r,\theta)} dr^{2} + r^{2} K(r,\theta) \left[ d\theta^{2} + \sin^{2}{\theta} \left[d\phi - \omega(r,\theta) dt \right]^{2}  \right]\,,
\ee
where $\nu(r,\theta)$, $\lambda(r,\theta)$, $K(r,\theta)$ and $\omega(r,\theta)$ are metric functions.

Even with these symmetries invoked, the resulting system of coupled and non-linear, partial differential equations is formidable, which forces us to face one of two options. The first is to find an ``exact'' solution by solving this system fully numerically, for example through a Green's function approach or through spectral methods. Such an approach has been implemented in the publicly available RNS~\cite{Stergioulas:1994ea} and LORENE/rostar~\cite{Bonazzola:1993zz,Bonazzola:1998qx} codes. The LORENE code solves the four elliptic-type equations that result from the Einstein equations~\cite{Gourgoulhon:2010ju} using multi-domain spectral methods and decomposing the metric functions in terms of Chebyshev polynomials. The RNS code solves the Einstein equations using a Green's function approach~\cite{Komatsu:1989zz,1992ApJ...398..203C}. Once a numerical solution has been obtained, the exterior multipole moments of the spacetime can be extracted through the evaluation of certain integrals~\cite{Yagi:2014bxa}. These codes are particularly well-suited to modeling rapidly rotating stars, since then the effect of rotation is large enough to be numerically resolvable. 

\begin{figure}[htb]
\begin{center}
\begin{tabular}{l}
\includegraphics[width=8.5cm,clip=true]{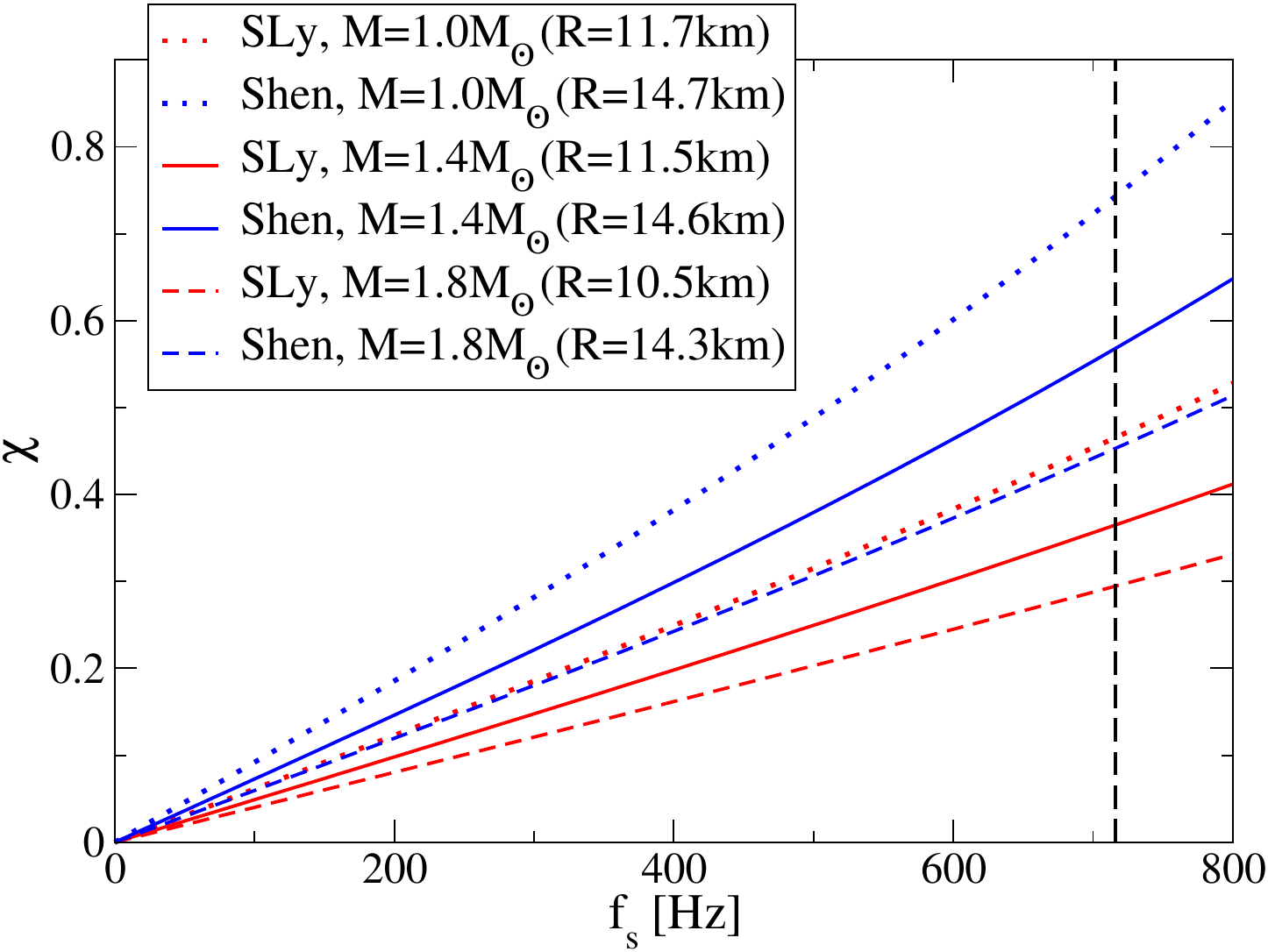} 
\end{tabular}
\caption{\label{fig:chi-f} Dimensionless spin parameter as a function of spin frequency in Hz for stars with different masses and different equations of state. The shaded regions correspond to stars with the same equation of state (red for SLy~\cite{Douchin:2001sv} and blue for Shen~\cite{Shen1,Shen2}). Observe that $\chi < 0.5$ for most stars, especially those with periods above $2$ ms (equivalent to $f_s < 500$ Hz), except those close to the spin frequency of the fastest millisecond pulsar observed~\cite{fastest-pulsar} (dashed black line). This figure is constructed with data obtained for slowly-rotating stars to third order in spin. Higher order terms in spin only affect the result by $\lesssim 3\%$. Each line has a fixed central density and the mass and radius in the legend are for a non-rotating configuration.
}
\end{center}
\end{figure}

But almost all neutron stars observed in Nature happen to be \emph{slowly-rotating}, which then allows for a different, semi-analytic approach. Although neutron stars are typically born with millisecond periods, they quickly spin-down due to magnetic braking and viscous damping~\cite{Shapiro:2000zh,Cook:2003ku,Liu:2003ay}. When in a binary, neutron stars may be spun-up or ``recycled'' through accretion, leading again to millisecond periods. The fastest millisecond pulsar observed,  J1939+2134~\cite{fastest-pulsar}, has a period of $1.5 \; {\rm{ms}}$, which is comparable to the rotational period of the blades in a professional kitchen blender. How can one then be justified in using a slow-rotation approximation? The reason is that although the raw angular velocity is large, it is much smaller than the Keplerian angular velocity of a test-particle at the surface (or the breakup angular velocity). Alternatively, one can introduce the dimensionless perturbation parameter $\chi = |\vec S|/M^{2}$ for neutron stars and quark stars (where $|\vec S| = I \Omega$ is the magnitude of the spin angular momentum, with $M$ the mass, $I$ the moment of inertia and $\Omega$ the angular velocity) and require it to be small. This quantity is plotted in Fig.~\ref{fig:chi-f} as a function of spin frequency for a few neutron star masses and equations of state. Observe that $\chi<0.5$ for a large portion of parameter space, especially those with periods above $2$ ms (equivalent to $f_s < 500$ Hz) 

The slow-rotation or ``Hartle-Thorne'' approximation~\cite{Hartle:1967he,Hartle:1968si} consists of expanding and solving the Einstein equations in $\chi \ll 1$. Because of the structure of these equations, the radial and angular sectors separate and the free metric functions can be product decomposed into Legendre polynomials of $\cos{\theta}$ and free functions of $r$. Moreover, axisymmetry requires that the metric functions that appear in the diagonal components of the metric be proportional to even powers of $\chi$, while the gravitomagnetic $(t,\phi)$ component be proportional to odd powers of $\chi$. The metric expanded in this form goes by the name of the Hartle-Thorne metric. The perturbed Einstein equations to ${\cal{O}}(\chi^{0})$ reduce to the well-known Tolman-Oppenheimer-Volkhoff equations (when using stress-energy tensor conservation) and their solution leads to the mass (the zeroth mass multipole moment) and the radius of the star (i.e.~the radius where the pressure vanishes). The solution to the perturbed Einstein equations to ${\cal{O}}(\chi)$ and ${\cal{O}}(\chi^{2})$ lead to the moment of inertia (related to the first current multipole moment) and the quadrupole moment (the second mass multipole moment) of the star. Yagi \et~\cite{Yagi:2014bxa} have extended the slow-rotation approximation to fourth-order in rotation and compared results to those obtained fully numerically~\cite{Yagi:2014bxa}. Throughout this review, ``slow-rotation limit'' will refer to calculations carried out to leading-order in the slow-rotation approximation.

Even within the slow-rotation approximation, the perturbed Einstein equations, however, still need to be solved numerically for most but the simplest equations of state. In practice, the differential system is simplified by changing radial coordinates such that the matter sector (the pressure and energy density) does not need to be perturbed~\cite{Hartle:1967he}. The perturbed Einstein equations are then numerically solved order by order in rotation, both inside and outside the star, imposing regularity at the stellar center and asymptotic flatness at spatial infinity. Matching the two numerical solutions at the stellar surface determines integration constants in the exterior metric tensor, whose asymptotic form at spatial infinity determines the multipole moments of the spacetime. For simple polytropic equations of state, like an $n=0$ polytrope and the TolmanVII equation of state~\cite{Tolman:1939jz,Lattimer:2000nx}, the perturbed Einstein equations to ${\cal{O}}(\chi^{0})$ can be solved analytically. 

An approximate analytic solution, e.g.~using an $n=0$ polytrope, can also be obtained to higher order in the dimensionless spin parameter through a weak-field or \emph{post-Minkowskian} approximation, i.e.~by expanding the perturbed Einstein equations in powers of the stellar compactness $C := M/R$, which determines the strength of the gravitational field at the stellar surface. Neutron stars are the most compact stars in Nature, but nonetheless $C \in (0.05,0.25)$, depending on their mass or central density. This implies that one may obtain an accurate solution to the perturbed Einstein equations if one carries out such an expansion to sufficiently high order in compactness, at any given (fixed) order in $\chi$. When one truncates the post-Minkowskian expansion to leading-order in compactness one obtains the so-called \emph{Newtonian limit}. Newtonian neutron stars are thus \emph{not} neutron stars constructed in Newtonian gravity, but rather neutron stars constructed in General Relativity to leading-order in an expansion about small compactness.  

Until now, we have mostly discussed the calculation of the gravitational field of isolated compact stars, but much of this review paper will deal with tidal deformations induced by companion bodies. When compact stars are in binaries, the gravitational field of the companion can distort the shape and the gravitational field of the primary star. These deformations are encoded in the so-called tidal deformabilities, related to the tidal Love numbers, of the star. All observed astrophysical neutron star binaries, including the most relativistic binary pulsar, are however very well-separated, with radial distances that exceed $10^{5}$ km. For stars so widely separated, one can employ a \emph{small-tide} approximation in which one solves the Einstein equations as an expansion in the tidal deformation about an isolated neutron star solution. In fact, such an approximation is valid even for the late-inspiral phase of neutron star binaries, which are also targets for ground-based gravitational wave interferometers, such as Adv.~LIGO. In practice, the calculation of the tidal deformation is similar to solving the perturbed Einstein equations in the slow-rotation approximation, except that here one sets the odd-parity (rotation-related) perturbations to zero and does not impose asymptotic flatness~\cite{Hinderer:2007mb,Damour:2009vw,Binnington:2009bb}. This allows for the extraction of not only the tidally-induced quadrupole moment, but also the strength of the tidal field, with which one can then calculate the tidal deformability.

%%%%%%%%%%%%%%%%%%%%%%%%%%%%%%%%%%%%%%%%%%%%%%%%%%%%%%
\subsection{I-Love-Q}
\label{sec:I-Love-Q}

The I-Love-Q relations are approximately-universal (i.e.~equation-of-state insensitive) inter-relations between the stellar moment of inertia $I$, the quadrupolar tidal deformability $\lambda_2$ and the (spin-induced) quadrupole moment $Q$. The moment of inertia quantifies how fast an object can spin with a given angular momentum. The quadrupole moment $Q$ determines the magnitude of the quadrupolar deformation of a star due to rotation. The tidal deformability $\lambda_{2}$ is defined as the ratio between the tidally-induced quadrupole moment and the strength of the external quadrupolar tidal field~\cite{Hinderer:2007mb}, and thus, it determines how easily an object can be deformed due to an external tidal field. The tidal deformability is related to the dimensionless tidal Love number (also known as the tidal apsidal constant) of the second kind $k_2$~\cite{love} via $k_2 =3 \lambda_2/(2 R^5)$~\cite{Hinderer:2007mb}.

%-----------------------------
\subsubsection{Newtonian Results}
\label{sec:I-Love-Q-Newton}

In the Newtonian limit, the tidal Love number $k_2$ is calculated as follows~\cite{1955MNRAS.115..101B,1959cbs..book.....K,Mora:2003wt,2014grav.book.....P}. The isodensity surfaces inside a star can be represented by the following function 
\be
r(\bar{r}, \theta, \phi) = \bar{r} \left[ 1 + \sum_{\ell, m} f_{\ell} (\bar{r}) Y_{\ell m} (\hat{\Omega}) Y_{\ell m} (\hat{n}) \right]\,,
\ee
where $\bar{r}$ is a radial parameter, $Y_{\ell m} (\hat{\Omega})$ are spherical harmonics in the $\hat{\Omega}$ direction and $\hat{n}$ denotes the principal direction of the perturbation. The dimensionless distortion function $f_\ell$ are obtained by solving the Clairaut-Radau equation, which comes from the Poisson and Euler equations, i.e.~the equations of structure of General Relativity in the Newtonian limit~\cite{2014grav.book.....P}:
\be
\bar r \frac{d\eta_\ell}{d \bar{r}} + 6 \frac{\rho}{\bar \rho} (\eta_\ell +1) + \eta_\ell (\eta_\ell -1) -\ell (\ell+1) =0\,,
\label{eq:Clairaut-Radau}
\ee
with the boundary condition $\eta_\ell(0) = \ell -2$, where $\eta_\ell (\bar{r}) \equiv {d \ln f_\ell}/{d \ln \bar{r}}$. In this equation, $\rho$ is the density of the undistorted configuration and $\bar \rho$ is the mean density of the undistorted star, where recall that the energy density is the same as the mass density in the Newtonian limit. The quadrupolar tidal Love number $k_2$ is given by $\eta_2$ via
\be
k_2 = \frac{3-\eta_2(R)}{2[2 + \eta_2(R)]}\,.
\label{k2N1}
\ee

In the Newtonian limit, the moment of inertia $I$ in the slow-rotation limit is given by~\cite{Hartle:1967he,2014grav.book.....P}
\be
I = \frac{8\pi}{3} \int^{R}_0 r^4 \rho (r)dr\,,
\label{I-Newton}
\ee
while the quadrupole moment $Q$ is given by~\cite{Ryan:1996nk,Laarakkers:1997hb}
\be
Q = 2\pi \int_0^{\pi} \int_0^{R(\theta)} \rho(r,\theta) r^4 P_2 (\cos \theta) \sin \theta \, dr \, d\theta\,,
\label{eq:Q-integ}
\ee
where $R(\theta)$ is the surface of the stellar ellipsoid, i.e.~the location of the surface when the star is rotating is a function of the angle $\theta$, where we note that $R(\theta) = R + {\cal{O}}(\chi^{2})$. One can also calculate $Q$ from $\lambda_2$ directly, using the fact that the quadrupolar \emph{rotational} deformability, given by $\lambda_{2}^{(\rm{rot})}=-Q/\Omega^2$ with $\Omega$  the stellar angular velocity~\cite{Mora:2003wt,Berti:2007cd}, is the same as $\lambda_2$ in the Newtonian limit~\cite{Mora:2003wt}.

The easiest way to see how the I-Love-Q relations come about is by calculating them analytically for an $n=0$ and an $n=1$ polytrope. For an isolated star, the density of a non-rotating configuration in these two cases is given analytically by~\cite{Yagi:2013awa}
\be
\rho^{(n=0)} = \frac{3}{4\pi} \frac{M}{R^3}\,, \qquad
\rho^{(n=1)}(r) = \frac{1}{4} \frac{M}{R^2} \frac{1}{r} \sin \left( \pi \frac{r}{R} \right)\,.
\ee
In the incompressible $(n=0)$ case, the density is obtained trivially by computing the mass of the star, while in the $n=1$ case it is obtained by solving the Tolman-Oppenheimer-Volkhoff equation and then using the polytropic equation of state [Eq.~\eqref{eq:polytropic-EoS}]. With the density at hand, one can then compute $(I,\lambda_{2},Q)$ as a function of compactness and then eliminate the compactness to find the following relations~\cite{Yagi:2013awa}:
\be
\label{eq:I-L-Q-Newt}
\bar{I} = C_{\bar{I} \bar{\lambda}_2}^{(n)} \bar{\lambda}_2^{2/5}\,, \qquad 
\bar{I} = C_{\bar{I} \bar{Q}}^{(n)} \bar Q^{2}\,, \qquad 
\bar Q = C_{\bar{Q} \bar{\lambda}_2}^{(n)}\bar{\lambda}_2^{1/5}\,,
\ee
where we have adimensionalized $(I,\lambda_{2},Q)$. The equation-of-state dependence of the I-Love-Q relations in Eq.~\eqref{eq:I-L-Q-Newt} is completely encoded in the coefficients $( C_{\bar{I} \bar{\lambda}_2}^{(n)}, C_{\bar{I} \bar{Q}}^{(n)},C_{\bar{Q} \bar{\lambda}_2}^{(n)})$, which depend on the polytropic index $n$. These coefficients, however, are approximately independent of $n$, which can be seen by taking the ratio between the coefficients in the $n=0$ and $n=1$ case~\cite{Yagi:2013awa}:
\ba
\frac{C_{\bar{I} \bar{\lambda}_2}^{(0)}}{C_{\bar{I} \bar{\lambda}_2}^{(1)}} &=& \frac{2^{2/5} 3^{3/5} \pi^{6/5} (15-\pi^2)^{2/5}}{5 \pi^2-30} \approx 1.002\,, \\
\frac{C_{\bar{I} \bar{Q}}^{(0)}}{C_{\bar{I} \bar{Q}}^{(1)}} &=& \frac{108 \pi^6 (\pi^2-15)^2}{3125 (\pi^2-6)^5} \approx 1.008\,, \\
\frac{C_{\bar{Q} \bar{\lambda}}^{(0)}}{C_{\bar{Q} \bar{\lambda}}^{(1)}} &=& \frac{25 (\pi^2-6)^2}{ 2^{4/5} 3^{6/5} \pi^{12/5} (15-\pi^2)^{4/5}} \approx 0.997\,.
\ea
The I-Love-Q relations are then roughly the same, with differences of about $0.2$--$0.8$\%, regardless of whether one constructs the Newtonian stars with two very different equations of state, an $n=0$ and an $n=1$ polytrope.

%-----------------------------
\subsubsection{Relativistic Results}
\label{sec:I-Love-Q-relativistic}

\begin{figure*}[htb]
\begin{center}
\begin{tabular}{l}
\includegraphics[width=8.6cm,clip=true]{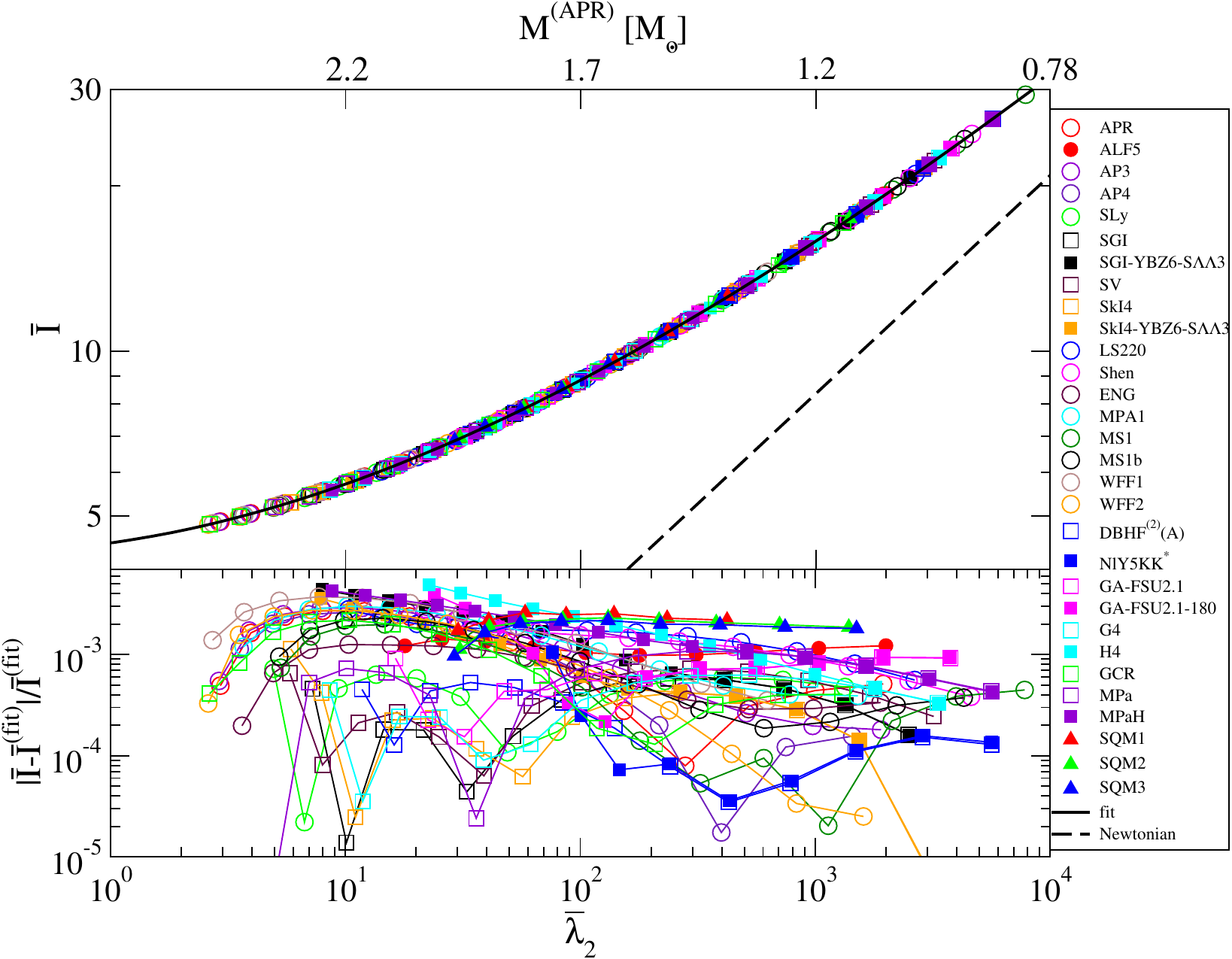} 
\includegraphics[width=7.5cm,clip=true]{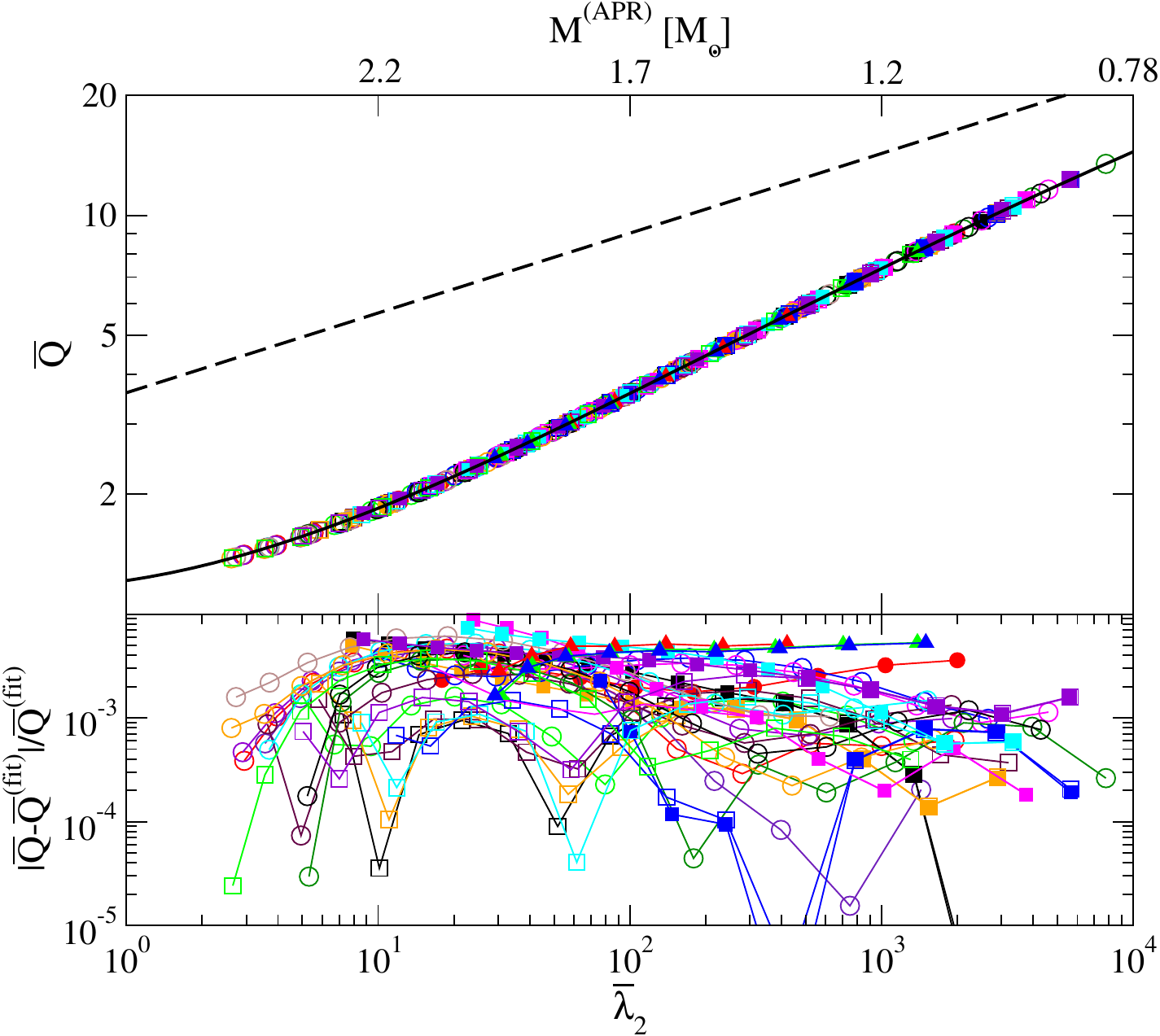} 
\end{tabular}
\caption{\label{fig:I-Love-Q}
(Top) The universal I-Love (left) and Q-Love (right) relations for slowly-rotating neutron stars and quark stars of $1M_\odot < M < M^{\mathrm{(max)}}$ with various equations of state. A single parameter along the curve is the stellar mass or compactness, which increases to the left of the plots. The solid curves show the fit in Eq.~\eqref{fit}. The top axis shows the corresponding stellar mass of an isolated, non-rotating configuration with the APR equation of state. (Bottom) Absolute fractional difference from the fit, while the dashed lines show the analytic Newtonian relations in Eq.~\eqref{eq:I-L-Q-Newt} with $n=0$. Observe that the relations are equation-of-state insensitive to $\mathcal{O}(1\%)$. 
}
\end{center}
\end{figure*}

Let us now consider the I-Love-Q relations in the full relativistic regime, outside of the Newtonian limit. As explained in Sec.~\ref{sec:Approxs}, this must be done numerically, either through a slowly-rotating and a tidally-deformed approximation or through a fully numerical analysis. Consider first the I-Love-Q relations in the slow-rotation approximation, as shown in the top panels of Fig.~\ref{fig:I-Love-Q} for various equations of state. The single parameter along the sequence of stellar configurations is the central density (high central density corresponds to low $\bar \lambda_2$, $\bar{Q}$ and $\bar{I}$), since the rotational frequency cancels out of the dimensionless quantities. We only show data with the stellar mass of an isolated, non-rotating configuration in the range $1M_\odot < M < M^{\mathrm{(max)}}$ with $M^{\mathrm{(max)}}$ representing the maximum mass for such a configuration. These numerical results can be validated against the analytic relations in the Newtonian limit presented in Sec.~\ref{sec:I-Love-Q-Newton} (dashed lines), where we see the former approach the latter as $\bar \lambda_2$ increases (as the compactness decreases).  Observe that the relations are insensitive to the equation of state, which allows one to construct a single fit (black solid curves) given by
\be
\ln y_i = a_i + b_i \ln x_i + c_i (\ln x_i)^2 + d_i (\ln x_i)^3+ e_i (\ln x_i)^4\,,
\label{fit}
\ee
with coefficients given in Table~\ref{table:coeff-I-Love-Q}. Such a fit corresponds to an updated version of that in Yagi and Yunes~\cite{Yagi:2013bca,Yagi:2013awa} due to a larger number of equations of state considered in this review. We stress that the fit is valid only within the range of $\bar I$, $\bar \lambda_2$ and $\bar Q$ shown in Fig.~\ref{fig:I-Love-Q}. The bottom panels of Fig.~\ref{fig:I-Love-Q} show the absolute fractional difference between all the data and the fit, which is at most $\sim 1\%$, slightly larger than the variation in the Newtonian limit of Sec.~\ref{sec:I-Love-Q-Newton}. These relations were first found in Yagi and Yunes~\cite{Yagi:2013bca,Yagi:2013awa} and later confirmed in Lattimer and Lim~\cite{Lattimer:2012xj} for different equations of state.

{\renewcommand{\arraystretch}{1.2}
\begin{table}
\begin{centering}
\begin{tabular}{cccccccc}
\hline
\hline
\noalign{\smallskip}
$y_i$ & $x_i$ &&  \multicolumn{1}{c}{$a_i$} &  \multicolumn{1}{c}{$b_i$}
&  \multicolumn{1}{c}{$c_i$} &  \multicolumn{1}{c}{$d_i$} &  \multicolumn{1}{c}{$e_i$}  \\
\hline
\noalign{\smallskip}
$\bar{I}$ & $\bar{\lambda}_2$ && 1.496 & 0.05951  & 0.02238 & $-6.953\times 10^{-4}$ & $8.345\times 10^{-6}$\\
$\bar{I}$ & $\bar Q$ && 1.393  & 0.5471 & 0.03028  & 0.01926 & $4.434 \times 10^{-4}$\\
$\bar Q$ & $\bar{\lambda}_2$ && 0.1940  & 0.09163 & 0.04812  & $-4.283 \times 10^{-3}$ & $1.245\times 10^{-4}$\\
\noalign{\smallskip}
\hline
\hline
\end{tabular}
\end{centering}
\caption{Updated numerical coefficients for the fitting formula of the I-Love, I-Q and Q-Love relations given in Eq.~\eqref{fit}.}
\label{table:coeff-I-Love-Q}
\end{table}
}

The I-Love-Q relations can also be obtained analytically in a full post-Minkowskian expansion (beyond the Newtonian limit) for special equations of state. Chan \textit{et al.}~\cite{Chan:2014tva} extended the results of Sec.~\ref{sec:I-Love-Q-Newton} through a post-Minkowskian expansion for incompressible stars to find analytic expressions for $\bar{I}$ and $\bar{\lambda}_{2}$ to sixth-order in compactness. Combining these expressions, they found the following I-Love relation for incompressible stars~\cite{Chan:2014tva}:
\ba
\label{eq:I-Love-PM}
\bar I &=& \frac{2}{5} \left( 2 \bar \lambda_2 \right)^{2/5} \left[1 + \frac{22}{7} \left( 2 \bar \lambda_2 \right)^{-1/5} + \frac{8726}{2205} \left( 2 \bar \lambda_2 \right)^{-2/5} + \frac{79840}{33957} \left( 2 \bar \lambda_2 \right)^{-3/5} + \frac{10621396}{21068775} \left( 2 \bar \lambda_2 \right)^{-4/5}  \right.
\nn \\
&&\left. -  \frac{495373192}{4866887025} \left( 2 \bar \lambda_2 \right)^{-1} - \frac{29520935754944}{286683980207625} \left( 2 \bar \lambda_2 \right)^{-6/5} + \mathcal{O} \left(  \frac{1}{\bar \lambda_2}  \right) \right]\,.
\ea
The expansion is here in powers of $\bar \lambda_{2}^{-1/5}$ because $\bar \lambda_{2} \propto C^{-5}$ for incompressible stars. One could attempt to resum this expansion, for example, through a Pad\'e approximation~\cite{Chan:2014tva}, but this is actually not necessary. The above equation is an excellent representation of the I-Love relation that cannot be visually distinguished from the fit used in Fig.~\ref{fig:I-Love-Q}, except for $\bar{\lambda}_{2} > 10$~\cite{Chan:2014tva}.~\footnote{We have attempted to determine the coefficients of Eq.~\eqref{eq:I-Love-PM} by fitting them to numerical data. We found that the first and second coefficients agree with those in Eq.~\eqref{eq:I-Love-PM} within 2\% and 12\% respectively, while higher coefficients disagree by 80\% or more. This is because Eq.~\eqref{eq:I-Love-PM} is an expansion in compactness, and thus, there is no reason to believe that the fit gives similar coefficients as in Eq.~\eqref{eq:I-Love-PM} for data that includes large compactnesses.} Chan \textit{et al.}~\cite{Chan:2015iou} extended the above analysis to self-bound stars, which include quark stars, and found that the I-Love relation is very similar to that for incompressible stars in the above equation. We will review the results of Chan \textit{et al.}~\cite{Chan:2015iou} in more detail in Sec.~\ref{sec:I-Love-Q-C-analytic}.

%%%%%%%%%%%%%%%%%%%%%%%%%%%%%%%%%%%%%%%%%%%%%%%%%%%%%%
\subsection{No-Hair Relations}

The no-hair relations are approximately-universal (i.e.~equation-of-state insensitive) inter-relations between the multipole moments of the exterior metric of a star. In General Relativity, there are two types of multipole moments: mass and current moments, associated with the energy density and the energy current density of the fluid respectively. These moments encode information about observable properties of the star (as measured by an observer at spatial infinity): the $\ell=0$ mass multipole moment is just the mass of the star, the $\ell=1$ current moment is the spin angular momentum of the star and the $\ell=2$ mass moment is the quadrupole moment. The odd mass moments and the even current moments vanish by symmetry considerations.

Two main definitions of multipole moments have been introduced in the literature. The Thorne multipole moments are defined at spatial infinity through an expansion of the metric tensor in a particular set of coordinates~\cite{Thorne:1980rm}. The Geroch-Hansen multipole moments are defined in terms of the conformal group~\cite{Geroch:1970cc,Geroch:1970cd}. These two sets of moments can be mapped into each other~\cite{1983GReGr..15..737G}, with either set capable of describing the exterior metric of a compact star~\cite{Ryan:1996nk,Backdahl:2005uz,Backdahl:2006ed}. In the Newtonian limit, both definition reduce to a single expression for the exterior multipole moments, modulo an overall constant. In the relativistic regime, we will mostly use the Geroch-Hansen moments in this article.

%-----------------------------
\subsubsection{Newtonian Results}
\label{sec:no-hair-Newton}

The mass $M_{\ell}$ and current $S_{\ell}$ multipole moments of a compact star in the Newtonian limit are~\cite{Ryan:1996nk}
\begin{align}
\label{eq:M-ell-integ}
M_\ell &= 2 \pi \int^\pi_0 \int^{R(\theta)}_0 \!\!\! \rho(r,\theta) \,  P_\ell (\cos \theta) \, \sin\theta  \, d\theta \, r^{\ell +2} dr\,, 
\\
\label{eq:S-ell-integ}
S_\ell &= \frac{4 \pi}{\ell+1}  \int^\pi_0 \int^{R(\theta)}_0 \!\!\!  \Omega \, \rho(r,\theta) \frac{d P_\ell (\cos \theta)}{d\cos\theta}  \,  \sin^3\theta  \, d\theta \, r^{\ell +3} dr\,,
\end{align}
where recall that $R(\theta)$ describes the stellar surface of an ellipsoidal star. Note that the above expression for $M_2$ is equivalent to the Newtonian expression for $Q$ in Eq.~\eqref{eq:Q-integ}. Note also that reflection symmetry about the equator, i.e.~that $M_{\ell}$ and $S_{\ell}$ remain invariant under $\theta \to \theta + \pi$, requires that $M_{2 \ell +1} = 0 = S_{2 \ell}$. 

In order to derive the no-hair like relations analytically, one must solve the above integrals to find closed-form expressions for the multipole moments in terms of the compactness. This is impossible without the use of further approximations due to the coupling of the radial and polar angle sectors in each integral. Therefore, in order to make analytic progress, one employs the \emph{elliptical isodensity approximation}~\cite{Lai:1993ve};
\begin{itemize}
\item[(i)] The stellar isodensity surfaces are self-similar ellipsoids with a fixed stellar eccentricity,
\item[(ii)] The density as a function of the isodensity radius $\tilde r$ for a rotating configuration matches that of a non-rotating configuration with the same volume,
\end{itemize}
where recall that the stellar eccentricity is defined via $e = \sqrt{1-a_3^2/a_1^2}$ with $a_1$ ($a_3$) representing the semi-major (semi-minor) axis of the ellipsoid. The isodensity radius $\tilde r$ is defined via $\tilde r \equiv r/\Theta(\theta)$~\cite{Stein:2014wpa} with
\begin{equation}
\Theta(\theta) \equiv \sqrt{\frac{1-e^2}{1-e^2 \sin^{2}{\theta}}}\,.
\end{equation}

\begin{figure}[ht]
\begin{center}
\includegraphics[width=7.cm,clip=true]{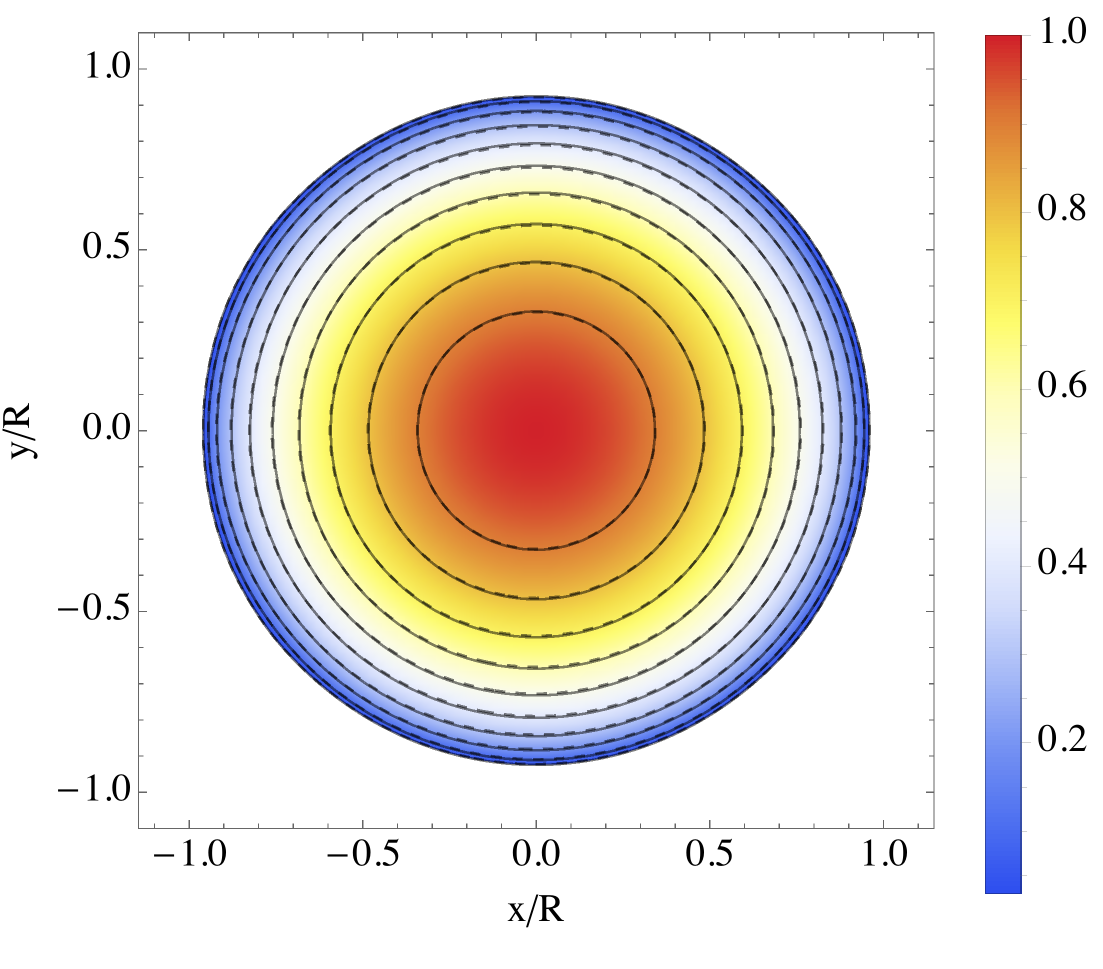} \qquad
\includegraphics[width=7.5cm,clip=true]{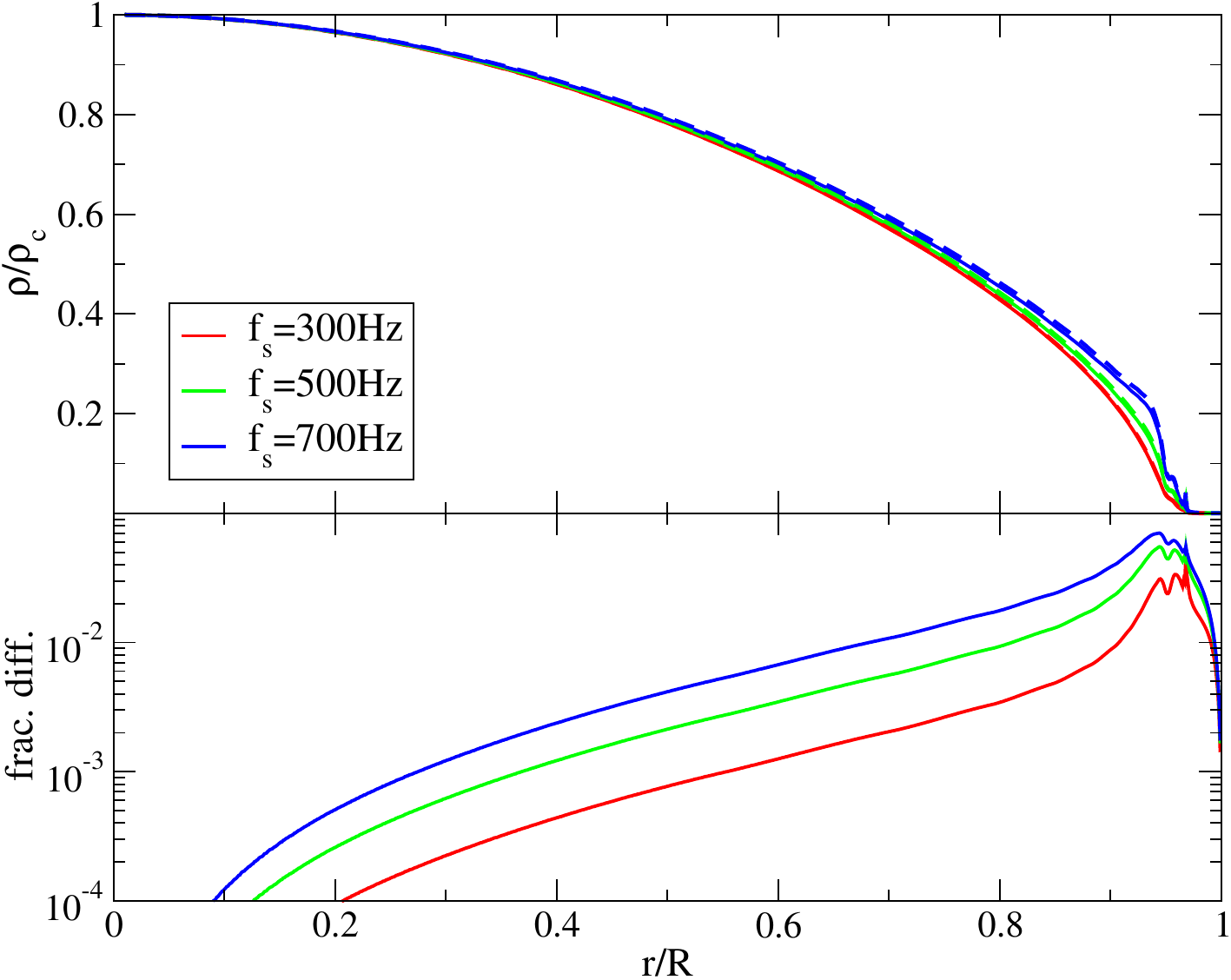}
\caption{\label{fig:isodensity}  (Left) Ratio of the stellar density to the central density (color gradient) and isodensity contours of a typical slowly-rotating neutron star with $M = 1.4 M_\odot$ and $f_s = 500$Hz, using the SLy equation of state~\cite{Douchin:2001sv}. The solid contours are obtained numerically by solving the slow-rotation equations of structure, while the dashed curves are obtained analytically using the elliptical isodensity approximation. (Top Right) The stellar density profile at the equator for stars rotating at various frequencies $f_s$, using the SLy equation of state. As in the left panel, the solid curves represent the profile obtained numerically, while the dashed curves represent that obtained analytically within the elliptical isodensity approximation. (Bottom 
Right) The fractional difference between the solid and dashed curves in the top panel for each $f_s$. Observe that the elliptical isodensity approximation accurately captures the realistic profile with an error of 7\% at most when $f_s \leq 700$Hz. The rapidly oscillatory behavior at $r/R \sim 0.95$ is due to a sudden change in the stellar density for an isolated, non-rotating configuration.
}
\end{center}
\end{figure}

The elliptical isodensity approximation is excellent when describing compact stars, such as neutron stars and quark stars, that rotate slowly. The left panel of Fig.~\ref{fig:isodensity} compares isodensity contours obtained numerically for slowly-rotating neutron stars (solid contours) and analytically\footnote{Strictly speaking, the elliptical isodensity result is obtained semi-analytically, as one needs to calculate the density profile for an isolated, non-rotating configuration numerically.} within the elliptical isodensity approximation (dashed contours), using the SLy equation of state and a stellar rotation frequency of $f_s = 500$Hz. We set the central density to $\rho_c = 1.0\times 10^{15}$g/cm$^3$, which gives a star with mass $M=1.4M_\odot$ for an isolated, non-rotating configuration. Observe that the two types of contours are indistinguishable. The top right panel of Fig.~\ref{fig:isodensity} shows the stellar density profile at the equator obtained numerically (solid) and analytically (dashed) for various frequencies $f_s$, with $f_s = 700$Hz roughly corresponding to the rotation frequency of the fastest spinning pulsar currently observed~\cite{fastest-pulsar}. The bottom panel shows the fractional difference between the numerical and analytic density profiles for each $f_s$. Observe that the elliptical isodensity approximation becomes less accurate, as expected, when one increases $f_s$. Observe also that the maximum fractional difference occurs close to the stellar surface. This is because the density profile changes significantly close to the surface, as shown in the top panel, and hence, a small error in isodensity contours leads to a relatively large difference between the two approaches. Notice, however, that the maximum fractional difference is 7\% at most for the largest frequency considered $f_s \leq 700$Hz. 

The elliptical isodensity approximation allows us to make analytic progress in the solution to the integrals for the multipole moments. The first assumption allows for the decomposition of the integrals as follows:
\be
M_{\ell}  = 2\pi \; I_{\ell,3} \; R_{\ell}\,,\qquad S_\ell = \frac{4 \pi \ell}{2 \ell +1} \Omega
\left(I_{\ell-1,5}-I_{\ell+1,3}\right) R_{\ell+1}\,,
\ee
where the radial and angular integrals are defined by
\be
\label{eq:R-I-def}
R_{\ell}  \equiv{} \int_{0}^{a_{1}} \!\!\! \rho(\tilde{r}) \tilde{r}^{\ell+2} d\tilde{r}\,,
\qquad
I_{\ell,k}  \equiv{} \int_{-1}^{+1} \!\!\! \Theta(\mu)^{\ell+k} P_{\ell}(\mu) d\mu\,.
\ee
Notice that $R_{\ell}$ can be written as a function of compactness and eccentricity, where the precise functional form will depend on the equation of state. On the other hand, $I_{\ell}$ is simply a function of eccentricity that is equation-of-state independent. In particular, the integrals $I_{\ell,3}$ and $I_{\ell,5}$ can be solved exactly as a function of $e$~\cite{2007tisp.book.....G}, but the radial integrals must be solved numerically for a generic equation of state. 

The second assumption of the elliptical isodensity approximation allows us to rewrite the radial integrals in terms of the Lane-Emden function $\vartheta_\LE (\xi)$, if we assume a polytropic equation of state. The Lane-Emden function, which is defined via $\vartheta_{\LE} = (\rho/\rho_{c})^{1/n}$ with $\rho_{c}$ representing the central density, is a solution to the Lane-Emden equation, which is nothing but the dimensionless form of the Poisson equation for the gravitational potential of a polytropic fluid~\cite{2014grav.book.....P}. The quantity $\xi$ is a dimensionless radial coordinate, with the stellar surface located at $\xi = \xi_1$. In terms of the Lane-Emden function, $M_\ell$ and $S_\ell$ are then given by~\cite{Stein:2014wpa}
\begin{align}
\label{eq:M2l2-CM}
M_{2 \ell + 2} =
\frac{\left(-\right)^{\ell+1}}{2 \ell + 3}
\frac{e^{2\ell+2}}{(1 - e^{2})^{\frac{\ell+1}{3}}}
\frac{{\cal{R}}_{n,2+2 \ell}}{\xi_{1}^{2 \ell + 4} |\vartheta'(\xi_{1})|}
\frac{M^{2 \ell+3}}{\bar C^{2\ell+2}}\,,
\qquad
S_{2 \ell + 1} =
\frac{(-)^{\ell}}{2 \ell + 3}
\frac{2 \Omega \; e^{2 \ell}}{(1 - e^{2})^{\frac{\ell+1}{3}}}
\frac{{\cal{R}}_{n,2+2\ell}}{\xi_{1}^{2 \ell + 4} |\vartheta'(\xi_{1})|}
\frac{M^{2 \ell+3}}{\bar C^{2\ell+2}}\,,
\end{align}
where 
\be
\label{eq:Rnl}
{\cal{R}}_{n,\ell} \equiv  \int_{0}^{\xi_{1}}
 \left[\vartheta_\LE (\xi)\right]^{n} \xi^{\ell+2} d \xi\,,
\ee
and where the geometric mean radius $\bar{R} \equiv a_1 (1 - e^2)^{1/6}$ was used in the definition of the compactness $\bar C \equiv M/\bar{R}$. 

With these analytic expressions for the multipole moments in terms of the eccentricity and the compactness, we can now compute the no-hair relations for compact stars in the Newtonian limit. Eliminating $\Omega$ and $\bar C$ using $S_1$ and $M_2$ respectively, one arrives at the three-hair relations~\cite{Stein:2014wpa}
\be
\label{eq:3-hair-dimensional}
M_{\ell} + i \frac{\tilde q}{a} S_{\ell} = \bar{B}_{n,\lfloor \frac{\ell-1}{2} \rfloor}  M (i \tilde q)^\ell\,,
\ee
or in terms of the dimensionless multipole moments~\cite{Stein:2014wpa}
\be
\label{eq:3-hair}
\bar{M}_{2\ell+2} + i \bar{S}_{2\ell+1} = \bar{B}_{n,\ell}  \bar{M}_2^{\ell} (\bar{M}_2 + i \bar{S}_1)\,.
\ee
The reduced quadrupole moment is given by $i\tilde q \equiv \sqrt{M_2/M}$ and $\lfloor x \rfloor$ corresponds to the largest integer not exceeding $x$. All of the equation-of-state dependence is encoded in the coefficients $\bar B_{n,\ell}$, which is defined as 
\be
\label{eq:Bbar-def}
\bar B_{n,\ell} \equiv \frac{3^{\ell+1}}{2\ell +3} \frac{\mathcal{R}_{n,0}^\ell \mathcal{R}_{n,2\ell+2}}{\mathcal{R}_{n,2}^{\ell+1}}\,,
\ee
with $\mathcal{R}_{n,0} = |\vartheta'(\xi_1)| \xi_1^2$ when using the Lane-Emden equation and $\bar B_{n,-1} = 1 = \bar B_{n,0}$. Observe that the three-hair relations for Newtonian polytropes resemble the black hole no-hair relations of Eq.~\eqref{eq:BH-no-hair}, although unlike in the latter, the multipole moments in the former depend not only on the mass and spin, but also on the quadrupole moment. Observe also that the coefficients $\bar{B}_{n,\ell}$ do not depend on $e$ at all, with all of the spin dependence factoring out through $\bar{M}_{2}$ and $\bar{S}_{1}$.

\begin{figure}[htb]
\begin{center}
\includegraphics[width=8.cm,clip=true]{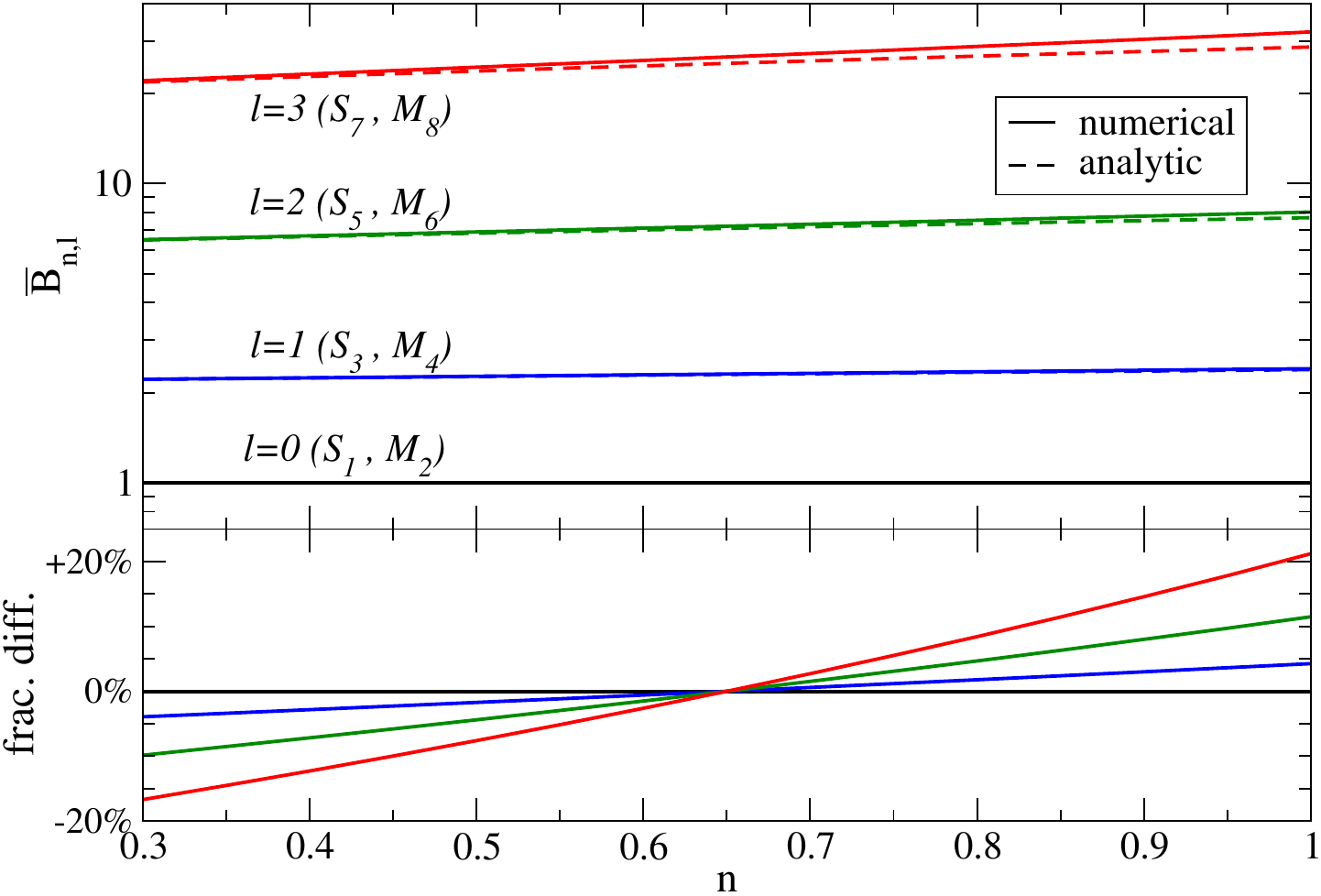}  
\caption{\label{fig:Bbar} (Top) The coefficients $\bar B_{n,\ell}$ against the polytropic index $n$ for various $\ell$ obtained numerically (solid) and analytically (dashed). The latter is obtained by solving the perturbed Lane-Emden equation about $n=0$. (Bottom) Fractional difference of the numerical results for various $n$ from the average $\langle n \rangle = 0.65$. Observe that the coefficients are universal to $\mathcal{O}(10\%)$.
This figure is taken and modified from Bretz \et~\cite{Bretz:2015rna}.
}
\end{center}
\end{figure}

Whether the no-hair relations are equation-of-state independent depends on how sensitive the $\bar{B}_{n,\ell}$ coefficients are to variations of the polytropic index. The solid curves in the top panel of Fig.~\ref{fig:Bbar} show $\bar B_{n,\ell}$ as a function of $n$ for various $\ell$, while the bottom panel shows the fractional difference with respect to $\bar B_{\langle n \rangle, \ell}$, where $\langle n \rangle = 0.65$ is the averaged polytropic index in $n \in [0.3,1]$. Observe that the equation-of-state variation increases as one increases $\ell$. Nonetheless, up to $\ell = 3$ (corresponding to the determination of $S_7$ and $M_8$ in terms of $M$, $S_{1}$ and $M_{2}$), the three-hair relations are equation-of-state universal to better than $\sim 20\%$. These results were first presented in Stein \et~\cite{Stein:2014wpa} and then extended to realistic equations of state (through piecewise polytropes) in Chatziioannou \et~\cite{Chatziioannou:2014tha}, which found similar levels of universality. 

Completely analytic three-hair relations can also be obtained if one studies stars with equations of state that are perturbations from $n=0$ polytropes~\cite{1978SvA....22..711S,Chatziioannou:2014tha}. Doing so, one finds that~\cite{Chatziioannou:2014tha,Bretz:2015rna}
\be
\label{eq:Bbar-analytic}
\bar B_{n,\ell} =  \frac{15 \cdot 5^\ell}{(2\ell+3) (2\ell+5)} \left\{1 - \frac{n}{15} \left[ 15 H \left( \ell + \frac{5}{2} \right) - 6\ell - 46 + 30\ln 2 \right]  \right\} + \mathcal{O}(n^2)\,,
\ee
where $H(\ell) \equiv \sum_{k=1}^{\ell}1/k$ is the $\ell$th harmonic number. Dashed lines in the top panel of Fig.~\ref{fig:Bbar} show the variability of $\bar{B}_{n,\ell}$ (as given in Eq.~\eqref{eq:Bbar-analytic}) with respect to  n. Observe that this analytic but approximate expression accurately describes the numerical results even when $n = 1$ (to $\mathcal{O}(1\%)$~\cite{Chatziioannou:2014tha}). Observe also that the $n$-dependence of Eq.~\eqref{eq:Bbar-analytic} is extremely weak, with $\partial \bar{B}_{n,\ell}/\partial n$ a very shallow function of $\ell$ for small $\ell$.

%-----------------------------
\subsubsection{Relativistic Results}
\label{sec:no-hair-relativistic}

The no-hair relations for compact stars hold, not only in the Newtonian limit, but also in full General Relativity. This can be shown to be the case through a numerical, slow-rotation treatment carried out in Yagi \et~\cite{Yagi:2014bxa}.  Let us begin by explaining how one can calculate the Geroch-Hansen multipole moments of a relativistic spacetime, where the integrals in Eqs.~\eqref{eq:M-ell-integ} and~\eqref{eq:S-ell-integ} are no longer valid. An obvious way is to simply follow the definition of such moments presented in~\cite{Hansen:1974zz}, but an easier approach was proposed in Pappas and Apostolatos~\cite{Pappas:2012ns}. Since the Geroch-Hansen moments are gauge invariant, one can extract them by calculating a gauge invariant quantity in two ways, in terms of (i) the given spacetime metric and (ii) the Geroch-Hansen moments, and then comparing the two expressions. Pappas and Apostolatos~\cite{Pappas:2012ns} considered a test particle around a stationary and axisymmetric spacetime in an equatorial, circular orbit and chose the energy change per logarithmic interval of the orbital frequency, $dE/d\ln\Omega_\orb$, as the gauge invariant quantity. The orbital angular velocity $\Omega_\orb$ and the energy per unit mass of a test particle $E$ are given in terms of the metric components as
\be
\Omega_{\orb} = \frac{-g_{t\phi,r} + \sqrt{(g_{t\phi,r})^2-g_{tt,r} g_{\phi\phi,r}}}{g_{\phi\phi,r}}\,, \quad 
E = - \frac{g_{tt}+g_{t\phi} \Omega_{\orb}}{\sqrt{-g_{tt}-2 g_{t\phi} \Omega_{\orb} - g_{\phi\phi} \Omega_{\orb}^2}}\,,
\ee
where commas represent partial derivatives. Substituting the metric for slowly-rotating compact stars valid to quartic order in spin in Yagi \et~\cite{Yagi:2014bxa} into the above equations, expanding about small velocity $v \equiv (M \Omega_\orb)^{1/3}$ and eliminating the radial coordinate $r$, one finds $dE/d\ln\Omega_\orb$ in a polynomial series of $v$. On the other hand, such an expression in terms of the Geroch-Hansen moments was also derived in Ryan~\cite{Ryan:1995wh}. Comparing the two term by term, one finds the Geroch-Hansen moments in terms of the integration constants in the metric~\cite{Yagi:2014bxa}.

\begin{figure*}[htb]
\begin{center}
\begin{tabular}{l}
\includegraphics[width=8.cm,clip=true]{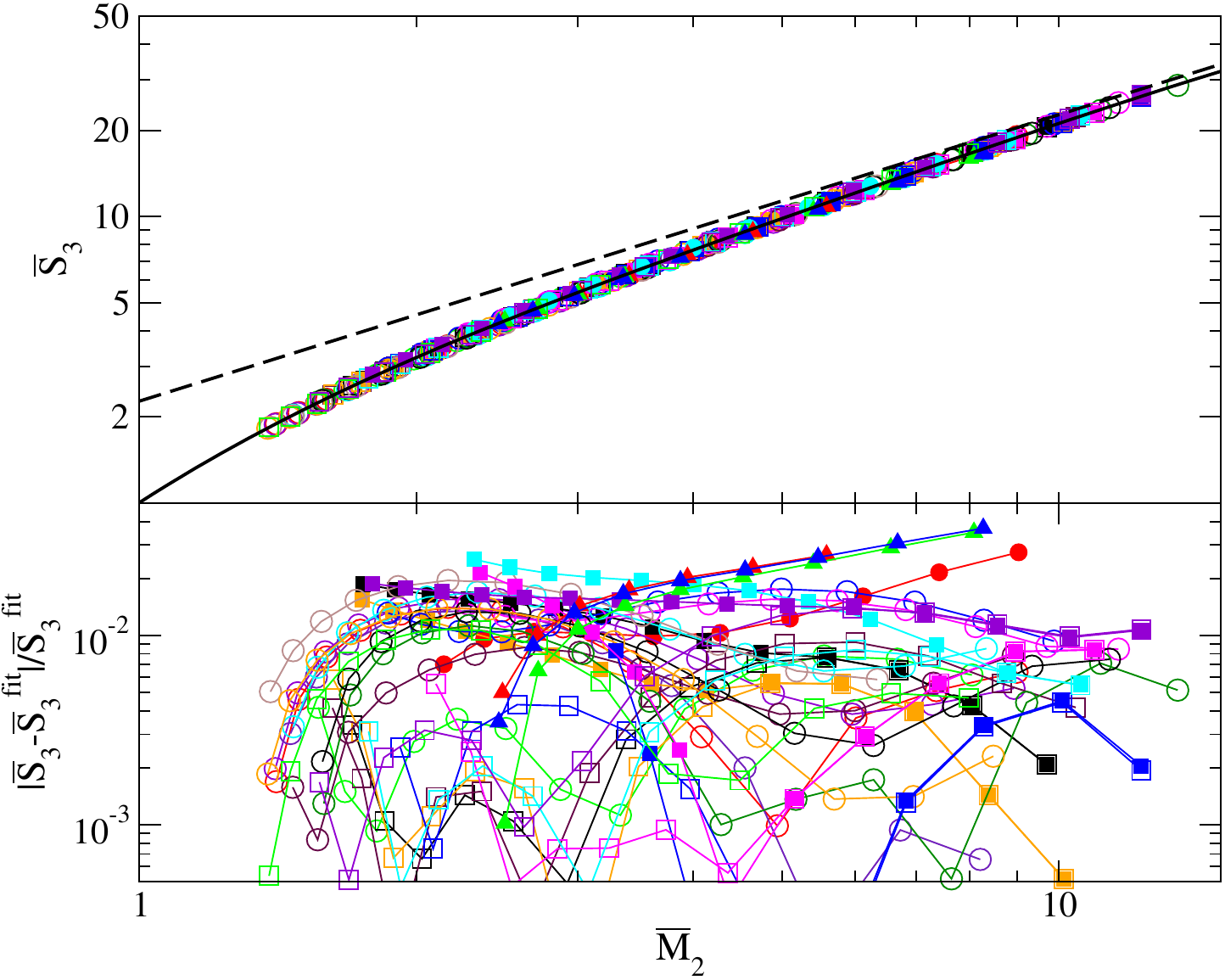} 
\includegraphics[width=8.cm,clip=true]{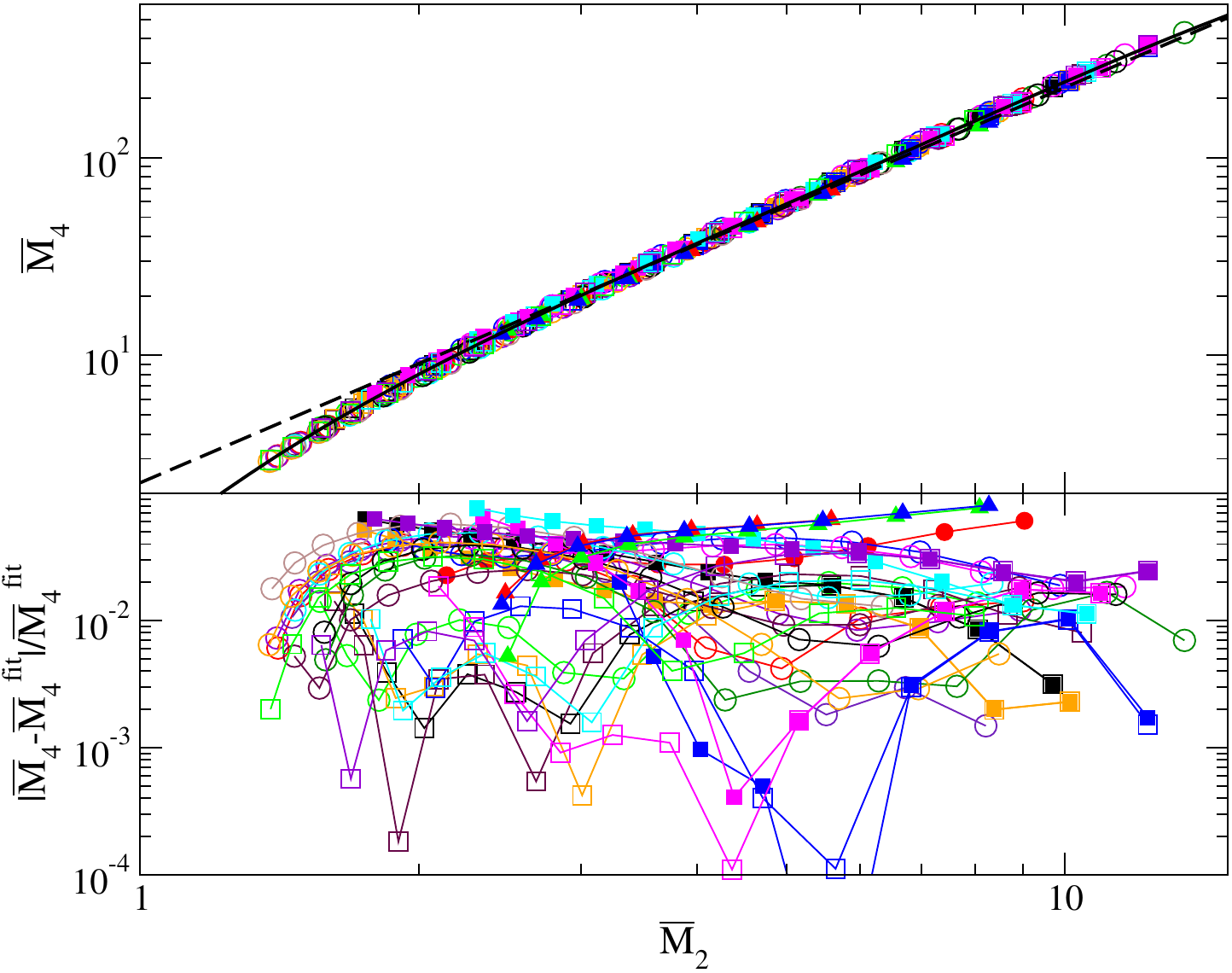} 
\end{tabular}
\caption{ \label{fig:no-hair}
(Top) The universal $\bar S_3$--$\bar M_2$ (left) and $\bar M_4$--$\bar M_2$ (right) relations for neutron stars and quark stars with various equations of state for $1M_\odot < M < M^{\mathrm{(max)}}$. The solid curves show a fit to all the numerical data given by Eq.~\eqref{fit}, while the dashed ones represent the Newtonian relations in Eq.~\eqref{eq:3-hair} with $n=0.5$. The meaning of each symbol is the same as in Fig.~\ref{fig:I-Love-Q}. (Bottom) Absolute fractional difference between the data and the fit. Observe that the relations are equation-of-state insensitive to $\sim 10\%$. 
}
\end{center}
\end{figure*}

With the multipole moments at hand, we now present the interrelations among them. The top panels of Fig.~\ref{fig:no-hair} show the $\bar S_3$--$\bar M_2$ and $\bar M_4$--$\bar M_2$ relations~\cite{Yagi:2014bxa} for slowly-rotating neutron stars and quark stars with various equations of state.  As before, the single parameter along each sequence is the central density, or equivalently, the stellar mass or compactness. First, observe that for large values of $\bar M_2$ (i.e.~as one approaches the Newtonian limit), the numerical results approach the semi-analytic Newtonian relations of Sec.~\ref{sec:no-hair-Newton} for an $n=0.5$ polytrope. Second, observe that the no-hair relations are approximately independent of the equation of state, which allows the construction of a single fitting function (black solid curves) of the form of Eq.~\eqref{fit} with coefficients in Table~\ref{table:coeff-no-hair} that is equation-of-state independent. As in the I-Love-Q case, we only present data for $1M_\odot < M < M^{\mathrm{(max)}}$ and the fit is only valid within the range of $\bar M_2$ shown in Fig.~\ref{fig:no-hair}. The bottom panels present the absolute fractional difference between the numerical data and the fit. Observe that the $\bar S_3$--$\bar M_2$ and $\bar M_4$--$\bar M_2$ relations are insensitive to the variation in the equation of state to $\sim 4\%$ and $\sim 10\%$ respectively. Such an amount of variation is somewhat larger than that found in the I-Love-Q relations of Fig.~\ref{fig:I-Love-Q}. The variation in the $\bar S_3$--$\bar M_2$ ($\bar M_4$--$\bar M_2$) relation is comparable to (larger than) that in the Newtonian no-hair relations in Fig.~\ref{fig:Bbar}. 

{\renewcommand{\arraystretch}{1.2}
\begin{table}
\begin{centering}
\begin{tabular}{cccccccc}
\hline
\hline
\noalign{\smallskip}
$y_i$ & $x_i$ &&  \multicolumn{1}{c}{$a_i$} &  \multicolumn{1}{c}{$b_i$}
&  \multicolumn{1}{c}{$c_i$} &  \multicolumn{1}{c}{$d_i$} &  \multicolumn{1}{c}{$e_i$}  \\
\hline
\noalign{\smallskip}
$\bar S_3$ & $\bar M_2$ && $3.131 \times 10^{-3}$  & 2.071 & $-0.7152$  & 0.2458 & $-0.03309$\\
$\bar M_4$ & $\bar M_2$ && $-0.02287$  & 3.849 & $-1.540$  & 0.5863 & $-8.337\times 10^{-2}$\\
\noalign{\smallskip}
\hline
\hline
\end{tabular}
\end{centering}
\caption{Numerical coefficients for the fitting formula of the no-hair like relations given in Eq.~\eqref{fit}.}
\label{table:coeff-no-hair}
\end{table}
}

%%%%%%%%%%%%%%%%%%%%%%%%%%%%%%%%%%%%%%%%%%%%%%%%%%%%%%%%
\section{Extensions of Universality in General Relativity}
\label{sec3:Extensions}

In this section, we review extensions to the analysis of the previous section. We begin by studying whether different ways of normalizing the multipole moments lead to a stronger universality in Sec.~\ref{sec:normalization}. We then relax the slow- and uniform-rotation approximations in Sec.~\ref{sec:rotation}. Sec.~\ref{sec:magnetized} deals with magnetized neutron stars. We introduce anisotropic pressure in Sec.~\ref{sec:anisotropy}. We then review the effect of non-barotropic equations of state in Sec.~\ref{sec:non-barotropic} and finally consider dynamical effects in the I-Love relation in Sec.~\ref{sec:dynamical-tides}. 

%%%%%%%%%%%%%%%%%%%%%%%%%%%%%%%%%%%%%%%%%%%%%%%%%%%%%%
\subsection{Different Normalizations to Strengthen Universality}
\label{sec:normalization}

Can we improve the strength of the universality in the I-Love-Q and the three-hair relations for the dimensionless multipole moments in Eq.~\eqref{eq:3-hair} by choosing a different normalization? This question was addressed by Majumder \et~\cite{Majumder:2015kfa} when they considered a new set of dimensionless moments of inertia and multipole moments:
\be
\label{eq:MM-new-def}
\bar{I}^{(a_{I})} \equiv \frac{\bar{I}}{C^{a_{I}}}\,,
\qquad
\bar{M}_{2\ell+2}^{(a_{M,2\ell+2})} \equiv \frac{\bar{M}_{2\ell+2}}{C^{a_{M,2 \ell+2}}}\,,
\qquad
\bar{S}_{2\ell+1}^{(a_{S,2\ell+1})} \equiv \frac{\bar{S}_{2\ell+1}}{C^{a_{S,2 \ell+1}}}\,.
\ee
Although these new definitions force the three-hair relations to depend in principle on both $\Omega$ and $e$, this dependence vanishes in the slow-rotation limit. One then finds that the fractional difference in the I--Q, $\bar M_4$--$\bar M_2$ and $\bar S_3$--$\bar M_2$ Newtonian relations between an $n=0$ and an $n=1$ polytrope is given by
\allowdisplaybreaks
\begin{align}
 \frac{\bar{I}^{(a_{I})}\vert_{n=0}-\bar{I}^{(a_{I})}\vert_{n=1}}{\bar{I}^{(a_{I})}\vert_{n=0}} &= 1 -4^{-\delta_0} 5^{\delta_0+1} \frac{\pi ^2-6}{3 \pi ^2}\,, \\
\label{frac24eqn}
 \frac{\bar{M}_4^{(a_{M,4})}\vert_{n=0}-\bar{M}_4^{(a_{M,4})}\vert_{n=1}}{\bar{M}_4^{(a_{M,4})}\vert_{n=0}} &= 1 - 4^{2-\delta_1} 5^{\delta_1-4}\frac{21 \left(120-20 \pi ^2+\pi ^4\right)}{\left(\pi ^2-6\right)^2}\,, \\
\label{frac23eqn}
 \frac{\bar{S}_3^{(a_{S,3})}\vert_{n=0}-\bar{S}_3^{(a_{S,3})}\vert_{n=1}}{\bar{S}_3^{(a_{S,3})}\vert_{n=0}} &= 1- 4^{1-\delta_2} 5^{\delta_2-3}\frac{21 \left(120-20 \pi ^2+\pi ^4\right)}{\left(\pi ^2-6\right)^2}\,,
\end{align}
with $\delta_0 \equiv (a_{I}+2)/(a_{M,2}+1)$, $\delta_1 \equiv (a_{M,4}+2)/(a_{M,2}+1)$ and $\delta_2 \equiv (a_{S,3}+1)/(a_{M,2}+1)$. Using these expressions, one can find a one-parameter family in $(a_{M,2},a_{I})$, $(a_{M,2},a_{M,4})$ and $(a_{M,2},a_{S,3})$ space such that the fractional differences vanish exactly. In other words, one can choose specific normalizations of the dimensionless multipole moments such that the I--Q, $\bar M_4$--$\bar M_2$ and $\bar S_3$--$\bar M_2$ relations become exactly the same for slowly-rotating Newtonian stars constructed with $n=0$ and $n=1$ polytropes. For realistic equations of state, there also exists a one-parameter family of normalization coefficients that improve the universality~\cite{Majumder:2015kfa}.

\begin{figure}[htb]
\begin{center}
\includegraphics[width=7.cm,clip=true]{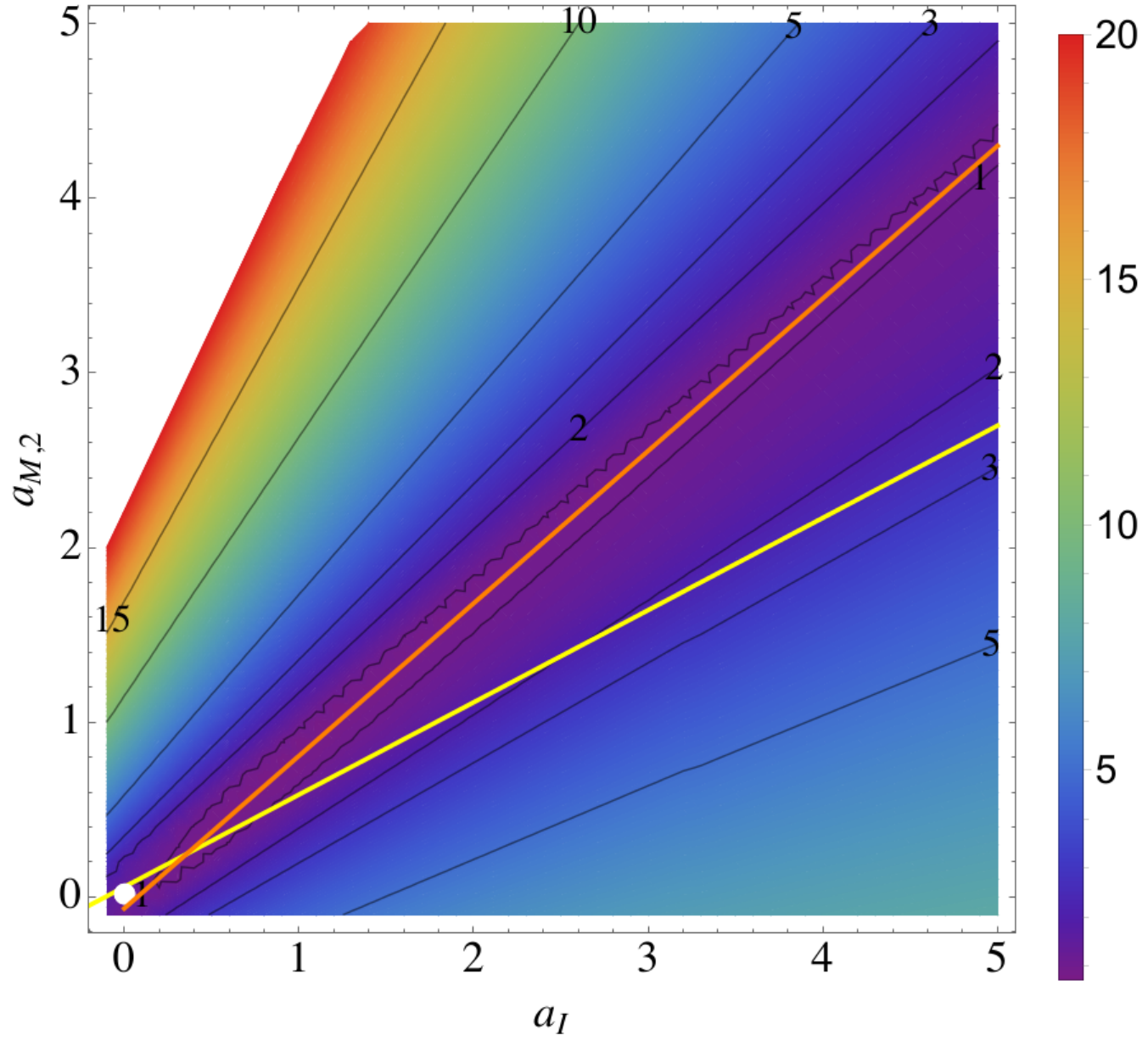}  \qquad
\includegraphics[width=7.cm,clip=true]{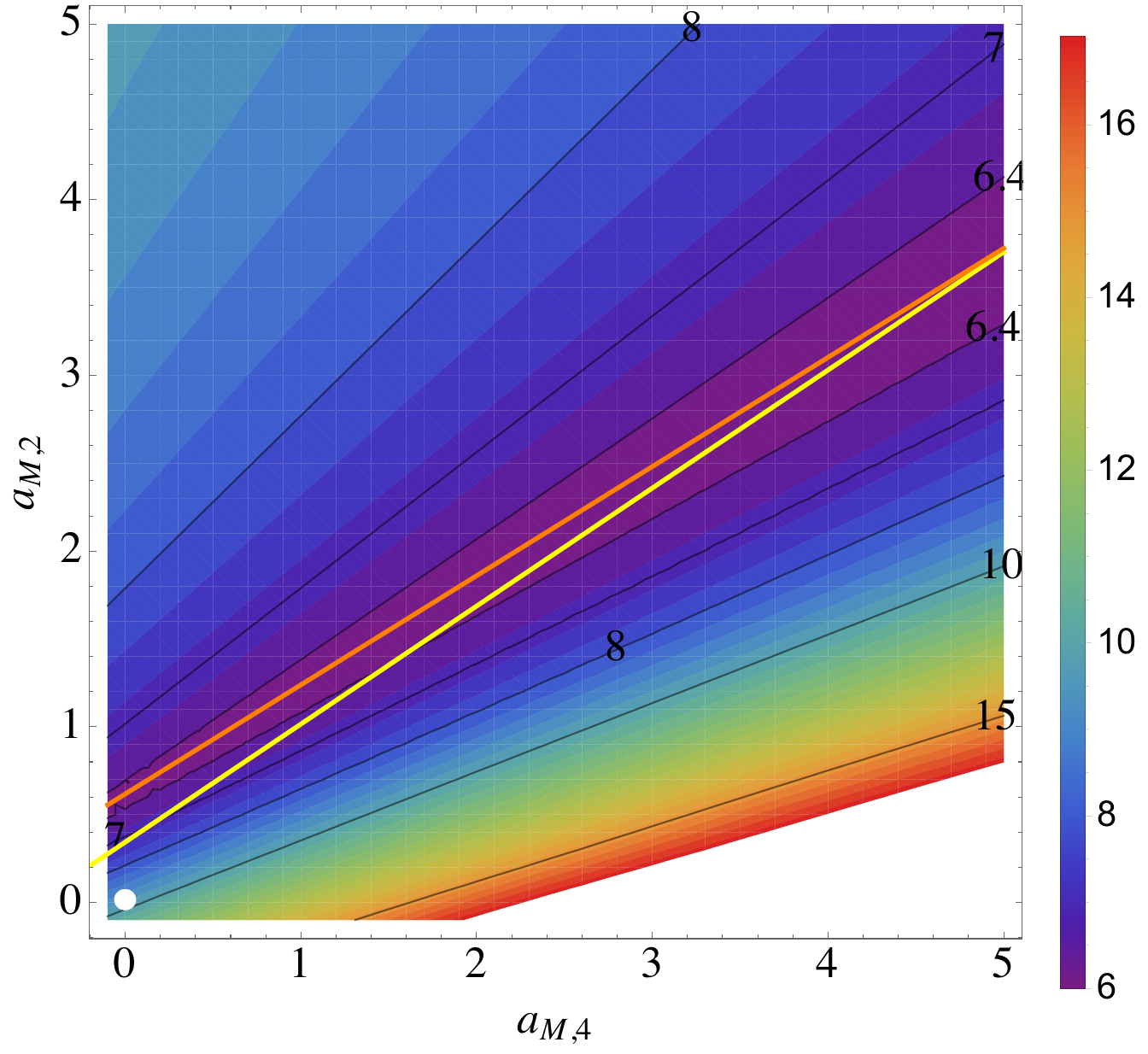}  
\caption{\label{fig:normalization} A maximum percent equation-of-state variation in the I--Q (left) and $\bar M_{4}$--$\bar M_{2}$ relations (right) for different values of the normalization constants introduced in Eq.~\eqref{eq:MM-new-def}. We also present the relations among these normalization constants that reduce to the least equation-of-state variation in the Newtonian limit (yellow) and in full General Relativity (orange). The original choice of normalization in~\cite{Yagi:2013bca,Yagi:2013awa,Stein:2014wpa,Yagi:2014bxa} is shown by white circles.
This figure is taken from Majumder \et~\cite{Majumder:2015kfa}.
}
\end{center}
\end{figure}

Figure~\ref{fig:normalization} presents the maximum equation-of-state variation in the I--Q and $\bar M_{4}$--$\bar M_{2}$ relations in full General Relativity for different choices of $(a_{M,2},a_{I})$ and $(a_{M,2},a_{M,4})$ from the relations with a fiducial equation of state (LS220~\cite{Lattimer:1991nc})~\cite{Majumder:2015kfa}, assuming slow-rotation. The original choice of normalization in~\cite{Stein:2014wpa,Yagi:2014bxa} is shown with white dots. The one-parameter family in each diagram that gives the least equation-of-state variation is shown with an orange line, while the yellow line shows the one-parameter family that gives zero variation between the relations for the $n=0$ and $n=1$ polytropes in the Newtonian limit. In the I--Q case, the Newtonian result for the least equation-of-state variation is quite different from the relativistic one, while the original normalization choice is close to the latter. In the $\bar M_{4}$--$\bar M_{2}$ case, the Newtonian result is close to the relativistic one, and the original universality can be improved by roughly a factor of two. The $\bar S_3$--$\bar M_2$ relation presents similar behavior to the I--Q relation, namely, the Newtonian expression for the least equation-of-state variation is quite different from the relativistic result, while the original normalization lies close to the latter~\cite{Majumder:2015kfa}. These results show that the amount of the universality in the I-Q and no-hair relations depends on how one normalizes the multipole moments. 

%%%%%%%%%%%%%%%%%%%%%%%%%%%%%%%%%%%%%%%%%%%%%%%%%%%%%%
\subsection{Rotation}
\label{sec:rotation}

In this section, we review how sensitive the universality is to the assumption of slow and uniform rotation of a relativistic star. We will relax these assumptions in turn by considering rapidly-rotating neutron stars and quark stars in Sec.~\ref{sec:rapid} and differentially-rotating stars in Sec.~\ref{sec:diff}.

%--------------------------------------------------
\subsubsection{Rapid Uniform Rotation}
\label{sec:rapid}

\begin{figure*}[htb]
\begin{center}
\begin{tabular}{l}
\includegraphics[width=7.5cm,clip=true]{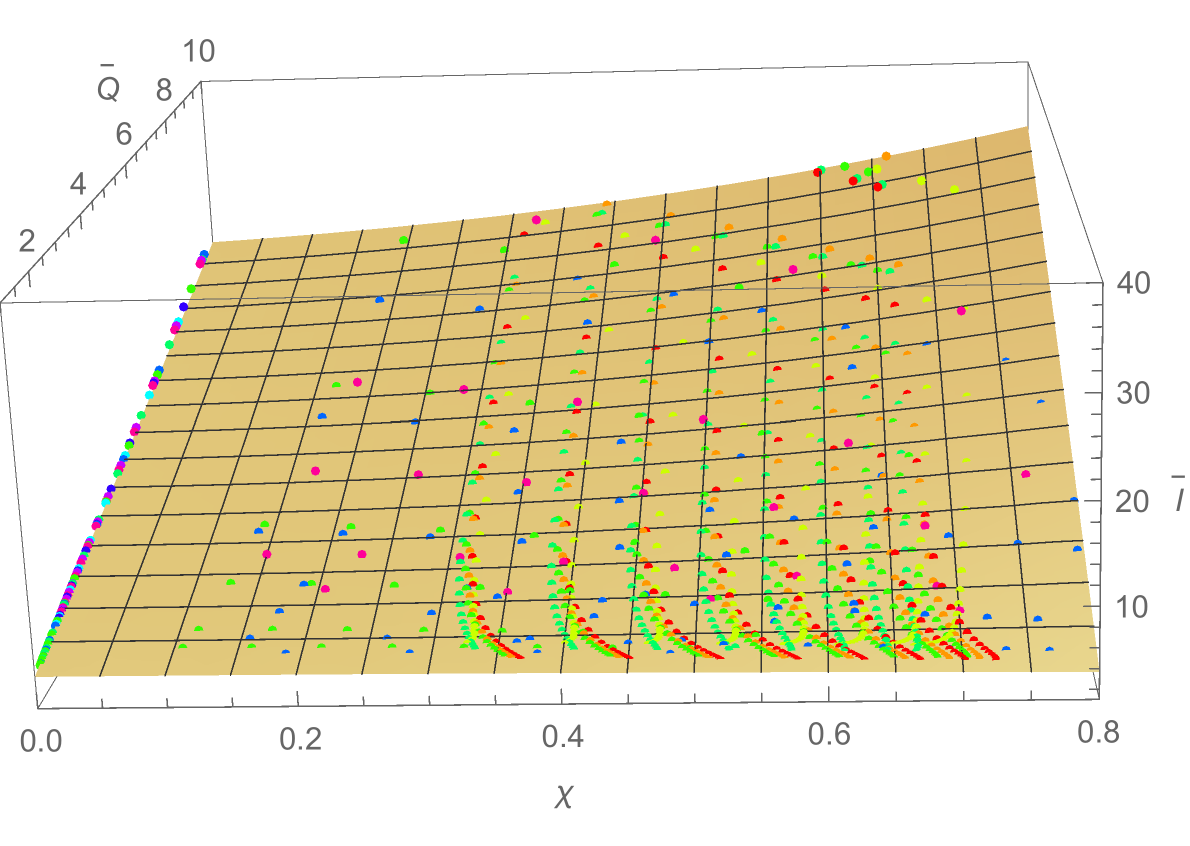} \qquad
\includegraphics[width=7.5cm,clip=true]{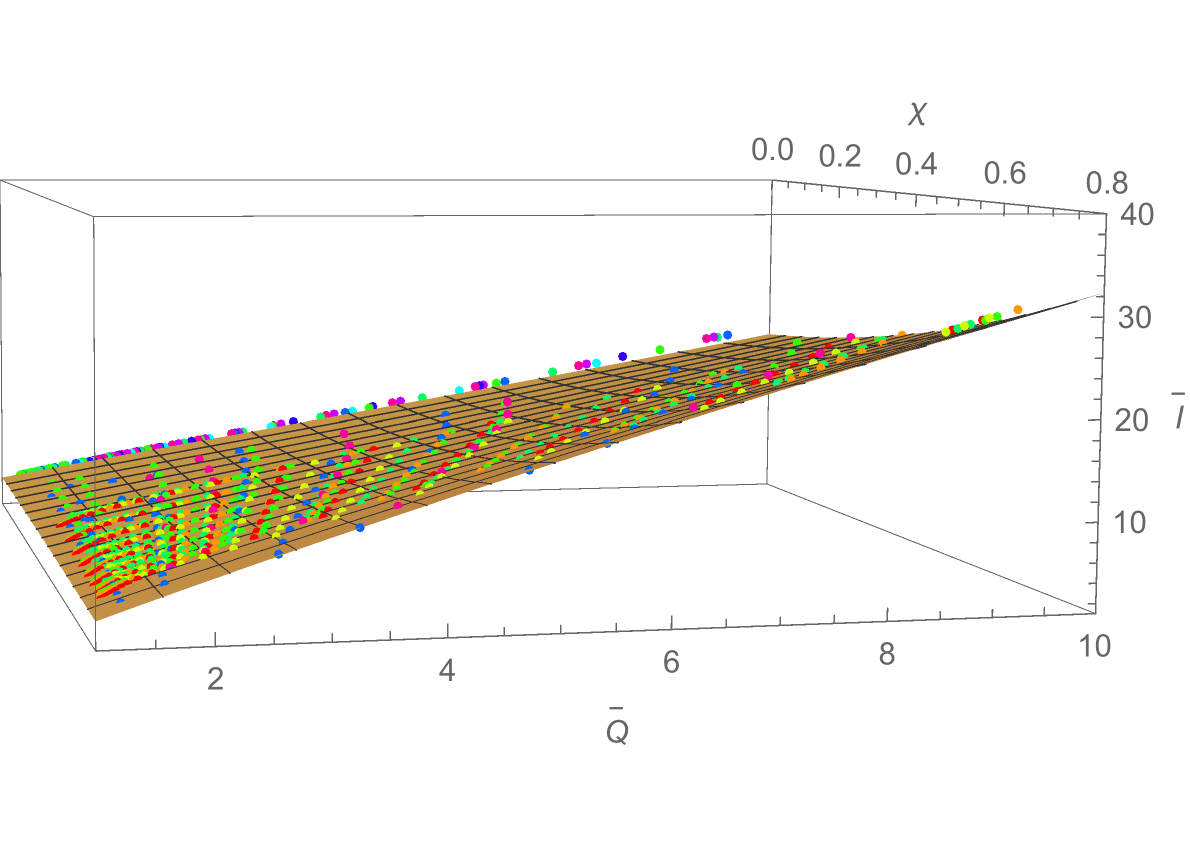} 
\end{tabular}
\caption{\label{fig:IQ-3D}
The universal I-Q relations in terms of the dimensionless spin parameter $\chi$ for both slowly-rotating and rapidly-rotating neutron stars and quark stars from two different viewpoints. Different colors correspond to different equations of state. Observe that the I-Q relation is a plane instead of a curve as in Fig.~\ref{fig:I-Love-Q}. Brown planes are the fit found in Pappas and Apostolatos~\cite{Pappas:2013naa}.
}
\end{center}
\end{figure*}

Although most neutron stars rotate slowly, the dimensionless spin of the fastest-spinning pulsar currently observed~\cite{fastest-pulsar} can be as large as $\chi \approx 0.65$ depending on its mass and radius (see Fig.~\ref{fig:chi-f}). Hence, it is important to investigate how spin corrections to the I-Love-Q and approximate no-hair relations affect the degree of universality. One can achieve this goal by numerically constructing rapidly- and uniformly-rotating neutron star and quark star solutions as described already in Sec.~\ref{sec:Approxs}.

The I-Q relation for rapidly-rotating neutron stars and quark stars was first studied in Doneva \et~\cite{Doneva:2013rha}. The authors investigated sequences of relativistic stellar solutions for different spin frequencies and found that the universality becomes significantly weaker as one increases the spin frequency. For example, the universality with a fixed spin frequency of $f_s =  800$Hz is $\sim 10\%$, an order of magnitude worse than the slow-rotation result ($f_s = 0$Hz). Thus, Doneva \et~\cite{Doneva:2013rha} concluded that the universality in the I-Q relation breaks for rapidly rotating stars when considering compact star sequences with fixed spin frequency. 

The quantity that is held constant in a neutron star sequence when considering universality can have a huge impact. Pappas and Apostolatos~\cite{Pappas:2013naa} found that the universality in the I-Q relation remains if one fixes the \emph{dimensionless} spin parameter $\chi$ instead of the dimensional spin frequency $f_s$ (for fixed $\chi$, the single parameter along the universal curve is the stellar mass or compactness, just like in the slow-rotation limit). Figure~\ref{fig:IQ-3D} presents the I-Q relation as a function of $\chi$ from two different viewpoints; observe that the numerical data lies approximately on a single universal plane, given by a fit in Pappas and Apostolatos~\cite{Pappas:2013naa}. Therefore, if one fixes $\chi$, the universality of the I-Q relations remains, but the universal relation is different from that found in the slow-rotation limit (which corresponds to the $\chi=0$ cross-section of Fig.~\ref{fig:IQ-3D}). Chakrabarti \et~\cite{Chakrabarti:2013tca} further found that a similar universal I-Q relation arises for rapidly rotating stars when one fixes other dimensionless spin combinations, such as $M f_s$ and $R f_s$. In these cases, the universality is valid to $\mathcal{O}(1\%)$, just like in the slow-rotation case. These studies make it clear that the I-Q relation does \emph{not} break for rapidly rotating stars, provided one fixes the appropriate spin parameter in the stellar sequence.

\begin{figure}[htb]
\begin{center}
\begin{tabular}{l}
\includegraphics[width=7.5cm,clip=true]{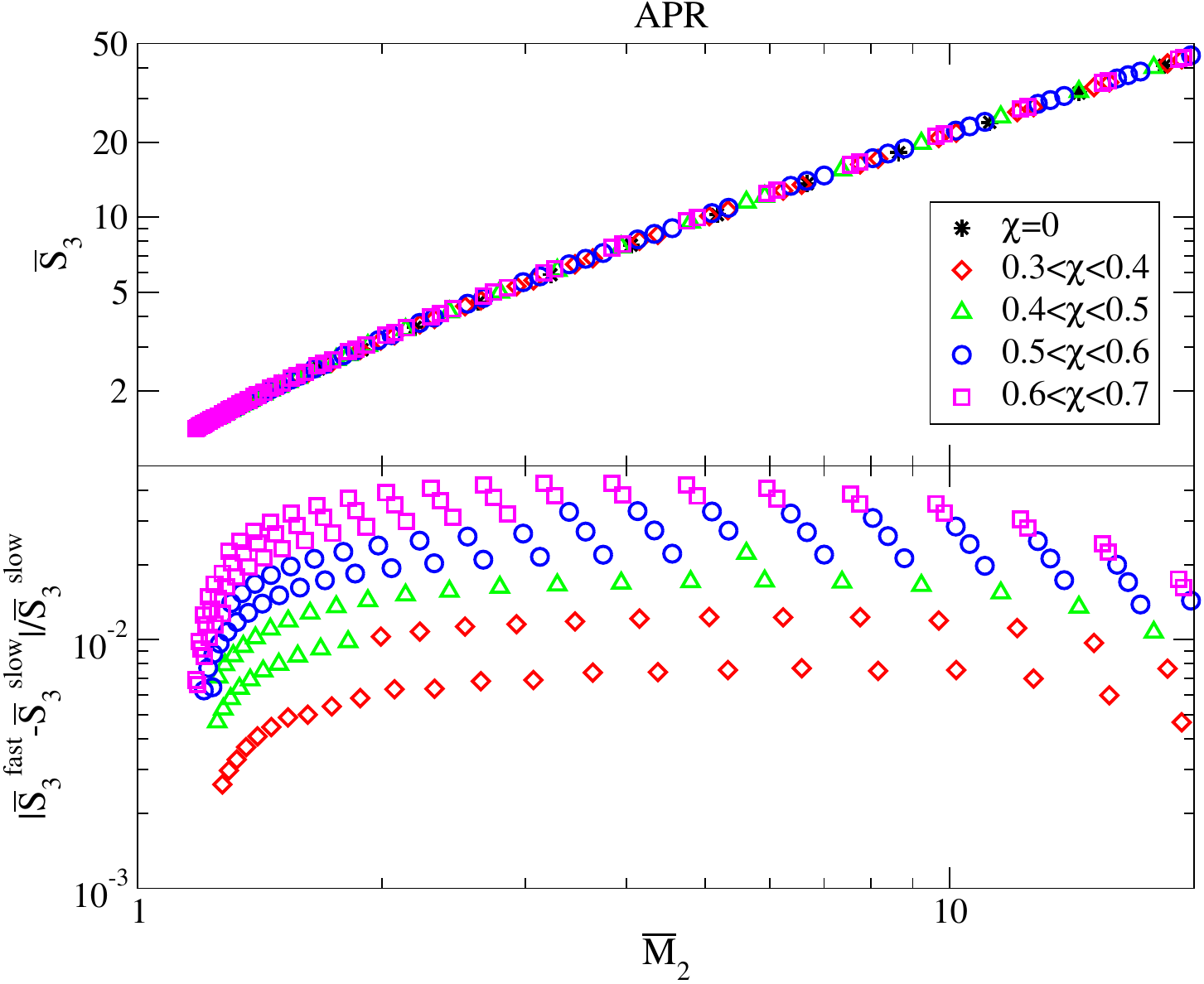} 
\includegraphics[width=7.5cm,clip=true]{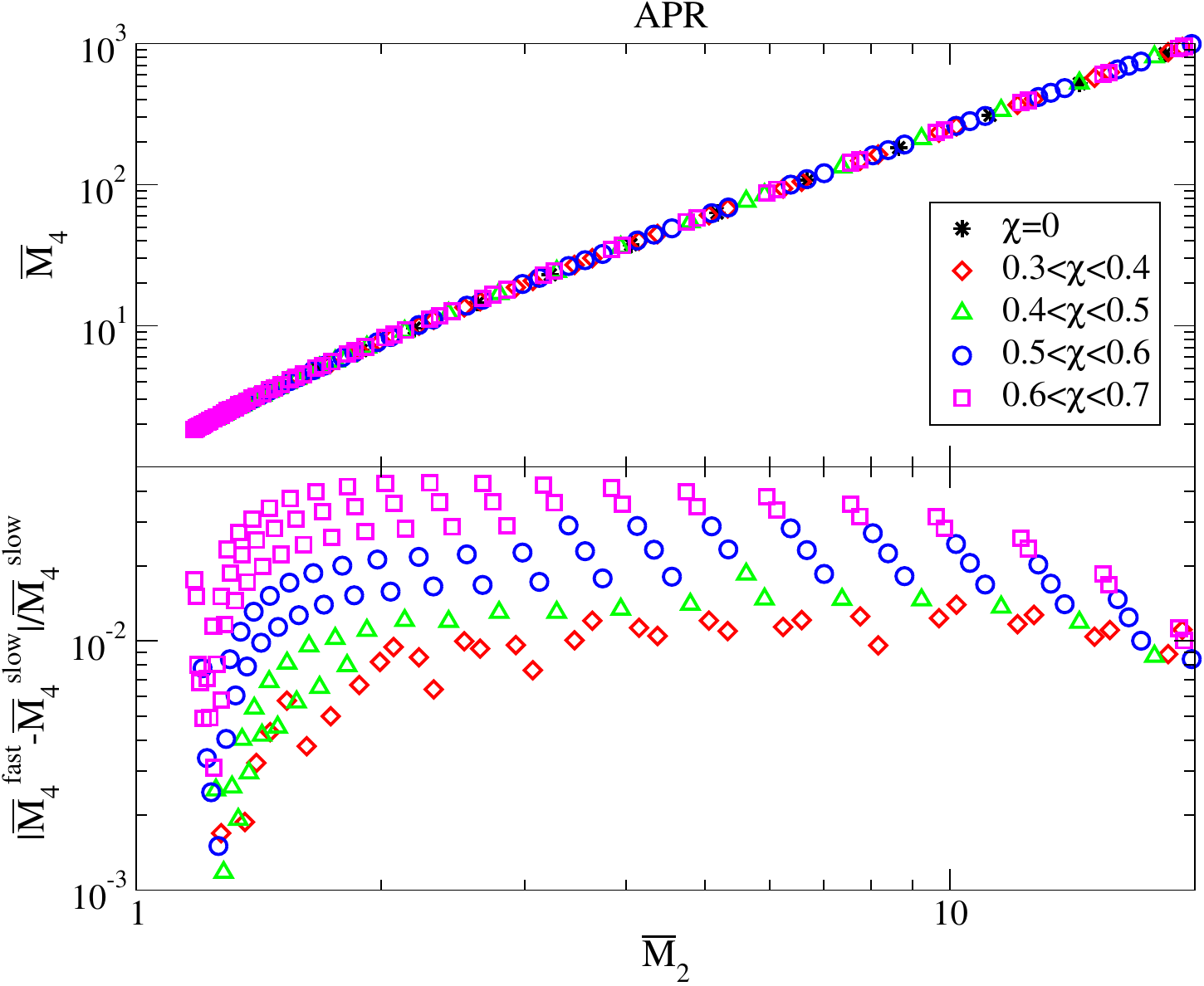} 
\end{tabular}
\caption{\label{fig:M4-M2-chidep}
(Top) Spin dependence of the $\bar{S}_3$--$\bar{M}_2$ (left) and $\bar{M}_4$--$\bar{M}_2$ (right) relations for an APR equation of state. (Bottom) Fractional difference between the relations for rapidly-rotating neutron stars using RNS and that in the slow-rotation limit with various dimensionless spin parameters. Observe that such a difference for the $\bar{S}_3$--$\bar{M}_2$ ($\bar{M}_4$--$\bar{M}_2$) relation is comparable to (smaller than) the equation-of-state variation in Fig.~\ref{fig:no-hair}. 
}
\end{center}
\end{figure}

Let us now discuss the approximate no-hair relations for rapidly-rotating neutron stars. The $\bar S_3$--$\bar M_2$ relation, first studied in Pappas and Apostolatos~\cite{Pappas:2013naa}, is shown in the top left panel of Fig.~\ref{fig:M4-M2-chidep} for an APR equation of state with different spins. Observe that although the fractional difference between the relation for rapidly-rotating and slowly-rotating stars (bottom panel) increases as one increases $\chi$, the relation is only sensitive to spin to $\sim 5$\%, which is comparable to the equation-of-state variation in Fig.~\ref{fig:no-hair}. The $\bar{M}_4$--$\bar{M}_2$ relation, shown in the right panel of Fig.~\ref{fig:M4-M2-chidep}, shows similar spin behavior relative to that of the $\bar{S}_3$--$\bar{M}_2$ relation, but now the spin variation is smaller than the equation-of-state variation. Observe also that the fractional difference in both the $\bar{S}_3$--$\bar{M}_2$ and $\bar{M}_4$--$\bar{M}_2$ relations decreases as one increases $\bar M_2$. This is because the spin dependence vanishes in the Newtonian limit within the elliptical isodensity approximation, as already explained in Sec.~\ref{sec:no-hair-Newton}. These findings allow us to conclude that the universality is also preserved in the approximate no-hair relations even for rapidly-rotating neutron stars and quark stars, provided one fixes the spin parameter appropriately in the stellar sequence.

%--------------------------------------------------
\subsubsection{Differential Rotation}
\label{sec:diff}

\begin{figure*}
\begin{center}
\includegraphics[width=8.cm,clip=true]{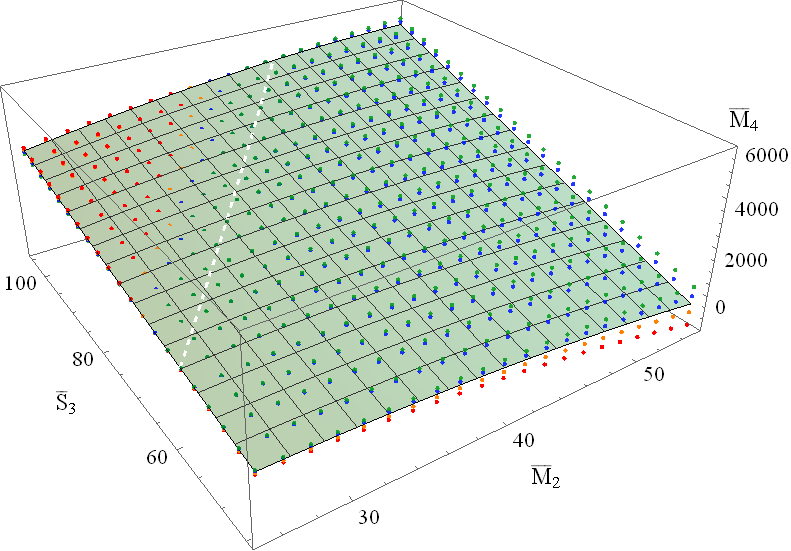}
\includegraphics[width=6.cm,clip=true]{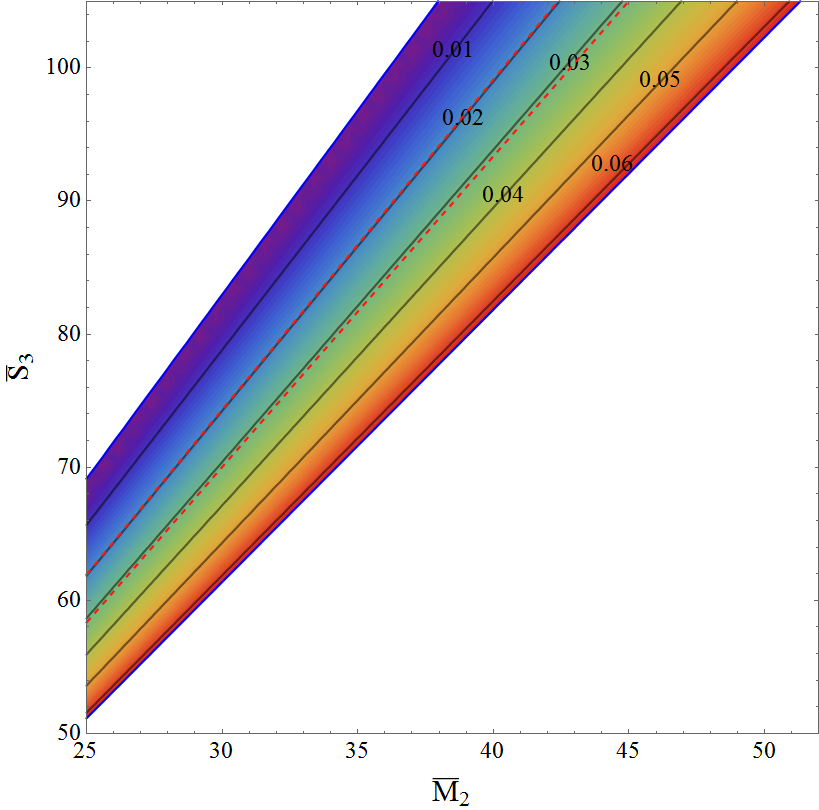}
\caption{\label{fig:diff-rot} (Left) The four-hair relations for $\bar{M}_4$ as a function of $\bar M_2$ and $\bar S_3$ for differentially-rotating stars in the Newtonian and slow-rotation limit (Eq.~\eqref{eq:4-hair-M4}) with $n= 0.3$ (green dots), 0.5 (blue dots), 0.8 (orange dots), 1 (red dots), and 0.65 (light green plane). Observe that the dots approximately lie on the plane. We also present the three-hair relation for uniformly-rotating stars in~\cite{Stein:2014wpa} (white dashed). (Right) Maximum fractional difference (color gradient and contours) in the four-hair relation in the left panel from the $n=0.65$ plane with index $n\in[0.3,1]$ within $|\gamma| < 0.5$, while the dashed red lines correspond to the region for $|\gamma| < 0.1$ with $n=1$. Observe that the maximum fractional difference is always less than $\sim 6\%$ and $3\%$ for $|\gamma| < 0.5$ and $|\gamma| < 0.1$ respectively.
This figure is taken from Bretz \et~\cite{Bretz:2015rna}.
}
\end{center}
\end{figure*}

We next study how the assumption of uniform rotation affects the universality by considering differentially rotating stars. Differential rotation is important in proto-neutron stars formed after supernova explosions~\cite{Dimmelmeier:2002bm,Ott:2003qg}, whose differential rotation may last for $\sim 1$ minute, and in hypermassive neutron stars formed after mergers of neutron star binaries~\cite{Shapiro:2000zh,Paschalidis:2012ff,Hotokezaka:2013iia}. Differential rotation may also arise due to nonlinear effects caused by r-mode oscillations~\cite{Rezzolla:1999he,Levin:2000vq,Stergioulas:2000vs,Sa:2003jw}. 

The first (and only) study of the no-hair relations in differentially-rotating stars is that of Bretz \et~\cite{Bretz:2015rna}, which restricts attention to the Newtonian limit and the small differential-rotation limit. Although the actual profile of rotation depends on the astrophysical situation one considers, the rotation law in the Newtonian limit can be parameterized in a generic way as~\cite{Galeazzi:2011nn}
\be
\frac{\Omega }{\Omega _c}=\left( 1-\alpha\gamma \frac{r^{2}}{{a_1}^{2}} \sin^{2}{\theta}\right)^{\frac{1}{\alpha}}\,,
\ee
where $\Omega_c$ is the stellar angular velocity at the center, $\gamma$ is a dimensionless parameter that controls the amount of differential rotation, and $\alpha$ determines the type of differential rotation. For example, the j-constant and v-constant laws, the Keplerian angular velocity profile and the rotation law for hypermassive neutron stars can be captured by choosing $\alpha= -1$, $-2$, $-4/3$ and $-4$ respectively. In the small differential rotation approximation, $|\gamma| \ll 1$. Keeping term to linear order in $\gamma$, the above equation can be expanded as  
\be
\label{eq:diff-rot-Omega-exp}
\frac{\Omega }{\Omega _c}=1-\gamma \frac{r^{2}}{{a_1}^{2}} \sin^{2}{\theta}+\mathcal{O}(\gamma^2)\,.
\ee
Notice that this equation does not depend on $\alpha$, and hence, it applies to a large class of differential rotation laws.

Consider now how differential rotation affects the three-hair relations for uniformly-rotating stars of Sec.~\ref{sec:no-hair-Newton}. One modification arises from the relation between $\Omega$ and $r$ in the current multipole moments of Eq.~\eqref{eq:S-ell-integ} and another from a change in the shape of the stellar surface. In Sec.~\ref{sec:no-hair-Newton}, we assumed that the latter is the same as that of a constant density star, which in rigid rotation is a spheroid. For differentially rotating stars, a constant density star deviates from a spheroid by $\mathcal{O}(\gamma)$ terms~\cite{Bretz:2015rna}, which affects both mass and current multipole moments. Taking these modifications into account, one can repeat the analysis in Sec.~\ref{sec:no-hair-Newton}, and keeping terms up to linear order in $\gamma$, one finds~\cite{Bretz:2015rna}
\be
\bar{M}_{2\ell+2}+i\bar{S}_{2\ell+1} = 
\bar{B}_{n,\ell}\bar{M}_2^{\ell} 
\left[ \bar{M}_2 \left(1+\gamma\,\alpha_{\ell}^{(3,\bar{M})} \right)
+ i \, \bar{S}_1 \left(1+\gamma\,\alpha_{n,\ell}^{(3,\bar{S})} \right)
\right] +  \mathcal{O}(\gamma^2)\,,
\ee
or equivalently,
\be
M_{\ell}+i\frac{q}{a}S_{\ell}= \bar{B}_{n,\lfloor\frac{\ell-1}{2}\rfloor}M_0( i \tilde q)^\ell \left( 1+\gamma\,\alpha_{n,\ell}^{(3)} \right) +  \mathcal{O}(\gamma^2)\,, \qquad 
\alpha_{n,\ell}^{(3)}=\left\{
\begin{array}{ll}
\alpha_{\frac{\ell-2}{2}}^{(3,\bar{M})}&\,\text{even}\,\ell \\
\alpha_{n,\frac{\ell-1}{2}}^{(3,\bar{S})}&\,\text{odd}\,\ell
\end{array}\right.\,,
\ee
where $\alpha_{\ell}^{(3,\bar{M})}$ and $\alpha_{n,\ell}^{(3,\bar{S})}$ are coefficients given by Eqs.~(B5) and~(B6) in Bretz \et~\cite{Bretz:2015rna}. The former depends only on $\ell$ and $e^2$, while the latter also depends on the polytropic index $n$. One can go one step further and replace $\gamma$ with the next independent multipole moment $S_3$. One then arrives at the \emph{four-hair} relations for differentially rotating Newtonian polytropes~\cite{Bretz:2015rna}:
\be
\bar{M}_{2\ell+2}+i\bar{S}_{2\ell+1} = \bar{B}_{n,\ell}\bar{M}_2^{\ell}
\left[ \bar{M}_2 \left(1+\alpha_{n,\ell}^{(4,\bar{M})} \right)
+ i \, \bar{S}_1 \left(1+\alpha_{n,\ell}^{(4,\bar{S})} \right) 
\right]+  \mathcal{O}(\gamma^2)\,,
\ee
where the coefficients $\alpha_{n,\ell}^{(4,\bar{M})}$ and $\alpha_{n,\ell}^{(4,\bar{S})}$ depend on $\bar M_2$ and $\bar S_3$ and are given by Eqs.~(B7) and~(B8) in Bretz \et~\cite{Bretz:2015rna} respectively.

In order to study the equation-of-state variation of the four-hair relations, let us now work in the slow-rotation limit. The mass hexadecapole $\bar M_4$ is then given by~\cite{Bretz:2015rna}
\be
\label{eq:4-hair-M4}
\bar{M}_4=\bar{M}_2^2 \bar{B}_{n,1}\left[ 1+\frac{49}{45}\frac{ \left(\bar{S}_3-\bar{M}_2 \bar{B}_{n,1}\right)}{ \bar{M}_2 \bar{B}_{n,1} \bar{C}_{n,1}}\right]+  \mathcal{O}(\gamma^2)\,, \qquad \bar{C}_{n,\ell} := \frac{{\mathcal{R}_{n,2 \ell+4}}}{\xi _1^2 \, {\mathcal{R}_{n,2 \ell+2}}}\,.
\ee
The left panel of Fig.~\ref{fig:diff-rot} shows this four-hair relation as a function of $\bar M_2$ and $\bar S_3$ with different polytropic indices. Observe that the data lies approximately on a single universal plane that is well approximated by the $n=0.65$ polytropic case. The three-hair relation for uniformly-rotating Newtonian polytropes is shown by the white dashed curve. The right panel of this figure shows the maximum fractional difference from the $n=0.65$ case for each fixed $\bar M_2$ and $\bar S_3$. We only show the region where $|\gamma| < 0.5$ to comply with the small differential rotation approximation, while the region enclosed by the red dashed lines correspond to $|\gamma| < 0.1$. Observe that the maximum fractional difference is at most $\sim 6\%$ ($3\%$) for $|\gamma| < 0.5$ ($|\gamma| < 0.1$). A 3\% equation-of-state variation is slightly smaller than the 4\% variation in  the uniformly-rotating three-hair case (see Fig.~\ref{fig:Bbar}). This is because one needs to fix both $\bar M_2$ and $\bar S_3$ in the relation for differentially-rotating stars, while one can only fix $\bar M_2$ for the uniformly-rotating case. Such an additional degree of freedom helps the relation be slightly more universal. The four-hair relations among $\bar S_5$, $\bar M_2$ and $\bar S_3$ present a similar behavior, though the equation-of-state variation is somewhat larger ($\sim 10\%$ with $|\gamma| < 0.1$)~\cite{Bretz:2015rna}.

%%%%%%%%%%%%%%%%%%%%%%%%%%%%%%%%%%%%%%%%%%%%%%%%%%%%%%
\subsection{Magnetic Fields}
\label{sec:magnetized}

In this section, we review how the I-Q relation for magnetized stars differ from that for unmagnetized stars~\cite{Haskell:2013vha}. For radio pulsars, the magnetic field strength at the surface is inferred to be $\lesssim 10^{12}$G, while that of magnetars can be $\sim 10^{15}$G. The internal magnetic field strength can be much larger than the surface value~\cite{Braithwaite:2008aw,Corsi:2011zi,Ozel:2012wu}. These magnetic fields give rise to a non-vanishing quadrupole moment, which may dominate the rotationally-induced one considered in Sec.~\ref{sec:I-Love-Q}. The toroidal component of the magnetic field forces the star to be prolate, which can change the I-Q relation from the unmagnetized case drastically.

Let us first look at a purely poloidal or toroidal configuration for simplicity. Based on the stellar ellipticity calculation of Haskell \et~\cite{Haskell:2007bh} in the Newtonian limit, the I-Q relation for magnetized Newtonian stars with an $n=1$ polytropic equation of state is approximately~\cite{Haskell:2013vha} 
\begin{align}
\bar Q &= 5 \bar{I}^{1/2} + 10^{-3} \bar I \left( \frac{B_p}{10^{12}\mathrm{G}} \right)^2 \left( \frac{P}{1\mathrm{s}} \right)^2\,, \qquad {\rm{poloidal}} \; {\rm{case}}\,, \\
\bar Q &= 5 \bar{I}^{1/2} - 3 \times 10^{-5} \bar I \left( \frac{\langle B \rangle}{10^{12}\mathrm{G}} \right)^2 \left( \frac{P}{1\mathrm{s}} \right)^2\,, \qquad {\rm{toroidal}} \; {\rm{case}}\,,
\end{align}
up to quadratic order in $(B \,  P) \sim 10^{-2} (B/10^{12}\mathrm{G}) (P/1\mathrm{s})$, where $P$ is the spin period, while $B_p$ and $\langle B \rangle$ are the magnetic field strength at the pole and the averaged field strength respectively. The first (second) term in the above equations corresponds to the rotationally- (magnetically-) induced quadrupole moment. The I-Q relation clearly acquires a correction due to the magnetic field, though the magnitude of such a correction is typically small, unless the magnetic field is of magnetar magnitude. Observe also that the magnetic field correction to $\bar Q$ for the toroidal configuration is negative, which means that it pushes the star toward a prolate configuration. 

In order to study the relativistic effect and the equation-of-state variation in the relation, Haskell \et~\cite{Haskell:2013vha} constructed magnetized neutron star solutions in full General Relativity using the LORENE code~\cite{Bocquet:1995je,Frieben:2012dz}. The authors found that although the I-Q relation deviates from the relation in the unmagnetized case as one increases $B$ or decreases $P$, the relation remains relatively equation-of-state insensitive if one constructs a neutron star sequence by fixing $B$ and $P$ for a purely poloidal or toroidal configuration. The effect of the magnetic field on the quadrupole moment, however, may be enhanced~\cite{1975Ap&SS..38....3J,1977PhRvD..16..275E} for a purely poloidal or toroidal configuration if one takes proton superconductivity in the outer core region into account.

One of the problems in such a simple poloidal or toroidal configuration is that it is dynamically unstable on an Alfv\'en timescale~\cite{1973MNRAS.163...77M,Ciolfi:2011xa,Kiuchi:2011yt,Lasky:2011un,Ciolfi:2012en,Lasky:2012ju}. Therefore, one needs to consider a more realistic and stable configuration, where both poloidal and toroidal components are present. In order to alleviate this problem, Haskell \et~\cite{Haskell:2013vha} considered a twisted-torus configuration by treating the magnetic field as a small perturbation to the non-rotating and unmagnetized background spacetime~\cite{2009MNRAS.397..913C,Ciolfi:2010td,Ciolfi:2013dta}. To construct a stellar solution with such a field configuration, one first needs to solve the Grad-Shafranov equation in the background spacetime for the azimuthal component of the vector potential $A_\phi$ with a regularity condition at the stellar center. Such a solution is then matched to an analytic, vacuum exterior solution, which is taken to be dipolar, namely $A_\phi \propto \sin^2 \theta$. One then solves the perturbed Einstein equations sourced by this vector potential, keeping terms up to quadratic order in $(B \, P)$. Haskell \et~\cite{Haskell:2013vha} found that the I-Q relation becomes sensitive to the equation of state with such a twisted-torus configuration if $P \gtrsim 10$s \emph{and} $B \gtrsim 10^{12}$G. These conditions are probably satisfied by slowly-rotating magnetars and isolated neutron stars, accreting pulsars and pulsars in high-mass X-ray binaries, but are typically not satisfied by millisecond or rotation-powered pulsars and pulsars in low-mass X-ray binaries~\cite{Harding:2013ij,Ho:2013rla}. The relation is also sensitive to magnetic field parameters such as the poloidal-to-toroidal field ratio. 

%%%%%%%%%%%%%%%%%%%%%%%%%%%%%%%%%%%%%%%%%%%%%%%%%%%%%%
\subsection{Anisotropic Pressure}
\label{sec:anisotropy}

Does the universality remain when we consider pressure anisotropy, i.e.~situations in which the radial pressure differs from the tangential one? (see Herrera and Santos~\cite{1997PhR...286...53H} for a review of anisotropy). Pressure anisotropy may be present in the stellar solid or superfluid cores of neutron stars~\cite{1990sse..book.....K,1996csnp.book.....G,Heiselberg:1999mq} and may arise due to strong magnetic fields~\cite{Yazadjiev:2011ks,Bocquet:1995je,Konno:1999zv,2001ApJ...554..322C,Ioka:2003nh,Ciolfi:2010td,Frieben:2012dz,Ciolfi:2013dta,2014MNRAS.439.3541P,2015MNRAS.447.3278B}, relativistic nuclear interactions~\cite{1972ARA&A..10..427R,1974ARA&A..12..167C}, phase transitions~\cite{Carter:1998rn}, pion condensation~\cite{Sawyer:1972cq} or crystallization of the core~\cite{Nelmes:2012uf}. Moreover, simple two-fluid models for normal and superfluid components can be well approximated by a single anisotropic fluid model~\cite{1980PhRvD..22..807L,1997PhR...286...53H}. Having said this, one may expect isotropy to be eventually restored inside a neutron star or a quark star due to its strong internal gravity, and thus, it is currently unclear whether compact stars have large anisotropic pressure. Here, we take an agnostic view and review how pressure anisotropy affects the degree of universality in the I-Love-Q and approximate no-hair relations.

One can achieve this goal by constructing tidally-deformed or slowly rotating neutron stars and quark stars with anisotropic pressure. Slowly rotating, anisotropic neutron stars to linear order in spin were constructed in Bayin~\cite{Bayin:1982vw} and Silva \et~\cite{Silva:2014fca} by extending the Hartle-Thorne formalism to anisotropic stars. Yagi and Yunes~\cite{Yagi:2015hda} extended these analyses to third order in spin and to tidally-deformed anisotropic stars, which allowed the authors to extract the tidal deformability and multipole moments up to octupole order. One way to construct anisotropic stars is to use the matter stress-energy tensor~\cite{Doneva:2012rd,Silva:2014fca}:
\be
T_{\mu \nu} = \rho u_{\mu} u_{\nu} + p k_\mu k_\nu + q_t(g_{\mu \nu}+u_\mu u_\nu-k_\mu k_\nu)\,,
\ee
where $p$ and $q_t$ are the radial and tangential pressures respectively, while $k^\mu$ is a unit radial vector that is spacelike $(k_\mu k^\mu = 1)$ and orthogonal to $u^\mu$ $(k_\mu u^\mu = 0)$.  

The choice of how $p$ differs from $q_t$ defines the anisotropy model. Yagi and Yunes~\cite{Yagi:2015hda} mainly considered the one proposed by Horvat \et~\cite{Horvat:2010xf} (H model), in which the difference between radial and tangential pressures in a spherically-symmetric background is given by~\cite{Horvat:2010xf,Doneva:2012rd,Silva:2014fca} 
\be
\sigma_0 \equiv p -q_t =  \lambda_\HH p \left( 1 - e^{-\lambda(r)} \right)\,, 
\ee
where $e^{\lambda(r)}$ is the $(r,r)$-component of the metric in the background spacetime. This model is constructed such that the effect of anisotropy in the hydrostatic equilibrium equation vanishes in the non-relativistic limit ($p \ll \rho$), and $\sigma_0$ vanishes at the stellar surface and at the stellar center; the latter is required to ensure the absence of singularities in the interior mass distribution~\cite{1974ApJ...188..657B}. The quantity $\lambda_\HH$ in the above equation controls the amount of anisotropy. The isotropic case is recovered when $\lambda_\HH=0$, and $\lambda_{\HH}$ can in principle be of order unity when anisotropy is produced by pion condensation~\cite{Sawyer:1972cq}. Alternatively, a Skyrme crystallization in the core predicts $-2 \leq \lambda_\HH \leq 0$ for a neutron star with mass $M \gtrsim 1.5 M_\odot$. Following Doneva and Yazadjiev~\cite{Doneva:2012rd}, Silva \et~\cite{Silva:2014fca} and Yagi and Yunes~\cite{Yagi:2015hda}, we focus on $-2 \leq \lambda_\HH \leq 2$. Higher order spin corrections to $\sigma_0$ are determined self-consistently by solving the perturbed Einstein equations order by order in the small-rotation expansion~\cite{Yagi:2015hda}.

\begin{figure*}[htb]
\begin{center}
\includegraphics[width=8.cm,clip=true]{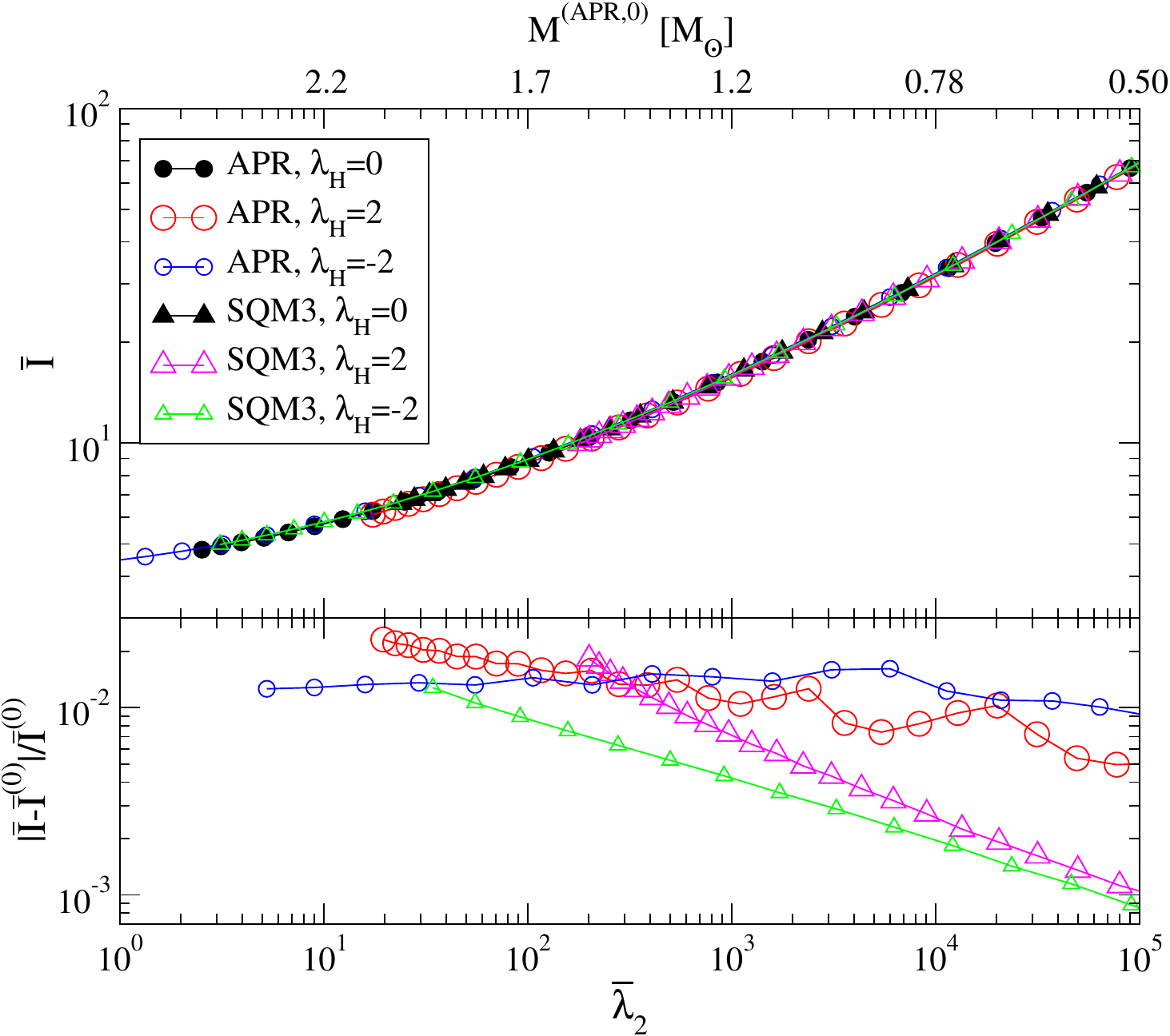}  
\includegraphics[width=8.3cm,clip=true]{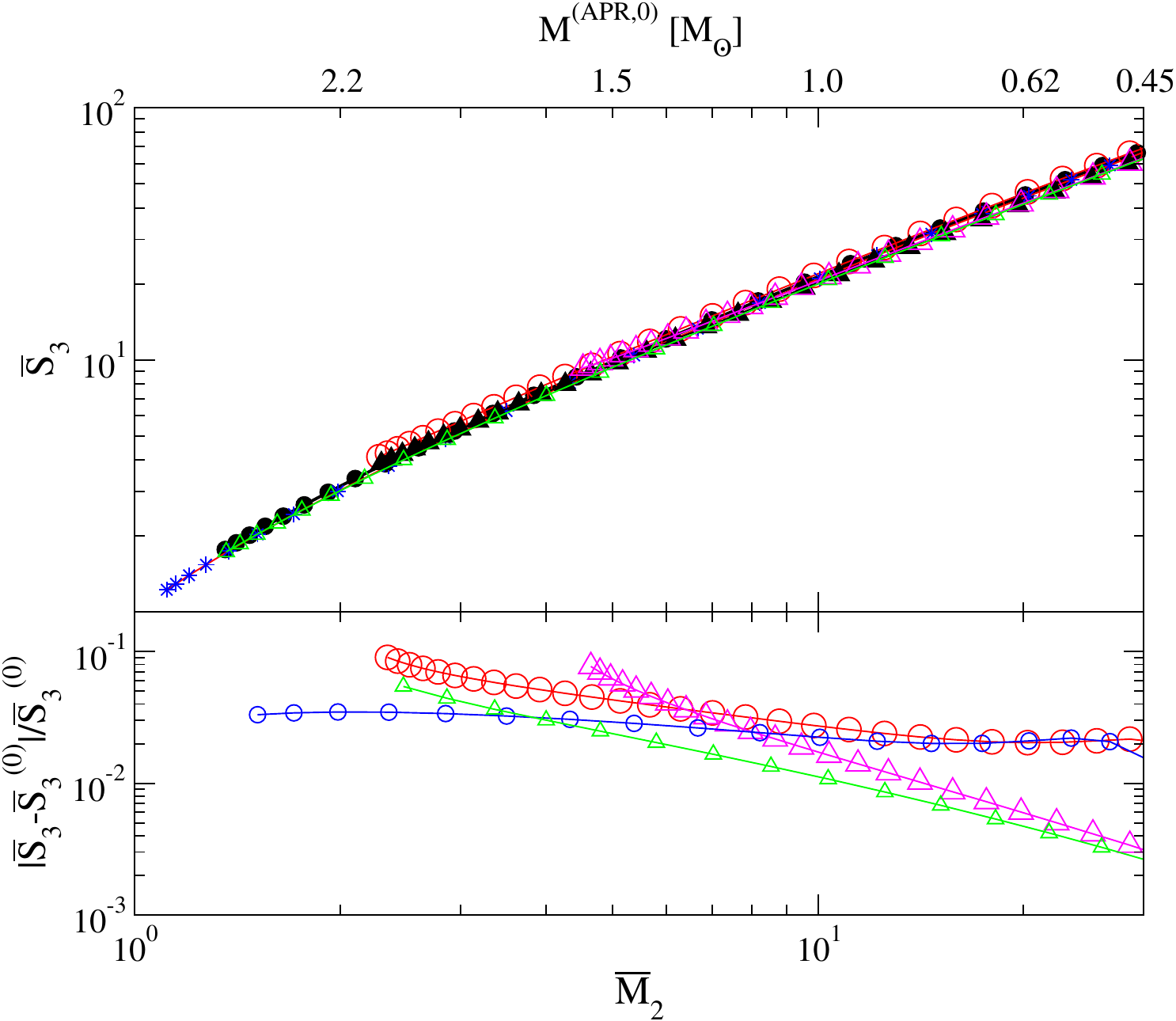}  
\caption{\label{fig:anisotropy} (Top) The approximately universal I-Love (left) and $\bar S_3$--$\bar M_2$ (right) relations for neutron stars (circle) and quark stars (triangle) with various anisotropy parameter $\lambda_\HH$. Top axes show the mass for isotropic neutron stars with the APR equation of state. 
(Bottom) Fractional difference from the isotropic relations. Observe that the anisotropy variation is comparable to the equation-of-state variation in Figs.~\ref{fig:I-Love-Q} and~\ref{fig:no-hair}.
 This figure is taken and edited from Yagi and Yunes~\cite{Yagi:2015hda}. 
}
\end{center}
\end{figure*}

The top panels of Fig.~\ref{fig:anisotropy} present the I-Love and $\bar S_3$--$\bar M_2$ relations for neutron stars (with an APR equation of state) and quark stars (with an SQM3 equation of state) for various values of $\lambda_\HH$. The bottom panels show the fractional difference of each anisotropic relation from the corresponding isotropic one. Observe that the difference from the isotropic case decreases as one decreases the mass or the compactness (moving to the right in each panel). This is because the effect of anisotropy vanishes from the hydrostatic equilibrium equation in the Newtonian limit. 

{\renewcommand{\arraystretch}{1.2}
\begin{table}[h]
\begin{center}
\begin{tabular}{ccccc}
\hline
\hline
\noalign{\smallskip}
Maximal Variability & I-Q &  I-Love &  Q-Love &  $\bar S_3$--$\bar M_2$ \\
\hline
with $\lambda_{\HH}$ (fixed APR equation of state) & 7 \% & 2 \% & 8 \%& 9 \% \\
with equation of state (fixed $\lambda_{\HH} = 2$) & 5 \% & 2 \% & 5 \% & 8 \%  \\
with equation of state (fixed $\lambda_{\HH} = 0$) & 2 \% & 0.7 \% & 2 \% & 5 \% \\
\noalign{\smallskip}
\hline
\hline
\end{tabular}
\end{center}
\caption{Maximum effect of anisotropy for fixed APR equation of state (first row), effect of equation-of-state variation for fixed, maximal anisotropy of $\lambda_{\HH} = 2$ (second row) and for isotropic pressure (third row) on I-Love-Q and approximate no-hair relations. Observe that anisotropy increases the amount of variation by a factor of 2--4 relative to the isotropic case. This table is taken from Yagi and Yunes~\cite{Yagi:2015hda}.
}
\label{table-summary}
\end{table}
}

Table~\ref{table-summary} compares the maximum amount of anisotropy and equation-of-state variation in different approximately universal relations. The first row shows the maximum variations of the relations with respect to anisotropy for a fixed (APR) equation of state. Such maximum variations for the I-Love and $\bar S_3$--$\bar M_2$ relations can be extracted from the bottom panels of Fig.~\ref{fig:anisotropy}. The second (third) row presents the maximum variation of the relations with respect to the equation of state for a fixed $\lambda_\HH = 2$ ($\lambda_\HH = 0$) anisotropy parameter\footnote{Maximum equation-of-state variations in the third row of Table~\ref{table-summary} are slightly larger than those extracted from Figs.~\ref{fig:I-Love-Q} and~\ref{fig:no-hair}, because the former include stars with masses less than $1M_\odot$.}. Observe that equation-of-state variability (in the first and second rows) is comparable and always less than 10\%, but larger than the variability for isotropic stars by a factor of 1.5--4. This result shows that the relations are approximately universal to variations in the equation of state and anisotropy to 10\% at most, but to a lesser degree than in the isotropic case.

How robust are these results to other choices of anisotropy models? To address this question, Yagi and Yunes~\cite{Yagi:2015hda} further studied the Bowers and Liang anisotropy model~\cite{1974ApJ...188..657B} (BL model):
\be
\sigma_0 =  \frac{\lambda_\BL}{3} (\rho + 3p) (\rho + p) e^{\lambda(r)} r^2\,,
\ee
where $\lambda_\BL$ is a constant that controls the amount of anisotropy, with $\lambda_\BL = 0$ reducing to the isotropic case. This model was constructed such that the modified Tolman-Oppenheimer-Volkoff equation for incompressible anisotropic stars can be solved analytically. However, the effect of anisotropy does not vanish in the hydrostatic equilibrium equation in the Newtonian limit. Such a feature seems unphysical if anisotropy originates from e.g.~the strain of nuclear matter at the core of compact stars. Notice also that $\sigma_0$ does not vanish at the surface if the stellar density is discontinuous there, like for constant density stars or quark stars. Because of these peculiarities, Yagi and Yunes~\cite{Yagi:2015hda} considered mainly the H model, using the BL model only as an auxiliary example to study how the universality depends on the models. 

Yagi and Yunes~\cite{Yagi:2015hda} found that the I-Love-Q and no-hair like relations are rather insensitive to the anisotropy model when considering neutron stars, but not when studying quark stars. In particular, the I-Love-Q relations for BL anisotropic quark stars can have large deviations from the relations for H anisotropic stars in the non-relativistic limit due to the pathologies of the BL model described above. On the other hand, the $\bar S_3$--$\bar M_2$ relation is similar in the two anisotropy models even for quark stars in the Newtonian limit. These results can be checked analytically by repeating the three-hair calculation in the Newtonian limit presented in Sec.~\ref{sec:no-hair-Newton} in the BL model. Doing so, one first finds that the Lane-Emden equation is modified to
\be
\label{eq:LE}
\frac{1}{\xi^2} \frac{d}{d\xi} \left[ \xi^2 \left( \frac{d\vartheta_\LE}{d\xi}  + \frac{\lambda_\BL}{6 \pi}  \xi \vartheta_\LE^n \right) \right]  = - \vartheta_\LE^n\,,
\ee
with the second term in square brackets on the left-hand side representing the anisotropic correction. One can solve such an equation analytically for incompressible stars with an $n=0$ polytropic index to find
\be
\vartheta_\LE^{(n=0)} (\xi) = 1 - \frac{\xi^2}{6} \left( 1 + \frac{\lambda_\BL}{2 \pi} \right)\,, \qquad \xi_1^{(n=0)} =\frac{2 \sqrt{3 \pi}}{\sqrt{2\pi + \lambda_\BL}}\,.
\ee
Substituting this solution in Eq.~\eqref{eq:Bbar-def}, one can show that $\bar B_{0,\ell}$ does not acquire any anisotropy corrections. Namely, the three-hair relations for anisotropic incompressible stars are the same as the isotropic ones. In order to see the anisotropy dependence with $n \neq 0$, one can further study the perturbed three-hair relations around $n=0$. Doing so, one arrives at Eq.~\eqref{eq:Bbar-analytic} even for anisotropic stars, which means that the anisotropy correction enters at $\mathcal{O}(n^2)$. Such analytic calculations mathematically explain why the $\bar S_3$--$\bar M_2$ relation for anisotropic Newtonian polytropes is similar to that for isotropic stars.

%%%%%%%%%%%%%%%%%%%%%%%%%%%%%%%%%%%%%%%%%%%%%%%%%%%%%%
\subsection{Non-barotropic Equations of State}
\label{sec:non-barotropic}

Can non-barotropic equations of state affect the degree of universality of the I-Love-Q relations~\cite{Martinon:2014uua}? So far, we have focused on barotropic equations of state, i.e.~those in which the stellar pressure is given purely in terms of the stellar density. This choice is suitable to model cold ($T \lesssim 10^9$K) and compact (relativistic) stars. However, the equation of state for hot ($T \gtrsim 10^{11}$K) proto-neutron stars is known to be non-barotropic. In particular, roughly 200--500ms after the bounce during gravitational collapse, a proto-neutron star enters a ``quasi-stationary'' phase, in which the stellar evolution can be modeled as a sequence of equilibrium configurations~\cite{Pons:1998mm,Pons:2000xf,Pons:2001ar,Fischer:2009af}. During this phase, the stellar radius contracts from 30--40km to 10--15km. Initially,  proto-neutron stars have large entropy gradients that become smoothed out by neutrino emission~\cite{Pons:1998mm}. During the quasi-stationary phase, dynamical and secular instabilities are suppressed due to the high temperature and slow rotation relative to the mass-shedding limit~\cite{Martinon:2014uua}. Non-perturbative equations of state for quark matter may also be non-barotropic in the low temperature regime, as suggested by Canfora \et~\cite{Canfora:2016xnc} based on lattice calculations~\cite{Bhagwat:2003vw,Parappilly:2005ei}.

The first study of the universal relations in proto-neutron stars was that of Martinon \et~\cite{Martinon:2014uua}. This reference extended the Hartle-Thorne formalism and constructed slowly-rotating or tidally-deformed proto-neutron star solutions in the quasi-stationary phase with non-barotropic equations of state. Such equations of state require pressure to be determined not only from the density, but also from the entropy and the lepton fraction, which need to be determined by solving transport equations~\cite{Camelio:2016fan}. Martinon \et~\cite{Martinon:2014uua} used GM3NQ non-barotropic profiles, which are a sequence of radial profiles of energy, pressure, lepton fraction and entropy~\cite{Pons:1998mm,Pons:2000xf,Pons:2001ar}. Such profiles are based on a non-barotropic equation of state derived using the mean-field approach with finite temperature, and are constructed by solving relativistic equations for neutrino transport and nucleon-meson coupling under a spherically symmetric background spacetime with a neutron-star baryonic mass of $1.6M_\odot$. 

Martinon \et~\cite{Martinon:2014uua} found that the I-Love-Q relations at the initial stage of the quasi-stationary phase of proto-neutron stars deviate from those for barotropic neutron stars by 20\% at most. However, just 2 seconds after the bounce, the deviations reduce to only $\sim 2\%$, showing how fast the universal relations approach the barotropic result during the quasi-stationary phase, which lasts about 1 minute. The authors also showed that the amount of deviation in these relations relative to the barotropic case is correlated to the magnitude of the radial gradient in the stellar entropy. Since Martinon \et~\cite{Martinon:2014uua} only studied a specific equation of state, it is currently unclear how the I-Love-Q relations for proto-neutron stars depend on different choices of non-barotropic equations of state.

%%%%%%%%%%%%%%%%%%%%%%%%%%%%%%%%%%%%%%%%%%%%%%%%%%%%%%
\subsection{Dynamical Tides}
\label{sec:dynamical-tides}

The final extension that we review in this section is the effect of dynamical tides on the I-Love relation. Until now, we have assumed that the external tidal field and the tidally-induced quadrupole moment are stationary, and hence, the tidal deformability is also time-independent. In a compact binary system, however, this assumption is valid only when the timescale of the stellar tidal deformation is much smaller than the orbital timescale, and thus, it becomes less and less valid as the binary inspirals.

The first study to consider the universal relations in dynamical scenarios was that of Maselli \et~\cite{Maselli:2013mva}. The authors applied the so-called post-Newtonian (PN) Affine approach~\cite{Maselli:2012zq,Ferrari:2011as} to compute the time-dependent tidal field and the induced quadrupole moment. As the name suggests, such an approach combines the PN and affine descriptions of a binary system. The former derives an approximate metric for a two-body system and the orbital evolution of such a system, assuming that the orbital velocity is much smaller than the speed of light. The latter treats the neutron stars as deformable ellipsoids, whose configuration is determined from the balance between self-gravity, pressure and the tidal field of the companion. The deformed neutron star is characterized by the three principal axes of the ellipsoid and two angles that define the orientation of the principal frame. These five dynamical variables are determined by solving a set of evolution equations together with the PN equations of motion. One can then derive the stellar tidal deformability in terms of the orbital frequency $f_\orb$ of the binary, with the zero frequency limit corresponding to the stationary case. One can also derive dynamical corrections to the moment of inertia using the PN Affine approach.

Maselli \et~\cite{Maselli:2013mva} found that the I-Love relation in a dynamical situation deviates more and more from the stationary result as the orbital frequency increases. For example, the relation at $f_\orb = 875$Hz differs from that at $f_\orb=0$Hz by $\sim 20\%$. However, for a given fixed $f_\orb$, the relation remains equation-of-state insensitive to $\lesssim 2\%$, with the equation-of-state variation increasing as $f_\orb$ increases. Maselli \et~\cite{Maselli:2013mva} derived a frequency-independent fit that can capture the relation for any $f_\orb$ to better than $\sim 5\%$. 

%%%%%%%%%%%%%%%%%%%%%%%%%%%%%%%%%%%%%%%%%%%%%%%%%%%%%%%%
\section{Connection to Other Universal Relations}
\label{sec4:Connections}

In this section, we review other approximately universal relations, different from the I-Love-Q and no-hair like ones, for neutron stars and quark stars. Section~\ref{sec:multipole-Love} begins by defining various tidal deformability parameters for a single star perturbed by some external environment, and then describing their inter-relations. Section~\ref{sec:binary-Love} refines this discussion by focusing on perturbations produced by a companion star in a binary system, and discussing the inter-relations among the deformabilities of the primary and the secondary star. Section~\ref{sec:oscillation} explains the connection between the I-Love-Q relations and the universal relations among stellar oscillation frequencies. Section~\ref{sec:I-Love-Q-C} concludes by reviewing the relation between the I-Love-Q trio and compactness, showing that the equation-of-state variation when using the compactness explicitly is much larger than that of the I-Love-Q relations. Finally, Sec.~\ref{sec:Darwin-Radau} reviews the Darwin-Radau relation (a relation between a different dimensionless version of I and Q) and compares it to the I-Q relation derived from the Newtonian 3-hair analysis in Sec.~\ref{sec:no-hair-Newton}.
 
%%%%%%%%%%%%%%%%%%%%%%%%%%%%%%%%%%%%%%%%%%%%%%%%%%%%%%
\subsection{Multipole Love Relations}
\label{sec:multipole-Love}

At the end of Sec.~\ref{sec:Approxs} and in Sec.~\ref{sec:I-Love-Q} we introduced the concept of tidal deformability by defining it as the response of a star (of a certain multipole order) to an external tidal perturbation. Back then, we concentrated on the $\ell=2$, electric-type tidal deformability $\lambda_{2}$, which is simply the (linear) quadrupolar response of a star to an even-parity perturbation. But of course, a small, external perturbation generically induces a linear response that can only be exactly recovered by summing up an infinite number of multipoles of both even (electric) and odd (magnetic) parity. In this section, we review the approximately equation-of-state independent inter-relations among the $\ell$th electric-type $(\lambda_\ell)$ and magnetic-type  $(\sigma_\ell)$ tidal deformabilities, as well as among the $\ell$th shape Love number $(h_\ell)$.

These $\lambda_{\ell}$ and $\sigma_{\ell}$ deformabilities are mathematically defined by~\cite{Damour:2009vw}
\be
\label{eq:lambda-sigma}
M_L = \lambda_\ell \, G_L\,, \quad S_L = \sigma_\ell \, H_L\,.
\ee
The quantities $M_L$ and $S_L$ are tidally-induced mass and current multipole moments, i.e.~the multipolar perturbations to the star's gravitational field or metric tensor. The quantities $G_L$ and $H_L$ are the gravito-electric and gravito-magnetic relativistic tidal moments, i.e.~the electric and magnetic parts of the external perturbation. Physically, one can think of $\lambda_\ell$ and $\sigma_\ell$ as parameters that quantify the linear response of the $\ell$th multipole moment due to an external perturbation characterized by the $\ell$th tidal moment. 

Similarly, $h_\ell$ is mathematically defined via~\cite{Damour:2009vw,Landry:2014jka}
\be
\label{eq:shape-Love}
\frac{\delta R_\ell}{R} = h_\ell \frac{U_\ell (R)}{C}\,, 
\ee
where $U_\ell (R)$ is the $\ell$th multipole coefficient of the external disturbing potential (in a Legendre decomposition at the stellar surface), while $\delta R_\ell / R$ is the $\ell$th fractional deformation of the stellar surface. One can thus think of $h_\ell$ as a coefficient that quantifies the linear response of the shape of the stellar surface due to the perturbation of an external potential. The shape Love number $h_\ell$ reduces to the first apsidal constant in the Newtonian limit. 

%%%%%%%%%%%%%%%%%%%%%%%%%%%%%%%%%%%%%%%%%%%%%%%%%%%%%%
\subsubsection{No Rotation}
\label{sec:no-rot-multipole-love}

Now that the tidal deformabilities have been mathematically defined and physically interpreted, let us present the simplest version of the multipole Love relations, working only in the Newtonian limit of General Relativity and neglecting any internal motions of the fluid for non-rotating neutron stars and quark stars. We begin by presenting and reviewing the \emph{electric multipole Love relations} (or just multipole Love relations for short), i.e.~approximately equation-of-state independent relations between the $\lambda_{\ell}$ parameters of different $\ell$ number (for example, the relations between $\lambda_{2}$ and $\lambda_{3}$). We then continue by discussing the relations between any given $\lambda_{\ell}$ (with $\ell$ fixed) and either $I$ or $Q$, by combining the electric multipole Love relations with the I-Love-Q relations. We conclude by commenting on the relations among the magnetic-type tidal deformability $\sigma_\ell$ and the shape Love number $h_\ell$, and we discuss how the fluid internal motion affects the universal relations.

In order to understand the multipole Love relations, we must first discuss how to calculate $\lambda_\ell$. The metric tensor of a star that has been disturbed by even-parity perturbations takes the form~\cite{1967ApJ...149..591T,1991PhRvD..43.1768I,Lindblom:1997un}
\be
ds^2 = ds_0^2 - Y_{\ell m}(\theta,\phi) \left[ e^{\nu(r)} H_{0,\ell} (r) dt^2 + 2 H_{1,\ell} (r) dt dr + e^{\lambda (r)} H_{2,\ell} (r) dr^2 +r^2 K_{\ell} (r) (d \theta^2 + \sin^2 \theta d\phi^2) \right]\,, 
\ee
where $ds_0^2$ is the background spacetime of Eq.~\eqref{eq:metric-ansatz} with $\nu(r,\theta) \to \nu(r)$, $\lambda(r,\theta) \to \lambda(r)$, $K(r,\theta) \to 1$ and $\omega(r,\theta) \to 0$, while $H_{0,\ell}$, $H_{1,\ell}$, $H_{2,\ell}$ and $K_{\ell}$ are perturbation functions. After perturbing the pressure and density in a similar way and substituting all of this into the linearized Einstein equations and the stress-energy conservation equations, one finds that $H_{1,\ell}=0$ and $H_{0,\ell} = H_{2,\ell} \equiv H_\ell$. Eliminating further $K_\ell$ from the remaining equations, one finds a \emph{master equation} for $H_\ell$~\cite{Lindblom:1997un}: 
\ba
\label{eq:master-multipole-love}
& & \frac{d^2 H_\ell}{dr^2} + \left\{ \frac{2}{r} + e^{\lambda} \left[ \frac{1-e^{-\lambda}}{r} + 4 \pi r (p-\rho) \right] \right\} \frac{dH_\ell}{dr} \nn \\
& & + \left\{ e^\lambda \left[ - \frac{\ell (\ell +1)}{r^2} + 4 \pi (\rho + p) \frac{d\rho}{dp} + 4 \pi (5 \rho + 9 p) \right] - \left( \frac{d\nu}{dr} \right)^2 \right\} H_\ell   = 0\,,
\ea
with the pressure $p$ and density $\rho$ taking their background values here. This equation can be solved numerically in the stellar interior by imposing regularity at the center, namely $H_\ell \propto r^{\ell}$ as $r \to 0$, while in the stellar exterior the solution is $H_\ell^{\mathrm{ext}} = a_\ell^P \, \hat{P}_{\ell 2} (x) + a_\ell^Q \, \hat{Q}_{\ell 2} (x)$, where $x \equiv r/M-1$ and $\hat{P}_{\ell 2}$ and $\hat{Q}_{\ell 2}$ are the normalized associated Legendre functions of the first and second kind respectively. The ratio between the integration constants $a_\ell^P$ and $a_\ell^Q$ is determined by matching the interior and exterior solutions of $y_\ell \equiv (r/H_\ell) (dH_\ell/dr)$ at the surface. This quantity is related to the dimensionless, $\ell$th-order, electric-type, tidal deformabilities $\bar \lambda_\ell \equiv \lambda_\ell / M^{2\ell +1}$ via~\cite{Damour:2009vw} 
\be
\label{eq:bar-lambda-ell}
\bar \lambda_\ell = \frac{1}{(2\ell - 1)!!} \frac{a_\ell^Q}{a_\ell^P} = - \frac{1}{(2\ell - 1)!!} \frac{\hat{P}'_{\ell 2} (x_c) - C \; y_{\ell}(R) \hat{P}_{\ell 2} (x_c)}{\hat{Q}'_{\ell 2} (x_c) - C \; y_{\ell}(R) \hat{Q}_{\ell 2} (x_c)}\,,
\ee
where we have defined $x_c \equiv C^{-1}-1$ and we recall that $C$ is the stellar compactness. Observe that $\bar \lambda_\ell$ depends both on the exterior solution ($\hat{P}_{\ell 2}$, $ \hat{Q}_{\ell 2}$) and on the interior solution ($y_\ell$) at the surface.

\begin{figure*}[htb]
\begin{center}
\includegraphics[width=8.cm,clip=true]{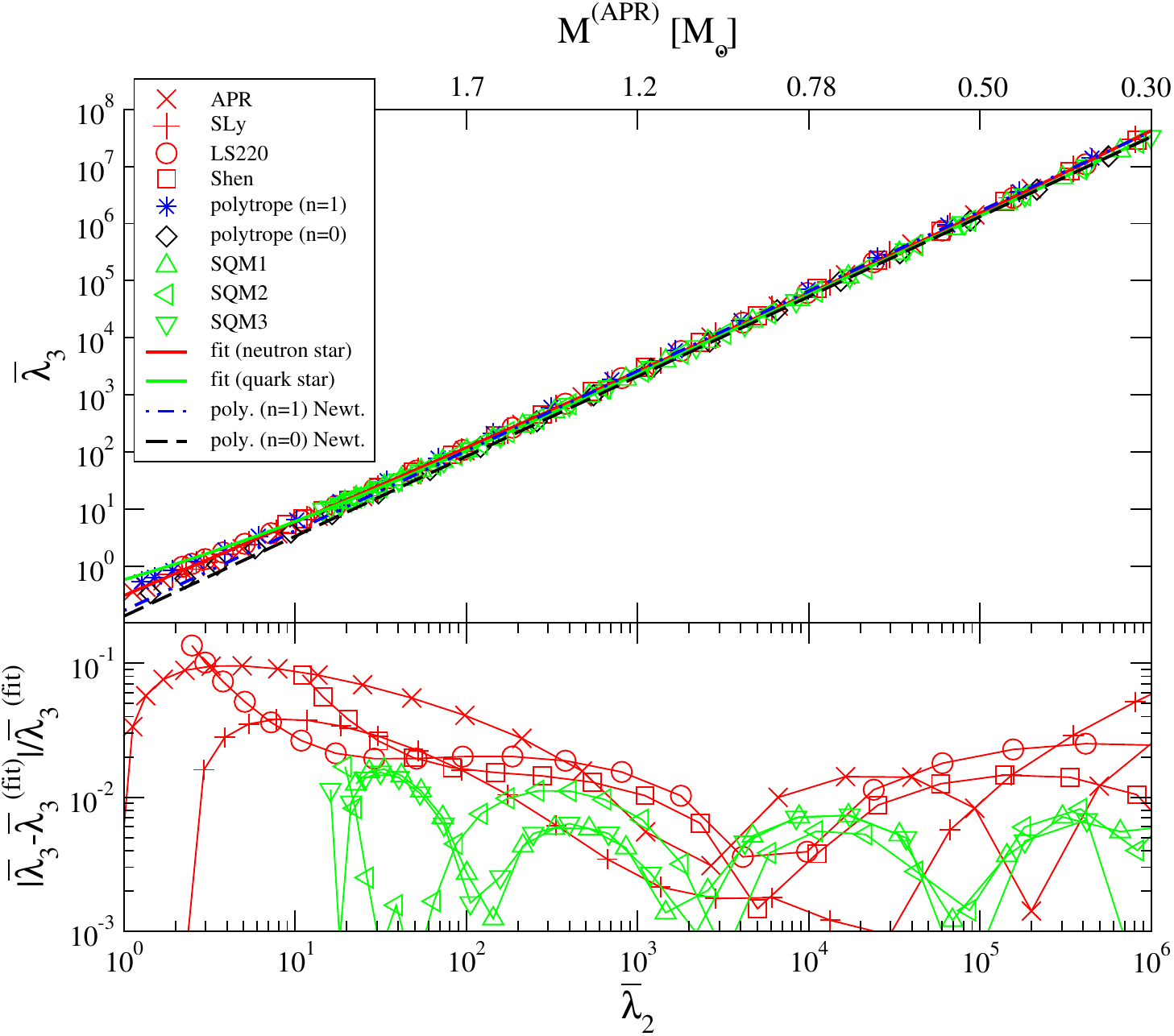}  
\includegraphics[width=8.cm,clip=true]{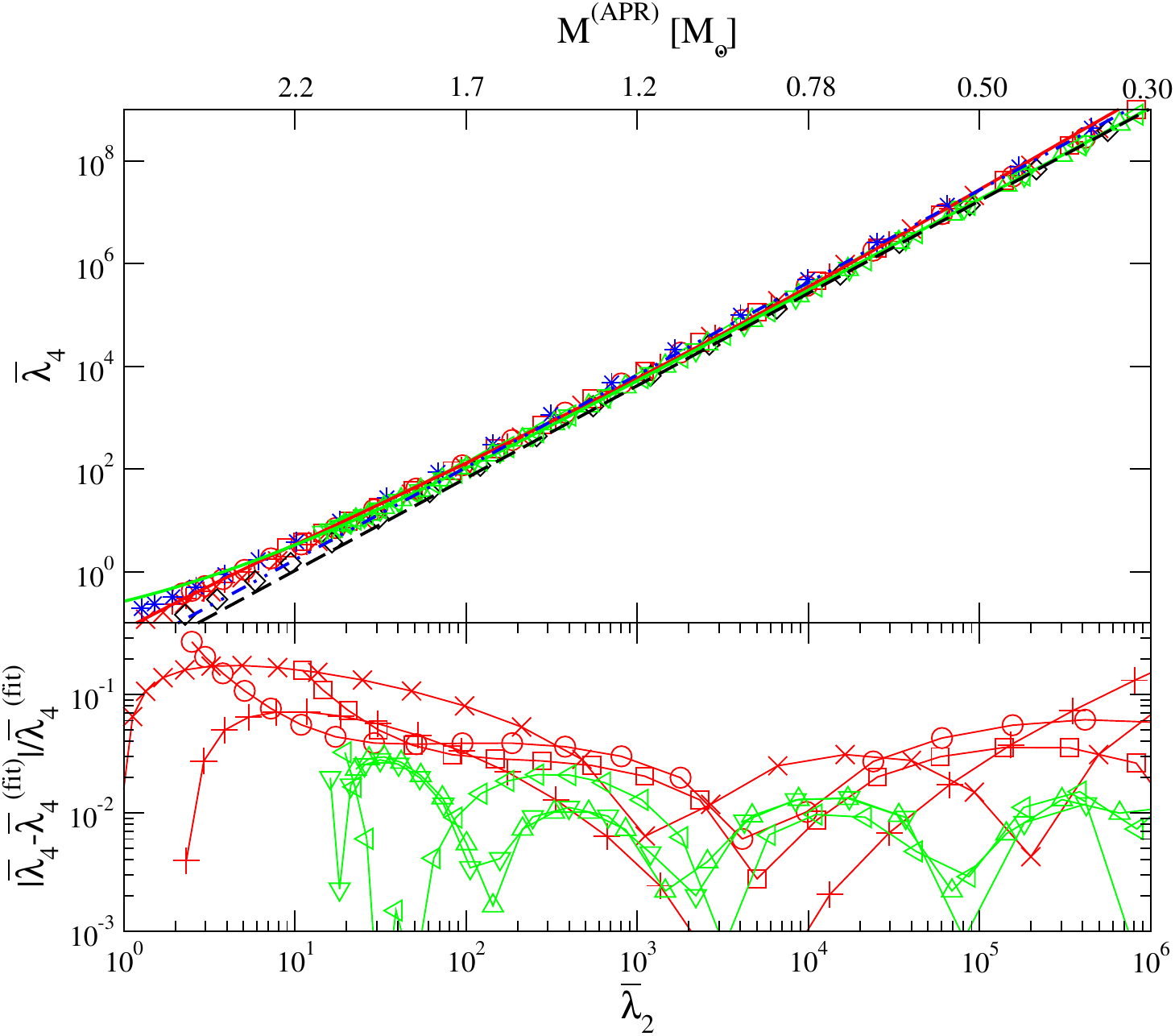}  
\caption{\label{fig:lambdabar34-lambdabar2} (Top) Universal $\bar{\lambda}_3$--$\bar{\lambda}_2$ (left) and $\bar{\lambda}_4$--$\bar{\lambda}_2$ (right) relations for neutron stars (red) and quark stars (green) with various realistic equations of state. We also present the relations with the $n=1$ (blue) and $n=0$ (black) polytropes and their Newtonian limit (dashed and dotted-dashed). Solid curves show the fit for each of the neutron star and quark star sequence. Top axes show the neutron star mass for the APR equation of state.
(Bottom) Fractional difference from the fit. Observe that the relations are universal to $\mathcal{O}(10\%)$.
This figure is taken and edited from Yagi~\cite{Yagi:2013sva}.
}
\end{center}
\end{figure*}

The top panels of Fig.~\ref{fig:lambdabar34-lambdabar2} show the multipole Love relations between $\bar \lambda_3$ and $\bar \lambda_2$ (left) and between $\bar \lambda_4$ and $\bar \lambda_2$ (right) for various equations of state appropriate to neutron stars (red) and quark stars (green). The figure also shows the relations computed with an $n=1$ (blue) and $n=0$ (black) polytropic equation of state. Observe that the relations are approximately equation-of-state independent, with the neutron star ones resembling the $n=1$ polytropic relation and the quark star ones resembling the $n=0$ relations, especially in the Newtonian limit. This behavior is expected given that quark star equations of state reduce to an $n=0$ polytropic one in the low pressure regime. The figure also includes fits to all the data, with the same form as in Eq.~\eqref{fit} but with different coefficients given in Yagi~\cite{Yagi:2013sva}. The bottom panels show the fractional difference between the numerical data and the fit. Observe that the relation is equation-of-state insensitive to $\mathcal{O}(10\%)$ ($\mathcal{O}(1\%)$) for the neutron star (quark star) sequence. 

Let us now attempt to understand these relations better through an analytical Newtonian study. In this limit, Eq.~\eqref{eq:master-multipole-love} reduces to~\cite{Yagi:2013sva}
\be
\label{eq:master-multipole-love-Newton}
\frac{d^2 H_\ell^\N}{dr^2} + \frac{2}{r} \frac{d H_\ell^\N}{dr} - \left( \frac{\ell (\ell + 1)}{r^2}  - 4 \pi \rho \frac{d\rho}{dp} \right) H_\ell^\N = 0\,,
\ee
where the superscript $N$ is to remind us that we are working in the Newtonian limit. Solving this equation in the exterior region, one finds that $\hat{P}_{\ell 2}$ and $\hat{Q}_{\ell 2}$ reduce to $(r/M)^{\ell}$ and $(r/M)^{-(\ell+1)}$ respectively. Thus, Eq.~\eqref{eq:bar-lambda-ell} reduces to
\be
\label{lambda-l-N}
\bar \lambda_\ell^\N = \frac{1}{(2 \ell -1)!!} \frac{\ell - y_\ell}{\ell + 1 - y_\ell} \frac{1}{C^{2 \ell + 1}}\,,
\ee
where we recall that $y_\ell$ is defined above Eq.~\eqref{eq:bar-lambda-ell}. In particular, for an $n=1$ polytrope, the solution to Eq.~\eqref{eq:master-multipole-love-Newton} is given in terms of Bessel functions $H_\ell^{\N,(n=1)} \propto \sqrt{R/r} \, J_{\ell+1/2}(\pi r/R)$~\cite{Hinderer:2007mb,Damour:2009vw}. For such polytropes, $\bar \lambda_\ell$ in the Newtonian limit is given by~\cite{Yagi:2013sva}
\be
\bar \lambda_2^\N{}^{, (n=1)} = \frac{15-\pi^2}{3 \pi^2} \frac{1}{C^5}\,, \quad
\bar \lambda_3^\N{}^{, (n=1)} = \frac{21-2 \pi^2}{9 \pi^2} \frac{1}{C^7}\,, \quad
\bar \lambda_4^\N{}^{, (n=1)} = -\frac{945 - 105 \pi^2 + \pi^4}{105 \pi^2 (\pi^2 -15)} \frac{1}{C^9}\,.
\ee
On the other hand, for an $n=0$ polytrope, $\bar \lambda_\ell$ in the Newtonian limit is given by~\cite{Damour:2009vw}
\be
\bar \lambda_\ell^\N{}^{, (n=0)} = \frac{3}{2 (\ell -1) (2 \ell -1)!!} \frac{1}{C^{2 \ell +1}}\,.
\ee
Combining these expressions, one can derive the multipole Love relations in the Newtonian limit as a function of the polytropic index $n$:
\begin{align}
\bar \lambda_3^\N &= C^{(n)}_{\bar \lambda_3 \bar \lambda_2} \left( \bar \lambda_2^{\N}\right)^{7/5}\,, \qquad
\bar \lambda_4^\N = C^{(n)}_{\bar \lambda_4 \bar \lambda_2} \left( \bar \lambda_2^{\N}\right)^{9/5}\,,
\end{align}
where $C^{(n)}_{\bar \lambda_3 \bar \lambda_2}$ and $C^{(n)}_{\bar \lambda_4 \bar \lambda_2}$ are constants that depend on $n$. 

These Newtonian multipole Love relations are shown in the top panels of Fig.~\ref{fig:lambdabar34-lambdabar2}. Observe that the numerical data agree with the analytic Newtonian relations in the large $\bar \lambda_2$ region, as expected. By taking the ratio between the $n=0$ and $n=1$ relations, one finds~\cite{Yagi:2013sva}
\ba
\label{eq:multipole-Love-Newton1}
\frac{C_{\bar \lambda_3 \bar \lambda_2}^{(n=0)}}{C_{\bar \lambda_3 \bar \lambda_2}^{(n=1)}} &=& \frac{2^{2/5} 3^{3/5} (15-\pi^2)^{7/5}}{10 \pi^{4/5} (21- 2 \pi^2 )} \approx 0.799\,, \\
\label{eq:multipole-Love-Newton2}
\frac{C_{\bar \lambda_4 \bar \lambda_2}^{(n=0)}}{C_{\bar \lambda_4 \bar \lambda_2}^{(n=1)}} &=& \frac{2^{4/5} (15-\pi^2 )^{14/5}}{3^{9/5} \pi^{8/5} (\pi^4 -105 \pi^2 +945)} \approx 0.616\,. 
\ea
This shows that the $\bar \lambda_3$--$\bar \lambda_2$  and the $\bar \lambda_4$--$\bar \lambda_2$ relations differ by 20\% and 40\% respectively when computed with an $n=0$ and an $n=1$ polytrope. Clearly, the multipole Love relations are not as equation-of-state insensitive as the I-Love-Q relations, whose equation-of-state variation in the Newtonian limit is only 0.2--0.8\% [see Sec.~\ref{sec:I-Love-Q-Newton}]. Note that the equation-of-state variation inferred from Eqs.~\eqref{eq:multipole-Love-Newton1} and~\eqref{eq:multipole-Love-Newton2} cannot be directly compared to that in the bottom panels of Fig.~\ref{fig:lambdabar34-lambdabar2}, as the former corresponds to the variation between the neutron star and quark star sequences, while the latter shows the variation \emph{within} each sequence.

The multipole Love relations can be combined with the I-Love-Q relations to obtain new, equation-of-state insensitive relations. For example, one can eliminate $\bar \lambda_2$ from these relations to find approximately universal relations between $\bar \lambda_{\ell \geq 3}$ and $\bar I$ or $\bar Q$, as first studied by Pani \et~\cite{Pani:2015nua}. This study revealed that the equation-of-state variation in the $\bar \lambda_3$--$\bar I$ and the $\bar \lambda_4$--$\bar I$ relations is at most $\mathcal{O}(10\%)$, which is much larger than the variation in the original I-Love relation. Clearly, the equation-of-state variation of the $\bar{\lambda}_{\ell \geq 3}$--$\bar{I}$ relation is dominated by that in the multipole Love relations. 

One can repeat the above analyses to derive relations between $\bar \lambda_\ell$ and the dimensionless, magnetic-type tidal deformabilities $\bar \sigma_\ell \equiv \sigma_\ell/M^{2\ell +1}$. Yagi~\cite{Yagi:2013sva} derived these relations by solving the master differential equation for odd-parity perturbations, given for example in Damour and Nagar~\cite{Damour:2009vw}. In particular, Yagi~\cite{Yagi:2013sva} studied the relation between $\bar \sigma_2$ and $\bar \lambda_2$ and found that the equation-of-state variation is similar to that in the $\bar \lambda_3$--$\bar \lambda_2$ relation. However, Pani \et~\cite{Pani:2015nua} pointed out later that the master equation in Damour and Nagar~\cite{Damour:2009vw} does not agree with that in Binnington and Poisson~\cite{Binnington:2009bb}, and the former may contain an error. Pani \et~\cite{Pani:2015nua} showed that the $\bar \sigma_2$--$\bar \lambda_2$ relation with $\bar \sigma_2$ derived from Damour and Nagar~\cite{Damour:2009vw} deviates from that with $\bar \sigma_2$ derived from Binnington and Poisson~\cite{Binnington:2009bb} in the relativistic regime, with the latter more universal (an equation-of-state variation of 3\% at most). Pani \et~\cite{Pani:2015nua} also studied the relation between $\bar \sigma_3$ and $\bar \lambda_3$ and between $\bar \sigma_4$ and $\bar \lambda_4$. In both cases, the equation-of-state variation is 0.7\% at most, and hence, the amount of universality in these cases is comparable to that in the I-Love-Q relations. The relations between $\bar \sigma_\ell$ and $\bar I$, on the other hand, have an equation-of-state variation of 5--20\%~\cite{Pani:2015nua}.

The above studies were carried out under a strict hydrostatic equilibrium condition, which requires the tidal fields to be stationary. Perhaps, a more realistic situation is to consider an \emph{irrotational} state, which takes into account internal motions of the fluid that arise from a time-dependent tidal environment due to the conservation of relativistic circulation within the fluid~\cite{Shapiro:1996up,Favata:2005da}. Landry and Poisson~\cite{Landry:2015cva} showed that the irrotational state does not affect $\lambda_\ell$, but $\sigma_\ell$ can become negative. Delsate~\cite{Delsate:2015wia} studied how the relations between $\bar \sigma_\ell$ and $\bar I$ are affected by taking such internal fluid motions into account, finding that the universality actually improves in the more realistic irrotational case. The author found that the maximum equation-of-state variation is reduced by a factor of two in the irrotational case relative to the hydrostatic equilibrium case.

Finally, one can also derive similar universal relations among dimensionless, shape tidal deformabilities $\bar \eta_\ell$ of different $\ell$ order. The latter are related to the $\ell$th shape Love number $h_\ell$ in Eq.~\eqref{eq:shape-Love} via $\bar \eta_\ell \equiv \{2/[(2\ell-1)!!]\} (h_\ell/C^{2\ell+1})$. Since the $\bar \eta_\ell$ quantities are electric-type tidal deformabilities, they are related to $\bar \lambda_\ell$ via~\cite{Yagi:2013sva,Landry:2014jka}
\be
\label{eq:eta-lambda}
\bar \eta_\ell = \left\{ \left[ 2 \bar \lambda_\ell \, \hat{Q}_{\ell 2} (x_c)  + \frac{2}{(2 \ell -1)!!} \hat{P}_{\ell 2} (x_c) \right]  [1+(\alpha_2 -2) C] + \left[ 2 \bar \lambda_\ell \, \hat{Q}'_{\ell,2}(x_c) + \frac{2}{(2 \ell -1)!!} \hat{P}'_{\ell,2}(x_c) \right] \alpha_1 \right\} \frac{1}{C^{\ell+1}}\,,
\ee
with
\be
\alpha_1 \equiv \frac{2 C}{(\ell -1) (\ell +2) }\,, 
\qquad
\alpha_2 \equiv \frac{1}{(\ell -1) (\ell +2)} \left[ \ell (\ell +1) + \frac{4 C^2}{1-2 C} - 2 (1-2C) \right]\,. 
\ee
In the Newtonian limit, Eq.~\eqref{eq:eta-lambda} simplifies to
\be
\bar \eta_\ell^\N = 2 \bar \lambda_\ell^\N + \frac{2}{(2 \ell -1)!!} \frac{1}{C^{2\ell +1}}\,.
\ee
Yagi~\cite{Yagi:2013sva} found that the relation between $\bar \eta_3$ and $\bar \eta_2$ is equation-of-state insensitive to 7--8\% for both neutron star and quark star sequences, which is similar to the degree of universality of the electric multipole Love relations.

%%%%%%%%%%%%%%%%%%%%%%%%%%%%%%%%%%%%%%%%%%%%%%%%%%%%%%
\subsubsection{Slow Rotation}

{\renewcommand{\arraystretch}{1.2}
\begin{table}
\begin{centering}
\begin{tabular}{ccc}
\hline
\hline
\noalign{\smallskip}
source & $\mathcal{O}(\chi^0)$ &$\mathcal{O}(\chi)$  \\
\hline
\noalign{\smallskip}
\multirow{2}{*}{$G_\ell$} & \multirow{2}{*}{$M_\ell$ $\left( \lambda_{\ell \ell}^{(+)} \right)$} & $S_{\ell-1}$  $\left( \sigma_{\ell-1 \ell}^{(-)} \right)$\\
\noalign{\smallskip}
 & & $S_{\ell+1}$  $\left( \sigma_{\ell+1 \ell}^{(-)} \right)$\\
\noalign{\smallskip}
\hline
\noalign{\smallskip}
\multirow{2}{*}{$H_\ell$} & \multirow{2}{*}{$S_\ell$ $\left( \sigma_{\ell \ell}^{(+)} \right)$} & $M_{\ell-1}$  $\left( \lambda_{\ell-1 \ell}^{(-)} \right)$\\
\noalign{\smallskip}
 & & $M_{\ell+1}$  $\left( \lambda_{\ell+1 \ell}^{(-)} \right)$\\
\noalign{\smallskip}
\hline
\hline
\end{tabular}
\end{centering}
\caption{Induced multipole moments at zeroth order (second column) and first order (third column) in spin, sourced by an external, axisymmetric tidal perturbation (first column). Corresponding non-vanishing tidal deformabilities are shown in brackets.}
\label{table:spin-Love}
\end{table}
}

Let us now review the universal relations among tidal deformabilities for \emph{slowly-rotating} neutron stars. To calculate the deformabilities in this case, one needs to consider tidal perturbations about slowly-rotating neutron star background solutions. Pani \et~\cite{Pani:2015nua} derived the spin corrections to the tidal deformabilities to linear order in spin for axisymmetric tidal perturbations following~\cite{Pani:2015hfa} (see also Landry and Poisson~\cite{Landry:2015zfa} for related work). The requirement of axisymmetry is satisfied when the tidal perturbation is additionally stationary. At zeroth order in spin, the $\ell$th electric-type (magnetic-type) tidal field only generates an  $\ell$th electric-type (magnetic-type) deformation to the stellar moments.  At linear order in spin, the $\ell$th electric-type (magnetic-type) tidal field sources an $\ell \pm 1$ magnetic-type (electric-type) deformation to the moments. Due to this mixing of parities, one needs to introduce new types of tidal deformabilities:
\ba
\label{eq:lambda-spin}
\lambda_{\ell \ell'}^{(+)} &\equiv& \frac{\partial M_\ell}{\partial G_{\ell'}}\,, \quad \lambda_{\ell \ell'}^{(-)} \equiv \frac{\partial M_\ell}{\partial H_{\ell'}}\,, \\
\label{eq:sigma-spin}
\sigma_{\ell \ell'}^{(+)} &\equiv& \frac{\partial S_\ell}{\partial H_{\ell'}}\,, \quad \sigma_{\ell \ell'}^{(-)} \equiv \frac{\partial S_\ell}{\partial G_{\ell'}}\,. 
\ea
Table~\ref{table:spin-Love} summarizes the induced multipole moments at zeroth and first order in spin for a given external tidal moment, together with the corresponding tidal deformabilities. At zeroth order in spin, the only non-vanishing tidal deformabilities are $\lambda_{\ell \ell}^{(+)}$ and $\sigma_{\ell \ell}^{(+)}$, which reduce to the $\lambda_\ell$ and $\sigma_\ell$ deformabilities of Eq.~\eqref{eq:lambda-sigma}. At linear order in spin, the only non-vanishing tidal deformabilities with axisymmetric perturbations are $\lambda_{\ell\pm1 \ell}^{(-)}$ and $\sigma_{\ell\pm1 \ell}^{(-)}$. Furthermore, equatorial symmetry forces the electric-type (magnetic-type) induced moments to have even (odd) values of $\ell$ in Eqs.~\eqref{eq:lambda-spin} and~\eqref{eq:sigma-spin}. For example, when $\ell,\ell' \leq 4$ with equatorially symmetric perturbations, the only non-vanishing deformabilities are $\lambda_{23}^{(-)}$, $\lambda_{43}^{(-)}$, $\sigma_{32}^{(-)}$ and $\sigma_{34}^{(-)}$. In this example, $\lambda_{23}^{(-)}$ is the linear response of the induced mass quadrupole moment due to the external tidal current octupole moment, and similarly for the other deformabilities.

\begin{figure*}[htb]
\begin{center}
\includegraphics[width=8.cm,clip=true]{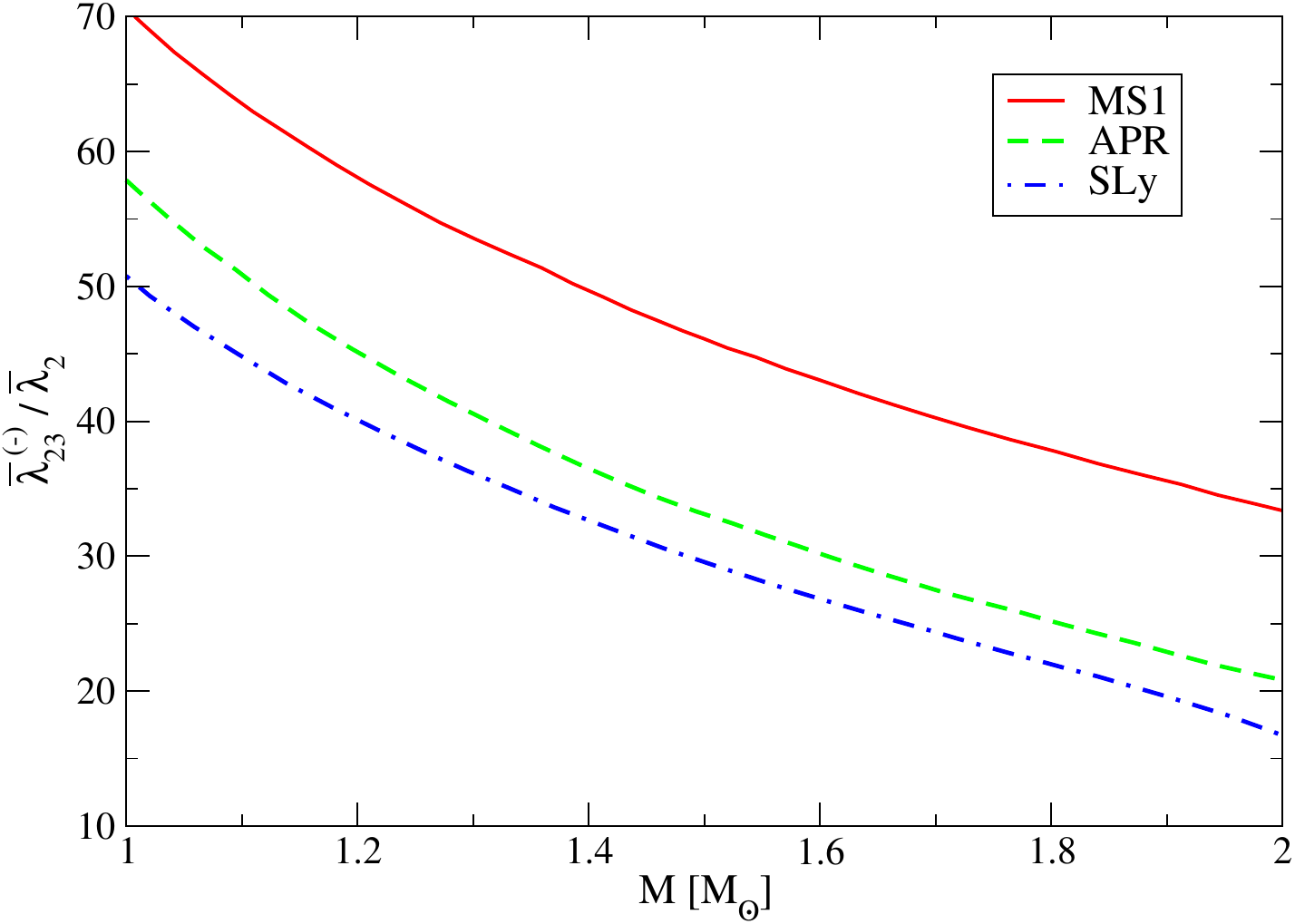}  
\includegraphics[width=8.cm,clip=true]{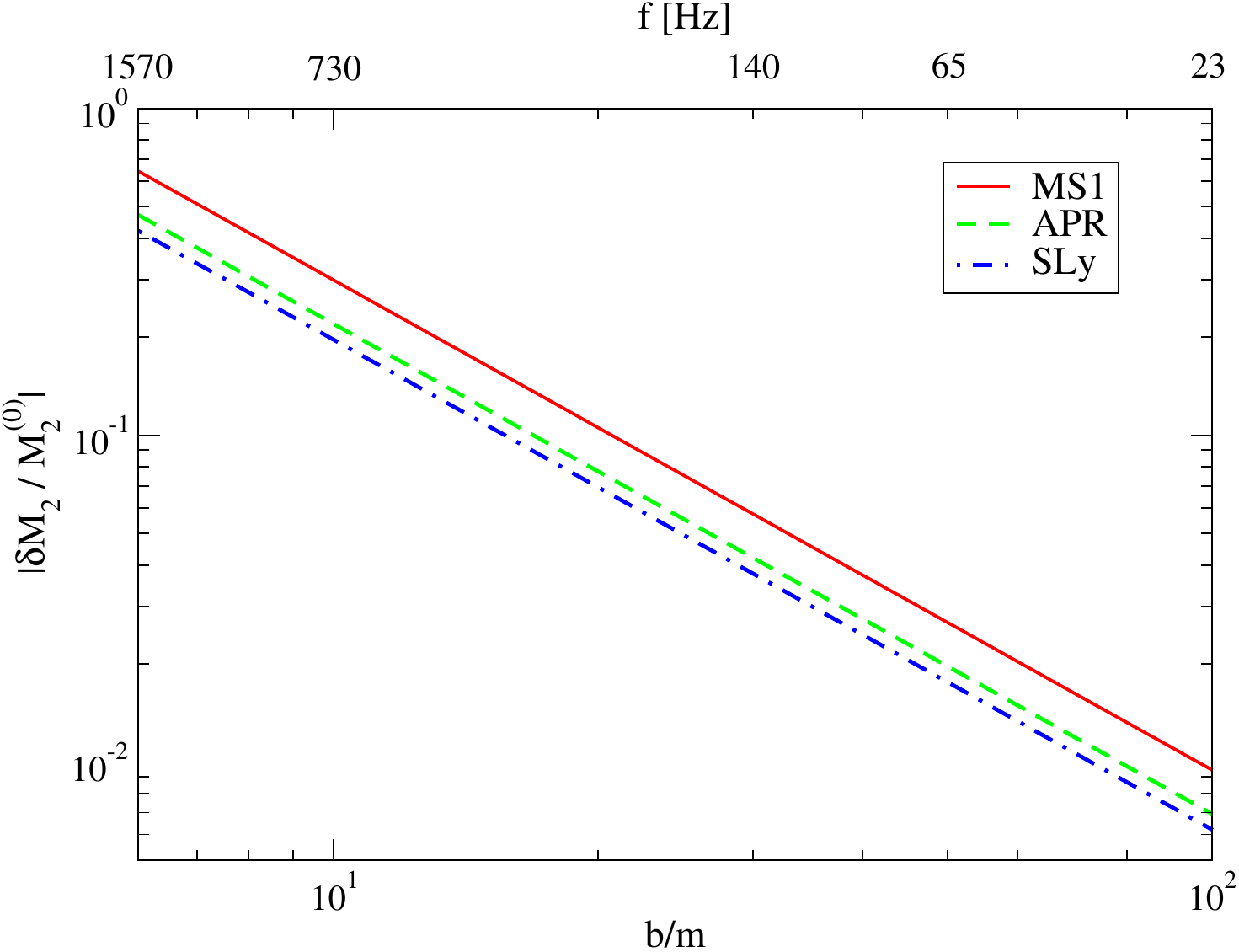}  
\caption{\label{fig:spin-Love} 
(Left) The ratio between $\bar \lambda_{23}^{(-)}$ at linear order in spin and $\bar \lambda_2$ at zeroth order in spin as a function of the neutron star mass for three representative equations of state. 
(Right) The absolute ratio between the spin correction to the induced quadrupole moment $\delta M_2$ and the induced moment at zeroth order in spin $M_2^{(0)}$ as a function of the binary separation normalized by the total mass. $b=6m$ corresponds to the location of the innermost stable circular orbit of a binary. We assumed an equal mass binary of $M=1.4M_\odot$ and the spin of the primary neutron star as $\chi = 0.05$. Observe that $\delta M_2$ is always smaller than $M_2^{(0)}$. The top axis shows the corresponding gravitational wave frequency calculated from the Newtonian relation $m/b = (\pi m f)^{2/3}$.
}
\end{center}
\end{figure*}

Let us now estimate the impact of the spin-corrected tidal deformabilities on the induced multipole moments relative to the non-rotating contribution. The most relevant spin-corrected tidal deformability to gravitational wave observations is $\lambda_{23}^{(-)}$ as others induce higher order multipole moments~\cite{Pani:2015hfa}. Thus, we will focus on the contribution of $\lambda_{23}^{(-)}$ to the induced quadrupole moment. To give a concrete example, let us assume the tidal field is created by a companion neutron star, whose mass is the same as the primary one, at a separation $b$ in a binary. The ratio between the linear spin correction to the induced quadrupole moment $\delta M_2$ and the induced moment for a non-rotating configuration $M_2^{(0)}$ is given by~\cite{Pani:2015hfa} 
\be
\label{eq:ratio-quadrupole}
\frac{\delta M_2}{M_2^{(0)}} = - \frac{9}{2} \sqrt{\frac{5}{7}}\, \frac{\bar \lambda_{23}^{(-)}}{\bar \lambda_2} \chi \left( \frac{m}{b} \right)^{3/2}\,,
\ee
where $m=2M$ is the total mass of a binary and $\bar \lambda_{23}^{(-)} \equiv \lambda_{23}^{(-)}/(M^{6} \chi)$ with $\bar \lambda_{23}^{(-)}/\bar \lambda_2$ shown in the left panel of Fig.~\ref{fig:spin-Love}. The factor of $b^{-3/2}$ arises from the ratio of $H_3 \propto b^{-9/2}$ to $G_2 \propto b^{-3}$, which sources $\delta M_2$ and $M_2^{(0)}$ respectively. The right panel of Fig.~\ref{fig:spin-Love} presents Eq.~\eqref{eq:ratio-quadrupole} with $M=1.4M_\odot$ and $\chi = 0.05$ as a function of $b/m$ for three representative equations of state. Observe that although the ratio $\bar \lambda_{23}^{(-)}/\bar \lambda_2$ easily exceeds unity, as shown in the left panel of Fig.~\ref{fig:spin-Love}, the contribution of this ratio to corrections to the quadrupole moment $\delta M_2/M_2^{(0)}$ is always much less than unity, as shown on the right panel of the figure. The contribution of $\lambda_{23}^{(-)}$ to the quadrupole moment becomes important only close to merger, when the spin is relatively large and the equation of state is rather stiff.

Pani \et~\cite{Pani:2015hfa} studied the relations between the moment of inertia $I$ and these tidal deformabilities at linear order in spin. They normalized $\lambda_{23}^{(-)}$ as given below Eq.~\eqref{eq:ratio-quadrupole} and defined $\bar \lambda_{43}^{(-)} \equiv \lambda_{43}^{(-)}/(M^{8} \chi)$, $\bar \sigma_{32}^{(-)} \equiv \sigma_{32}^{(-)}/(M^{6} \chi)$ and $\bar \sigma_{34}^{(-)} \equiv \sigma_{34}^{(-)}/(M^{8} \chi)$. They found that the equation-of-state variation in these relations is much larger than in the original I-Love relation for a non-rotating configuration. For example, the variation in the relation between $\bar \lambda_{23}^{(-)}$ and $\bar I$ can be as large as 50\%, and that between $\bar \sigma_{32}^{(-)}$ and $\bar I$ can exceed 100\%. This study suggests that the universality is completely lost when considering rotating stars and the relatively higher $\ell$ order deformabilities, relative to the $\bar \lambda_2$ case for non-rotating stars. 

%%%%%%%%%%%%%%%%%%%%%%%%%%%%%%%%%%%%%%%%%%%%%%%%%%%%%%
\subsection{Binary Love Relations}
\label{sec:binary-Love}

\begin{figure*}[thb]
\begin{center}
\includegraphics[width=7.3cm,clip=true]{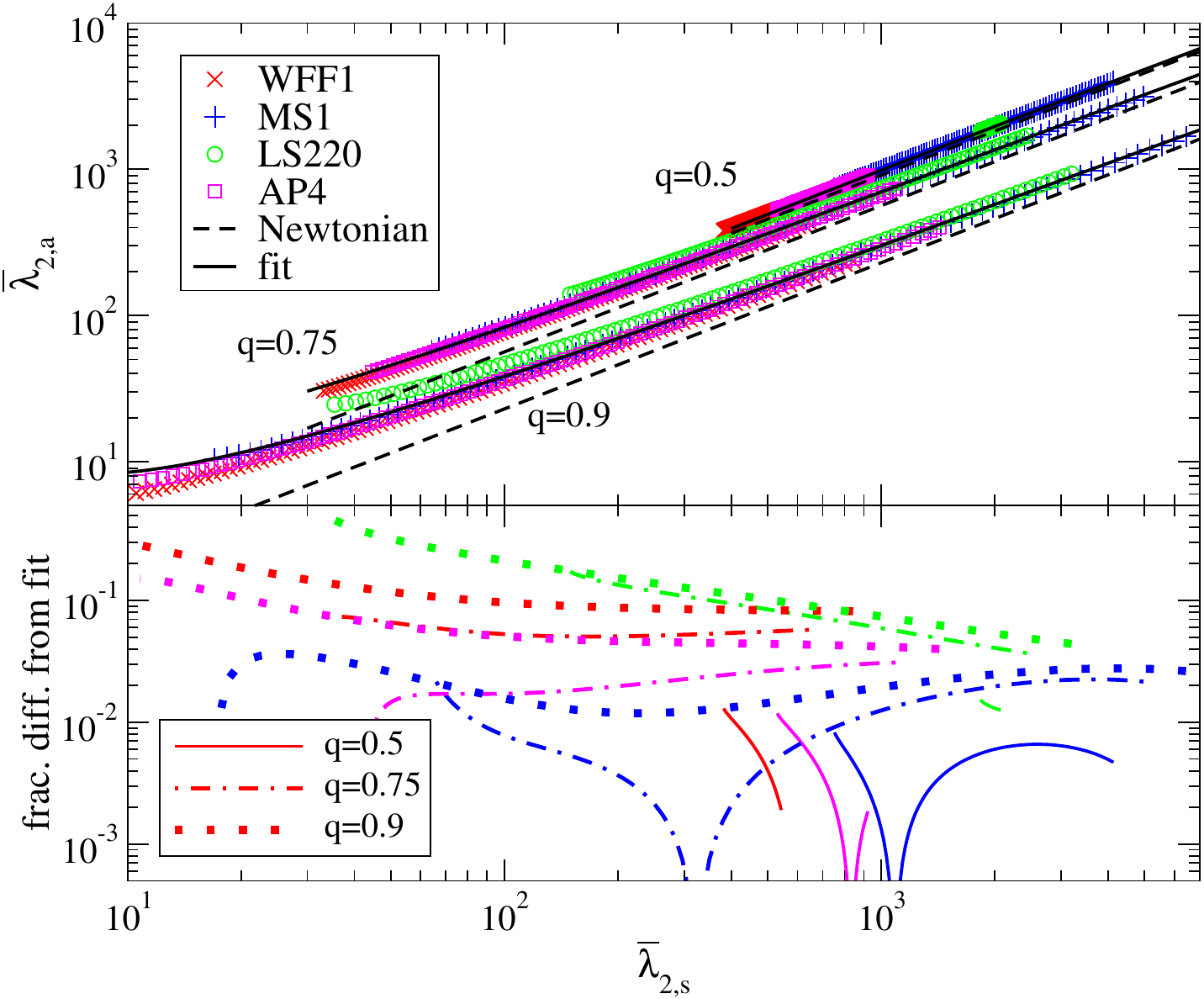}  
\includegraphics[width=7.5cm,clip=true]{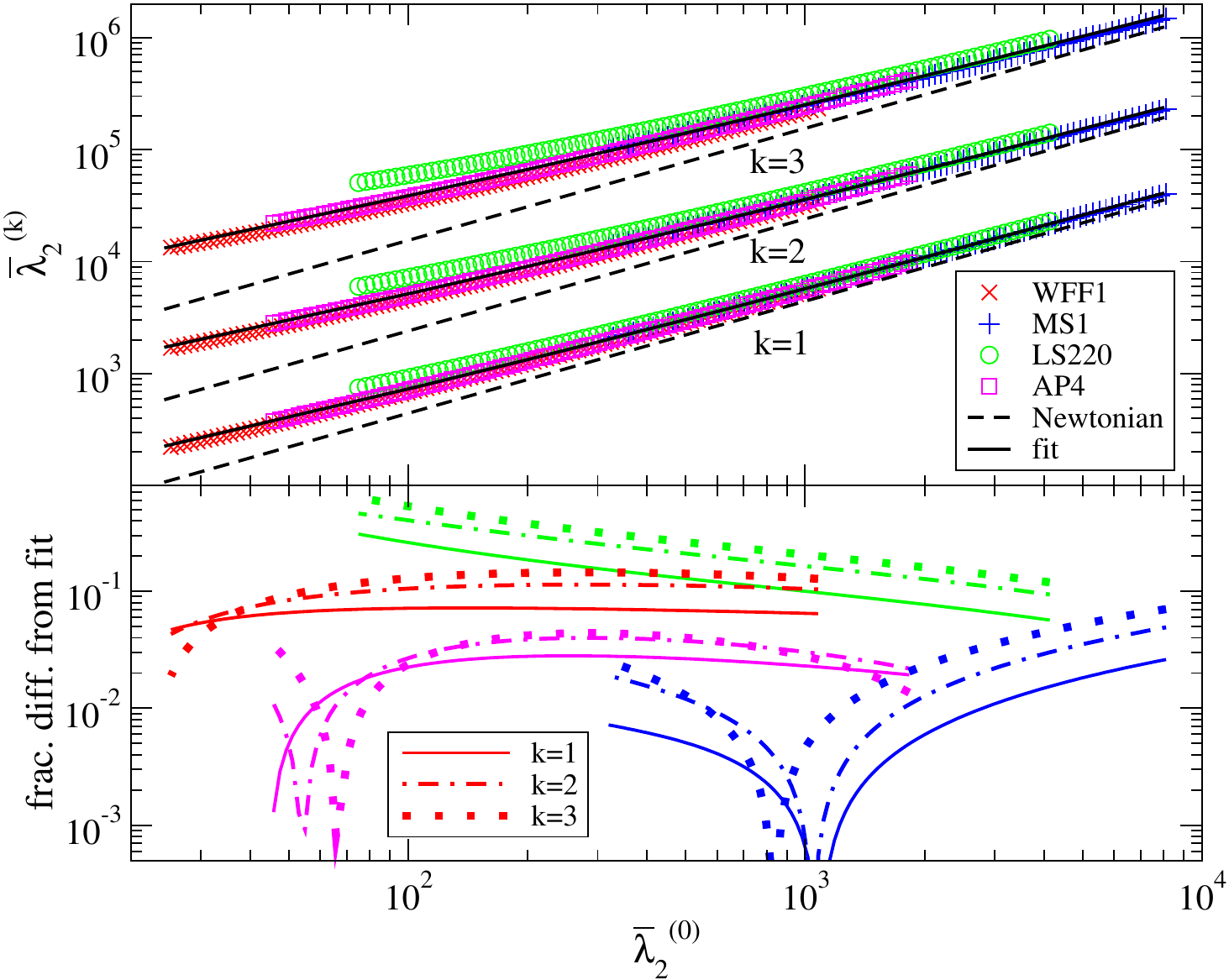}  
\caption{\label{fig:lambdas-lambdaa} 
The $\bar \lambda_{2,s}$--$\bar \lambda_{2,a}$ relation for binaries with mass ratio $q = 0.5$, $0.75$ and $0.9$ (top left) and the $\bar \lambda_2^{(0)}$--$\bar \lambda_2^{(k)}$ relations (top right) for neutron stars with four realistic equations of state. The solid lines represent fits to the numerical data, while the dashed lines are analytic relations in the Newtonian limit. (Bottom) Absolute fractional difference between the numerical data and the fits. This figure is taken and edited from Yagi~\cite{Yagi:2013sva}.} 
\end{center}
\end{figure*}

We have so far focused on approximately universal relations satisfied among stellar quantities that characterize a single neutron star or quark star. Let us now consider approximately universal relations between tidal deformability parameters that characterize both components of a binary system. Consider then a binary system in which star 1 is perturbed by the external field of star 2 (leading to a linear, quadrupolar, even-parity response parametrized by $\lambda_{2,1}$) and star 2 is perturbed by the external field of star 1 (leading to a linear, quadrupolar, even-parity response parametrized by $\lambda_{2,2}$). In particular, consider the symmetric and antisymmetric combinations of these deformabilities,
\be
\label{eq:lambdas-lambfaa-def}
\bar \lambda_{2,s} \equiv \frac{\bar \lambda_{2,1} + \bar \lambda_{2,2}}{2}, \quad \bar \lambda_{2,a} \equiv \frac{\bar \lambda_{2,1} - \bar \lambda_{2,2}}{2}\,,
\ee
and study whether there exist \emph{binary Love relations}, i.e.~approximately equation-of-state independent relations between $\lambda_{\ell,s}$ and $\lambda_{\ell,a}$. The top left panel of Fig.~\ref{fig:lambdas-lambdaa} shows this relation when $\ell=2$ calculated with various equations of state at a fixed mass ratios $q \equiv m_1/m_2$ with $m_1 \leq m_2$. We only show data with $m_1 \geq 1M_\odot$ (a $q$-dependent fit to this data was also constructed in Yagi and Yunes~\cite{Yagi:2013sva,Yagi:2016qmr}, which is also shown in Fig.~\ref{fig:lambdas-lambdaa}). The bottom left panel shows the absolute fractional difference between the numerical data and the fit. Observe that the universality worsens as one increases $q$ (and approaches the equal-mass limit) and as one approaches the relativistic limit (small $\bar \lambda_{2,s}$ limit). The equation-of-state variation can be as large as 50\%, but if one restricts attention to neutron star masses smaller than $1.8M_\odot$, the approximate universality is better than 20\%.  

In order to better understand the approximate universality of the binary Love $\bar \lambda_{2,s}$--$\bar \lambda_{2,a}$ relation, let us carry out analytic calculations in the Newtonian limit~\cite{Yagi:2016qmr}. Using the fact that a Newtonian polytrope satisfies the relations $m_A \propto C_A^{(3-n)/2}$ and $\bar \lambda_{2,A} \propto C_A^{-5}$ with $m_A$ and $C_A$ the mass and compactness of the $A$th body, one finds that $\bar \lambda_{2,s} \propto 1 + q^{10/(3-n)}$ and $\bar \lambda_{2,a} \propto 1 - q^{10/(3-n)}$, which then imply~\cite{Yagi:2016qmr} 
\be
\bar \lambda_{2,a} = F_n(q) \bar \lambda_{2,s}\,, \quad F_n(q) \equiv \frac{1 - q^{10/(3-n)}}{1 + q^{10/(3-n)}}\,.
\ee
This Newtonian relation is also shown in the top panel of Fig.~\ref{fig:lambdas-lambdaa}, which agrees with the numerical data in the large $\bar \lambda_{2,s}$ region (in the Newtonian limit). The above relation reduces to $\bar \lambda_{2,a} = \bar \lambda_{2,s}$ in the small mass ratio limit $(q \to 0)$, while it reduces to $\bar \lambda_{2,a} = 0$ in the equal-mass limit $(q\to 1)$. The absolute fractional difference of the relation from that with an averaged polytropic index $\bar n$ near $q=1$ is given by
\be
\left| \frac{F_n - F_{\bar n}}{F_{\bar n}} \right| = \frac{n-\bar n}{3-\bar n} + \mathcal{O}\left[(1-q)^2\right] \lesssim 0.13\,.
\ee
This analytic calculation explains mathematically why the universality deteriorates as one increases $q$.  

Another parameterization of tidal deformabilities that is useful in gravitational wave observations of neutron-star binaries is~\cite{Favata:2013rwa,Wade:2014vqa} 
\ba
\bar \Lambda &=& \frac{16}{13} \left[ (1 + 7 \eta - 31 \eta^2) \bar \lambda_{2,s}  - \sqrt{1- 4 \eta} (1 + 9 \eta - 11 \eta^2) \bar \lambda_{2,a} \right]\,, \\
\delta \bar \Lambda &=& \sqrt{1-4\eta} \left( 1-\frac{13272}{1319} \eta + \frac{8944}{1319} \eta^2 \right) \bar \lambda_{2,s}  - \left( 1 - \frac{15910}{1319} \eta + \frac{32850}{1319} \eta^2 + \frac{3380}{1319} \eta^3 \right) \bar \lambda_{2,a}\,,
\ea
where $\eta$ is the symmetric mass ratio defined by
\be
\label{eq:eta}
\eta \equiv \frac{m_1 m_2}{(m_1 + m_2)^2} = \frac{q}{(1+q)^2}\,. 
\ee
Yagi and Yunes~\cite{Yagi:2016qmr} found that the relation between $\bar \Lambda$ and $\delta \bar \Lambda$ also depends on $q$ and has a similar equation-of-state variation as that of the $\bar \lambda_{2,s}$--$\bar \lambda_{2,a}$ relation. 

Yet, another useful parameterization is to Taylor expand $\bar \lambda_{2,A}(m_A)$ about a fiducial mass $m_0$~\cite{Yagi:2013sva}:
\be
\label{eq:Taylor-expand-Love}
\bar \lambda_{2,A} (m_A) \equiv  \sum_{k=0} \frac{\bar \lambda_2^{(k)}}{k!} \left(1 - \frac{m_A}{m_0}  \right)^k\,, \quad
\bar \lambda_2^{(k)} \equiv  (-1)^k m_0^k  \frac{d^k \bar \lambda_{2,A}}{d (m_A)^k} \bigg|_{m_A=m_0}\,,
\ee
which is similar to the parameterization proposed in Messenger and Read~\cite{Messenger:2011gi}, Damour \et~\cite{Damour:2012yf}, Del Pozzo \et~\cite{DelPozzo:2013ala} and Agathos \et~\cite{Agathos:2015uaa}. The top right panel of Fig.~\ref{fig:lambdas-lambdaa} shows the relation between $\bar \lambda_2^{(k)}$ and $\bar \lambda_2^{(0)}$ for various values of $k$ and equations of state. The single parameter along each curve is $m_0$. A fit to the numerical data, shown in the figure, was constructed in Yagi~\cite{Yagi:2013sva}, and the absolute fractional difference between this fit and the numerical data is shown in the bottom right panel of Fig.~\ref{fig:lambdas-lambdaa}. Observe that the relation with $k=1$ is equation-of-state insensitive to $\sim 30\%$, while the variation increases as one increases $k$. The top right panel also shows the Newtonian relation, $\bar \lambda_2^{(k)} = \bar \lambda_2^{(0)} \Gamma [k+10/(3-n)]/\Gamma [10/(3-n)]$ with $\Gamma(x)$ the Gamma function. Observe that the numerical data approaches the Newtonian relations as one increases $\bar \lambda_2^{(0)}$.

%%%%%%%%%%%%%%%%%%%%%%%%%%%%%%%%%%%%%%%%%%%%%%%%%%%%%%
\subsection{I-Love-Q and Oscillation Frequencies}
\label{sec:oscillation}

Let us proceed by reviewing the connection between the I-Love-Q relations and the universal relations among the f-mode oscillation frequencies. The latter are relations associated with the real and imaginary frequencies of the modes of an oscillating star, i.e.~the central frequency and the damping time of an oscillation mode. These frequencies and certain combinations of the stellar mass and radius satisfy approximately universal relations that were discovered in the 1990s~\cite{Andersson:1996pn,Andersson:1997rn,Benhar:1998au,Benhar:2004xg,Tsui:2004qd}. Lau \et~\cite{Lau:2009bu} (see also Chirenti \et~\cite{Chirenti:2015dda}) found that the amount of universality in these relations improves if one uses the moment of inertia instead of the radius for non-rotating configurations. This is because $\sqrt{I/M}$ serves as an effective radius that measures the average size of a star weighted by its mass distribution, and is more relevant to the stellar dynamics than the geometric radius. Doneva and Kokkotas~\cite{Doneva:2015jba} extended the relation to rapidly-rotating neutron stars and to higher-mode oscillation frequencies. Together with the I-Love relation, one can easily see that a universal relation exists between the f-mode oscillation frequency and the tidal deformability.

Chan \et~\cite{Chan:2014kua} studied the relation between the $\ell$th central oscillation frequency $\bar \omega_\ell$ (normalized by the stellar mass) and the $\ell'$th electric-type tidal deformability $\bar \lambda_{\ell'}$. Interestingly, they found that the equation-of-state variation becomes smallest when $\ell = \ell'$ (smaller than 1\%), while the variation increases when $\ell \neq \ell'$. To better understand this behavior, the authors carried out an analytic calculation in the Newtonian limit. They considered a generalized Tolman model, whose density profile is given by $\rho (r) = \rho_c (1-\delta \, r^2/R^2)$, where $\rho_c$ is the central density of the star, while $0 \leq \delta \leq 1$ corresponds to an effective polytropic index near the center ($\delta = 0$ corresponds exactly to an incompressible star profile, while $\delta =1$ reduces to the original Tolman model). Using such a profile, one can perturbatively solve the Tolman-Oppenheimer-Volkhoff equation about $\delta = 0$ to find the pressure profile. One can then calculate the f-mode oscillation frequency, given by~\cite{1964ApJ...139..664C}
\be
\label{eq:omega_ell}
\omega_\ell = \frac{2 \ell (\ell-1)(2\ell-1)}{2\ell+1} \frac{\int_0^R p(r) \, r^{2\ell-2} \, dr}{\int_0^R \rho(r) \, r^{2\ell} \, dr}\,,
\ee
which depends on $\rho_c$. Using the relations
\ba
M &=& 4 \pi \rho_c \int_0^R \left( 1 - \delta \frac{r^2}{R^2} \right) r^2 dr = \frac{4\pi}{3}  \rho_c R^3 \left(1 - \frac{3}{5} \delta \right)\,, \\
I &=& \frac{8 \pi}{3} \rho_c \int_0^R \left( 1 - \delta \frac{r^2}{R^2} \right) r^4 dr = \frac{8\pi}{15}  \rho_c R^5 \left(1 - \frac{5}{7} \delta \right)\,,
\ea
one can eliminate the radius $R$ and find $\rho_c (M, I)$. Substituting this in Eq.~\eqref{eq:omega_ell}, one finds
\be
\label{eq:omega-l2}
\bar \omega_\ell^2 \equiv (M \omega_\ell)^2 \propto \bar I^{-\frac{3}{2}} \left[ 1 + \frac{12(2-\ell)}{35(2\ell+3)} \delta + \mathcal{O}\left(\delta^2 \right) \right]
\ee
near $\delta = 0$. On the other hand, following the calculation in Sec.~\ref{sec:no-rot-multipole-love}, one finds that the tidal deformability is given by
\be
\label{eq:lambda-l}
\bar \lambda_{\ell'} \propto \bar I^{\frac{2 \ell'+1}{2}} \left[ 1 + \frac{4 (2\ell'+1) (\ell'-1) (\ell'-2)}{35(\ell'-1) (2\ell'+3)} \delta + \mathcal{O}\left( \delta^2 \right) \right]\,.
\ee
Eliminating $\bar I$ from Eqs.~\eqref{eq:omega-l2} and~\eqref{eq:lambda-l}, one arrives at
\be
\bar \omega_\ell^2 \propto \bar \lambda_{\ell'}^{-\frac{3}{2\ell'+1}} \left[ 1 - \frac{12 (\ell-\ell')}{5(2\ell+3) (2\ell'+3)} \delta + \mathcal{O}\left( \delta^2 \right) \right]\,.
\ee
This shows that when $\ell \neq \ell'$, the equation-of-state variation is of ${\cal{O}}(\delta)$, while it is of ${\cal{O}}(\delta^{2})$ when $\ell = \ell'$. In other words, the relation becomes \emph{stationary} (the linear dependence vanishes) around $\delta = 0$ in the latter case, suppressing the variation. Similarly, observe that by setting $\ell' = 2$ in Eq.~\eqref{eq:lambda-l}, the I-Love relation in the Newtonian limit depends on $\delta$ at quadratic order, which explains mathematically why the universality holds so strongly in this relation~\cite{Chan:2014kua}. In particular, this also shows mathematically why the I-Love relation is more equation-of-state insensitive than the 3-hair ones, since the latter depends linearly in the deviation from incompressible stars [see Eq.~\eqref{eq:Bbar-analytic}].

%%%%%%%%%%%%%%%%%%%%%%%%%%%%%%%%%%%%%%%%%%%%%%%%%%%%%%
\subsection{I-Love-Q and Compactness}
\label{sec:I-Love-Q-C}

Approximate universal relations between each of the I-Love-Q trio and the stellar compactness were found before the discovery of the I-Love-Q relations~\cite{1994ApJ...424..846R,Bejger:2002ty,Lattimer:2004nj,Urbanec:2013fs,Baubock:2013gna,Maselli:2013mva,Breu:2016ufb,Staykov:2016mbt}. One might then naively think that the existence of the universal I-Love-Q relations is trivial. For example, one can combine the I-C and the C-Love relations to obtain the I-Love relation and expect the latter to be as universal as the former two. The aim of this subsection is to show explicitly that the equation-of-state variation in the relation between the compactness and any of the I-Love-Q trio members is much larger than the variation in the I-Love-Q relations. We will also here review a mathematical explanation for why this equation-of-state variation is suppressed in the I-Love relations, while it is not in the I-C and C-Love relations~\cite{Chan:2015iou}.  

%--------------------------------------------
\subsubsection{Numerical Analysis}

Let us begin by reviewing various I-C relations. Ravenhall and Pethick~\cite{1994ApJ...424..846R} discovered the first approximate universal relation between $C$ and $\bar I C^2 e^{\lambda(R)}$ in the 1990s (where recall that $e^{\lambda(r)}$ is the $(r,r)$ component of the metric, given in Eq.~\eqref{eq:metric-ansatz}) for slowly-rotating neutron stars to first-order in slow-rotation. A decade later, Bejger and Haensel~\cite{Bejger:2002ty} and then Lattimer and Schutz~\cite{Lattimer:2004nj} studied the relation between $C$ and $\bar I C^2 = I/(M R^2)$ (motivated by the fact that $I \propto M R^2$ for Newtonian polytropes). These studies revealed that the I-C relation of a neutron star sequence is quite different from that of a quark star or incompressible star sequence. Baubock \et~\cite{Baubock:2013gna} studied the relation between $C$ and $\bar I C^{3/2}$ using 4 different neutron-star equations of state and showed that the universality holds to $\sim 4\%$ at most when $C \gtrsim 0.15$. Breu and Rezzolla~\cite{Breu:2016ufb} and Staykov \et~\cite{Staykov:2016mbt} investigated the relations both between $\bar I C^2$ and $C$ and between $\bar I$ and $C$ with many different neutron-star equations of state, finding that the latter is slightly more universal than the former. The maximum equation-of-state variation in the latter was found to be ${\cal{O}}(5\%)$ when $C \gtrsim 0.07$~\cite{Breu:2016ufb,Staykov:2016mbt}. Breu and Rezzolla~\cite{Breu:2016ufb} and Staykov \et~\cite{Staykov:2016mbt} also studied the relations for rapidly-rotating neutron stars and found that they can deviate from the relations for non-rotating stars, with a variation of ${\cal{O}}(20\%)$ when $\chi = 0.6$~\cite{Breu:2016ufb}; this variation is larger than that induced by differences in the equation of state in the slow rotation limit.

\begin{figure*}[htb]
\begin{center}
\includegraphics[width=8.1cm,clip=true]{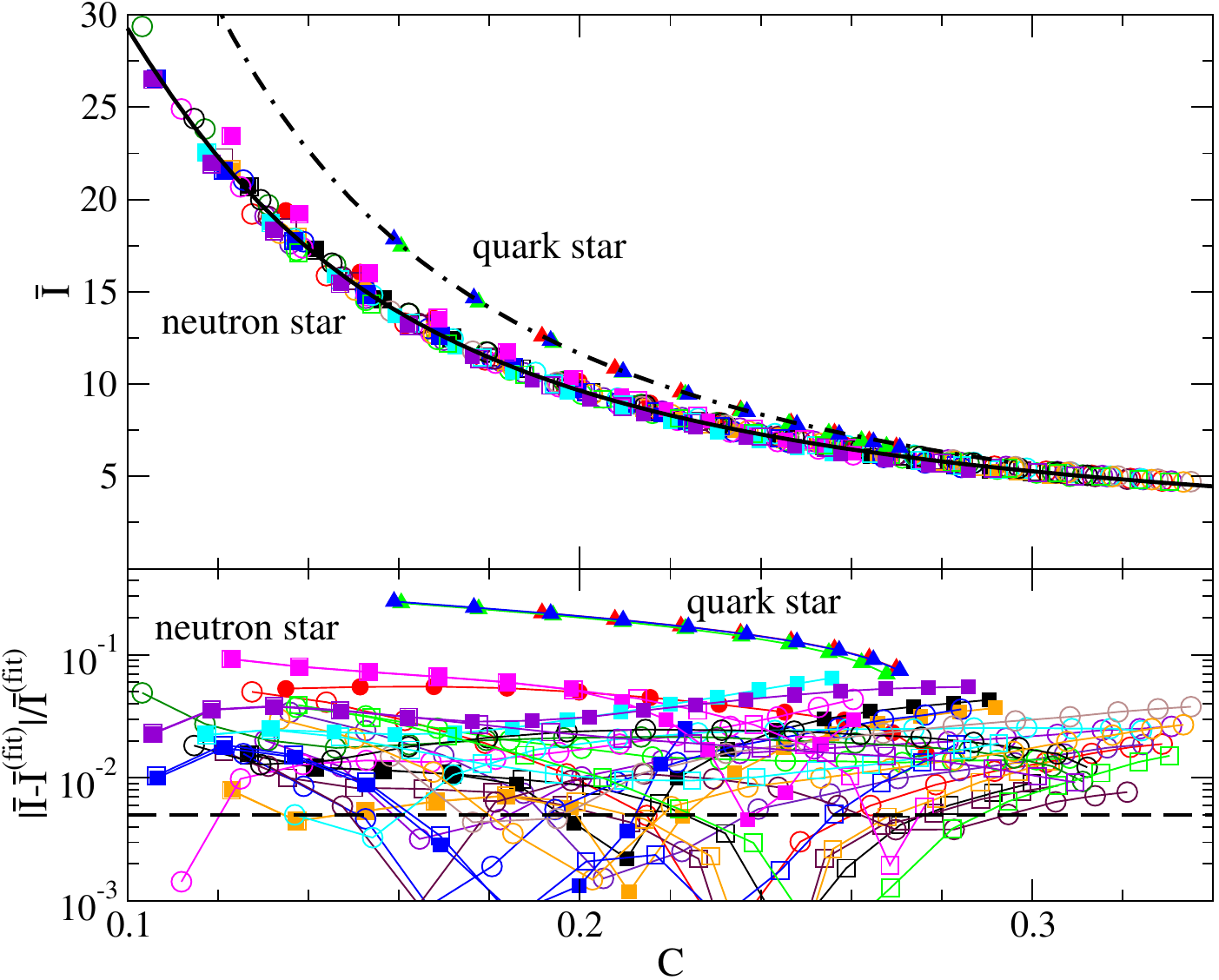}  
\includegraphics[width=8.cm,clip=true]{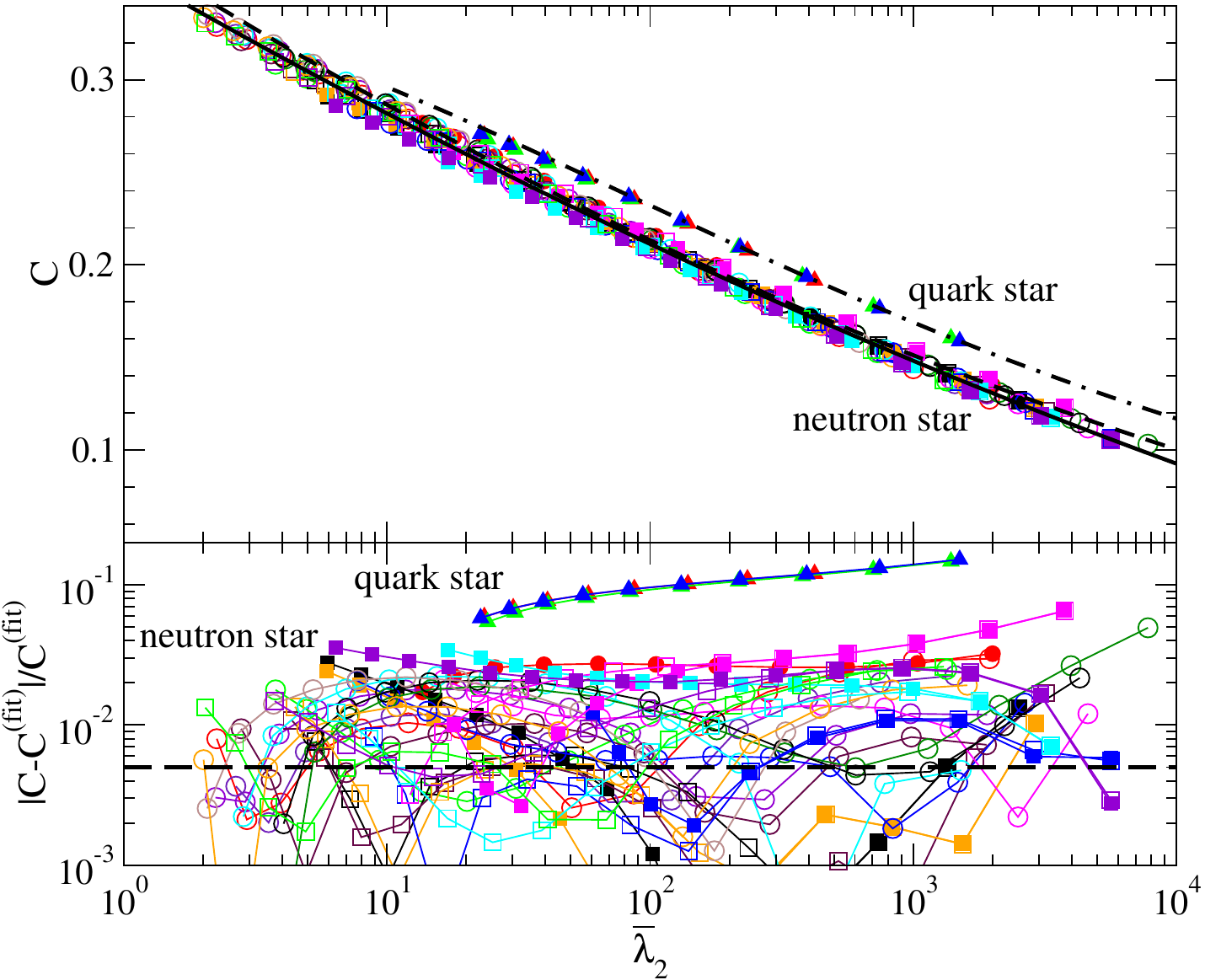}  
\caption{\label{fig:I-Love-C}
(Top) I-C (left) and C-Love (right) relations for slowly-rotating neutron stars and quark stars with masses larger than $1M_\odot$ and the meaning of symbols being same as in Fig.~\ref{fig:I-Love-Q}. We also show the fit in Eqs.~\eqref{eq:I-C-fit} and~\eqref{eq:C-Love-fit} as solid curves, constructed without including quark stars. The dashed curve in the C-Love plane represents the fit created in Maselli \et~\cite{Maselli:2013mva} among 3 different equations of state with masses $1.2 M_\odot < M < 2M_\odot$. Dotted-dashed curves are analytic expressions for quark stars derived within the post-Minkowskian approximation~\cite{Chan:2015iou}. (Bottom) Fractional difference from the fit. Observe that such a fractional difference, in particular for quark stars (triangles), is much larger than that of the I-Love relation in Fig.~\ref{fig:I-Love-Q}, whose maximum equation-of-state variation of $\sim 0.5\%$ is shown by horizontal dashed lines.
}
\end{center}
\end{figure*}

The top left panel of Fig.~\ref{fig:I-Love-C} presents the relation between $\bar I$ and $C$ for slowly-rotating neutron stars and quark stars with the same equations of state used in Fig.~\ref{fig:I-Love-Q}. As also presented in Ravenhall and Pethick~\cite{1994ApJ...424..846R}, Bejger and Haensel~\cite{Bejger:2002ty} and Lattimer and Schutz~\cite{Lattimer:2004nj}, this figure shows that the I-C relation of quark stars (triangles) is quite different from that of neutron stars (circles), as also found in Staykov \et~\cite{Staykov:2016mbt}. Following Breu and Rezzolla~\cite{Breu:2016ufb} and Staykov \et~\cite{Staykov:2016mbt}, we construct a polynomial fit given by
\be
\label{eq:I-C-fit}
y = \sum_{k=1}^4 a_k \; C^{-k}\,,
\ee
where $y = \bar I$ and the best fitted coefficients (without using the quark star data) are $a_1=1.317$, $a_2=-0.05043$, $a_3=0.04806$ and $a_4=-0.002692$, which is consistent with that found in Breu and Rezzolla~\cite{Breu:2016ufb} and Staykov \et~\cite{Staykov:2016mbt}. The bottom left panel of Fig.~\ref{fig:I-Love-C} shows the relative fractional difference of the numerical data and the fit. Observe that the maximum fractional difference for the neutron star sequence is at most $9\%$, which is consistent with Breu and Rezzolla~\cite{Breu:2016ufb} but slightly larger than that found in Baubock \et~\cite{Baubock:2013gna} and Staykov \et~\cite{Staykov:2016mbt} (due to the use of more equations of state in this review). The quark star sequence deviates from the neutron star one by $27\%$ at most. These deviations can be understood by studying Newtonian polytropes, where the dimensionless moment of inertia when $n=0$ and $n=1$ is given by $\bar I_{\N}^{(n=0)} = (2/5) C^{-2} = 0.4 C^{-2}$ and $\bar I_{\N}^{(n=1)} = 2 (\pi^2 - 6)/(3 \pi^2) C^{-2} \sim 0.26 C^{-2}$ respectively~\cite{Yagi:2013awa}, leading to a fractional difference between the two coefficients of $\sim 53\%$.  

We next move on to the Q--C relation, the approximate universality of which was discussed in Urbanec \et~\cite{Urbanec:2013fs}. As in the I-C case, the authors found that the Q-C relation of quark stars is quite different from that of neutron stars. We do not present the Q-C relation here because it is qualitatively similar to the $\bar{I}$ and $C$ relation, shown in the left panel of Fig.~\ref{fig:I-Love-C}. A good fit to such a Q-C relation (using neutron star data only) is given by Eq.~\eqref{eq:I-C-fit} with $y = \bar Q$ and the coefficients $a_1=-0.2588$, $a_2=0.2274$, $a_3=0.0009528$ and $a_4=-0.0007747$. The maximum fractional difference of the neutron star data from the fit is $9\%$ at most, while the maximum deviation between the quark star and neutron star sequences is $29\%$. 

Let us finally look at the C-Love relation, which was first studied by Maselli \et~\cite{Maselli:2013mva}. The authors investigated three different neutron star (including hybrid star) equations of state with neutron star masses $1.2 M_\odot \leq M \leq 2M_\odot$, and found that the relation is approximately universal to within $\sim 2\%$. The top right panel of Fig.~\ref{fig:I-Love-C} shows the C-Love relation with various equations of state. A fit of the form~\cite{Maselli:2013mva}
\be
\label{eq:C-Love-fit}
C = \sum_{k=0}^2 a_k \left( \ln \bar \lambda_2 \right)^k\,,
\ee
with $a_0^{\mathrm{M}} = 0.371$, $a_1^{\mathrm{M}} = -0.0391$ and $a_2^{\mathrm{M}}= 0.001056$ obtained in~\cite{Maselli:2013mva} is also shown in this figure with a dashed curve. Using the wider set of equations of state considered in this review, the best-fit coefficients change slightly to $a_0^{\mathrm{YY}} = 0.360$, $a_1^{\mathrm{YY}} = -0.0355$ and $a_2^{\mathrm{YY}} = 0.000705$, which is shown in the figure with a solid curve. The bottom right panel shows the fractional difference between the data and the second fit. The maximum difference for the neutron star sequence is 6.5\% (which is more than 3 times larger than that in Maselli \et~\cite{Maselli:2013mva}), while the maximum deviation in the quark star sequence relative to the neutron star one is 15\%.

We now compare the equation-of-state variation in the I-C, Q-C and C-Love relations with that in the I-Love-Q relations. The maximum equation-of-state variation in the I-Love relation of Fig.~\ref{fig:I-Love-Q} is shown by horizontal dotted-dashed lines in the bottom panels of Fig.~\ref{fig:I-Love-C}. Observe that this variation is more than one order of magnitude smaller than that in the I-C and C-Love relations. In particular, the I-Love relation of neutron stars is essentially indistinguishable from that of quark stars. These results show that the degree of universality in the I-Love relation cannot be simply explained by that in the I-C and C-Love relations. Once one combines the two relations, the equation-of-state variation in one relation must partially cancel that in the other, leading to a suppressed overall variation in the combined I-Love case. 

%--------------------------------------------
\subsubsection{Analytic Investigation}
\label{sec:I-Love-Q-C-analytic}

Can we understand this suppression mechanism analytically? Chan \et~\cite{Chan:2015iou} showed mathematically that the equation-of-state variation is suppressed in the I-Love relation relative to that in the I-C and C-Love relations for self-bound stars. The energy density of such stars does not vanish in the pressureless limit, but rather $\rho = \sum_{k=0} c_k \; p^k$, where $c_0 > 0$ and $c_1 = (d\rho/dp)_{p=0} \geq 0$ is a measure of the stellar compressibility at $p=0$. Incompressible stars correspond to the $c_0 > 0$ and $c_{k\geq 1}=0$ case, while quark stars (modeled with a simple MIT bag model~\cite{Johnson:1975zp,Alcock:1986hz,Witten:1984rs}) corresponds to the $c_0 = 4B$, $c_1 = 3$ and $c_{k\geq 2}=0$ case, where $B$ is the bag constant. Chan \et~\cite{Chan:2015iou} extended the analysis of incompressible stars in~\cite{Chan:2014tva} (see Eq.~\eqref{eq:I-Love-PM}) to self-bound stars and proved that the $c_j$ coefficients first enter at $\mathcal{O}(C^j)$ relative to the leading order term in compactness in the I-C and C-Love relations, while it first enters at higher order in compactness in the I-Love relation. For example, with a general linear equation of state with $c_{k\geq 2}=0$, the I-C and C-Love relations do not depend on $c_0$, but their dependence on $c_1$ is 
\be
\frac{1}{\bar I} \left( \frac{\partial \bar I}{\partial c_1} \right)_C = -0.05715 C + \mathcal{O}\left( C^2 \right)\,,
\qquad
\frac{1}{\bar \lambda_2} \left( \frac{\partial \bar \lambda_2}{\partial c_1} \right)_C = -0.1428 C + \mathcal{O}\left( C^2 \right)\,,
\ee
respectively. On the other hand, the dependence of $c_1$ of the I-Love relation is 
\ba
\frac{1}{\bar I} \left( \frac{\partial \bar I}{\partial c_1} \right)_{\bar \lambda_2} &=& -0.002061 \bar \lambda_2^{-2/5} +  \mathcal{O}\left( \bar \lambda_2^{-3/5} \right) \nn \\
&=& -0.002720 C^2 + \mathcal{O}\left( C^3 \right)\,.
\ea
Therefore, the $c_1$ dependence on the I-Love relation is suppressed compared to that of the I-C and C-Love relations by a factor of $\mathcal{O}(C)$ and since $C < 1/2$ for all self-bound stars, the former is more universal than the latter two.

With this argument at hand, we can now better understand why the I-Love relation is more universal than the I-C and C-Love relations. Chan \et~\cite{Chan:2015iou} showed that 
\be
\label{eq:I-Love-Chan}
\bar I^{(\SBS)} \left( C^{(\SBS)} \right) = \bar I^{(\IS)} \left( C^{(\IS)} \right) \left[1 + \mathcal{O}\left(C^{(\IS)} \right) \right]\,,
\quad 
\bar \lambda_2^{(\SBS)} \left( C^{(\SBS)} \right) = \bar \lambda_2^{(\IS)} \left( C^{(\IS)} \right)\left[1 + \mathcal{O}\left(C^{(\IS)} \right) \right]\,,
\ee
where the superscripts (SBS) and (IS) refer to self-bound stars and incompressible stars respectively, i.e.~one can adjust the compactness of an incompressible star $C^{(\IS)}$ such that both $\bar I$ and $\bar \lambda_2$ match those of a self-bound star with a given compactness $C^{(\SBS)}$ in the low compactness limit. Since the I-Love relation is obtained by eliminating the compactness, the difference between $C^{(\SBS)}$ and $C^{(\IS)}$ is irrelevant in the I-Love relation, and such a relation for self-bound stars agrees with that of incompressible stars to leading-order in compactness. Thus, the equation-of-state variation of the I-Love relation enters at $\mathcal{O}(C)$ higher than that of the I-C and C-Love relations. Equation~\eqref{eq:I-Love-Chan} is realized due to (i) the similarity in the response of $\bar I$ and $\bar \lambda_2$ to variations in the stellar compressibility and (ii) the proper normalization of $\bar I$ and $\bar \lambda_2$ (see also Sec.~\ref{sec:normalization} for the latter point)~\cite{Chan:2015iou}.

%%%%%%%%%%%%%%%%%%%
\subsection{Darwin-Radau Relation}
\label{sec:Darwin-Radau}

In this subsection, we will compare the Newtonian I-Q relation obtained from the 3-hair analysis in Sec.~\ref{sec:no-hair-Newton} with the Darwin-Radau relation. In 1885 and later in 1899, Radau~\cite{radau} and Darwin~\cite{1899MNRAS..60...82D} found that a certain dimensionless version of the quadrupole moment and of the moment of inertia, which we shall call $J_{2}$ and $\tilde{I}$, satisfy a quadratic relation that depends on the stellar eccentricity. This relation was found by making a series of assumptions: (i) Newtonian gravity (since General Relativity did not exist yet), (ii) slow rotation, (iii) hydrostatic equilibrium of the stellar interior, and (iv) an equation of state that deviates only mildly from constant density. These assumptions happen to be excellent in planetary astrophysics, where the Darwin-Radau relation has been used to infer the rotation rate of an extrasolar planet from the oblateness measurement via transit photometry~\cite{Barnes:2003yw} (assuming \emph{a priori} knowledge of the planet's internal structure). For planets in the Solar System, the rotation rate has been measured and one can check that it is accurately recovered from the Darwin-Radau relation within a few percent error~\cite{Barnes:2003yw} using the model-dependent moment of inertia of Hubbard and Marley~\cite{1989Icar...78..102H}.

Let us then begin by explaining how to derive the Darwin-Radau relation. Our starting point is the moment of inertia given by Eq.~\eqref{I-Newton}. Instead of integrating the dipole moment $r^{2} \rho(r)$ from the core to the stellar surface, we will integrate up to the mean radius $\bar R$ defined below Eq.~\eqref{eq:Rnl}. Rewriting the resulting expression in terms of the mean density $\bar \rho$, one finds
\be
\label{eq:I-DR}
I = \frac{8\pi}{9} \left( \bar \rho(\bar R)\, \bar R^5 - 2 \int_0^{\bar R} \bar \rho (\bar r )\, \bar r^4\, d\bar r \right)\,.
\ee
Let us next return to the Clairaut-Radau equation of Eq.~\eqref{eq:Clairaut-Radau}, which determines the tidal Love number $k_{2}$ through the solution for $\eta_{2}$ (the logarithmic derivative of the distortion function):
\be
\frac{d}{d\bar r} \left( \bar \rho\, \bar r^5\, \sqrt{1+\eta_2} \right) = 5 \bar \rho\, \bar r^4\, \psi(\eta_2)\,,
\qquad
\psi (\eta_2) \equiv \frac{10+5\eta_2-\eta_2^2}{10\sqrt{1+\eta_2}}\,.
\ee
Using the approximation $\psi (\eta_2) \approx 1$, which is exact for an $n=0$ polytrope, we can solve the above equation to find
\be
\int_0^{\bar R} \bar \rho (\bar r)\, \bar r{}^4\, d\bar r = \frac{\bar \rho ( \bar R )\, \bar R^5} {5} \sqrt{1+\eta_2 ( \bar R )}\,.
\ee
Substituting this and $\bar \rho = 3M/(4\pi \bar R^3)$ in Eq.~\eqref{eq:I-DR} one finds, 
\be
\tilde I \equiv \frac{I}{M \bar R^2} = \frac{2}{3} \left( 1 - \frac{2}{5} \sqrt{1+\eta_2 (\bar R )} \right)\,.
\ee
Using further the relation $\eta_2 (\bar R ) = 3 \left[ 1 - 5 J_2/(2 f_e) \right]$, with $J_2 \equiv - Q/\left(M \bar R^2\right)$ and $f_e \equiv 1-a_3/a_1$, where recall that $a_{3}$ and $a_{1}$ are the semi-major and semi-minor axis of the stellar ellipsoid, one arrives at the Darwin-Radau relation~\cite{radau,1899MNRAS..60...82D,2000ssd..book.....M}:
\be
\label{eq:Darwin-Radau}
\frac{J_2}{f_e} = -\frac{3}{10} + \frac{5}{2} \tilde I - \frac{15}{8} \tilde I^2\,.
\ee
Observe how the new dimensionless quadrupole moment $J_{2}$ depends quadratically on the new dimensionless moment of inertia $\tilde{I}$ with an overall constant of proportionality $f_{e}$ that depends on the stellar eccentricity.

Let us now compare the Darwin-Radau relation with the I-Q relation obtained from the 3-hair relation in Sec.~\ref{sec:no-hair-Newton} within the elliptical isodensity approximation. Setting $\ell = 0$ in Eq.~\eqref{eq:M2l2-CM}, one finds
\be
\label{eq:I-Q-3-hair}
Q = - \frac{1}{3} \frac{\mathcal{R}_{n,2}}{\xi_1^4 |\vartheta'(\xi_1)|} \frac{e^2}{\left( 1-e^2 \right)^{1/3}} M \bar R^2\,,
\qquad
I = \frac{2}{3} \frac{\mathcal{R}_{n,2}}{\xi_1^4 |\vartheta'(\xi_1)|} \frac{1}{\left( 1-e^2 \right)^{1/3}} M \bar R^2\,.
\ee
Using further that the ellipticity $e^2 \sim 2 f_e$ in the slow-rotation approximation, one finds
\be
\label{eq:Darwin-Radau-3-hair}
\frac{J_2}{f_e} = \tilde I\,.
\ee
Observe that the I-Q relation obtained from the 3-hair relations in the Newtonian limit differs from the Darwin-Radau relation in Eq.~\eqref{eq:Darwin-Radau}, with the former being linear in the moment of inertia instead of quadratic.

Interestingly, both the Darwin-Radau and 3-hair I-Q relations can be obtained from a certain limit of the so-called \emph{core-ocean model}~\cite{1979Icar...37..310D,1979Icar...37..575D}. In this model, one treats a star as an two-fluid ellipsoid: a constant-density core of mean radius $A$ and a constant-density ocean of mean radius $B$. The relation between $J_2/f_e$ and $\tilde I$ in the core-ocean model is given by~\cite{2000ssd..book.....M}
\be
\label{eq:core-ocean}
\frac{J_2}{f_e} = \frac{2}{3} + \frac{5 \tilde I - 2 r_{\AB}^2}{5 \left(1-r_\AB^2 \right)} + \frac{8-20r_{\AB}^2+10 \tilde I \left( 5r_\AB^3-2 \right)}{12 \left( r_{\AB}^5-1 \right)+15 \tilde I \left( 2-5 r_\AB^3+3r_\AB^5 \right)}\,,
\ee
where $r_\AB \equiv A/B$ is the ratio of the mean radii. One recovers the Darwin-Radau relation (Eq.~\eqref{eq:Darwin-Radau}) in the \emph{shallow-ocean limit} $r_\AB \to 1$, while the 3-hair I-Q relation (Eq.~\eqref{eq:Darwin-Radau}) is recovered in the \emph{point-core limit} $r_\AB \to 0$. In particular, in the latter limit, $\tilde I$ and $J_2/f_e$ are given by~\cite{1979Icar...37..575D} $\tilde I =(2/5) (M_\mrm{B}/M) = J_2/f_e$, where $M$ is the total mass of the star and $M_B$ is the mass of the ocean. Thus, although $\tilde I$ and $J_2/f_e$ in the point-core mode differ from the 3-hair Eq.~\eqref{eq:I-Q-3-hair}, the relation between these quantities is identical. Note also that when $0<r_\AB<1$, the relation in Eq.~\eqref{eq:core-ocean} lies between the Darwin-Radau and 3-hair I-Q curve.
 
\begin{figure}[htb]
\begin{center}
\includegraphics[width=8.cm,clip=true]{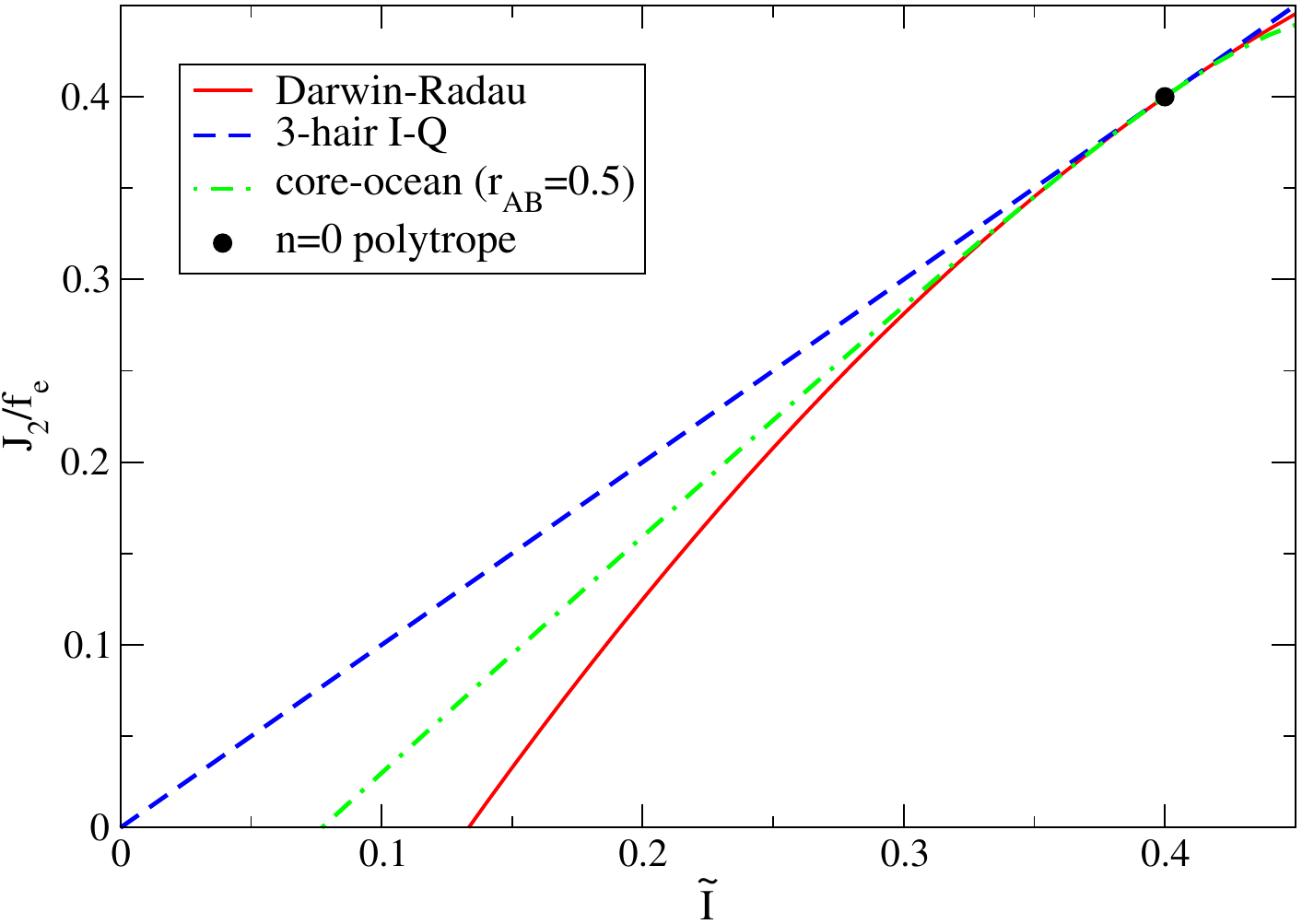}  
\caption{\label{fig:Darwin-Radau}
The Darwin-Radau relation and the 3-hair I-Q relation in the Newtonian limit.  Both relations become exact for an $n=0$ polytrope shown by the black dot. We also show the relation in the core-ocean model with $r_\AB = 0.5$ for reference, while the Darwin-Radu and 3-hair I-Q relations are obtained as a limit $r_\AB \to 1$ and $r_\AB \to 0$ respectively.
}
\end{center}
\end{figure}
Figure~\ref{fig:Darwin-Radau} compares the Darwin-Radau relation of Eq.~\eqref{eq:Darwin-Radau} with the 3-hair I-Q relation of Eq.~\eqref{eq:Darwin-Radau-3-hair} and the core-ocean relation of Eq.~\eqref{eq:core-ocean}. Observe that they all agree at a point corresponding to an $n=0$ polytrope, since all relations become exact in this limit. On the other hand, the relations deviate from each other as one moves away from an $n=0$ polytrope, as each curve relies on a different set of approximations. 

Let us conclude this subsection by stressing that the Darwin-Radau relation is \emph{different} from the I-Q relation discussed in Sec.~\ref{sec:I-Love-Q-relativistic}. One reason is that the former is only valid for Newtonian (non-compact) stars, while the latter holds for relativistic (compact) stars, like neutron stars and quark stars. Another reason is the difference in the normalizations of the moment of inertia and quadrupole moment, which can affect the universality drastically as shown in Sec.~\ref{sec:normalization}. Indeed, the universality seems to be lost for relations between $J_2/f_e$ and $\tilde I$ when considering neutron stars and quark stars. A more detailed study on the comparison between the Darwin-Radau and I-Q relations is currently in progress.

%%%%%%%%%%%%%%%%%%%%%%%%%%%%%%%%%%%%%%%%%%%%%%%%%%%%%%%%
\section{Why Universality Holds}
\label{sec5:Why}

After the initial discovery of the I-Love-Q relations, several possible hypotheses for the origin of the universality were proposed, but none of these survived detailed scrutiny. The first proposal for the origin of the universality was rooted in the similarity of the different equations of state in the outer layers of neutron stars,  which is required to ensure agreement with experimental constraints at low densities. The idea here was that the moment of inertia and quadrupole moment are mostly determined by these outer layers, and since the equations of state must all be similar to each other in that region, then the moment of inertia and the quadrupole moment should be fairly equation-of-state independent. This explanation was shown to be unsatisfactory in Yagi \et~\cite{Yagi:2014qua}, since the moment of inertia and the quadrupole moment receive their largest contributions from matter between $50\%$ and $95\%$ from the center of the star, and at these densities (the slope of) the equations of state can differ by as much as 30\%. 

A second proposal for the origin of the universality was based on the idea that the original I-Love-Q papers only considered ``simple'' compact stars, e.g.~stars that rotate slowly and have no magnetic fields. The implication here was that if more complicated neutron stars were considered, such as those with quark condensates in their inner cores or stars that rotate very rapidly, then the universality would be lost. As explained in Sec.~\ref{sec3:Extensions}, however, the universality was shown to remain in rapid rotation, in differential rotation and in weakly-magnetized stars, as well as for a very large set of equations of state that include phase transitions. 

A third proposal for the origin of the universality was related to their relativistic or \emph{black hole} limit. Indeed, the black hole no-hair theorems guarantee that the exterior gravitational field of isolated, stationary and axisymmetric black holes depends only on their mass, spin and charge. All higher multipole moments of the exterior field are then determined entirely in terms of these three quantities with exactly zero sensitivity to the interior composition of the black hole or the nature of the singularity. Neutron stars and quark stars, however, are not nearly as compact as black holes, and in fact, no continuous stationary sequence exists to map between these two sets of compact objects. Thus, it was not clear at first whether one could establish a concrete relation between the no-hair theorems of black holes and the approximate no-hair relations of compact stars.

If none of these explanations hold, then what is the origin of the universality? This is the topic of this section, which will begin with a study of the relations in the Newtonian limit (Sec.~\ref{sec:Why-I-Love-Q-Newton}) and then in the black hole limit (Sec.~\ref{sec:BH-limit}). The section concludes with a discussion of the only phenomenological model that seems capable of explaining the universality in terms of symmetry considerations (Sec.~\ref{sec:emergent-symmetry}). 

%%%%%%%%%%%%%%%%%%%%%%%%%%%%%%%%%%%%%%%%%%%%%%%%%%%%%%
\subsection{The Newtonian Limit}
\label{sec:Why-I-Love-Q-Newton}

Given the complexity of the I-Love-Q relations, perhaps one should first study the Newtonian approximate no-hair relations of neutron stars to gain some insight. Upon doing so, one quickly discovers that these relations rely heavily on the elliptical isodensity approximation (see Sec.~\ref{sec:no-hair-Newton}), and thus, perhaps this approximation is responsible for the universality. In view of this, Yagi \et~\cite{Yagi:2014qua} considered breaking each of the assumptions that go into this approximation separately to see which one has the largest effect in the universal relations. The first possibility is to break the \emph{elliptical condition} (i.e.~that the stellar isodensity surfaces are self-similar ellipsoids), for example by requiring that the isodensity contours take a self-similar spherical shape or a self-similar peanut shape. Yagi \et~\cite{Yagi:2014qua}, however, found that this only modifies the angular integrals $I_{\ell,k}$ in Eq.~\eqref{eq:R-I-def}, which are equation-of-state independent, thus leaving the universality unaffected. The second possibility is to break the \emph{spherical density profile condition} (i.e.~that the density as a function of the isodensity radius for a rotating configuration equals that of a non-rotating configuration with the same volume), for example by replacing the Lane-Emden functions $\vartheta_{\LE}$ (valid in spherical symmetry) by some other function. Yagi \et~\cite{Yagi:2014qua}, however, found that this modification changes the structure of the radial integrals $R_{\ell}$ only by a small amount, provided the function that replaces $\vartheta_{\LE}$ is a small deformation from the Lane-Emden function. 

The last possibility that remains is to break the self-similar isodensity condition and force the eccentricity to be a function of the isodensity radius $\tilde r$. Indeed, realistic compact stars are more spherical toward the center than close to the surface, as discussed in Sec.~\ref{sec:no-hair-Newton}. One can model this breakage of self-similarity through the replacement $e \to e(\tilde r) = e_0 f(\tilde r/a_1)$, where $e_0$ is the eccentricity at the surface, $f(\tilde r/a_1)$ is an arbitrary function with $f(1)=1$ and recall that $a_1$ is the semi-major axis. The multipole integral in Eqs.~\eqref{eq:M-ell-integ} and~\eqref{eq:S-ell-integ} can still be separated into radial and angular integrals~\cite{Yagi:2014qua}
\be
\label{eq:M_S_ecc}
M_{\ell}  = 2\pi \; I_{\ell,3}^{(0)} \; R_{\ell}^{(M)}\,,\qquad S_\ell = \frac{4 \pi \ell}{2 \ell +1} \Omega
\left(I_{\ell-1,5}^{(0)}-I_{\ell+1,3}^{(0)}\right) R_{\ell+1}^{(S)}\,,
\ee
where the superscript (0) refers to replacing $e$ with $e_0$ in the angular integrals of Eq.~\eqref{eq:R-I-def}, and where 
\be
R_{\ell}^{(A)}  \equiv  \int_0^{a_1} \rho (\tilde{r}) \, \tilde{r}^{\ell +2} \left[f \left( \frac{\tilde{r}}{a_1} \right) \right]^{n_A} \sqrt{\frac{1- e_0^2 \left[ f \left( \tilde{r}/a_1 \right)  \right]^2}{1-e_0^2}} d\tilde{r}\,, 
\ee
with $A = (M,S)$, $n_M = \ell$ and $n_S = \ell-2$. Observe that $R_{\ell}^{(A)}$ reduces to $R_\ell$ in Eq.~\eqref{eq:R-I-def} when $f(\tilde r/a_1) =1$. Repeating the calculation in Sec.~\ref{sec:no-hair-Newton} with Eq.~\eqref{eq:M_S_ecc}, one finds that the three-hair relations still have the form of Eq.~\eqref{eq:3-hair-dimensional} or~\eqref{eq:3-hair} but with~\cite{Yagi:2014qua}
\be
\label{eq:Bbar-ecc}
\bar B_{n,\ell} = \frac{3^{\ell+1}}{2\ell +3} \left( \frac{\mathcal{R}_{n,0}^{(M)}}{\mathcal{R}_{n,2\ell+2}^{(M)}} \right)^{\ell} \left( \frac{\mathcal{R}_{n,2\ell+2}^{(S)}}{\mathcal{R}_{n,2}^{(S)}} \right)^{\ell+1}\,,
\ee
where
\be
\mathcal{R}_{n,\ell}^{(A)} \equiv  \int_0^{\xi_1} \vartheta_\LE^n  \, \xi^{\ell +2} \left[f \left( \frac{\xi}{\xi_1} \right) \right]^{n_A} \sqrt{\frac{1- e_0^2 \left[ f \left( \xi/\xi_1 \right)  \right]^2}{1-e_0^2}} d\xi\,.
\ee
Observe again that Eq.~\eqref{eq:Bbar-ecc} reduces to Eq.~\eqref{eq:Bbar-def} when $f(\tilde r/a_1) =1$.

\begin{figure*}[t]
\begin{center}
\includegraphics[width=8.cm,clip=true]{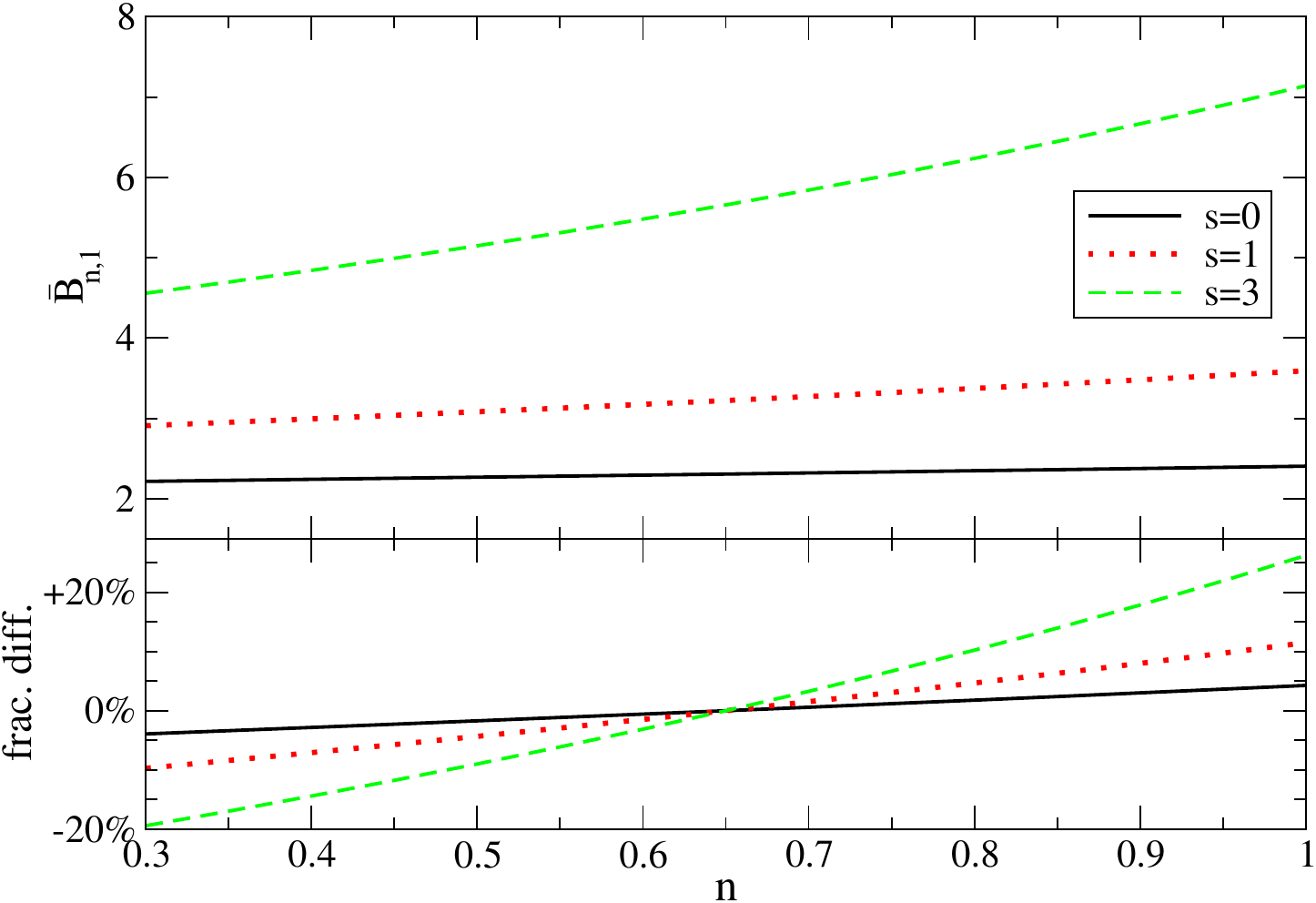}  
\includegraphics[width=8cm,clip=true]{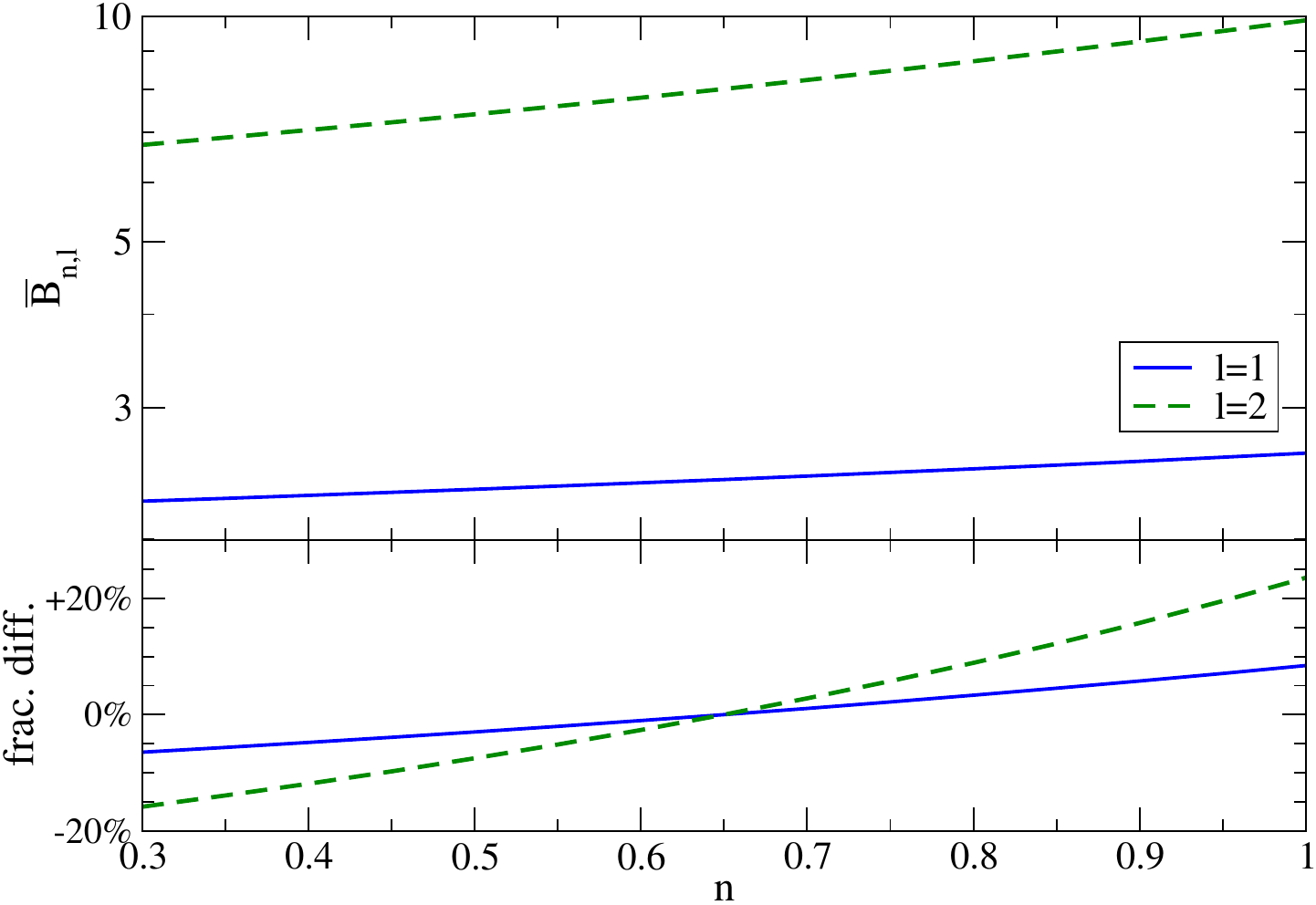}  
\caption{\label{fig:self-similar-break} (Left) Coefficient $\bar{B}_{n,1}$ (Eq.~\eqref{eq:Bbar-ecc}) for different values of the exponent $s$ in Eq.~\eqref{eq:ecc-dep-0} as a function of the polytropic index $n$ (top) and fractional difference between the coefficient and its value at $n=0.65$ (bottom). Observe how the universality deteriorates with large $s$. 
(Right) Coefficient $\bar{B}_{n,\ell}$ for different values of $\ell$ for the case where the eccentricity is given by Eq.~\eqref{eq:ecc-dep} (top), and fractional difference between the coefficient and its value at $n=0.65$ (bottom). Observe that in the $\ell=1$ case the fractional difference is very small, unlike in the left panel of this figure.
}
\end{center}
\end{figure*}

The degree of universality that is preserved depends sensitively on the functional form of $e(\tilde r)$ or equivalently $f(\tilde r/a_{1})$. Let us first review the case
\be
\label{eq:ecc-dep-0}
e(\tilde{r}) = e_{0} \left(\frac{\tilde{r}}{a_{1}}\right)^{s}\,,
\ee
where $s$ is a constant. When $s > 1$, the eccentricity is small near the core, growing rapidly as $\tilde r \gtrsim a_{1}/2$, while when $s < 1$ the eccentricity grows very rapidly right outside the core. The left panel of Fig.~\ref{fig:self-similar-break} shows the coefficient $\bar{B}_{n,\ell}$ of Eq.~\eqref{eq:Bbar-ecc} for the $\ell=1$ case and different values of $s$. Observe how the universality deteriorates greatly with large $s$, leading to 30\% variability in the relations when $s=3$. 
The model of the eccentricity profile in Eq.~\eqref{eq:ecc-dep-0}, however, is not realistic. A much more suitable model is
\be
\label{eq:ecc-dep}
e(\tilde{r}) = e_{0} \left\{1 - \delta e \left[ \left(\frac{\tilde{r}}{a_{1}}\right)^{s}-1 \right]\right\}\,,
\ee
where this time the constants $s$ and $\delta e$ are determined by fitting the above expression to the eccentricity profile of numerically constructed neutron stars with a given equation of state; for Newtonian polytropic stars with $n \in (0,1)$, $s \in (3,3.5)$ and $\delta e \in (0,0.2)$. The latter implies that the eccentricity profile only changes by $\lesssim 20\%$ inside neutron stars, since $\delta e$ represents the fractional difference in the eccentricity profile between that at the center and at the surface. The right panel of Fig.~\ref{fig:self-similar-break} shows the coefficient $\bar{B}_{n,\ell}$ for two different values of $\ell$ using the eccentricity expression of Eq.~\eqref{eq:ecc-dep}. Observe that this time the fractional difference is much smaller than when the eccentricity was prescribed via Eq.~\eqref{eq:ecc-dep-0}. This is as expected, since the profile of Eq.~\eqref{eq:ecc-dep} is fitted to numerically constructed neutron star solution, which we already know is approximately constant throughout the star. 

%%%%%%%%%%%%%%%%%%%%%%%%%%%%%%%%%%%%%%%%%%%%%%%%%%%%%%
\subsection{The Black Hole Limit}
\label{sec:BH-limit}

Another method to study the origin of the universality of the I-Love-Q relations is to consider the relativistic or black hole limit of the approximate no-hair relations. We define here the black hole limit as the limit in which the \emph{compactness} of the star reaches the critical value $C_{\rm BH} = 1/2$. The compactness is here defined as $C_{\rm e} \equiv M/R_{\rm e}$, where $R_{\rm e}$ is the equatorial radius of the star. For a black hole, this quantity can be defined via $R_{\rm e} = r_{\rm H} + a^{2}/r_{\rm H}$, where $r_{\rm H}$ is the location of the horizon in Boyer-Lindquist coordinates and $a$ is the Kerr spin parameter, which thus leads to $C_{\rm BH} = 1/2$ irrespective of the spin~\cite{Kleihaus:2016dui}. Alternatively, one can use Thorne's Hoop Conjecture\footnote{The Hoop Conjecture stipulates that if a body of mass $M$ is contained inside a circumference of $4 \pi M$, then the object will form a black hole~\cite{Thorne:1972ji,Gibbons:2009xm}.} to argue that the equatorial radius of relevance must be $2 M$ for a black hole to form, such that $C_{\rm BH} = 1/2$. 

This black hole limit is difficult to consider when studying a \emph{stationary} sequence of neutron star solutions. In General Relativity, it is well-known that neutron stars that are modeled as a perfect fluid achieve a maximum compactness that is below the black hole limit, i.e.~$C_{\rm NS} < C_{\rm BH}$. For example, for an incompressible fluid equation of state, the maximum compactness of a neutron star is $C_{\rm incomp} = 4/9$, at which point the central pressure diverges. For other equations of state, the maximum compactness is smaller than this number, i.e. any regular and thermodynamically stable perfect fluid star must have $C < 4/9$. This is called the Buchdahl limit~\cite{Buchdahl:1959zz,Bondi:1964zz}.

The natural way to approach the black hole limit is to relax the requirement of stationarity. One can imagine numerically evolving the collapse of a neutron star into a black hole, starting from an initially unstable configuration, and monitoring the I-Love-Q relations or the approximate no-hair relations dynamically. One possible problem is the definition of multipole moments in a dynamical spacetime. The Geroch-Hansen moments~\cite{Geroch:1970cc,Geroch:1970cd}, for example, are only defined in stationary spacetimes and typically require resolving the metric near spatial infinity. It is probably for this reason that this line of study has not yet been pursued. 

Another way to approach the black hole limit is to relax the requirement that neutron stars be described by a perfect fluid. This can be achieved by considering \emph{anisotropic stars}, i.e.~stars with a stress-energy tensor that deviates from the perfect fluid model through the introduction of a diagonal anisotropic tensor that allows for different pressures along different principal directions of the star's ellipsoidal shape (see Sec.~\ref{sec:anisotropy}). For realistic equations of state, static stable stars have a compactness of $C \lesssim 2/5$ even with strongly anisotropic pressure ($\lambda_\BL \approx - \pi$). On the other hand, incompressible static stars in the BL model have $C \to 1/2$ as $\lambda_{\BL}  \to -2 \pi$. 

\begin{figure*}[t]
\begin{center}
\includegraphics[width=8cm,clip=true]{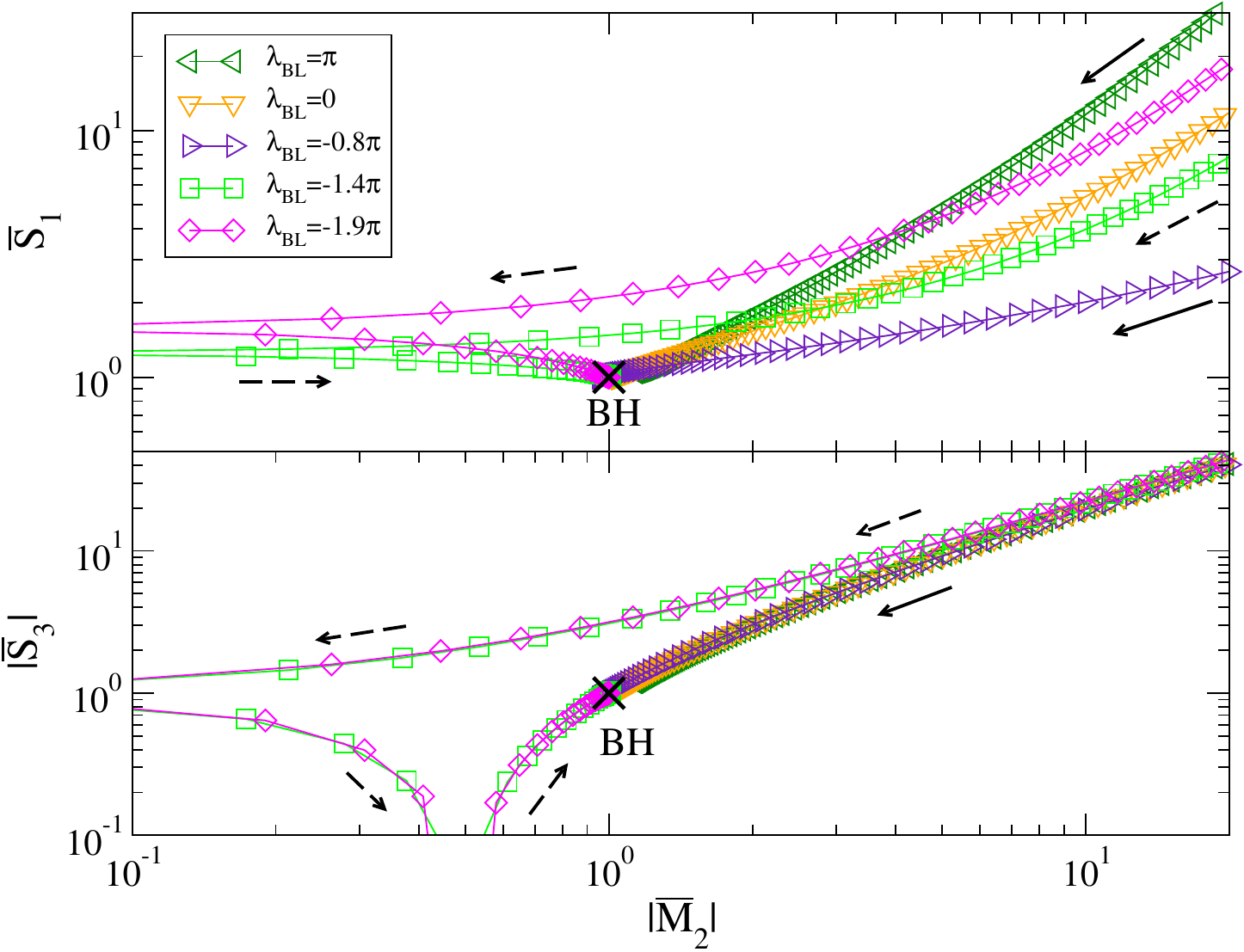}  
\includegraphics[width=8.1cm,clip=true]{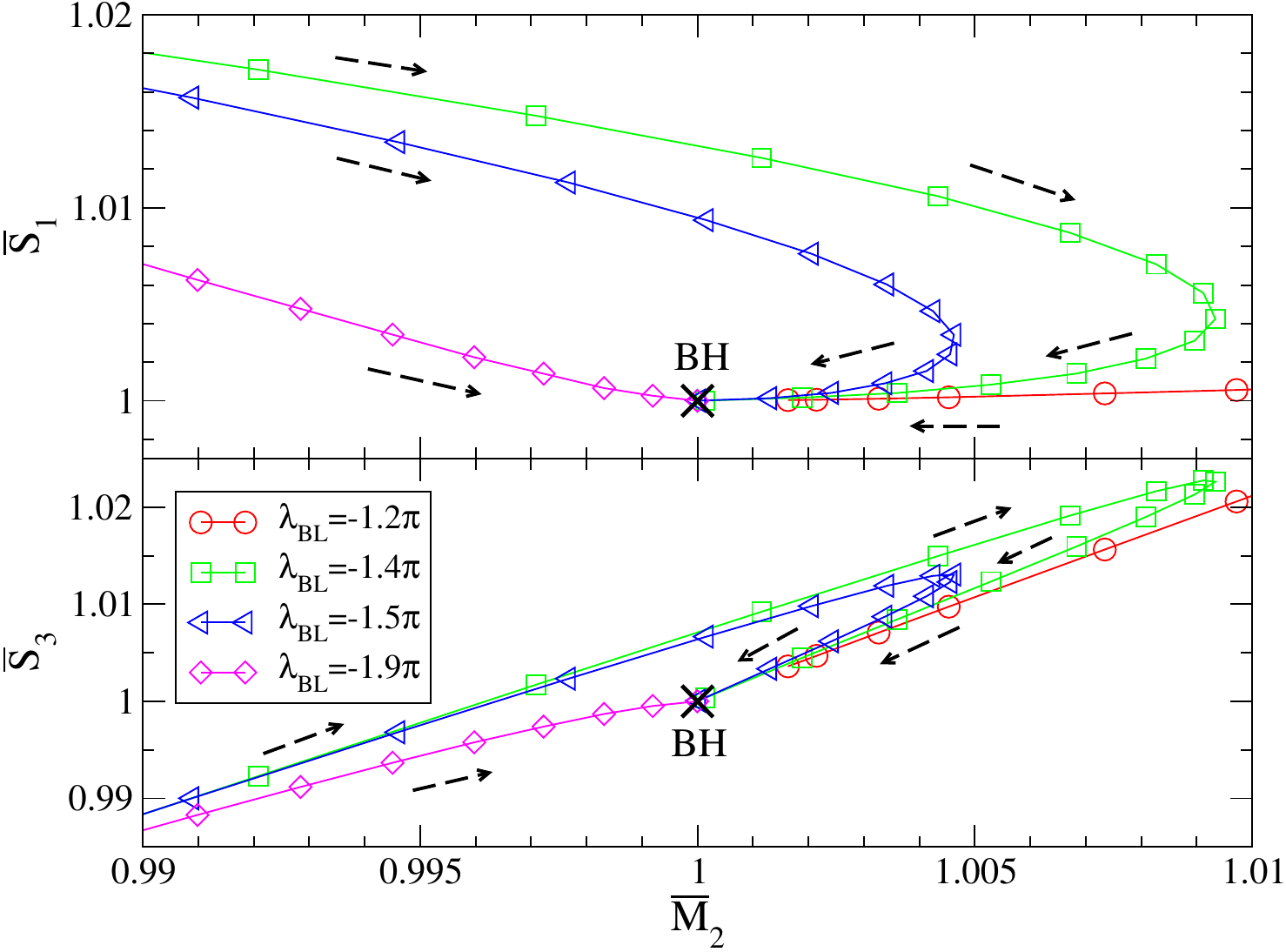}  
\caption{\label{fig:BH-limit} (Left) $\bar S_1$--$\bar M_2$ (top) and $\bar S_3$--$\bar M_2$ (bottom) relations for compact stars with constant density and various anisotropy parameters $\lambda_\BL$. The black crosses represent the black hole values. Arrows show the direction of increasing compactness, with $\lambda_\BL \leq -0.8 \pi$ (dashed) and $\lambda_\BL > -0.8 \pi$ (solid). 
(Right) Zooming into the region close to the black hole limit in the left panels. Observe how each sequence approaches the limit in a nontrivial way.
This figure is taken and edited from Yagi and Yunes~\cite{Yagi:2015upa}.
}
\end{center}
\end{figure*}

Yagi and Yunes~\cite{Yagi:2015upa,Yagi:2016ejg} considered anisotropic compact stars with constant density as a toy model to study the I-Love-Q and no-hair relations in the approach to the black hole limit. Figure~\ref{fig:BH-limit} shows some of the results of this study by considering the $\bar S_1$--$\bar M_2$ and $\bar S_3$--$\bar M_2$ relations in the black hole limit for different anisotropic magnitudes. Observe how the approximate no-hair relations for compact stars approach the exact black hole no-hair relations (marked with a cross) as the compactness increases (indicated by arrows). In particular, the approach to this limit is sensitive to the magnitude of the anisotropy parameter. Observe also that the behavior of the approach is different between the $\lambda_\BL \leq -0.8 \pi$ and the $\lambda_\BL > -0.8 \pi$ case. This is because the sign of $\bar M_2$ and $\bar S_3$ changes in the former case as one increases the stellar compactness.

One can analytically prove that $S_1$ \emph{reaches} the black hole value in the limit $C \to 1/2$ for a strongly-anisotropic constant density star with $\lambda_\BL = -2\pi$~\cite{Yagi:2015upa,Yagi:2016ejg}. In such a case, one can solve analytically the Einstein equations at linear order in spin both in the interior and exterior regions modulo integration constants. One then matches these solutions at the surface continuously and smoothly to determine these integration constants. In particular, the current dipole moment is directly related to the integration constant in the exterior region, which one finds depends on hypergeometric functions via~\cite{Yagi:2015upa,Yagi:2016ejg} 
\be
\label{eq:crit-exp-I}
\bar S_1 (C)  =  \frac{18\,{\it C}\,
{\mbox{$_2$F$_1$}(\frac{5}{2},{\frac {13}{4}};\,\frac{7}{2};\,2\,{\it C})}-9\,
{\mbox{$_2$F$_1$}(\frac{5}{2},{\frac {13}{4}};\,\frac{7}{2};\,2\,{\it C})}+5\,
{\mbox{$_2$F$_1$}(\frac{3}{2},\frac{9}{4};\,\frac{5}{2};\,2\,{\it C})} } {  8 C^2 \left[ 18\,{{
\it C}}^{2}
{\mbox{$_2$F$_1$}(\frac{5}{2},{\frac {13}{4}};\,\frac{7}{2};\,2\,{\it C})}-9\,{\it 
C}\,{\mbox{$_2$F$_1$}(\frac{5}{2},{\frac {13}{4}};\,\frac{7}{2};\,2\,{\it C})}+15\,{
\it C}\,{\mbox{$_2$F$_1$}(\frac{3}{2},\frac{9}{4};\,\frac{5}{2};\,2\,{\it C})}-5\,
{\mbox{$_2$F$_1$}(\frac{3}{2},\frac{9}{4};\,\frac{5}{2};\,2\,{\it C})} \right] }\,, 
\ee
where $\bar S_1 \equiv S_1/S_{1,\BH}$.
Expanding this about $C=C_\BH = 1/2$, one finds $\bar S_1 (C)  =  1 - 10 (C-C_\BH) + 56 (C-C_\BH)^2  +  \mathcal{O}\left[  (C-C_\BH)^3 \right]$. Thus, $\bar S_1(C \to 1/2) = 1$, which agrees with the black hole result.

\begin{figure}[t]
\begin{center}
\includegraphics[width=8.2cm,clip=true]{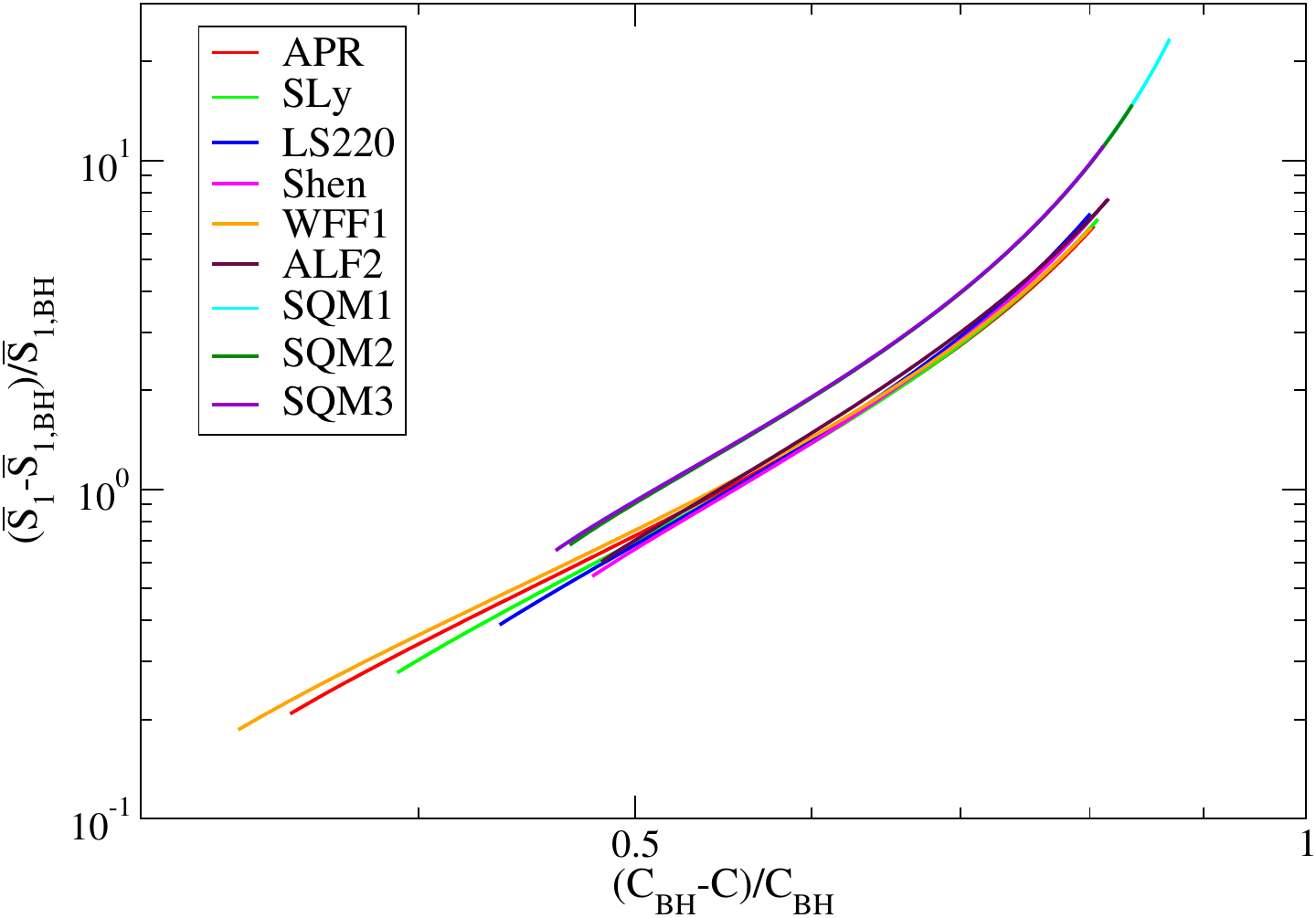}  
\includegraphics[width=8.cm,clip=true]{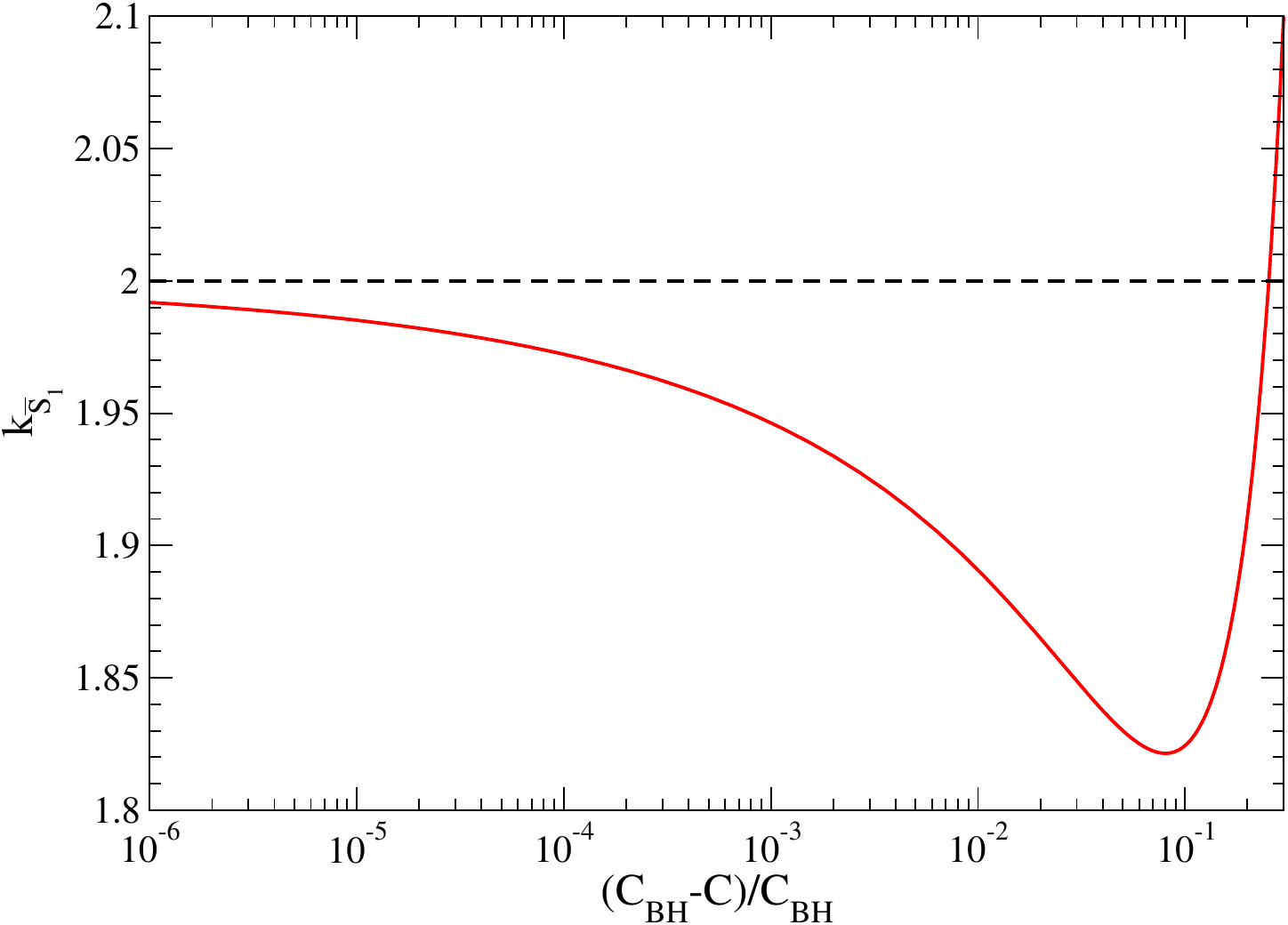}  
\caption{\label{fig:I-C-crit-iso} (Left) Fractional difference in $\bar S_1$ from the black hole value as a function of the fractional difference in the stellar compactness for isotropic neutron stars with various equations of state. Observe that slopes of these curves are similar for $(C_\BH-C)/C_\BH \lesssim 0.7$.
(Right) Scaling exponent of $\bar S_1$ defined in Eq.~\eqref{eq:exponent} as a function of the fractional difference in the compactness from the black hole value for a constant density, anisotropic star with $\lambda_\BL = -2 \pi$. Such a scaling exponent is extracted from the analytic expression in Eq.~\eqref{eq:crit-exp-I}. Observe that the exponent approaches 2 (black dashed) as one approaches the black hole limit ($\tau \to 0$). The right panel of this figure is taken and edited from Yagi and Yunes~\cite{Yagi:2016ejg}.
}
\end{center}
\end{figure}

The approximate universality of the no-hair relations is preserved in the presence of anisotropy (see Sec.~\ref{sec:anisotropy} and Fig.~\ref{fig:anisotropy}), and given that the black hole limit is a single point in phase space, it is possible that the approach to this critical point also presents universality. Yagi and Yunes~\cite{Yagi:2015upa,Yagi:2015hda} studied this possibility for isotropic stars and found that indeed the slope at which the universality is approached seems to be fairly independent of the equation of state. The left panel of Fig.~\ref{fig:I-C-crit-iso} shows the approach of the dipole moment to the black hole limit for various equations of state (similar behavior is observed for higher order multipole moments)~\cite{Yagi:2015upa}. Although the black hole limit cannot be cleanly resolved for isotropic stars, observe that all equations of state seem to approach the black hole limit with roughly the same slope. One can quantify this behavior further by defining the slope via
\be
\label{eq:exponent}
k_{\bar S_1} \equiv \frac{d \ln \Delta \bar S_1}{d \ln \tau}\,, \qquad \tau \equiv \frac{C_\BH - C}{C_\BH}\,, \quad \Delta \bar S_1 \equiv \frac{\bar S_1 - \bar S_{1,\BH}}{\bar S_{1,\BH}}\,,
\ee
in which case one finds that $k_{\bar S_1} \approx -3.9$ with a 10\% equation-of-state variation~\cite{Yagi:2015upa}. The right panel of Fig.~\ref{fig:I-C-crit-iso} presents $k_{\bar S_1}$ as a function of $\tau$ for a strongly-anisotropic, constant density star with $\lambda_\BL = -2 \pi$, calculated analytically from Eq.~\eqref{eq:crit-exp-I}. Observe that $k_{\bar S_1} \to 2$ as $C \to C_\BH$, which is different from the isotropic value. Such universal behavior in the approach of a critical point suggests an analogy with phase transitions and with critical behavior in the collapse of a compact star to a black hole, although much more work is needed to establish such an analogy rigorously.  

%%%%%%%%%%%%%%%%%%%%%%%%%%%%%%%%%%%%%%%%%%%%%%%%%%%%%%
\subsection{An Approximate Emergent Symmetry}
\label{sec:emergent-symmetry}

The two sets of analyses described above, in the Newtonian and in the black hole limits, begin to paint a physical, albeit phenomenological, picture postulated in Yagi \et~\cite{Yagi:2014qua} for the reason the universality holds. Imagine a phase space spanned by quantities that characterize stars, e.g.~their compactness, their temperature, their differential rotation rate, their equation of state, etc. In principle, this is a space of very large dimension, but as one considers more and more compact stars, a subspace of much smaller dimension becomes more and more important\footnote{We do not mean that the phase space of solutions changes dimensionality as the star becomes more compact. Rather, we mean that some physical variables, such as temperature, affect the structure of compact stars less than the structure of non-compact stars.}.

\begin{figure}[tb]
\centering
\includegraphics[width=8.cm,clip=false]{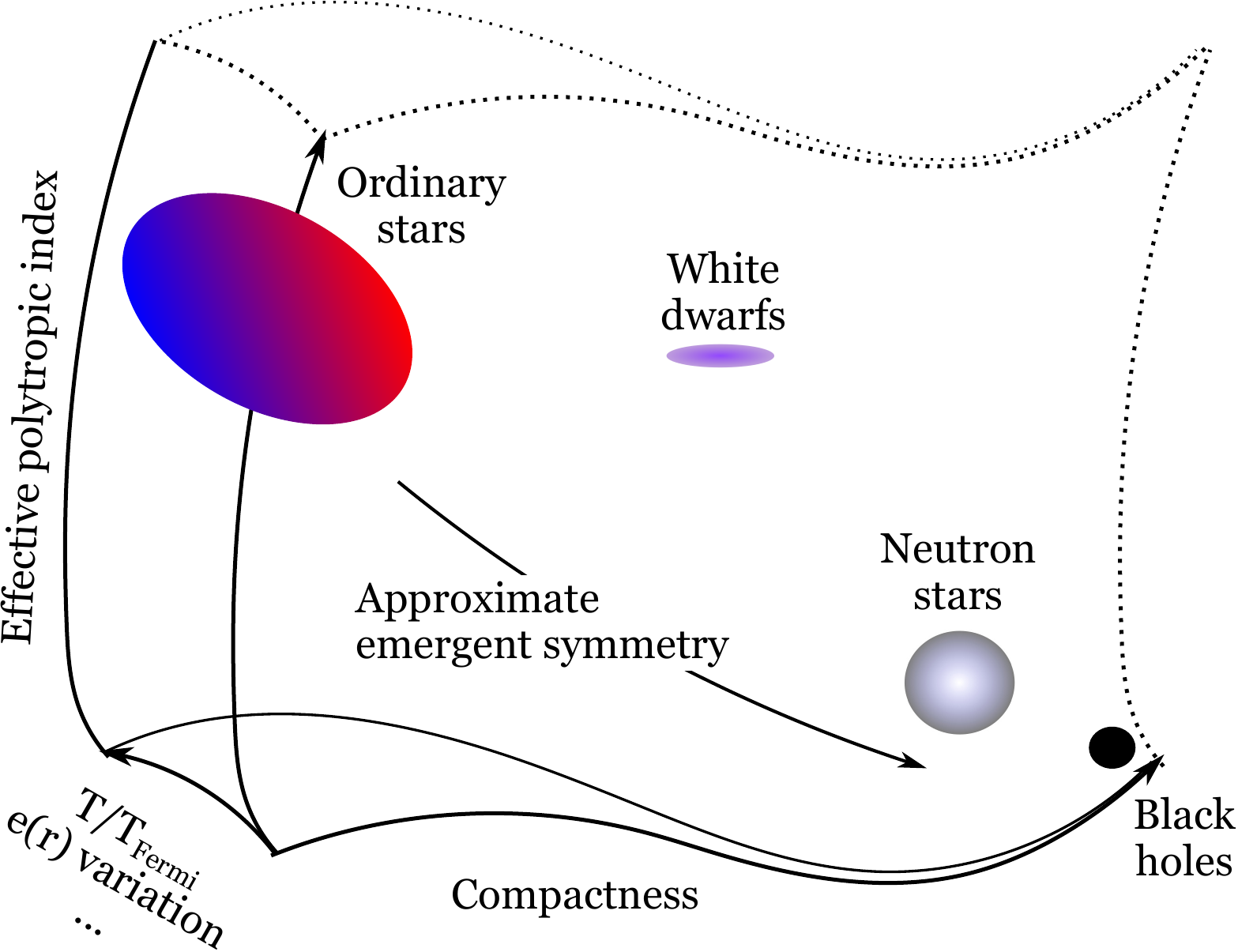}
\caption{ 
\label{fig:emergent-symmetry}
Schematic diagram of the stellar phase space. Compact objects live in one corner of this space, while non-compact stars live in another corner. As one flows from the latter to the former, degrees of freedom other than the polytropic index and compactness are suppressed. Then, an approximate self-similarity in isodensity contours emerges, which is responsible for the universality in no-hair relations for compact stars. This figure is taken from Yagi \et~\cite{Yagi:2014qua}.
}
\end{figure}

Let us then imagine the two-dimensional subspace spanned by the compactness and an effective equation of state polytropic index $n$, as shown in Fig.~\ref{fig:emergent-symmetry}. Realistic equations of state are obviously not polytropic, but they can be effectively represented as piecewise polytropes~\cite{Read:2008iy}. Regular non-compact stars, such as supergiants with $C \sim 10^{-8}$ and temperatures of around $10^4$K, exist in the small compactness region. The interior density of these stars is subnuclear, and thus, the equation of state can be captured through computer simulations, experimental data and  helio-seismological observations, all indicating an effective polytropic index of $n>1$. Neutron stars, on the other hand, exist in the large compactness $C \in (0.1,0.4)$ and $n \in (0.5,1)$ region, with temperatures much smaller than their Fermi temperature. Isolated neutron stars typically rotate rigidly due to the absence of external perturbations, i.e.~vorticity and differential rotation are unsourced in the barotropic limit.

As one flows from the non-compact stellar region to the neutron star region, the other dimensions of the space become less and less important, effectively shrinking as shown in Fig.~\ref{fig:emergent-symmetry}. These other dimensions are related to the microphysics of the neutron star, which effectively efface away at high compactness, influencing the main characteristics of the neutron star only mildly. It is because of this effacing that the equation of state of neutron stars can be treated as barotropic and modeled through a set of piecewise polytropes. 

As this effacement occurs, the radial eccentricity profiles become less and less variable, becoming nearly constant in the region of the star's interior that contributes the most to the low-order multipole moments. This occurs because such a nearly constant eccentricity profile minimizes the energy of the system~\cite{Lai:1993ve}. Thus, as one considers more and more compact stars, an approximate symmetry, isodensity self-similarity, emerges. The emergence of an approximate symmetry is most probably what causes the approximate universal behavior in the approximate no-hair relations for compact stars. 

\begin{figure}[tb]
\centering
\includegraphics[width=8.4cm,clip=false]{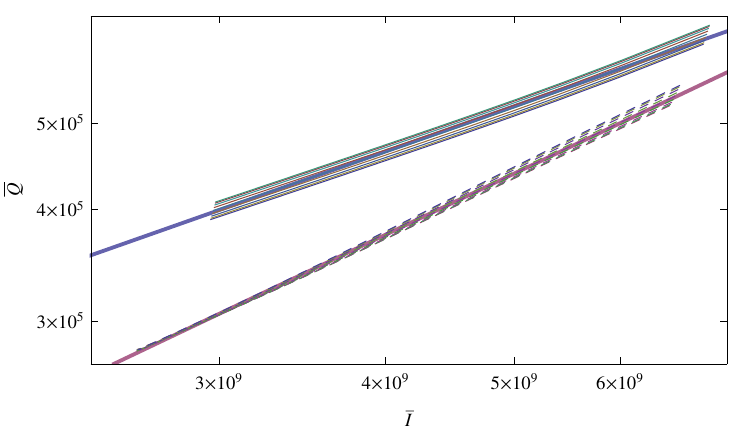}
\includegraphics[width=7.7cm,clip=false]{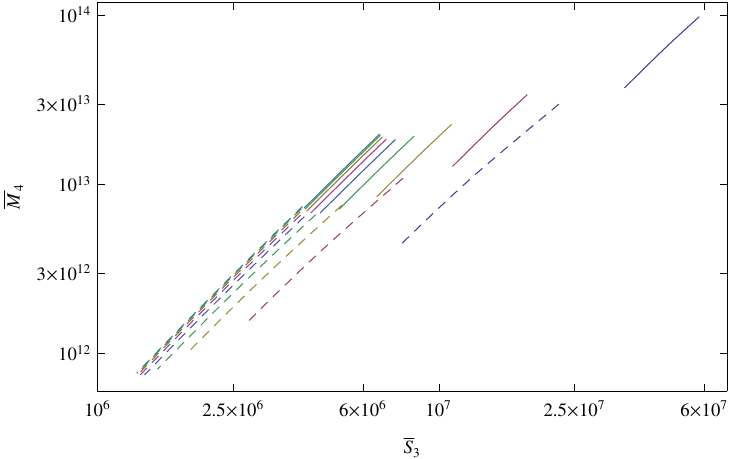}
\caption{ 
\label{fig:noncompact}
(Left) The I-Q relation for rotating noncompact stars with Kramer's (thin solid) and OPAL (thin dashed) opacity law, and $M \in (2,10)M_\odot$. Within each opacity family, each curve corresponds to the stellar rotation of $\Omega_s/\Omega_\mrm{bk}$ from 0.1 to 0.9 in increments of 0.1 that corresponds to $\chi \in (3,17.5)$, where $\Omega_s$ and $\Omega_\mrm{bk}$ are spin angular velocities at the surface and for breakup respectively. Thick solid curves are the fit to the relations within each opacity family. Observe that the relation depends sensitively on the opacity law, while it is less sensitive to spin.
(Right) The $\bar M_4$--$\bar S_3$ relation for non-compact stars with Kramer's (thin solid) and OPAL (thin dashed) opacity law. The stellar rotation of $\Omega_s/\Omega_\mrm{bk}$ increases from 0.1 to 0.9 in increments of 0.1 from left to right. Observe that the universality is lost for rapidly-rotating noncompact stars.
This figure is taken from Yagi \et~\cite{Yagi:2014qua}.
}
\end{figure}

Such an approximate symmetry is not present in non-compact stars, as the eccentricity can vary drastically in their interiors. Yagi \et~\cite{Yagi:2014qua} constructed rotating non-compact stellar solutions with the publicly-available ESTER code~\cite{Rieutord:2012rt,Lara:2012wv} and found that the eccentricity varies by 300--600\% within the stars, a variation much larger than 20\% found in neutron stars (see Sec.~\ref{sec:Why-I-Love-Q-Newton}). Figure~\ref{fig:noncompact} presents the I-Q and $\bar M_4$--$\bar S_3$ relations for such rotating, non-compact stars with two opacity laws, which effectively corresponds to changing equations of state. Observe that these relations are very sensitive to the opacity law. For example, within each opacity family, the I-Q relation only varies by a few \% from the fit, while the two fits differ by 40\%. Such a finding is consistent with the approximate emergent symmetry argument; the amount of universality deteriorates as the eccentricity variation increases.

When one pushes beyond the neutron star region of subspace toward the black hole region during gravitational collapse, the approximate symmetry becomes more and more accurate, leading to exact universal relations. The latter are simply a manifestation of the black hole no-hair theorems discussed earlier. This extension to the black hole limit, however, is subtle because black holes are vacuum solutions, and thus there is no meaning in isodensity profiles in their interior. One is here imagining the process of gravitational collapse, in which matter is present in the black hole interior right after a dynamical horizon has formed, although this matter will continue to contract toward the singularity.    

This physical picture is consistent with the mathematical reasoning recently presented in Chatziioannou \et~\cite{Chatziioannou:2014tha}, Sham \et~\cite{Sham:2014kea} and Chan \et~\cite{Chan:2015iou} and that we reviewed in Secs.~\ref{sec:no-hair-Newton},~\ref{sec:oscillation} and~\ref{sec:I-Love-Q-C-analytic}. These references showed that the I-Love and 3-hair relations are perturbatively insensitive with respect to changes in the equation of state about the incompressible limit (polytropic index $n=0$). 
This mathematical result supports the physical model described above because isodensity self-similarity becomes exact in the incompressible limit. Some care must be taken, however, since realistic equations of state cannot be well-modeled in the incompressible limit, as e.g.~showed in Read \et~\cite{Read:2008iy}. 

The physical picture described above is also consistent with recent numerical simulations of proto-neutron stars~\cite{Martinon:2014uua}, as already discussed in Sec.~\ref{sec:non-barotropic}. This reference studied the evolution of proto-neutron stars shortly after their birth, during which there are large thermodynamical gradients and strong neutrino emission. During this highly dynamical phase lasting roughly one second, Martinon \et~\cite{Martinon:2014uua} found that the eccentricity profiles are not approximately constant and the I-Love-Q relations are different from those of cold neutron stars with barotropic equations of state. However, soon after the entropy gradients relax to a smooth configuration, the eccentricity profiles become nearly constant and the original I-Love-Q relations are re-established. 

%%%%%%%%%%%%%%%%%%%%%%%%%%%%%%%%%%%%%%%%%%%%%%%%%%%%%%%%
\section{I-Love-Q in Modified Gravity and Exotic Compact Objects}
\label{sec6:ILQ-Mod-Grav}

Up until now, we have focused on universal relations valid in General Relativity, where the action is given by
\be
S = S_\EH + S_\mat\,, \qquad S_\EH \equiv \kappa \int d^4 x \; \sqrt{-g} \; \mathcal R\,,
\ee
with $S_\EH$ the Einstein-Hilbert action, $g$ and $\mathcal R$ the metric determinant and the Ricci scalar, $\kappa \equiv (16 \pi)^{-1}$, and $S_\mat$ the action for the matter fields. In this section we review what is known about the I-Love-Q relations in theories other than General Relativity and for exotic compact objects within General Relativity (see e.g.~\cite{Yunes:2013dva,Berti:2015itd} for recent reviews of modified gravity).

As we shall see, the I-Love-Q relations in modified gravity remain approximately equation-of-state insensitive and differ from their General Relativistic counterpart by an amount proportional to the coupling constants of the modified theory. For theories that have already been stringently constrained by Solar System, binary pulsar or cosmological observations, such as certain scalar-tensor theories and quadratic gravity models, the modified gravity I-Love-Q relations will be very close to their General Relativistic counterpart. For other theories that are only weakly constrained by Solar System experiments, such as certain parity-violating gravity models, the modified gravity I-Love-Q relations will differ significantly from their General Relativistic counterpart, even when saturating current constraints. We will also see that the I-Love-Q relations in exotic compact objects, such as gravastars, are both qualitatively and quantitatively different from those for neutron stars and quark stars.

%-----------------------------
\subsection{Dynamical Chern-Simons Gravity}
\label{sec:dCS}

Dynamical Chern-Simons (dCS) gravity is a theory that introduces parity-violation into the gravitational interaction. Motivated from heterotic superstring theory~\cite{Polchinski:1998rq,Polchinski:1998rr}, loop quantum gravity~\cite{Alexander:2004xd,Taveras:2008yf,Calcagni:2009xz} and effective field theories of inflation~\cite{Weinberg:2008hq}, the dCS action is given by~\cite{Jackiw:2003pm,Smith:2007jm,Alexander:2009tp}
\be
S = S_\EH + S_\Pont + S_\vartheta + S_\mat\,, 
\qquad 
S_\Pont \equiv \frac{\alpha_\CS}{4} \int d^4 x \sqrt{-g} \vartheta \pont\,, 
\quad 
S_\vartheta \equiv -\frac{\beta}{2} \int d^4 x \sqrt{-g} \left[ \nabla_\mu \vartheta \nabla^\mu \vartheta + 2 V(\vartheta) \right]\,,
\ee
where $\vartheta$ is a pseudo-scalar field, $\alpha_\CS$ and $\beta$ are coupling constants, $V(\vartheta)$ is a potential for $\vartheta$ and $\pont \equiv \mathcal R_{\nu \mu \rho \sigma} {}^* \mathcal R^{\mu\nu\rho\sigma}$ is the Pontryagin density, which depends on the dual Riemann tensor ${}^* \mathcal R^{\mu\nu\rho\sigma} \equiv (1/2) \epsilon^{\rho \sigma \alpha \beta} \mathcal R^{\mu \nu}{}_{\alpha\beta}$, defined in terms of the Levi-Civita tensor $\epsilon^{\rho \sigma \alpha \beta}$. 

The current most stringent constraints on dCS gravity come from Solar System observations of frame-dragging with LAGEOS~\cite{Ciufolini:2004rq} and Gravity Probe B~\cite{Everitt:2011hp,Everitt:2015qri} and from table top experiments~\cite{Kapner:2006si}. The simplest way to quantify these constraints is to assume $\vartheta$ is dimensionless, such that $\alpha_\CS$ and $\beta$ have units of (length)$^2$ and (length)$^0$ respectively, and we  can define the characteristic length scale $\xi_\CS^{1/4} \equiv [\alpha_\CS^{2}/(\kappa \beta)]^{1/4}$. DCS gravity reduces to General Relativity in the $\xi_\CS \to 0$ limit. Current observations then constrain $\xi_\CS^{1/4} \lesssim 10^8$km~\cite{AliHaimoud:2011fw,Yagi:2012ya}.

DCS gravity arises from more fundamental theories of gravitation in a truncated low-curvature/low-energy expansion, and thus, it must be understood as an effective field theory. The dCS action is defined by including up to quadratic order terms in the curvature, and thus, one must require that the coupling constants be such that these terms lead to \emph{small deformations} from General Relativistic predictions in any relevant observation. This means that such an effective action is not valid in regimes of extreme curvature, where all curvature terms are equally important. To ensure we are within the regime of validity of the effective field theory description, most studies have worked in the \emph{small coupling approximation}: $\xi_\CS M^2/R^6 \ll 1$~\cite{AliHaimoud:2011fw,Yagi:2013mbt}. Such a treatment ensures the well-posedness of the initial value problem of the theory~\cite{Delsate:2014hba}. 

Before proceeding, one must typically choose a potential for the scalar field. From a heterotic string theory viewpoint, the field $\vartheta$ is an axion, one of the many moduli fields in string theory that has a potential with many flat directions. If the potential is flat, then the field is massless and one can just set $V(\vartheta)=0$. If the potential is zero, then the field is \emph{shift-symmetric}, i.e.~the Pontryagin density can be rewritten as a total derivative~\cite{Alexander:2009tp}, and hence, $S_\Pont$ becomes shift invariant upon integration by parts.

\begin{figure}[htb]
\begin{center}
\begin{tabular}{l}
\includegraphics[width=8.1cm,clip=true]{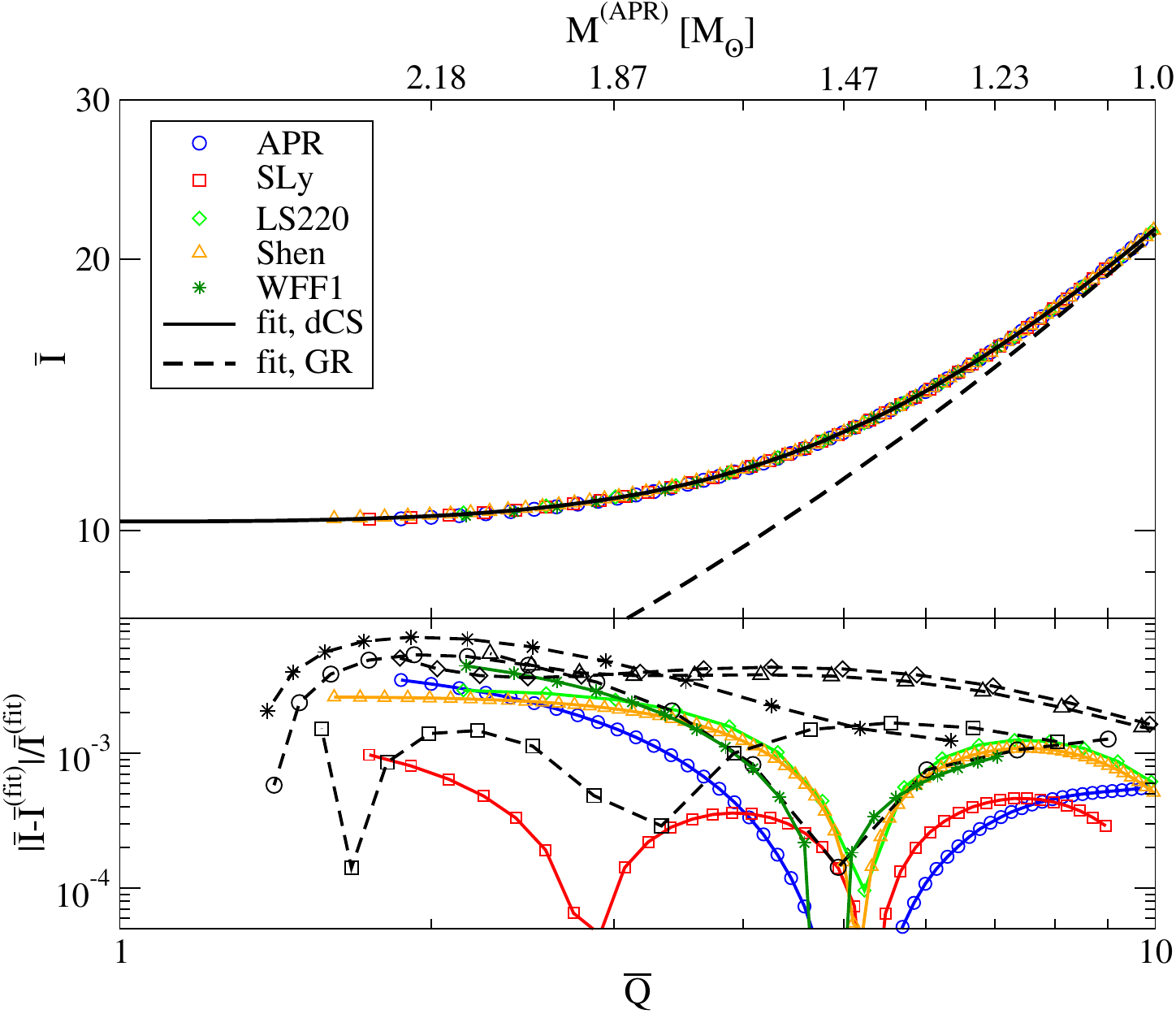} 
\includegraphics[width=8.cm,clip=true]{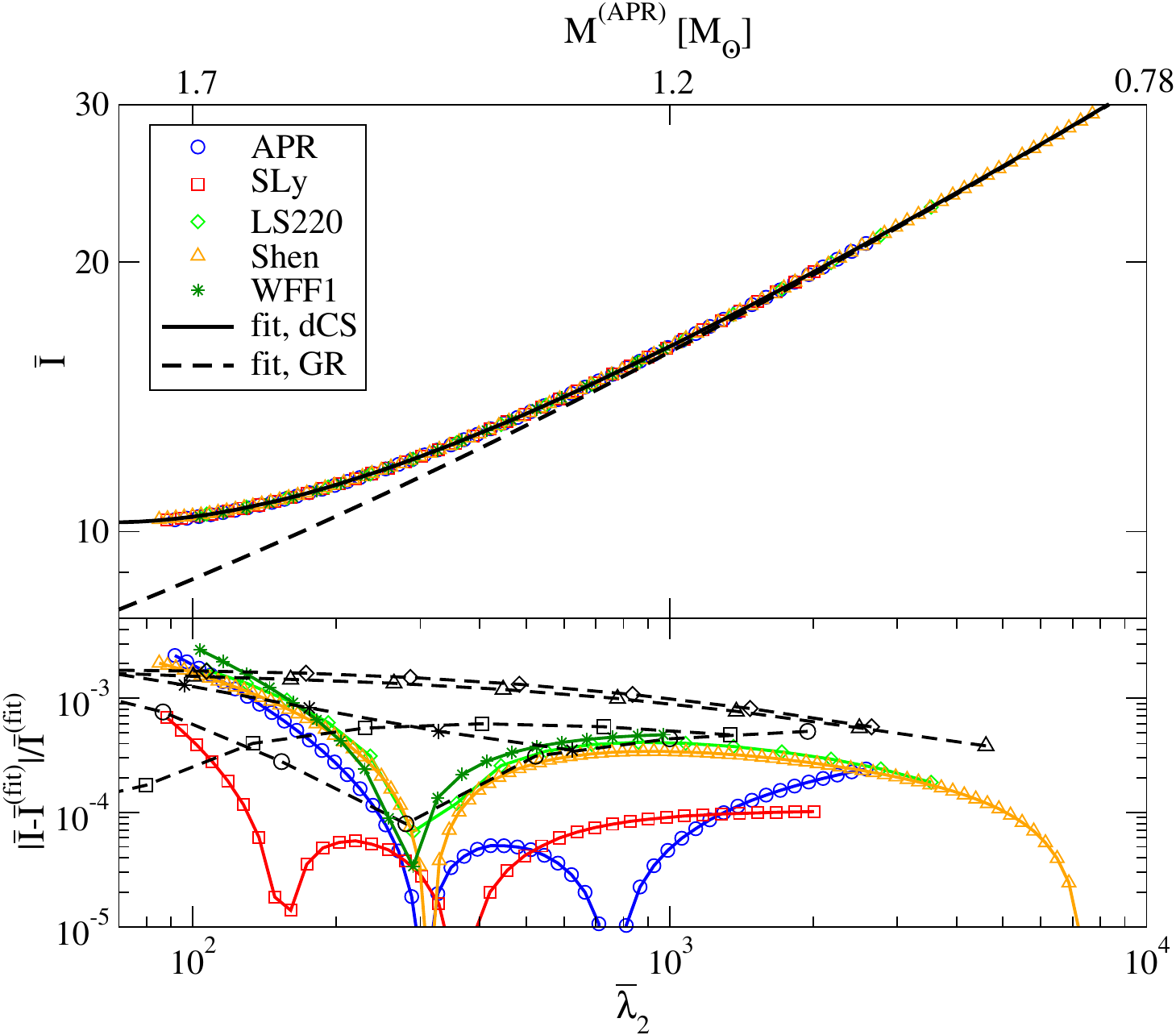} 
\end{tabular}
\caption{\label{fig:I-Love-Q-CS} 
(Top) The I-Q (left) and I-Love (right) relations in dCS gravity with a coupling constant fixed to $\xi_\CS/M^4 = 10^3$ for various equations of state. We also show a fit to the dCS data, together with the fit to the General Relativistic data given in Sec.~\ref{sec:I-Love-Q-relativistic}. The relations in dCS are different from those in General Relativity, but they remain approximately universal within each theory. For reference, we show the neutron star mass in General Relativity with an APR equation of state on the top axis. (Bottom) Absolute fractional difference from the fit. Black dashed curves correspond to the General Relativistic case. Observe that the degree of universality in dCS is comparable to that in General Relativity. These figures are taken from Majumder \et~\cite{barun}.
}
\end{center}
\end{figure}

Let us now look at the I-Love-Q relations in dCS gravity. The neutron star moment of inertia and quadrupole moment, first calculated in Yunes \et~\cite{Yunes:2009ch}, Ali-Haimoud and Chen~\cite{AliHaimoud:2011fw} and Yagi \et~\cite{Yagi:2013mbt}, are different from the General Relativistic prediction. On the other hand, the electric-type tidal deformability is the same as in General Relativity due to parity considerations~\cite{Yagi:2011xp}. The top panels of Fig.~\ref{fig:I-Love-Q-CS} present the I-Q and I-Love relations in dCS gravity for a set of representative equations of state studied in Yagi and Yunes~\cite{Yagi:2013bca,Yagi:2013awa} and Majumder \et~\cite{barun}. The coupling parameter is fixed to $\xi_\CS/M^4 = 10^3$, which is allowed by current bounds. Observe that the dCS relations deviate from the General Relativistic ones in the relativistic regime (small $\bar Q$ or $\bar \lambda_2$). Such a large deviation allows one to place strong constraints on the theory with future observations, as we will review in Sec.~\ref{sec:application-gravitational}. The bottom panels show the amount of equation-of-state universality in both dCS and General Relativity. Observe that the degree of universality is comparable between the two theories. However, whether the relations remain universal depends on how one fixes the dimensionless coupling parameter. For example, if one fixes $\xi_\CS M^2/R^6$ instead of $\xi_\CS/M^4$, the amount of equation-of-state variation increases significantly~\cite{barun}.

%-----------------------------
\subsection{Einstein-Dilaton Gauss-Bonnet Gravity}

Einstein-Dilaton Gauss-Bonnet (EDGB) gravity is a theory that introduces quadratic-curvature modifications to the Einstein-Hilbert action. Motivated from heterotic superstring theory~\cite{Polchinski:1998rq,Polchinski:1998rr}, the EDGB action we consider here is given by~\cite{Jackiw:2003pm,Smith:2007jm,Alexander:2009tp}
\begin{align}
& \qquad \qquad \qquad\qquad S = S_\EH + S_{\EDGB} + S_\phi + S_\mat\,, 
\\\ 
S_{\EDGB} &\equiv \alpha_\EDGB \kappa \int d^4 x \sqrt{-g} e^{-\gamma_{\EDGB} \phi} \mathcal R_{\EDGB}^{2}\,, 
\quad 
S_\phi \equiv -\frac{\kappa}{2} \int d^4 x \sqrt{-g} \nabla_\mu \phi \nabla^\mu \phi \,,
\end{align}
where $\phi$ is a scalar field (the dilaton), $\alpha_\EDGB$ and $\gamma_{\EDGB}$ are coupling constants, $V(\phi)$ is a potential for $\phi$ and $\mathcal R_{\EDGB}^{2} = \mathcal R_{\mu \nu \rho \sigma} \mathcal R^{\mu\nu\rho\sigma} - 4 \mathcal R_{\mu \nu} \mathcal R^{\mu \nu} + \mathcal R^{2}$ is the Gauss-Bonnet density, which depends on the Riemann tensor, the Ricci tensor and the Ricci scalar. Notice that we have not included here a potential for the scalar field, following the same reasoning as in dCS gravity. 

\begin{figure}[htb]
\begin{center}
\begin{tabular}{l}
\includegraphics[width=8.cm,clip=true]{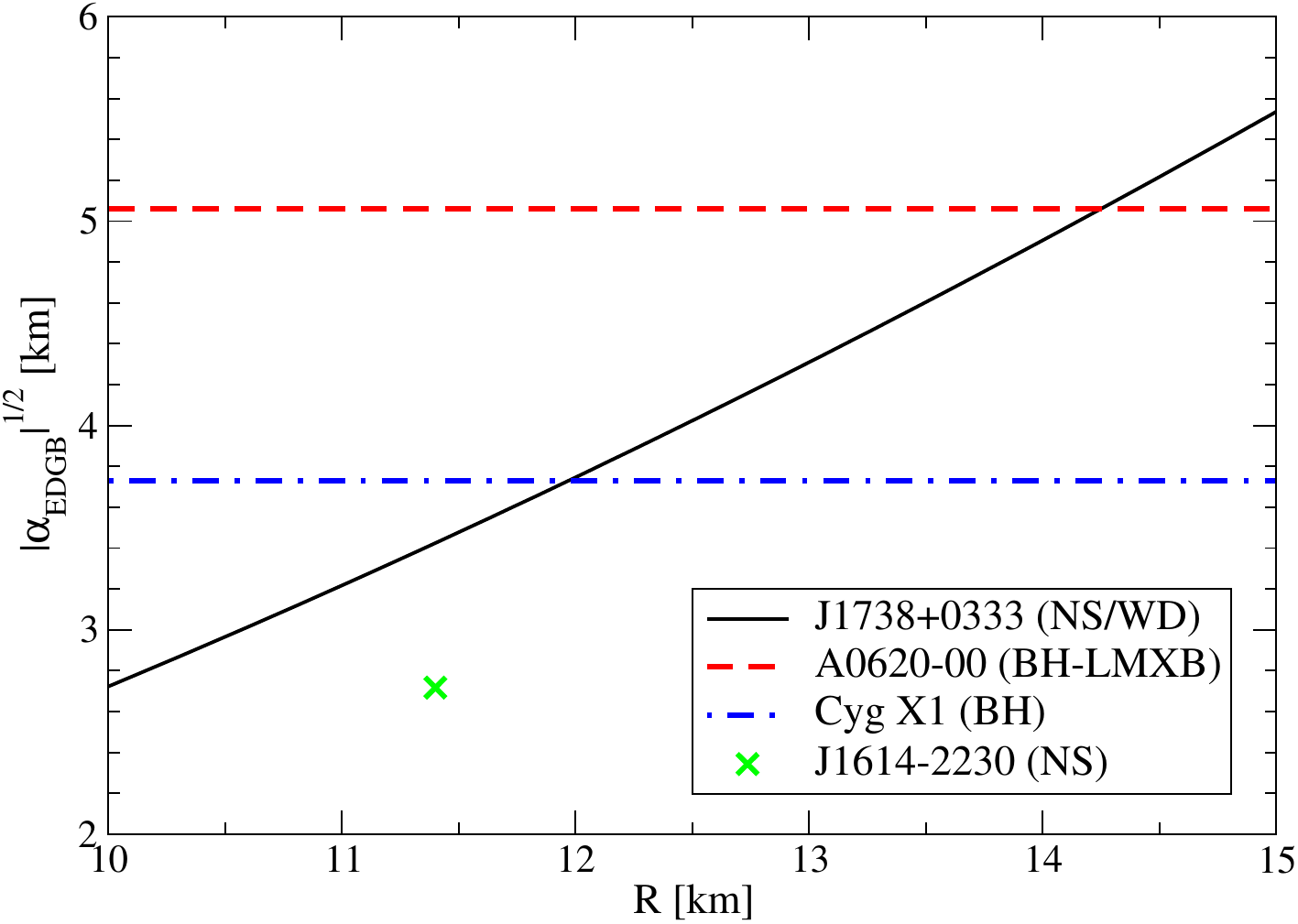} 
\end{tabular}
\caption{\label{fig:EDGB-constr}
Current constraints on the characteristic length scale of EDGB gravity with $\gamma_\EDGB=1$ coming from the observation of the neutron star/white dwarf binary J1738+0333~\cite{Freire:2012mg} (black solid) as a function of the unknown neutron star radius. The Tolman VII equation of state was assumed here, which can be used to effectively model realistic equations of state~\cite{Tolman:1939jz,Lattimer:2000nx}. Constraints from the black hole LMXB A0620-00~\cite{Yagi:2012gp} (red dashed) and the Cyg X1 black hole~\cite{Pani:2009wy,Pani:2011xm} (blue dotted-dashed) are also shown. The green cross corresponds to the bound derived from the existence of neutron star J1614-2230~\cite{1.97NS} obtained in Pani \et~\cite{Pani:2011xm} assuming the APR equation of state. 
}
\end{center}
\end{figure}

Current stringent constraints on EDGB gravity come from observations of the rate of change of the orbital period in a low-mass X-ray binary (LMXB) system~\cite{Yagi:2012gp} and the existence of stellar-mass black holes~\cite{Pani:2009wy}. An equation-of-state dependent constraint using the maximum mass of neutron stars was also derived in Pani \et~\cite{Pani:2011xm}. The simplest way to quantify this constraint is to set $\gamma_\EDGB = 1$ and to assume $\phi$ is dimensionless, such that $\alpha_\EDGB$ has units of (length)$^2$, and thus, the characteristic length scale is given by $|\alpha_\EDGB|^{1/2}$. EDGB gravity reduces to General Relativity in the $\alpha_\EDGB \to 0$ limit. Current constraints on $|\alpha_\EDGB|^{1/2}$ are summarized in Fig.~\ref{fig:EDGB-constr}, where observe that they are all $|\alpha_\EDGB|^{1/2} \lesssim (2.5-5.5)$km. 

As in the dCS gravity case, EDGB gravity also arises from more fundamental theories through a truncated low-curvature/low-energy expansion, and thus, it should also be understood as an effective field theory. Working in the small coupling approximation $\alpha_\EDGB^2 M^2/R^6 \ll 1$, one then seeks small deformations from General Relativistic solutions, and thus, one expands the exponential in $S_{\EDGB}$ via $\exp(-\gamma_{\EDGB} \phi) \sim 1 - \gamma_{\EDGB} \phi$. The first term in this expansion does not contribute to the field equations because the Gauss-Bonnet density is a topological invariant. The second term, however, does contribute and the modifications to the field equations are proportional to $\gamma_{\EDGB} \,\alpha_{\EDGB}$. Since there is no way to decouple this product, one typically sets $\gamma_{\EDGB} = 1$ and allows $\alpha_{\EDGB}$ to control the magnitude of the deformation. The field equations of EDGB gravity are manifestly second order even if the theory is taken to be exact, though the well-posedness and initial value problem of the theory remain mostly unexplored.

\begin{figure}[htb]
\begin{center}
\begin{tabular}{l}
\includegraphics[width=8.cm,clip=true]{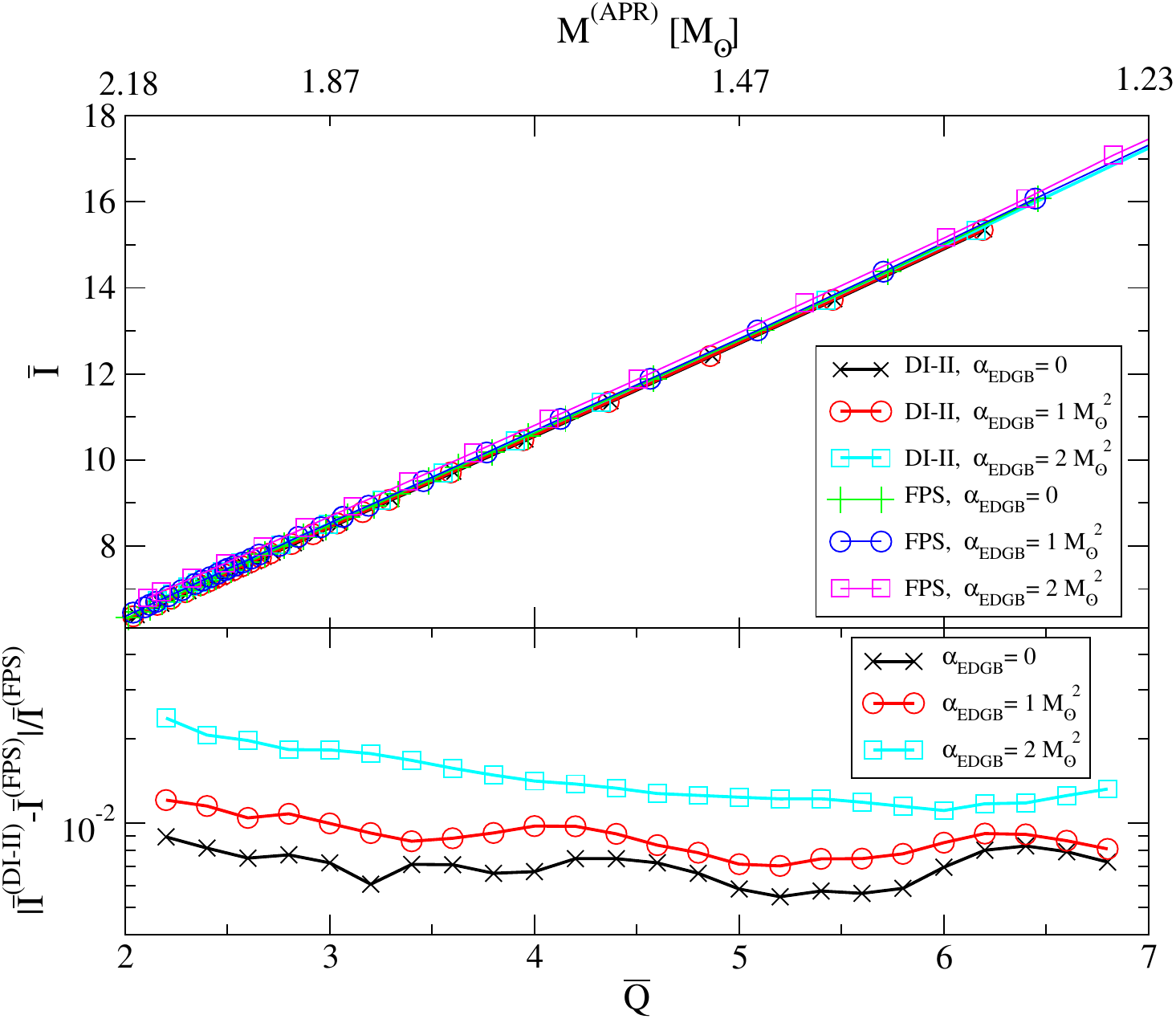} 
\end{tabular}
\caption{\label{fig:I-Love-EDGB}
(Top) The I-Q relation in EDGB gravity with a a couple of fixed EDGB coupling constants. For reference, the top axis shows the neutron star mass in General Relativity using the APR equation of state. The $\alpha_\EDGB=0$ line corresponds to the I-Q relation in General Relativity. (Bottom) Absolute fractional difference between the EDGB I-Q relations with an FPS equation of state and a DI-II equation of state for different fixed $\alpha_\EDGB$. Observe that the EDGB I-Q relation is still approximately universal, although the degree of universality is worse than in General Relativity. This figure is adapted and modified from the data presented in Kleihaus \et~\cite{Kleihaus:2014lba,Kleihaus:2016dui}.
}
\end{center}
\end{figure}

Let us now look at the I-Love-Q relations in EDGB gravity. The neutron star moment of inertia and quadrupole moment were calculated in Kleihaus \et~\cite{Kleihaus:2014lba,Kleihaus:2016dui}, while the electric-type tidal deformability has not yet been calculated. The top panel of Fig.~\ref{fig:I-Love-EDGB} shows the I-Q relation in EDGB gravity for two equations of state~\cite{Kleihaus:2014lba,Kleihaus:2016dui}: FPS~\cite{1989ASIB..205..103P} (derived using the variational method, also adopted to derive e.g.~AP4 discussed in Sec.~\ref{sec:Preliminaries}) and DI-II~\cite{1985ApJ...291..308D} [a polytropic-like equation of state with $n \approx 0.75$]. The I-Q relation is shown for three different values of EDGB coupling constants: $\alpha_{\EDGB} = 0$ (General Relativistic limit), $\alpha_{\EDGB} = M_{\odot}^{2} \approx 2.16 \; {\rm{km}}^2$ and $\alpha_{\EDGB} = 2 M_{\odot}^{2} \approx 4.32 \; {\rm{km}}^2$. Observe that the EDGB relation is quite similar to that in General Relativity because EDGB gravity has been stringently constrained by current observations; for larger values of $\alpha_{\EDGB}$, the EDGB I-Q relation would differ much more from the General Relativistic relation. The bottom panel shows the degree of universality of the I-Q relation in EDGB gravity and in General Relativity. As in the dCS case, observe that the relation is still equation-of-state insensitive, although the variation is larger than that in General Relativity. 

%-----------------------------
\subsection{Scalar-Tensor Theories}
\label{sec:ST}

Scalar-tensor (ST) gravity is a class of theories where the gravitational interaction mediated by both a massless graviton and a fundamental scalar field. Motivated from high-energy modifications of Einstein's theory~\cite{Sotiriou:2015lxa} and certain inflation models~\cite{Steinhardt:1990zx,Martin:2013tda}, the ST action in the \emph{Jordan frame} is given by~\cite{Pani:2014jra,Doneva:2014faa}
\be
\label{eq:ST-action-Jordan}
S = S_{\delta \EH} + S_\Phi + S_\mat\,, 
\qquad 
S_{\delta \EH} \equiv \kappa \int d^4 x \sqrt{-g} \, F(\Phi) \, \mathcal R \,, 
\quad 
S_\Phi \equiv - \kappa \int d^4 x \sqrt{-g} \, Z(\Phi) \left[ \nabla_\mu \Phi \nabla^\mu \Phi + V(\Phi) \right]\,,
\ee
where recall that $\kappa = (16 \pi)^{-1}$, $\Phi$ is the scalar field, $F(\Phi)$ and $Z(\Phi)$ are coupling functions, and $V(\Phi)$ is a potential for the scalar field. ST gravity is a class of metric theories, where the scalar field couples to matter only indirectly, through the metric tensor. 

The field equations derived from the action in Eq.~\eqref{eq:ST-action-Jordan} are quite involved, but one can always perform the following transformation into the \emph{Einstein frame} 
\begin{align}
g^{\E}_{\mu \nu} = F(\Phi) \; g_{\mu \nu}\,,
\qquad
A(\Phi^{\E}) = F^{-1/2}(\Phi)\,,
\qquad
V^{\E}(\Phi^{\E}) = \frac{V(\Phi)}{F^{2}(\Phi)}\,,
\qquad
\Phi^{\E} = \int \frac{d\Phi}{\sqrt{4 \pi}} \sqrt{\frac{3}{4} \frac{F'(\Phi)^{2}}{F(\Phi)^{2}} + \frac{1}{2} \frac{Z(\Phi)}{F(\Phi)}}\,,
\end{align}
to recast the action in Einstein-like form
\begin{align}
S = S^{\E}_{\delta \EH} + S^{\E}_\Phi + S^{\E}_\mat\,, 
\qquad 
S^{\E}_{\delta \EH} \equiv \kappa \int d^4 x \sqrt{-g^{\E}} \, \mathcal R^{\E} \,, 
\quad 
S_\Phi \equiv - \int d^4 x \sqrt{-g^{\E}} \,  \left[  \frac{1}{2}  \nabla^{\E}_\mu \Phi_{\E} \nabla_{\E}^\mu \Phi_{\E} + \kappa V^{\E}(\Phi^{\E}) \right]\,,
\end{align}
where $S^{\E}_{\mat}$ is the matter action, which now couples to the combination $A^{2}(\Phi_{\E}) g_{\mu \nu}$. As one would expect, the field equations derived from the Einstein frame action are significantly simpler. When computing physical observables, however, one must always map all calculations back to Jordan frame quantities.

The choice of $[F(\Phi),Z(\Phi),V(\Phi)]$ in the Jordan frame define the particular ST theory under consideration, and here, we will first focus on the (massless) Damour/Esp\'osito-Farese (DEF) model, defined via 
\begin{align}
F(\Phi) = \Phi\,,
\qquad
Z(\Phi) &= \frac{\omega(\Phi)}{\Phi}\,,
\qquad 
\omega(\Phi) := -\frac{3}{2} - \frac{1}{8 \kappa \beta \ln{\Phi}}\,,
\qquad
V(\Phi) = 0\,,
\nn \\
A(\Phi^{\E}) &= e^{\frac{\beta}{2} \Phi_{\E}^{2}}\,,
\qquad
V^{\E}(\Phi^{\E}) = 0\,.
\end{align}
This model is controlled by two quantities: the coupling constant $\beta$ and the asymptotic value of the scalar field at spatial infinity $\Phi_{0}^{\infty}$. The latter is related to the former and to the Brans-Dicke coupling parameter via
\be
\Phi_{0}^{\infty} = e^{-\frac{4 \pi}{\beta} \frac{1}{3 + 2 \omega_{\BD}}}\,,
\qquad
\Phi_{\E,0}^{\infty} = \frac{2 \sqrt{\pi}}{|\beta| \sqrt{3 + 2 \omega_{\BD}}}\,,
\ee
and thus, it is constrained by Solar System experiments via $\omega_{\BD} > 4 \times 10^{4}$~\cite{Bertotti:2003rm}. Upon cosmological evolution, N\"ordtvedt and Damour~\cite{Damour:1992kf,Damour:1993id} showed that when $\beta > 0$, General Relativity is an attractor and $\Phi$ evolves toward zero exponentially, automatically satisfying Solar System constraints. On the other hand, when $\beta < 0$, Sampson \et~\cite{Sampson:2014qqa} were the first to show that $\Phi$ evolves away from zero polynomially, maximally violating Solar System constraints today. 

In spite of this problem with Solar System constraints, ST theories with $\beta<0$ have received special attention due to the non-linear process of \emph{scalarization}~\cite{Damour:1993hw,Damour:1996ke}. This phenomenon is analogous to ferromagnetism: below a certain critical binding energy stellar solutions are as in General Relativity, but above this critical energy, new solutions appear that possess a non-trivial scalar field and are energetically favorable~\cite{Damour:1993hw,Damour:1996ke}. When considering neutron stars in isolation, the process is called \emph{spontaneous scalarization} and the critical energy is a function of the central density only. When considering neutron stars in binary systems, the process is called \emph{induced or dynamical scalarization} (depending on whether the companion star is spontaneously scalarized or not) and the critical energy is the binary's binding energy~\cite{Barausse:2012da,Palenzuela:2013hsa,Shibata:2013pra,Taniguchi:2014fqa}. Such scalarized solutions are constrained by binary pulsar observations~\cite{Freire:2012mg,Wex:2014nva,Berti:2015itd,Kramer:2016kwa}, since the activation of a non-trivial scalar field unavoidably leads to scalar dipole radiation, which is not observed in binary pulsars. This observations require that $\beta \gtrsim -4.5$, where the approximate sign is because the precise value at which neutron stars scalarize depends on its equation of state (although the dependence is very mild~\cite{Harada:1998ge,Silva:2014fca}).

\begin{figure}[htb]
\begin{center}
\begin{tabular}{l}
\includegraphics[width=8.1cm,clip=true]{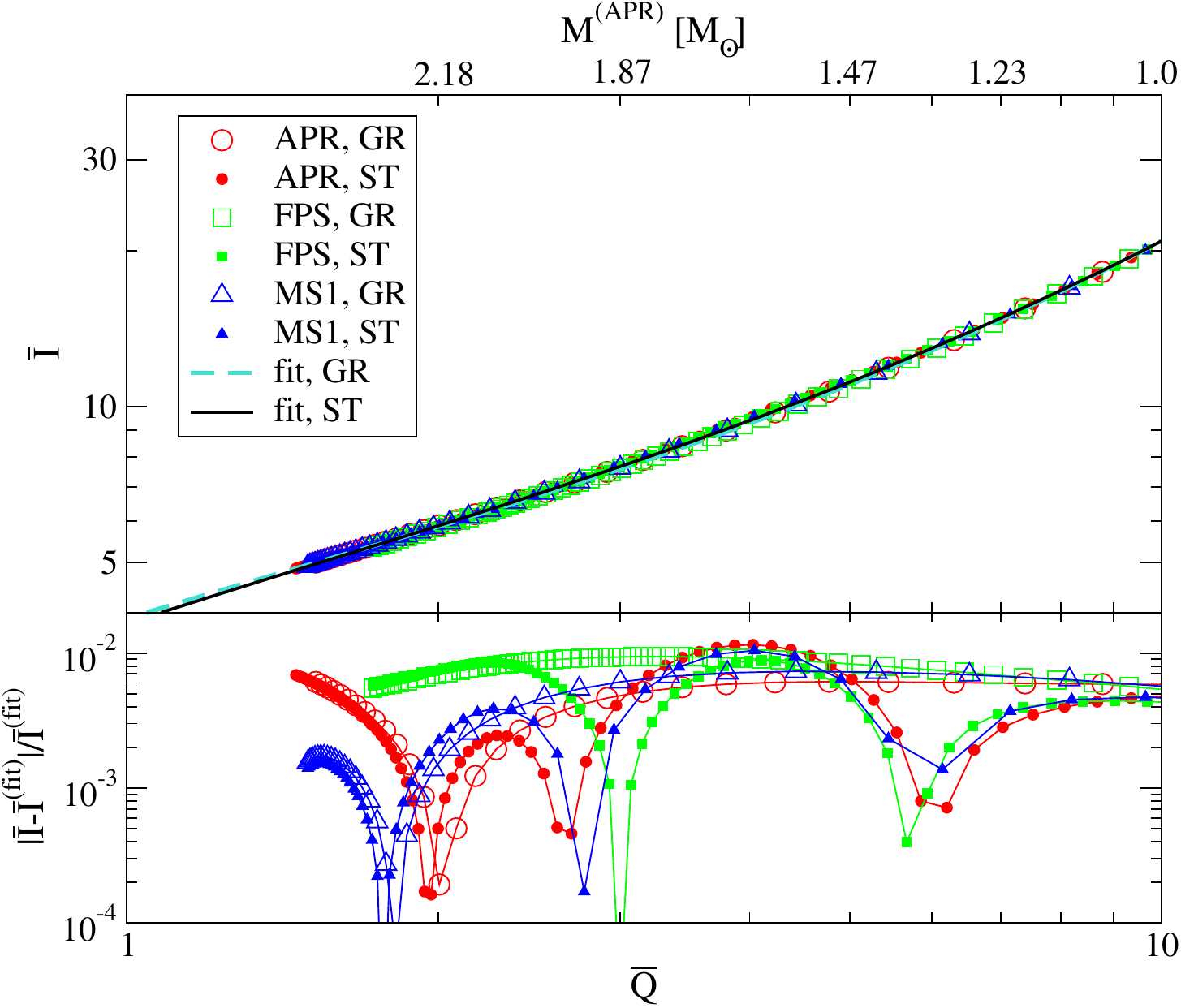} 
\includegraphics[width=7.9cm,clip=true]{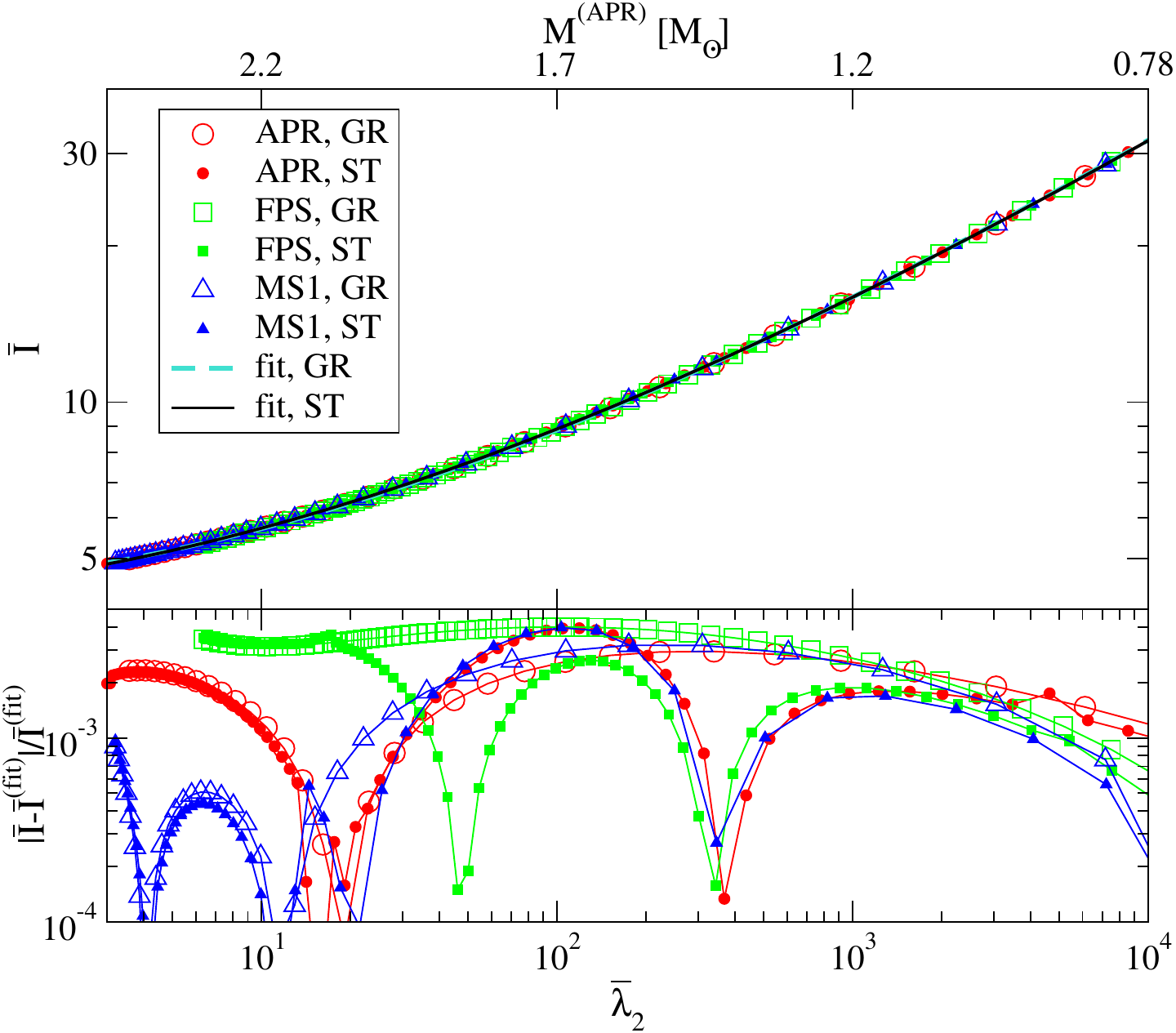} 
\end{tabular}
\caption{\label{fig:I-Love-ST}
(Top) The I-Q (left) and I-Love (right) relations in ST gravity with $\beta = -4.5$ and $\Phi_{\E,0}^{\infty} = 10^{-3}$. For reference, the top axis shows the neutron star mass in General Relativity using the APR equation of state. (Bottom) The absolute fractional difference between the relations with different equations of state with respect to the fit. Observe that the ST I-Love-Q relations are still approximately universal to roughly the same degree as in General Relativity. Observe also that the ST relations are very close to the General Relativistic I-Love-Q relation because of the choice of ST coupling parameters. These figures are adapted from the data presented in Pani and Berti~\cite{Pani:2014jra}.
}
\end{center}
\end{figure}

Let us now look at the I-Love-Q relations in the DEF model of ST gravity. The first analysis of the I-Q relation for rapidly rotating neutron stars in DEF ST gravity was carried out by Doneva \et~\cite{Doneva:2014faa}, while the study of the I-Love-Q relations was completed in Pani and Berti~\cite{Pani:2014jra} for slowly-rotating neutron stars. The top panels of Fig.~\ref{fig:I-Love-ST} show the I-Q and I-Love relations in DEF ST gravity, for three representative equations of state, as well as the relation in General Relativity for comparison. The figure fixes the coupling constants to $\beta = -4.5$ and $\Phi_{\E,0}^{\infty} = 10^{-3}$, which saturates binary pulsar constraints. Observe that the ST relations are quite similar to the General Relativistic ones because of the strength of the binary pulsar constraints. The bottom panels show the degree of universality of the I-Love-Q relation in ST gravity and in General Relativity\footnote{The fit in General Relativity was constructed with the data only for the three equations of state shown in the figure, which is slightly different from that presented in Sec.~\ref{sec:I-Love-Q-relativistic}, which used a larger number of equations of state.}. Observe that the relations are still equation-of-state insensitive, with variations comparable to those in General Relativity.

Very recently, Doneva and Yazadjiev~\cite{Doneva:2016xmf} studied the I-Q relation in massive DEF theory, in which $V^{\E}(\Phi^{\E}) = m_s^2 \left( \Phi^{\E} \right)^2/2$ with $m_s$ is the mass of the scalar field. Spontaneous scalarization in such a theory was first studied in~\cite{popchev,Ramazanoglu:2016kul} for non-rotating neutron stars, and was later extended to slowly-rotating~\cite{Yazadjiev:2016pcb} and rapidly-rotating~\cite{Doneva:2016xmf} neutron stars. The advantage of considering a massive theory is that if the scalar field mass is sufficiently large, non-GR modifications are screened and one is likely to evade Solar System and binary pulsar bounds on $\beta$ for the massless theory. Ramazanoglu and Pretorius~\cite{Ramazanoglu:2016kul} found that neutron stars (and not white dwarves) can scalarize and evade these observational bounds at the same time when $-10^{-3} \lesssim \beta \lesssim -3$ and $10^{-16} \mrm{eV} \ll m_s \lesssim 10^{-9} \mrm{eV}$. Doneva and Yazadjiev~\cite{Doneva:2016xmf} found that the universality in the I-Q relation is worse than that in GR, though the former remains equation-of-state universal within a few percent for a fixed $f_s M$ (where we recall $f_s$ is the stellar spin frequency). They also found that such a relation deviates from that in GR by 20\% at most for $\beta = -6$ and $\Phi_{\E,0}^{\infty} = 0$. 

%-----------------------------
\subsection{$f(\mathcal R)$ Gravity}

This modified theory replaces the Ricci scalar $\mathcal R$ in the Einstein-Hilbert action by an arbitrary function of $\mathcal R$:
\be
\label{eq:action-fR}
S = S_{f(\mathcal R)} + S_\mat\,, \quad S_{f(\mathcal R)} \equiv \kappa \int d^4x \sqrt{-g} \; f(\mathcal R)\,.
\ee
The arbitrary function $f(\mathcal R)$ needs to satisfy $d^2f/d\mathcal R^2 \geq 0$ and $df/d\mathcal R \geq 0$ to avoid tachyonic instabilities and ghost modes. Such a theory is motivated mainly from cosmology, as it can explain inflation or the accelerating expansion of the Universe without introducing an inflaton or dark energy (see e.g. Sotiriou and Faraoni~\cite{Sotiriou:2008rp} and De Felice and Tsujikawa~\cite{DeFelice:2010aj} for reviews on $f(\mathcal R)$ gravity). 

$f(\mathcal R)$ gravity is equivalent to Brans-Dicke theory with $\omega_\BD=0$ and a non-vanishing scalar field potential. To see this explicitly, one starts with an action with an auxiliary field $\psi$ given by
\be
\label{eq:action-fR2}
S = \kappa \int d^4x \sqrt{-g} \left[ f(\psi) + f'(\psi) (\mathcal R-\psi) \right] + S_\mat\,.
\ee
Varying this action with respect to $\psi$, one finds $f''(\psi) (\mathcal R-\psi) = 0$. Thus, $\psi=\mathcal R$ provided $ f''(\psi) \neq 0$ and one recovers Eq.~\eqref{eq:action-fR} if one integrates out $\psi$ from Eq.~\eqref{eq:action-fR2}. We thus say that Eq.~\eqref{eq:action-fR2} is dynamically equivalent to Eq.~\eqref{eq:action-fR}. Introducing further a new field $\Phi = f'(\psi)$, one can rewrite Eq.~\eqref{eq:action-fR2} in Brans-Dicke form in Jordan frame with $\omega_\BD=0$ and a potential:
\be
S = \kappa \int d^4x \sqrt{-g} \left[ \Phi \, \mathcal R -V(\Phi) \right] + S_\mat\,,
\ee
where $V(\Phi) \equiv \psi (\Phi) \; f' \left[ \psi(\Phi) \right] - f\left[ \psi(\Phi) \right]$. Notice that the observation of the Shapiro time delay with signals sent by the Cassini spacecraft~\cite{Bertotti:2003rm} places the bound $\omega_\BD > 4 \times 10^4$ on Brans-Dicke theory \emph{without} a potential; therefore, such a bound does not rule out the above $f(\mathcal R)$ gravity model with $\omega_\BD=0$. Thanks to this formulation, one can construct neutron star solutions in the same way as in scalar-tensor theories (see e.g. Yazadjiev \et~\cite{Yazadjiev:2014cza}).

Universal relations in $f(\mathcal R)$ gravity have been studied in Doneva \et~\cite{Doneva:2015hsa} (I-Q relation) and Staykov \et~\cite{Staykov:2015cfa} (relation between the f-mode frequency and the moment of inertia) with $f(\mathcal R) = \mathcal R + a \mathcal R^2$, where $a$ is not to be confused with the Kerr spin parameter, but rather it is a coupling constant with unit of length squared. Such a function was first introduced by Starobinsky~\cite{Starobinsky:1980te} to explain inflation and can also explain the accelerated expansion of the Universe, while remaining consistent with cosmic microwave background observations. With this choice of the $f(\mathcal R)$ function, the potential becomes $(\Phi-1)^2/(4a)$, and the theory reduces to a massive scalar-tensor theory. The current strongest bound on $a$ is obtained from table-top experiments, which require $a \lesssim 10^{-10}\mrm{m}^2$~\cite{Naf:2010zy}. Solar system bounds further require $a \lesssim 5 \times 10^{11}\mrm{m}^2$ from the Gravity Probe B experiment~\cite{Naf:2010zy} and $a \lesssim 6 \times 10^{17}\mrm{m}^2$ from the observation of the perihelion precession of Mercury~\cite{Berry:2011pb}. All of these constraints, however, can be avoided if one invokes a chameleon mechanism~\cite{Khoury:2003aq,Khoury:2003rn} that forces $f(\mathcal R)$ gravity to reduce rapidly to General Relativity in regions with relatively high density.

\begin{figure}[htb]
\begin{center}
\begin{tabular}{l}
\includegraphics[width=8.5cm,clip=true]{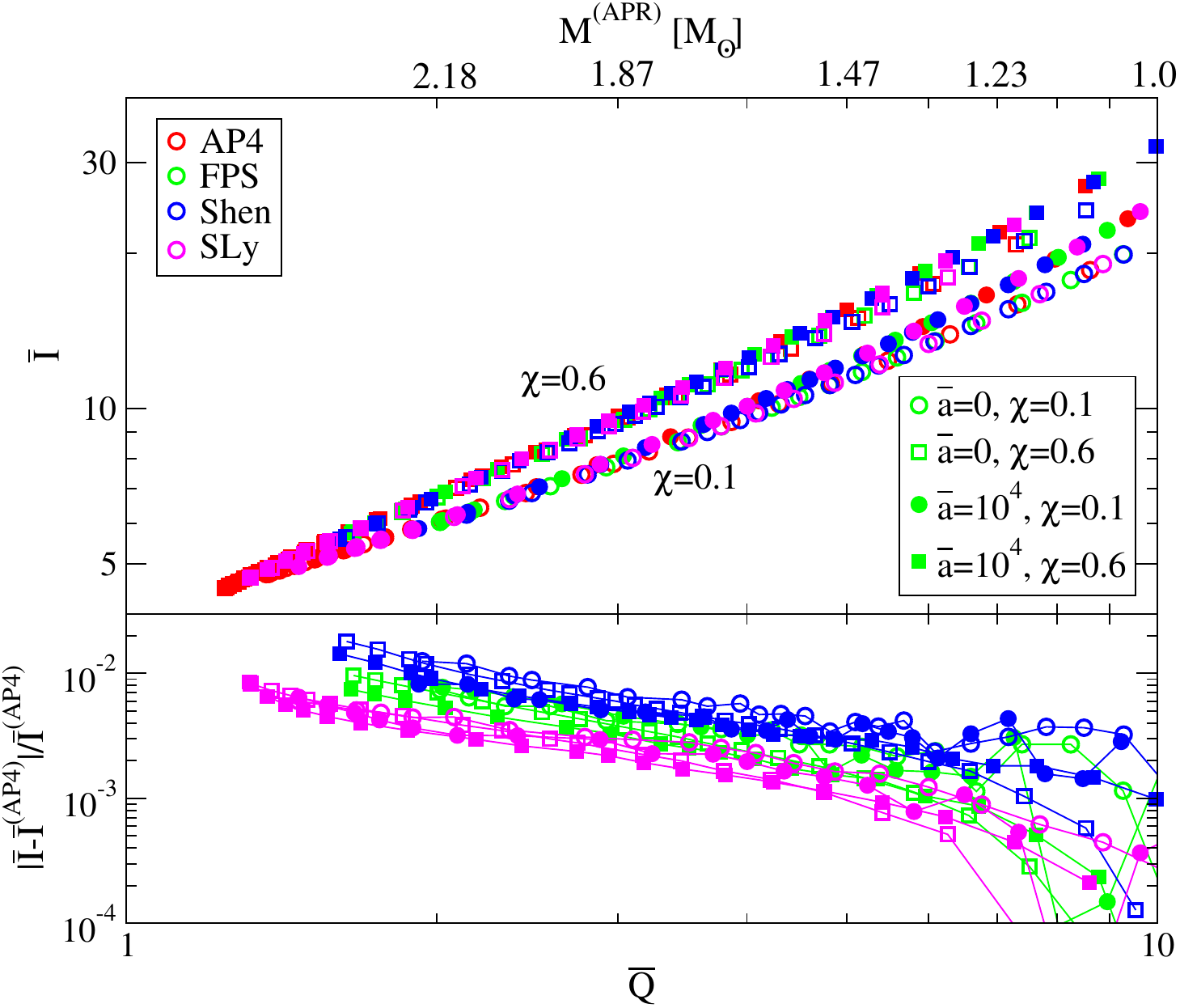} 
\end{tabular}
\caption{\label{fig:I-Q-fR}
(Top) I-Q relations in $f(\mathcal R)$ gravity with four representative equations of state, two dimensionless spins $\chi$ and two coupling constants $\bar a$. The top axis shows the neutron star masses with the APR equation of state in General Relativity.
(Bottom) Fractional difference from the relations with AP4. Observe that the universality becomes better than General Relativity. 
These figures are adapted from the data presented in Doneva \et~\cite{Doneva:2015hsa}.
}
\end{center}
\end{figure}

Figure~\ref{fig:I-Q-fR} presents the I-Q relation in $f(\mathcal R)$ gravity for rotating neutron stars with various equations of state, $\chi = 0.1$ and $0.6$ and $\bar a \equiv a/M_\odot^2 = 0$ (General Relativistic case) and $10^4$~\cite{Doneva:2015hsa} without imposing the slow-rotation approximation, as well as the absolute fractional difference from the AP4 equation of state. Notice that $\bar a = 10^4$ is completely ruled out by laboratory experiments since these require $\bar a \leq 5 \times 10^{-17}$, while it satisfies Solar System bounds. Observe first that the relations depend more sensitively on $\chi$ than on $\bar a$, i.e.~the sequence of squares and circles in the figure cluster separately. Observe also that for a fixed $\chi$ the relations remain approximately equation-of-state universal to roughly ${\cal{O}}(1 \%)$, i.e.~each cluster of squares or circles lies approximately on the same curve, irrespective of the equation of state symbolized by different colors. Finally, observe that the General Relativistic relations differ from the $f(\mathcal R)$ relations the most for large $\bar{Q}$ (small mass) stars, i.e.~the filled symbols are farthest from the unfilled symbols for large $\bar{Q}$. This is because black hole solutions in $f(\mathcal R)$ gravity are the same as in General Relativity; thus, the universal relations in the two theories approach the same point in the I-Q plane, forcing deviations to become smaller as one increases the compactness (or decreases $\bar Q$). The closeness of the General Relativistic and $f(\mathcal R)$ I-Q sequences is merely an artifact of the size of $\bar{a}$ chosen and the stringent constraints on this theory from observations and experiments; obviously, the relation in $f(\mathcal R)$ becomes essentially indistinguishable from that in General Relativity if one saturates the bound obtained from laboratory experiments $(\bar a \leq 5 \times 10^{-17})$. 

%-----------------------------
\subsection{Eddington-Inspired Born-Infeld Gravity}

Eddington-Inspired Born-Infeld (EiBI) gravity modifies the coupling between gravity and matter in a nonlinear way. The starting point is an alternative action to $S_\EH$ proposed by Eddington~\cite{1923mtr..book.....E}, who suggested that the fundamental field should be the connection $\Gamma^\alpha_{\beta \gamma}$ instead of the metric $g_{\mu\nu}$. Eddington's action is given by 
\be
S = 2 \kappa \kappa_\Edd \int d^4x \sqrt{-|\bar{\mathcal{R}}|}\,,
\ee
where the determinant $|\bar{\mathcal R}|$ of the (symmetric) Ricci tensor $\bar{\mathcal{R}}_{\mu\nu}$ is constructed solely from the connection (in contrast to $\mathcal R_{\mu\nu}$, which is constructed from the metric) and $ \kappa_\Edd$ is the coupling parameter of Eddington's theory with dimensions of length squared. One can show that $S_\Edd$ is exactly equivalent to $S_\EH$ by calculating the field equations, which are exactly equivalent to the Einstein equations. Ba\~nados and Ferreira~\cite{Banados:2010ix} extended Eddington's theory by introducing $g_{\mu\nu}$ and matter fields in the action in a manner similar to what is done in Born-Infeld nonlinear electrodynamics~\cite{Born:1934gh} to eliminate the divergence of the electron self-energy. The resulting theory, EiBI gravity, is then defined in the Palatini formalism by~\cite{Banados:2010ix,Berti:2015itd}
\be
\label{eq:EiBI-action}
S = \frac{2 \kappa}{\kappa_\EiBI } \int d^4x \left( \sqrt{-|g_{\mu\nu}+\kappa_\EiBI \bar{\mathcal{R}}_{\mu\nu}|} - \sqrt{-g} \right) + S_\mat\,,
\ee
where we set the cosmological constant to zero, $\kappa_\EiBI$ is the coupling parameter of EiBI theory with dimensions of length squared and the matter action only depends on $g_{\mu\nu}$ and the matter fields. Unlike Eddington's theory, EiBI gravity is not equivalent to General Relativity and the former reduces to the latter only in the limit $\kappa_\EiBI \to 0$. This can be seen more explicitly by expanding Eq.~\eqref{eq:EiBI-action} about $\kappa_\EiBI = 0$:
\be
\label{eq:EiBI-action-exp}
S = \kappa \int d^4x \sqrt{-g} \, \bar{\mathcal R} \left[ 1 + \frac{\kappa_\EiBI}{4  \bar{\mathcal R}} \left( \bar{\mathcal R}^2 - 2 \bar{\mathcal R}_{\mu\nu} \bar{\mathcal R}^{\mu\nu}  \right) + \mathcal{O}\left( \kappa_\EiBI^2 \bar{\mathcal R}^2 \right) \right] + S_\mat \,.
\ee
Clearly, as one takes the $\kappa_{\EiBI} \to 0$ limit of the above equation, one recovers General Relativity in the Palatini formulation, which is equivalent to Einstein's theory in the metric formulation. The metric formulation found by Deser and Gibbons~\cite{Deser:1998rj} contains ghosts unless one introduces additional terms to eliminate them. Vollick~\cite{Vollick:2003qp,Vollick:2005gc,Vollick:2006qd} considered a theory similar to EiBI gravity, but the matter fields were introduced in a different way. EiBI gravity avoids singularities in early cosmology~\cite{Banados:2010ix} and in the non-relativistic gravitational collapse of non-interacting particles~\cite{Pani:2011mg,Pani:2012qb}.
 
Although EiBI gravity, as defined in Eq.~\eqref{eq:EiBI-action}, looks quite different from General Relativity, one can show that such a theory is equivalent to Einstein's theory with a modified matter sector~\cite{Delsate:2012ky}. Although Eq.~\eqref{eq:EiBI-action-exp} contains a higher-curvature correction, the Palatini formalism ensures that the field equations do not contain higher derivatives of the metric. Varying Eq.~\eqref{eq:EiBI-action} with respect to the metric and connection, one finds
\be
\label{eq:field-eq-EiBI}
\sqrt{-q}q^{\mu\nu} = \sqrt{-g} \left( g^{\mu\nu} - 8\pi \kappa_\EiBI T^{\mu\nu} \right)\,, \quad
\Gamma^\alpha_{\beta\gamma} = \frac{1}{2} q^{\alpha \sigma} \left( \partial_\gamma q_{\sigma \beta} + \partial_\beta q_{\sigma \gamma} - \partial_\sigma q_{\beta \gamma} \right)\,,
\ee
where $T^{\mu\nu}$ is the matter stress-energy tensor and $q_{\mu\nu}$ is an auxiliary metric, defined by $q_{\mu\nu} = g_{\mu\nu} + \kappa_\EiBI \bar{\mathcal R}_{\mu\nu}$, that is compatible with the connection. Notice that in vacuum $(T^{\mu\nu}=0)$, $q_{\mu\nu} = g_{\mu\nu}$ and $\bar{\mathcal R}_{\mu\nu} = \mathcal R_{\mu\nu} = 0$, which shows that EiBI gravity is equivalent to General Relativity in the absence of matter. Combining the above field equations, one can rewrite them such that they resemble Einstein equations:
\be
\bar{\mathcal R}^{\mu}{}_{\nu} - \frac{1}{2} \delta^{\mu}{}_{\nu}  \bar{\mathcal R} =  8 \pi \tilde T^{\mu}{}_{\nu}\,, \quad 
\tilde T^{\mu}{}_{\nu} \equiv  \tau \; T^{\mu}{}_{\nu} - \left( \frac{1 - \tau}{8\pi \kappa_\EiBI} + \frac{\tau}{2}T \right) \delta^{\mu}{}_{\nu}\,,
\ee
where $\mathcal R^{\mu}{}_{\nu} = \mathcal R_{\nu\rho} q^{\mu\rho}$, $T^{\mu}{}_{\nu} = T^{\mu\rho} g_{\nu\rho}$ and $\tau \equiv 1/\sqrt{|\delta^{\mu}{}_\nu - 8 \pi \kappa_\EiBI T^\mu{}_\nu|}$. Observe that the above modified Einstein equations have a nonlinear matter coupling on the right-hand side, whereas the Einstein tensor is linearly proportional to the matter stress-energy tensor in General Relativity. In the case of perfect fluid matter [see Eq.~\eqref{eq:Tmunu}], one finds that $\tau = [(1 + 8\pi\kappa_\EiBI \rho)(1 - 8\pi \kappa_\EiBI p)]^{-1/2}$ and $\tilde T_{\mu \nu}$ can be rewritten as~\cite{Delsate:2012ky} 
\be
\tilde T^{\mu}{}_{\nu} = \left(\tilde \rho + \tilde p \right) \tilde u^{\mu} \tilde u_{\nu} + \tilde p \; \delta^{\mu}{}_{\nu}\,,
\ee
with
\be
\tilde \rho \equiv \tau \rho - \frac{\tau - 1}{8 \pi \kappa_\EiBI} - \frac{(3p-\rho)\tau}{2}\,, \quad \tilde p \equiv \tau p + \frac{\tau - 1}{8 \pi \kappa_\EiBI} - \frac{(3p-\rho)\tau}{2}\,,
\ee
and $\tilde u^\mu$ satisfying $\tilde u^\mu \tilde u^\nu q_{\mu \nu} = -1$ and $\tilde u^\mu \tilde u_\nu = u^\mu u_\nu$. The last equation does not necessarily imply that $\tilde u^\mu$ is equivalent to $u^\mu$ since the indices of the former (latter) are raised and lowered with $q_{\mu\nu}$ $(g_{\mu\nu})$. One then sees that EiBI gravity is nothing but General Relativity but with a different equation-of-state description for matter~\cite{Delsate:2012ky}.

Let us comment on constraints on the theory and possible pathologies. The most stringent (although approximate and equation-of-state dependent) constraint on EiBI gravity comes from the existence of neutron stars, which requires $\zeta_\EiBI \equiv 8\pi \kappa_\EiBI \rho_c \lesssim 0.1$~\cite{Avelino:2012ge}, where $\rho_c \; (\gtrsim 10^{14}$ g/cm$^3$) is the central energy density of a neutron star. Such a constraint on $\kappa_\EiBI$ is seven and nine orders of magnitude stronger than the bound from solar observations~\cite{Casanellas:2011kf} and big bang nucleosynthesis~\cite{Avelino:2012ge} respectively. Constraints from Solar System experiments have not been derived since PN solutions do not fit into the standard parameterized PN framework~\cite{Pani:2013qfa}. EiBI gravity may suffer from curvature singularities if the energy density is discontinuous at the surface of stars, as is the case e.g.~for stars constructed with certain polytropes~\cite{Pani:2012qd} or with equations of state with phase transition layers in their interior~\cite{Sham:2013sya}; such singularities may be avoided by back-reaction of gravity onto the matter sector~\cite{Kim:2013nna}.

\begin{figure}[htb]
\begin{center}
\begin{tabular}{l}
\includegraphics[width=8.1cm,clip=true]{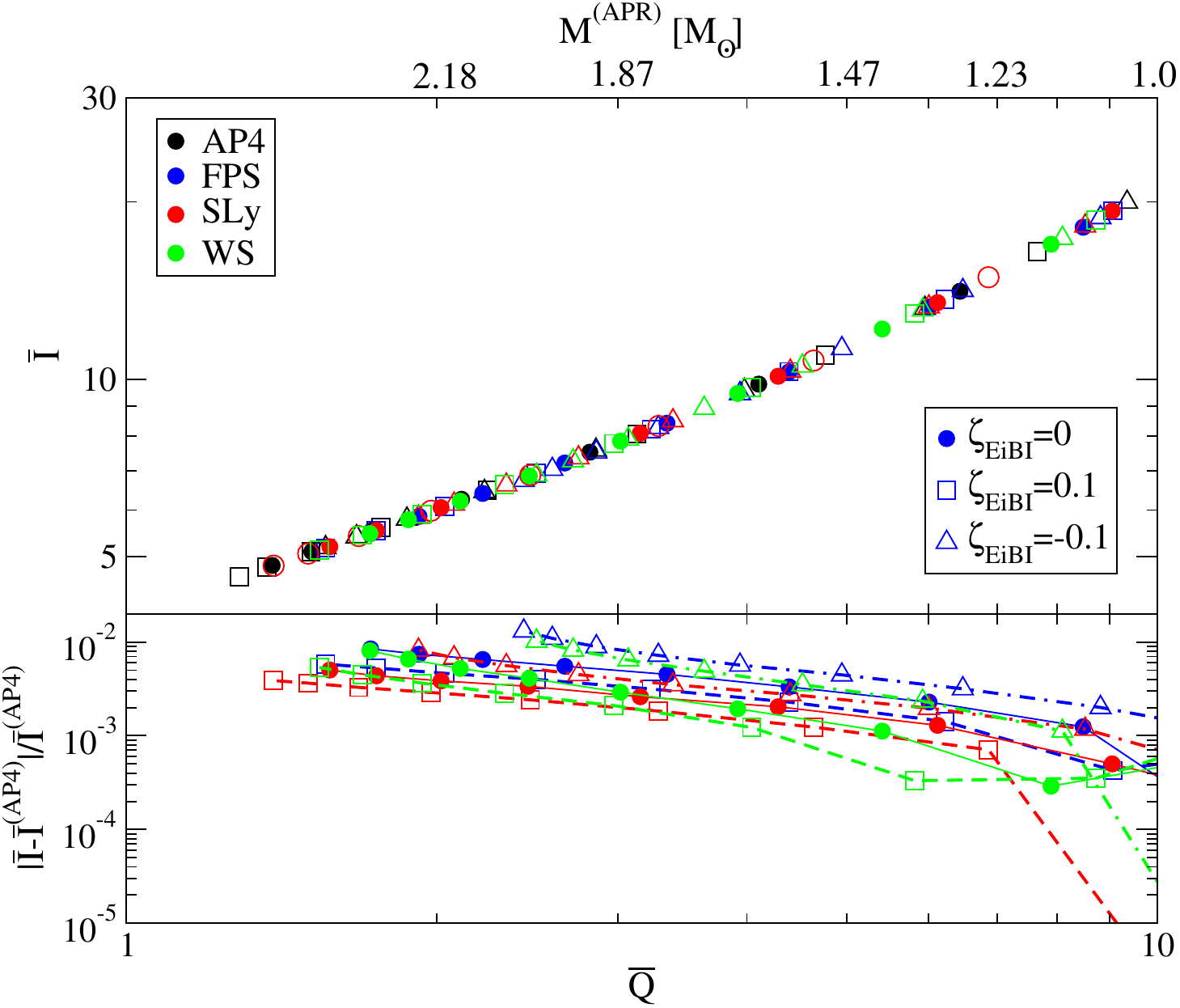} 
\includegraphics[width=8.cm,clip=true]{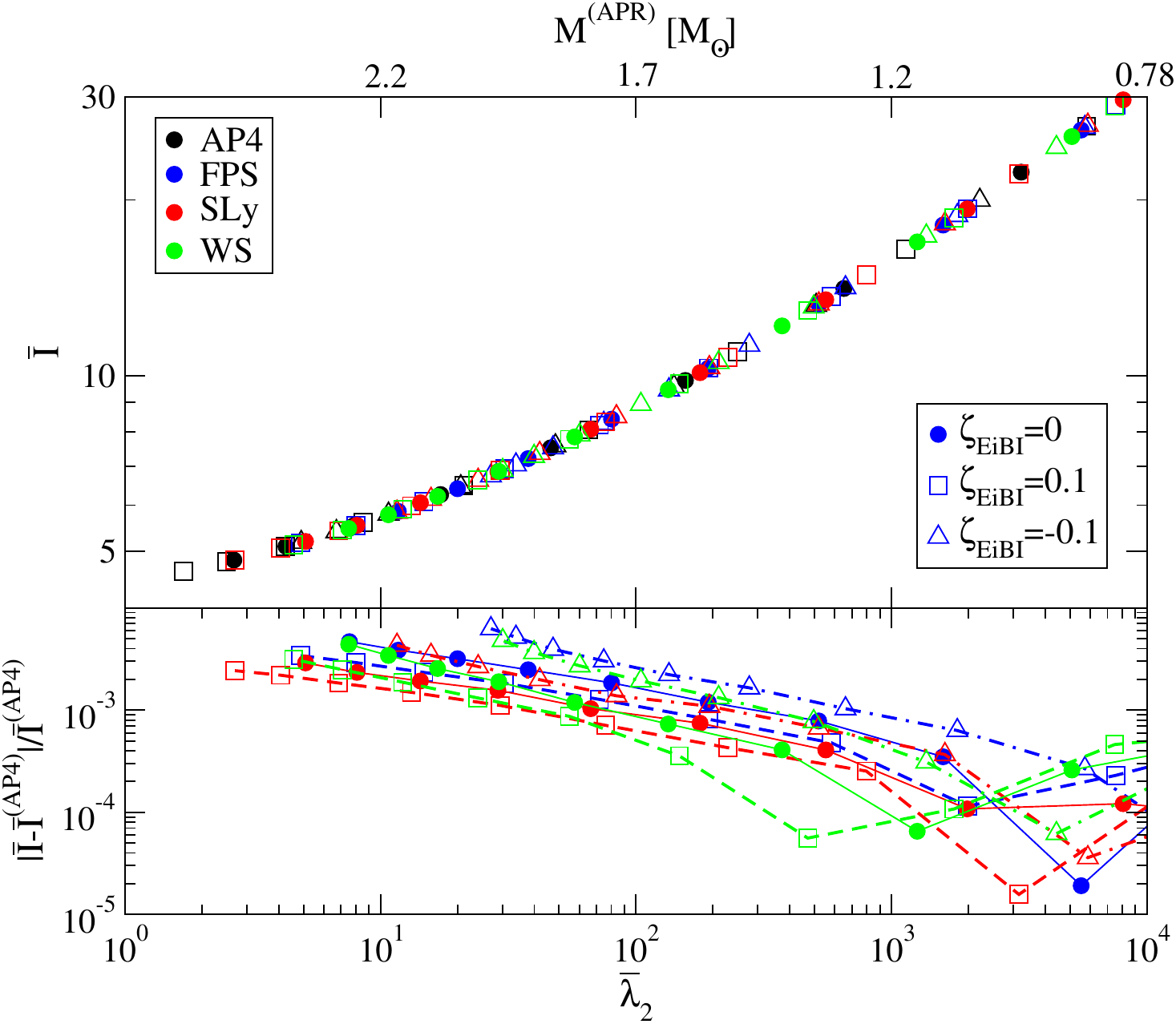} 
\end{tabular}
\caption{\label{fig:I-Love-Q-EiBI}
(Top) I-Q (left) and I-Love (right) relations in EiBI gravity with four representative equations of state and three coupling constants $\zeta_\EiBI$. The top axis shows the neutron star masses with the APR equation of state in General Relativity.
(Bottom) Fractional difference from the relations with AP4. Observe that the universality becomes better (worse) than the General Relativistic one for positive (negative) $\zeta_\EiBI$. 
These figures are adapted from the data presented in Sham \et~\cite{Sham:2013cya}.
}
\end{center}
\end{figure}

Universal I-Love-Q relations (and relations involving f-mode frequencies) of slowly-rotating and tidally-deformed neutron stars were considered in Sham \et~\cite{Sham:2013cya}\footnote{Other universal relations studied in EiBI gravity include those between the radius of a 0.5$M_\odot$ neutron star and the neutron skin thickness of ${}^{208}$Pb~\cite{Sotani:2014goa}.}. Thanks to the apparent equation-of-state formulation of Delsate and Steinhoff~\cite{Delsate:2012ky}, one can calculate the I-Love-Q trio in the same way as in General Relativity. The top panels of Fig.~\ref{fig:I-Love-Q-EiBI} show the I-Q and I-Love relations for various equations of state with $\zeta_\EiBI = 0$ (the General Relativistic limit),~$\zeta_\EiBI = 0.1$ and $\zeta_\EiBI = -0.1$, while the bottom panels show the fractional difference from the relation with the AP4 equation of state. Observe first that the relations remain universal for $\zeta_{\EiBI} \neq 0$ with respect to variation of the equation of state. Observe also that the universal relations with $\zeta_{\EiBI} \neq 0$ are very similar to those obtained in the General Relativistic limit. This is not surprising, given that (i) EiBI gravity has already been stringently constrained by neutron star observations and (ii) EiBI gravity is identical to General Relativity with a modified equation of state~\cite{Delsate:2012ky}. Observe, nonetheless, that the degree of universality improves (deteriorates) for positive (negative) $\zeta_\EiBI$ relative to the General Relativistic case. 

%-----------------------------
\subsection{Gravastars}

Let us now shift gears and review the I-Love-Q relations in exotic compact objects, focusing on gravitational vacuum condensate stars or \emph{gravastars}~\cite{Mazur:2001fv,Mazur:2004fk,Mottola:2011ud}. Such compact objects are constructed by matching an exterior black hole spacetime to a de Sitter interior spacetime at a would-be horizon through a ``matter'' shell with equation of state $p=-\rho$. The motivation behind such a construction is that quantum effects, such as the trace anomaly of the stress-energy tensor in a curved spacetime, backreacts onto the spacetime near the horizon, replacing the latter with a quantum phase boundary layer. 

Such exotic compact objects are known to be stable both thermodynamically~\cite{Mazur:2001fv} and dynamically~\cite{Visser:2003ge,Chirenti:2007mk,Pani:2009ss} in their non-rotating configuration, to linear order in the metric perturbation and with reasonable equations of state in the boundary layer. On the other hand, very compact, horizonless objects with a light ring may be nonlinearly unstable~\cite{Keir:2014oka,Cardoso:2014sna}. Moreover, such ultracompact objects with large spin are known to suffer from an \emph{ergoregion instability}~\cite{1978RSPSA.364..211C,1978CMaPh..63..243F,1996MNRAS.282..580Y} (see Brito \et~\cite{Brito:2015oca} for a recent review), as considered for example by Cardoso \et~\cite{Cardoso:2007az} for slowly rotating gravastars and Chirenti and Rezzolla~\cite{Chirenti:2008pf} for rapidly rotating gravastars. These studies showed that one can construct rotating gravastars without an ergoregion, and hence, these objects do not suffer from the ergoregion instability. 

Gravastar models have been compared against astrophysical data of black hole candidates in Broderick and Narayan~\cite{Broderick:2007ek} and Chirenti and Rezzolla~\cite{Chirenti:2016hzd} (see also the related work by Giudice \et~\cite{Giudice:2016zpa}). The former showed that XTE J1118+480 and Sgr A$^{*}$ are inconsistent with a gravastar model in the non-General Relativistic theories studied in Chapline \et~\cite{Chapline:2000en} unless modifications from General Relativity are sub-Planckian. The latter showed that the final remnant of the source of the recently-detected gravitational wave event GW150914~\cite{Abbott:2016blz} is not consistent with a rotating gravastar if the latter had the same mass and spin inferred by assuming that the remnant is a Kerr black hole~\cite{Abbott:2016blz,TheLIGOScientific:2016wfe}. The GW150914 event, however, describes the full late inspiral and merger of a compact binary, thus placing stringent constraints on the \emph{dynamics} of the compact objects that coalesced. Thus, whether gravastars need to be taken seriously as a source of GW150914 is questionable at best, given that such exotic objects do not have a sound theoretical underpinning to describe the dynamics of a merger event in the first place~\cite{Yunes:2016jcc}.

Because of the lack of matter in their interior, one cannot study the equation-of-state dependence of the I-Love-Q relations in gravastars in the same manner as when studying neutron stars. The only possible equation-of-state dependence arises from the matter in the thin shell that connects the interior to the exterior vacuum regions. Gravastars with vanishing energy density in the thin shell, however, can have compactnesses as large as those of black holes, greatly exceeding the isotropic, perfect fluid limit of $4/9$ for a non-rotating configuration~\cite{Buchdahl:1959zz}. In this sense, such gravastars resemble the anisotropic compact stars of Sec.~\ref{sec:BH-limit}, and thus, they can also be used to study how the I-Love-Q relations approach the black hole limit. 

The I-Love-Q relations have been studied for slowly-rotating/tidally-deformed gravastars with an infinitesimally thin shell through an extension of the Hartle-Thorne formalism in Pani~\cite{Pani:2015tga} and in Uchikata \et~\cite{Uchikata:2016qku}, based on non-rotating configurations constructed by Mazur and Mottola~\cite{Mazur:2001fv,Mazur:2004fk} and Visser and Wiltshire~\cite{Visser:2003ge}. The exterior solutions are formally the same as the Hartle-Thorne ones, and hence, the only differences appear in the interior solutions and the matching conditions at the boundary layer, with the latter determined through the Israel junction conditions. 

Pani~\cite{Pani:2015tga} studied the I-Love-Q relations \emph{analytically} for gravastars by setting the energy density of the thin shell to zero. Such a condition forces $\nu$ and $\lambda$ in Eq.~\eqref{eq:metric-ansatz} to be continuous at the shell and $L^2/M^2 = 1/(2 C^3)$, where $L$ is the de Sitter horizon radius in the interior. At zeroth order in spin, the interior solutions are given by $e^{\nu} = e^{-\lambda} = 1-2Cr^2/R^2$, while at linear order in spin, the interior solution for $\omega$ is simply $\omega = \Omega$ and the dimensionless moment of inertia is given by $\bar I = 1/C^2$. Observe that when one expands $\bar I$ for a slowly-rotating gravastar about $C = C_\BH = 1/2$, one finds that $\bar I = 4 + \mathcal{O}(C-1/2)$, and hence, $\bar I$ agrees with that of a black hole in the $C \to C_\BH$ limit. At second order in spin, the interior solutions are given by a combination of polynomial and hyperbolic arctangent functions. The dimensionless quadrupole moment is then given by
\begin{align}
\bar Q &= 1 - \frac{4 C^5}{5 \Delta_{\bar Q}}\left[ 2\sqrt{C} \left( 16C^2-6C-9 \right) -9 \sqrt{2} \left( 4 C^2 - 1 \right) \tanh^{-1} \left( \sqrt{2 C}\right) \right]\,,
\end{align}
with 
\begin{align}
\Delta_{\bar Q} & \equiv  2 \sqrt{C} \left[ C \left(9 - 12 C + 9 C^2 + 8 C^3 \right) - 
    3 \left(3 - 7 C + 6 C^2 \right) \tanh^{-1}\left( \frac{C}{1 - C} \right) \right] \nn \\
 &  -  3 \sqrt{2} \tanh^{-1} \left( \sqrt{
   2 C} \right) \left[ C \left(3 - 6 C - 5 C^2 + 6 C^3 \right) - 
    3 \left(1 - 3 C + 4 C^3 \right) \tanh^{-1} \left( \frac{C}{1 - C} \right) \right]\,.
\end{align}
In the $C \to 1/2$ limit, $\bar Q \to 1 +(8/45)[\ln(1-2C)]^{-1}$, and hence, $\bar Q$ approaches the black hole value like an inverse logarithm. On the other hand, in the $C \to 0$ limit, $\bar Q \to -3$, which suggests that Newtonian, slowly-rotating gravastars are prolate due to the negative sign. This resembles slowly-rotating, strongly-anisotropic incompressible Newtonian stars~\cite{Glampedakis:2013jya,Yagi:2015upa,Yagi:2016ejg}, which are also prolate in the Newtonian limit. Regarding tidal perturbations, the interior solutions are given by hypergeometric functions and the dimensionless tidal deformability is given by\footnote{This corrects the expressions in the original version of Pani~\cite{Pani:2015tga} and now agrees with the results in Uchikata \et~\cite{Uchikata:2016qku} and Cardoso \et~\cite{Cardoso:2017cfl}.}
\begin{align}
%\bar \lambda_2 &=
%\frac{56 \left(1 - 2 C \right)^2}{\Delta_{\bar \lambda}} \left[10 \sqrt{ C} \left(72 C^3 - 108 C^2 + 34 C - 3\right) + 
 %  3 \sqrt{2} \left(1 - 2 C \right)^2 \left(36 C^2 - 40 C + 5 \right) \tanh^{-1}\left( \sqrt{2 C} \right) \right]\,, \nn \\
   \bar \lambda_2 =\frac{16}{15} \frac{\Delta_{\bar \lambda}^{(n)}}{\Delta_{\bar \lambda}^{(d)}}\,,
\end{align}
with
\begin{align}
%\Delta_{\bar \lambda} & \equiv    210 C^{3/2} \left(720 C^6 + 440 C^5 - 5084 C^4 + 6590 C^3 - 3210 C^2 + 
  %   645 C - 45 \right) \nn \\
  %   &  + 
 % \frac{315}{2} \left(1 - 2 C \right)^2 \left\{ 10 \sqrt{
  %    C} (6 C - 1) \left(12 C^2 - 16 C + 3 \right) \ln (1 - 2 C) \right. \nn \\
   %   &  \left. + 
   %  \sqrt{2} \tanh^{-1} \left(\sqrt{
    %   2 C} 
    % \right) \left[2 C \left(72 C^5 + 72 C^4 - 474 C^3 + 488 C^2 - 
     %      165 C + 15\right) \right. \right. \nn \\
       %    &  \left. \left. + 
      %  3 \left(1 - 2 C\right)^2 \left(36 C^2 - 40 C + 5\right) \ln(1 - 2 C)\right] \right\}\,.
\Delta_{\bar \lambda}^{(n)} & \equiv  -2 C \left(120
   C^4-536 C^3+130 C^2+156 C-45\right) \nn \\
   &+6 \sqrt{2C} \left(16 C^5+152 C^4-204 C^3-2 C^2+62
   C-15\right) \tanh ^{-1}\sqrt{2C} \nn \\
   &-9 \left(4 C^3+6 C^2-6 C-5\right) (1-2 C)^3 \left(\tanh ^{-1}\sqrt{2C}\right)^2\,, \\
\Delta_{\bar \lambda}^{(d)} & \equiv 2 C (10 C-3) \left[\left(-36 C^3+150 C^2+6
   C-45\right) \log (1-2 C)+2 C \left(32 C^4+32 C^3+96
   C^2-39 C-45\right)\right] \nn \\
   &+9 (1-2 C)^2 \left[2 C \left(-16
   C^3-46 C^2+3 C+15\right)+3 \left(8 C^4+8 C^3-18
   C^2-4 C+5\right) \log (1-2 C)\right] \nn \\
   &\times \left(\tanh^{-1} \sqrt{2C} \right)^2+6 \sqrt{2C} \left[2 C \left(64 C^6+64
   C^4-460 C^3+120 C^2+141 C-45\right) \right.  \nn \\
   & \left.+3 \left(16 C^5+152
   C^4-204 C^3-2 C^2+62 C-15\right) \log (1-2 C)\right] \tanh
   ^{-1}\sqrt{2C}\,. 
\end{align}
In the $C \to 1/2$ limit, $\bar \lambda_2 \to (32/15) [23-6\ln 2 + 9 \ln(1-2C)]^{-1}$~\cite{Cardoso:2017cfl}, and thus, $\bar \lambda_2$ reaches the black hole value in the black hole limit. The tidal deformability $\bar \lambda_2$ for gravastars is always negative, which again resembles the case of strongly-anisotropic, incompressible stars~\cite{Glampedakis:2013jya,Yagi:2015upa,Yagi:2016ejg}. 

\begin{figure}[htb]
\begin{center}
\begin{tabular}{l}
\includegraphics[width=8.1cm,clip=true]{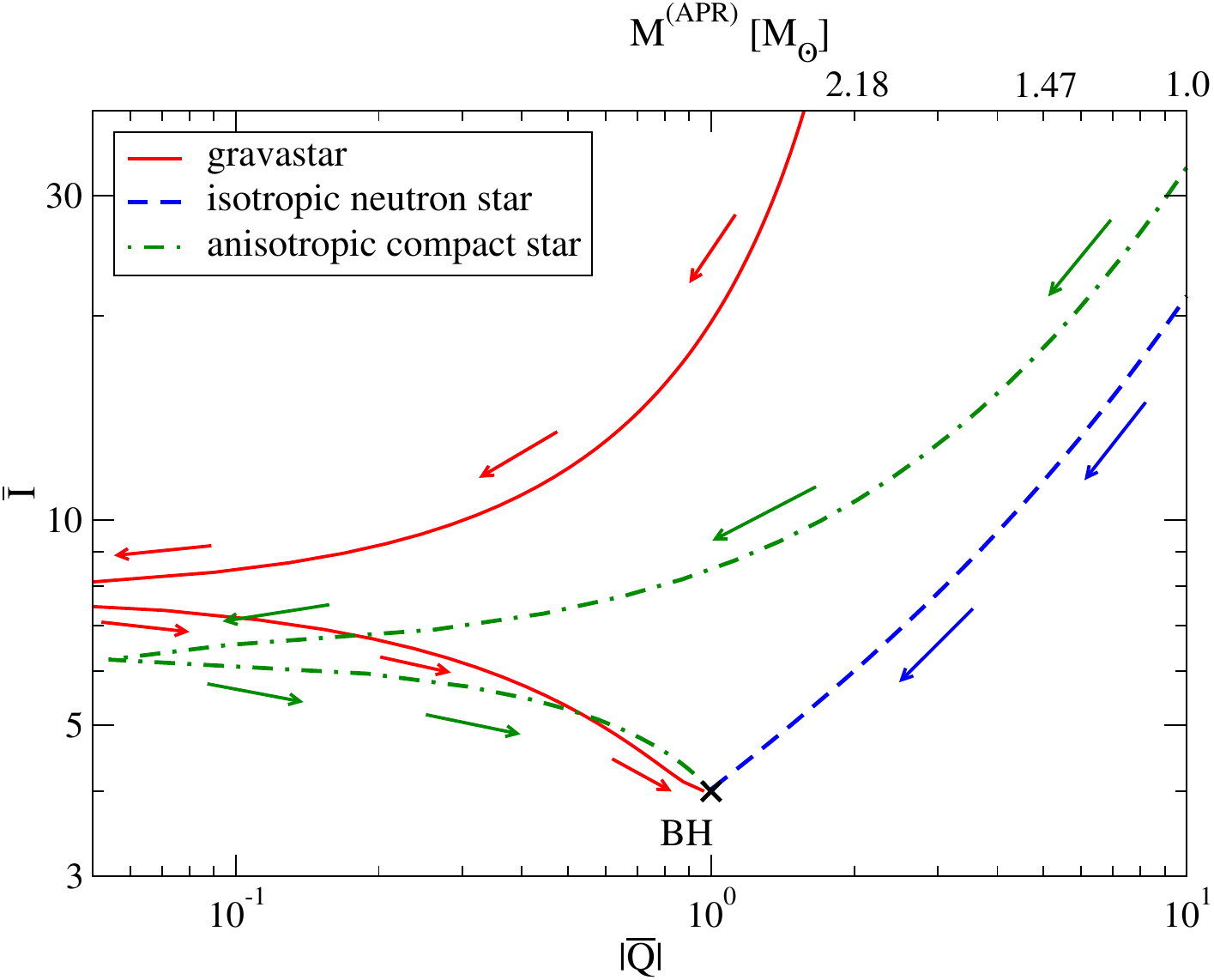} 
\includegraphics[width=8.cm,clip=true]{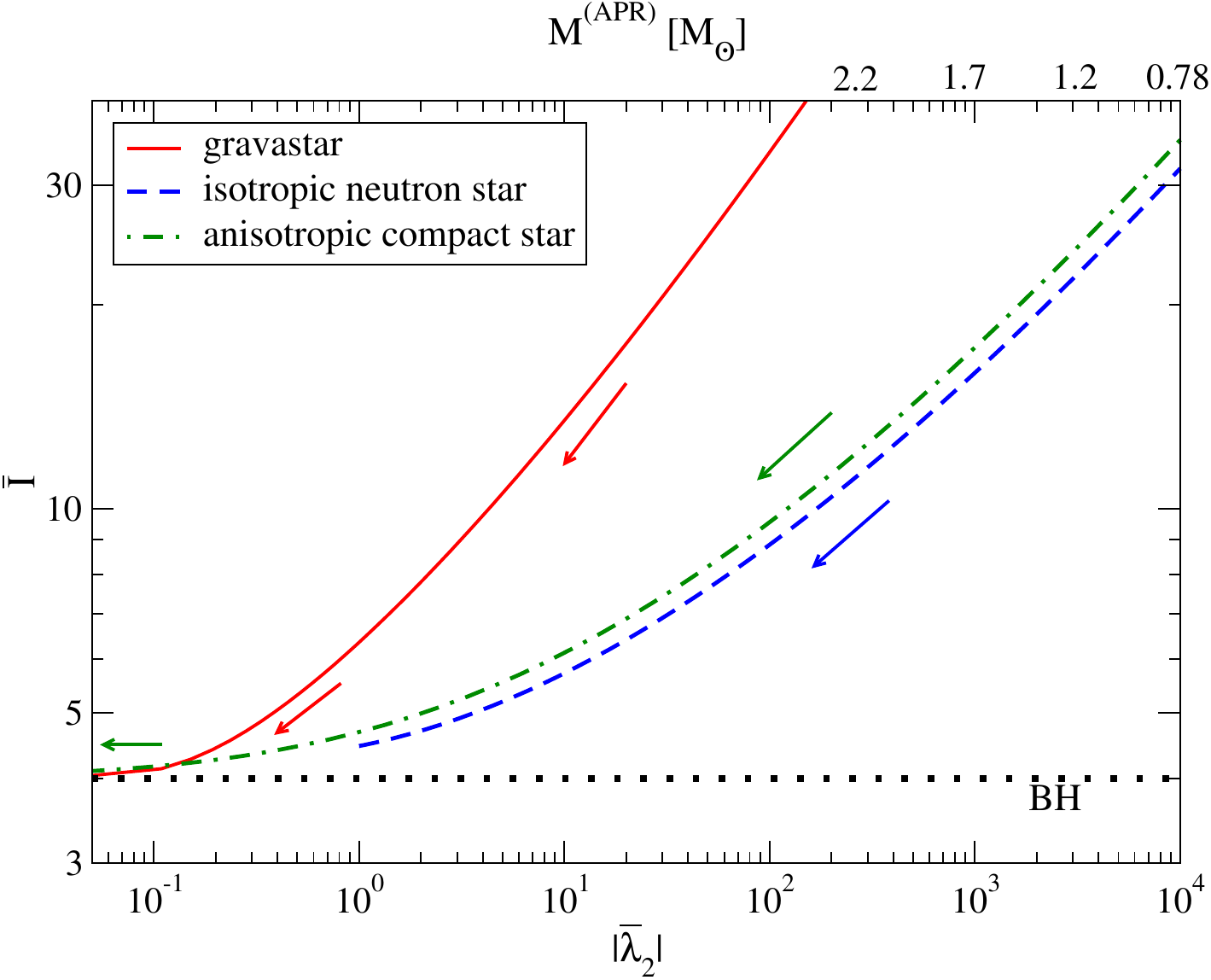} 
\end{tabular}
\caption{\label{fig:I-Love-Q-gravastar}
I-Love (left) and I-Q (right) relations for thin-shell gravastars with vanishing shell energy density~\cite{Pani:2015tga}. For reference, we also show the relations for isotropic neutron stars and anisotropic constant density stars~\cite{Yagi:2015upa,Yagi:2016ejg}, with the mass of the former for an APR equation of state shown in the top axes. The black cross and dotted horizontal line show $\bar I$ and $\bar Q$ for a black hole. The arrows show the direction of increasing compactness. Observe that the gravastar relations are quite different from those of isotropic neutron stars, but the way the former approach the black hole limit is qualitatively similar to the anisotropic compact star case.
}
\end{center}
\end{figure}

With these analytic expressions at hand, one can calculate the interrelations between $\bar I$, $\bar Q$ and $\bar \lambda_2$, which are shown in Fig.~\ref{fig:I-Love-Q-gravastar}, where we take the absolute values of $\bar Q$ and $\bar \lambda_2$ as they can be negative. Observe that the I-Love-Q relations for thin-shell gravastars are quite different from those for isotropic neutron stars, but qualitatively similar to the relations for strongly-anisotropic, incompressible stars (see Sec.~\ref{sec:BH-limit}). Quantitatively, however, the relations for gravastars approach the black hole limit in a different manner. For example, as explained in the previous paragraph, $\bar Q$ approaches this limit as an inverse logarithm for gravastars but as a polynomial for strongly-anisotropic stars~\cite{Yagi:2015upa,Yagi:2016ejg}. 

Uchikata \et~\cite{Uchikata:2016qku}, on the other hand, studied the equation-of-state variation of the I-Love-Q relations \emph{numerically} by considering a polytropic equation of state for the shell, based on slowly-rotating gravastars constructed in~\cite{Uchikata:2015yma}. The authors found that, just as for neutron stars and gravastars with vanishing energy density, the I-Love-Q relations for polytropic gravastars approach the black hole limit as one increases the compactness, but how they approach this limit is sensitive to the equation of state. In fact, the relations for gravastars are not only sensitive to the polytropic index $n$, but also to the overall magnitude $K$ in Eq.~\eqref{eq:polytropic-EoS}, while the I-Love-Q relations for neutron stars and quark stars are only sensitive to $n$. The thin-shell gravastar model reviewed in this subsection is the simplest one used to describe gravastars;  it may be interesting to study the I-Love-Q relations for other gravastar models, such as those with a finite-thickness boundary layer~\cite{Chirenti:2007mk,Chirenti:2008pf}, with the interior region supported by dark energy instead of a cosmological constant (phantom gravastars or dark energy stars)~\cite{Bilic:2005sn,Lobo:2005uf}.

%%%%%%%%%%%%%%%%%%%%%%%%%%%%%%%%%%%%%%%%%%%%%%%%%%%%%%%%
\section{Applications}
\label{sec7:Applications}

Universal relations have various useful applications in astrophysics. For example, if one can measure any member of the I-Love-Q trio, one automatically obtains the remaining two quantities without having to know the correct equation of state in nature \emph{a priori}. The moment of inertia, for example, is expected to be measured to $\lesssim 10\%$ accuracy from the spin-orbit coupling effect in the rate of periastron advance of the double binary pulsar J0737-3039, using e.g.~SKA~\cite{Lattimer:2004nj,Kramer:2009zza,Kehl:2016mgp}. Combining such a measurement with the I-Love-Q relations, one can obtain the tidal deformability and the quadrupole moment of the primary pulsar in J0737-3039, which would not be easily measured in any other way. On the other hand, the tidal Love number is expected to be measured from future gravitational wave observations of neutron star mergers~\cite{Flanagan:2007ix,Hinderer:2009ca,Lackey:2011vz,Damour:2012yf,Lackey:2013axa,Read:2013zra,Favata:2013rwa,Yagi:2013baa,DelPozzo:2013ala,Wade:2014vqa,Lackey:2014fwa,Agathos:2015uaa}. Again, by combining this measurement with the I-Love-Q relations, one can obtain the moment of inertia and the quadrupole moment of neutron stars in a binary system, which would also be difficult to measure from gravitational wave observations. The remainder of this section reviews different applications of the universal relations to nuclear physics with gravitational wave observations (Sec.~\ref{sec:nuclear-GW}) and X-ray observations (Sec.~\ref{sec:nuclear-Xray}), to gravitational wave astrophysics (Sec.~\ref{sec:GW-astro}), to experimental relativity (Sec.~\ref{sec:application-gravitational}) and to cosmology (Sec.~\ref{sec:cosmology}). 

%%%%%%%%%%%%%%%%%%%%%%%%%%%%%%%%%%%%%%%%%%%%%%%%%%%%%%
\subsection{Nuclear Physics through Gravitational Wave Observations }
\label{sec:nuclear-GW}

Recently, Adv.~LIGO detected gravitational waves from black hole binaries~\cite{Abbott:2016blz,Abbott:2016nmj}, and it is likely that it will detect gravitational waves from neutron star binaries in near future. Such observations will be very useful in probing nuclear physics as the waveform depends on the internal structure of neutron stars through their tidal interactions~\cite{Flanagan:2007ix,Hinderer:2009ca,Lackey:2011vz,Damour:2012yf,Lackey:2013axa,Read:2013zra,Favata:2013rwa,Yagi:2013baa,DelPozzo:2013ala,Wade:2014vqa,Lackey:2014fwa,Agathos:2015uaa,Hinderer:2016eia,Hotokezaka:2016bzh}. 

Let us begin by reviewing the structure of the gravitational waves produced by inspiraling neutron star binaries in the Fourier domain. The response of an interferometer to an impinging gravitational wave is simply the contraction of the metric perturbation very far from the source onto a detector response tensor. In the Fourier domain, this response can be schematically written as $\tilde h (f) = \mathcal{A}(f) \exp [i\Psi(f)]$, where $\mathcal{A}(f)$ is a slowly-varying Fourier amplitude and $\Psi(f)$ is a rapidly-varying Fourier phase. In practice, the metric perturbation is decomposed in spherical harmonics, each of which contributes to the response with a structure similar to that presented above, but for quasi-circular inspiraling neutron stars typically a single harmonic dominates~\cite{Blanchet:2002av}. The Fourier phase $\Psi (f)$ is more important than the Fourier amplitude $\mathcal{A}(f)$ in interferometric gravitational wave observations due to the number of cycles that are typically in the sensitive band of the detector; this is especially so for parameter estimation purposes, when a set of models is compared to the signal via a matched filtering analysis~\cite{Cutler:1994ys}. 

One can decompose the Fourier phase into a \emph{point-particle} part $\Psi_\mathrm{pp}(f)$ and a \emph{tidal} part $\Psi_\mathrm{tidal} (f)$: $\Psi = \Psi_\mathrm{pp} + \Psi_\mathrm{tidal}$. When the binary separation is large and the orbital velocities are small relative to the speed of light, the PN framework\footnote{A perturbative solution to the field equations as an expansion in weak-field and slow-velocities. A term is said to be of $N$th PN order if it is proportional to $(v/c)^{2N}$ relative to its leading-order, controlling factor. See Blanchet~\cite{Blanchet:2002av} for a review of PN theory.} is a good approximation to model the orbital and gravitational wave dynamics. In this framework, the Fourier phase becomes a series in $x \equiv (\pi m f)^{2/3}$, where $m \equiv m_1+m_2$ is the total mass; we recognize $x$ as effectively the relative orbital velocity of the binary. The point-particle contribution to the phase $\Psi_\mathrm{pp}$ for non-spinning, quasi-circular binaries is known up to 3.5PN order~\cite{Arun:2004hn}. The tidal contribution $\Psi_\mathrm{tidal}$ can be decomposed via $\Psi_\mathrm{tidal} = \sum_{\ell=2} \left( \Psi_{\bar \lambda_\ell} + \Psi_{\bar \sigma_\ell} \right)$, where $\Psi_{\bar \lambda_\ell}$ and $ \Psi_{\bar \sigma_\ell}$ are terms that depend on $\bar \lambda_\ell$ and $\bar \sigma_\ell$ respectively (recall that $\bar{\lambda}_{\ell}$ ($\bar{\sigma}_{\ell}$) is the $\ell$th-order multipole, dimensionless electric-type (magnetic-type) tidal deformability). Both of these contributions can be calculated in the PN approximation~\cite{Yagi:2013sva}\footnote{We neglect terms that represent nonlinear interactions between different tidal deformabilities, as these are much more difficult to extract with a gravitational wave observation. We also neglect terms associated with dynamical tides characterized by the f-mode oscillation frequency of a neutron star~\cite{Flanagan:2007ix,Hinderer:2016eia}.}:
\ba
\Psi_{\bar{\lambda}_\ell} &=& - \sum_{A=1}^2 \left[ \frac{5}{16} \frac{(2\ell -1)!! (4\ell +3) (\ell +1)}{(4\ell -3) (2\ell -3)}  \bar{\lambda}_{\ell,A} X_A^{2\ell -1} x^{2 \ell -3/2} +  \frac{9}{16} \delta_{\ell 2}  \bar{\lambda}_{2,A} \frac{X_A^4}{\eta} x^{5/2} \right] + \mathcal{O} \left( x^{2\ell -1/2} \right)\,, \nn \\
\\
\Psi_{\bar{\sigma}_2} &=& \sum_{A=1}^2 \frac{5}{224}  \bar{\sigma}_{2,A} \frac{X_A^5}{\eta} ( X_A- X_B) x^{7/2} + \mathcal{O}(x^{9/2})\,,
\ea
where $X_A \equiv m_A/m$. The $\Psi_{\bar{\lambda}_2}$ contribution was first derived by Flanagan and Hinderer~\cite{Flanagan:2007ix}, while higher PN terms in $\Psi_{\bar{\lambda}_2}$ can be found in Vines \et~\cite{Vines:2011ud} and Damour \et~\cite{Damour:2012yf}. Given that the leading PN order, Newtonian term in $\Psi_\mathrm{pp}$ is proportional to $x^{-5/2}$, the contributions of $\bar \lambda_\ell$ and $\bar \sigma_\ell$ enter first at $2\ell+1$ and $3\ell$ PN order respectively.

In order to determine which terms are important in future gravitational wave observations, one can calculate the number of gravitational wave cycles.
The \emph{useful} number of gravitational wave cycles, weighted by the noise spectral density $S_n$, was introduced by Damour \et~\cite{Damour:2000gg}. Sampson \et~\cite{Sampson:2014qqa} improved on these by introducing the \emph{effective} number of gravitational wave cycles, defined by 
\be
\label{eq:useful}
\mathcal{N}_\mrm{eff}  \equiv  \frac{1}{2\pi} \left( \int_{f_\mrm{min}}^{f_\mrm{max}} \frac{|\tilde h(f)|^2}{S_n(f)} df \right)^{-1/2} \min_{\Delta t,\Delta\phi} \left[\left( \int_{f_\mrm{min}}^{f_\mrm{max}} \frac{|\tilde h(f)|^2 \Delta \Phi^2}{S_n(f)} df  \right)^{1/2}  \right]\,.
\ee
Here, $|\tilde h(f)| \propto f^{-7/6}$ is the waveform amplitude in the Fourier domain while $\Delta \Phi \equiv \Delta \Psi + 2\pi f \Delta t - \Delta \phi$, with $\Delta t$ and $\Delta \phi$ an arbitrary time and phase shift. One can calculate $\mathcal{N}_\mrm{eff}$ for $\bar \lambda_\ell$ and $\bar \sigma_\ell$ by setting $\Delta \Psi = \Psi_{\bar \lambda_\ell}$ and $\Delta \Psi = \Psi_{\bar \sigma_\ell}$ respectively. The lower limit of integration $f_\mrm{min}$ is associated with the minimum frequency at which the detector has a non-negligible sensitivity, while $f_\mrm{max}$ is the minimum between the frequency at contact and the innermost stable circular orbit of a point particle in a Schwarzschild spacetime with its mass equivalent to the total mass of a binary (roughly the frequency at which the PN approximation breaks down). The advantage of using $\mathcal{N}_\mrm{eff}$ is that it is related to the Bayes factor (BF) between a model with $\Delta \Psi \neq 0$ and one with $\Delta \Psi =0$~\cite{Sampson:2014qqa}:
\ba
\label{eq:BF}
\ln \mrm{BF} & \approx & \min_{\theta^i}  \left[ 2 \pi^2 \left( \int_{f_\mrm{min}}^{f_\mrm{max}} \frac{|\tilde h(f)|^2}{S_n(f)} df \right) \mathcal{N}_\mrm{eff} ^2 \right]  :=   \min_{\theta^i} \left( \ln \mrm{BF}^{(u)} \right)\,,
\ea
where $\theta^i$ are the waveform parameters, while $\mrm{BF}^{(u)}$ is the upper bound on the BF without minimizing. For equal priors between two competing models, the BF corresponds to the odds that the data favors one model over the other. For example,  a BF of 10 means that one model is 10 times more likely to be correct than the other given the data. Since $\mrm{BF}^{(u)}$ is the \emph{upper bound} of the BF, if it exceeds a threshold BF, the corresponding phase ($\Delta \Psi$) may or may not be important. On the other hand, if $\mrm{BF}^{(u)}$ is below this threshold, then the model with the corresponding phase is disfavored, suggesting that such a term is not important in data analysis. 

\begin{figure}[htb]
\begin{center}
\includegraphics[width=8.5cm,clip=true]{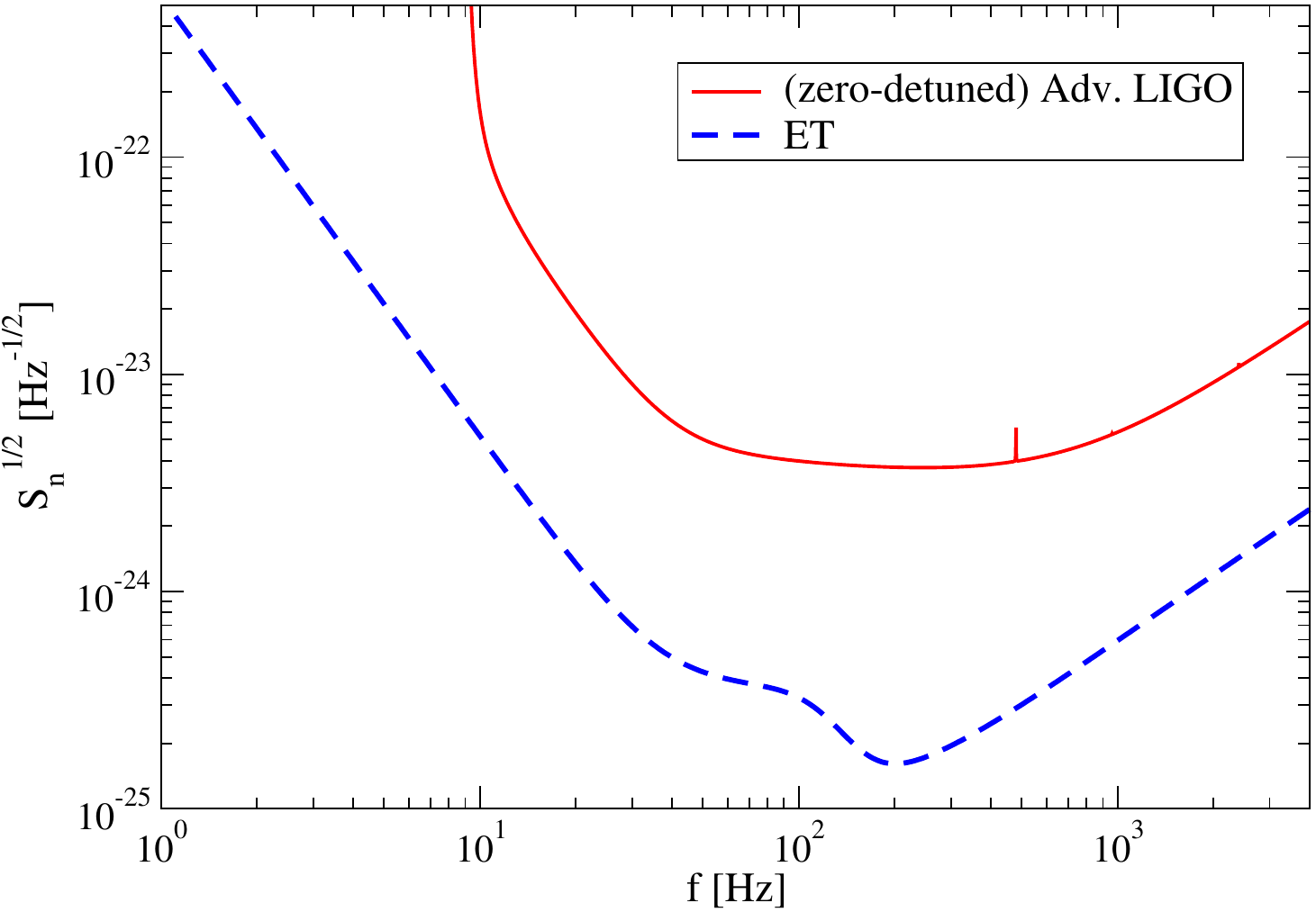}  
\caption{\label{fig:noise} 
Noise spectral density for Adv.~LIGO with zero-detuned configuration~\cite{Adhikari:2013kya} (red solid) and ET with the ``B'' configuration~\cite{Hild:2008ng,Mishra:2010tp,Sathyaprakash:2009xs} (blue dashed). This figure is taken and edited from Yagi~\cite{Yagi:2013sva}.
}
\end{center}
\end{figure} 

\begin{figure}[thb]
\begin{center}
\includegraphics[width=8.cm,clip=true]{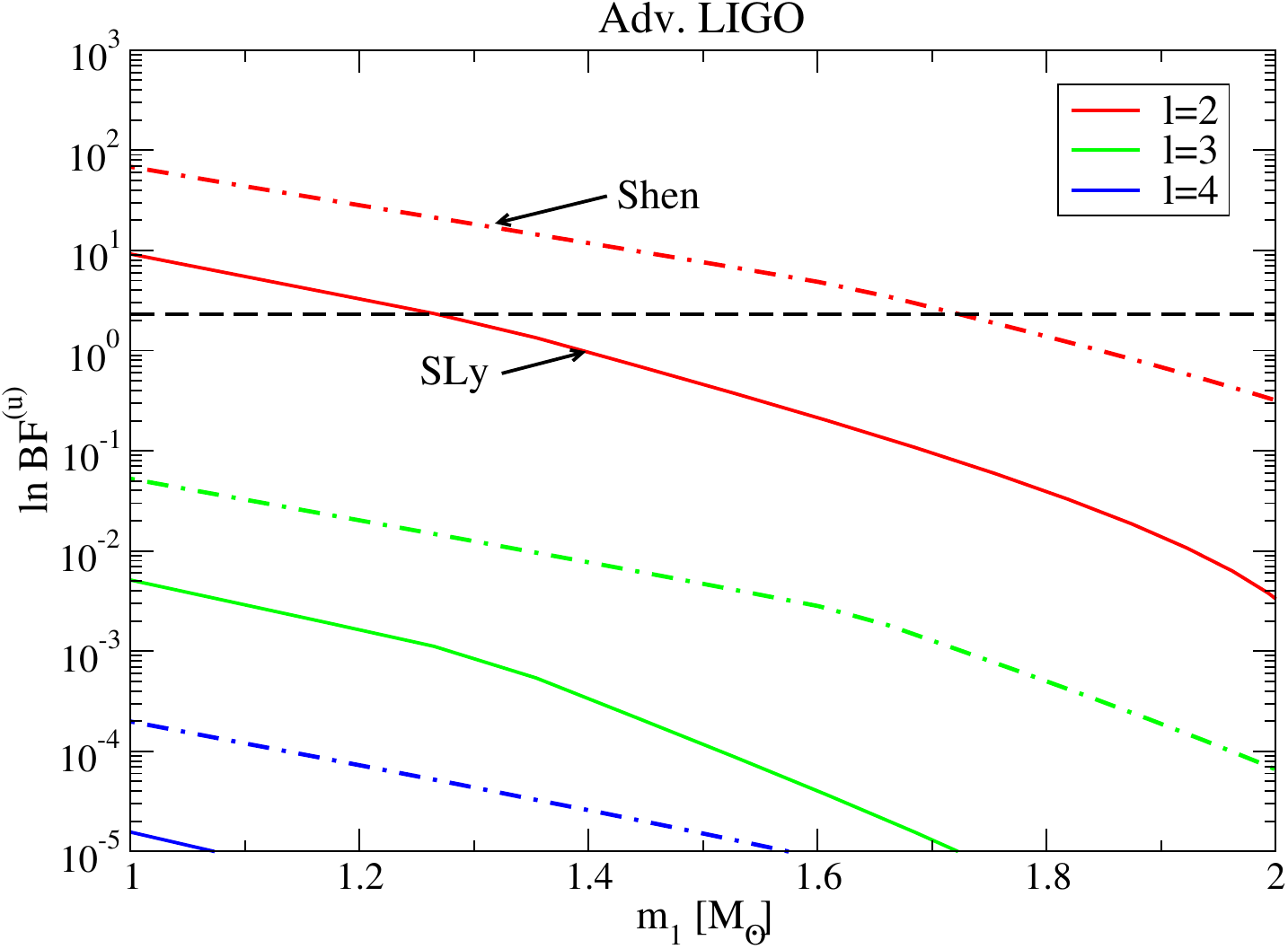}  
\includegraphics[width=8.cm,clip=true]{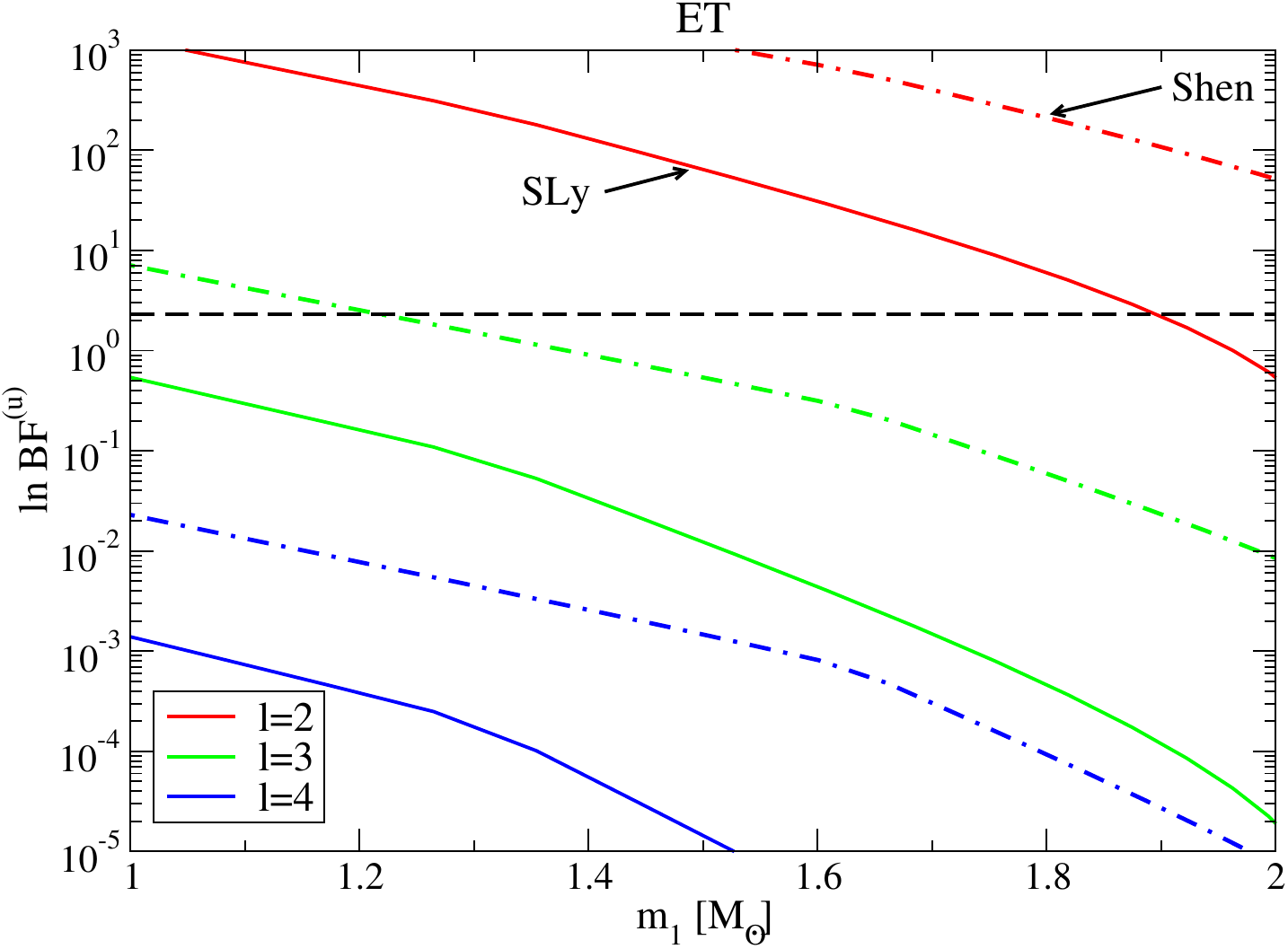}  
\caption{\label{fig:multipole-love-useful} 
Upper bound on the Bayes factor (defined via Eq.~\eqref{eq:BF}) between the waveform with $\bar \lambda_{\ell} \neq 0$ and $\bar \lambda_{\ell} = 0$ as a function of the neutron star mass for various $\ell$ and two representative equations of state. We assume that Adv.~LIGO (left) and ET (right) detects gravitational wave signals emitted from equal-mass, non-spinning neutron star binaries at 100Mpc. The dashed horizontal line shows BF~$=10$, which is the threshold BF chosen in this analysis. The contribution $\bar \lambda_\ell$ to the waveform phase may be important if the upper bound BF is larger than this threshold. Observe that higher-order tidal deformabilities are not important for Adv.~LIGO, while $\bar \lambda_3$ can be important for ET observations of low-mass neutron star binaries.
}
\end{center}
\end{figure}

Let us now estimate the importance of the tidal contributions to the Fourier phase through $\mrm{BF}^{(u)}$, assuming a future observation of a binary neutron star inspiral with Adv.~LIGO at a high-power/zero-detuned configuration in design sensitivity~\cite{AdvLIGO-noise,Ajith:2011ec}, and with an upgrade to the Einstein Telescope (ET) in the ``B'' configuration~\cite{Hild:2008ng,Mishra:2010tp,Sathyaprakash:2009xs}; the noise spectral densities of these detectors are shown in Fig.~\ref{fig:noise}. Figure~\ref{fig:multipole-love-useful} presents the upper bound of the Bayes factor with $\Delta \Psi = \Psi_{\bar \lambda_\ell}$ for an equal-mass neutron star binary at 100Mpc with Adv.~LIGO (left) and ET (right); $\Psi_{\bar \sigma_2} = 0$ for such equal-mass systems. In this figure, we use the Shen (SLy) equation of state as a representative member of the stiff (soft) equation-of-state class. As explained above, a BF analysis instructs us to take into account $\Psi_{\bar \lambda_\ell}$ if the corresponding $\mrm{BF}^{(u)}$ is larger than the threshold BF, which we take here to be 10 (horizontal dashed lines). Observe that $\Psi_{\bar \lambda_2}$ is the only term that seems to be relevant for Adv.~LIGO, while both $\Psi_{\bar \lambda_2}$ and $\Psi_{\bar \lambda_3}$ need to be taken into account for ET observations. Such a finding is consistent with Yagi~\cite{Yagi:2013sva} who calculated the \emph{useful} number of gravitational wave cycles~\cite{Damour:2000gg} and compared it with the inverse of the signal-to-noise ratio.

\begin{figure*}[thb]
\begin{center}
\includegraphics[width=7.5cm,clip=true]{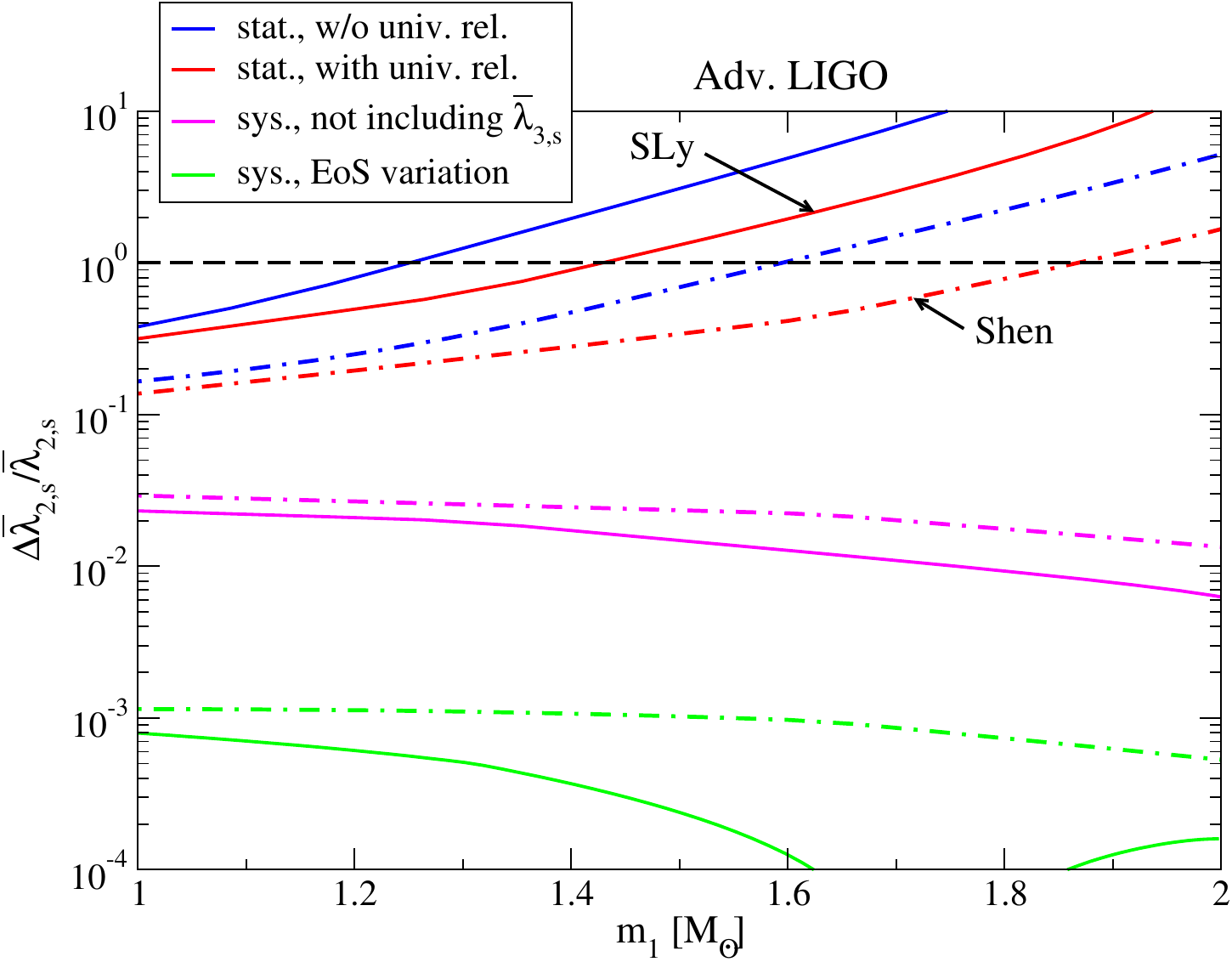}  
\includegraphics[width=7.5cm,clip=true]{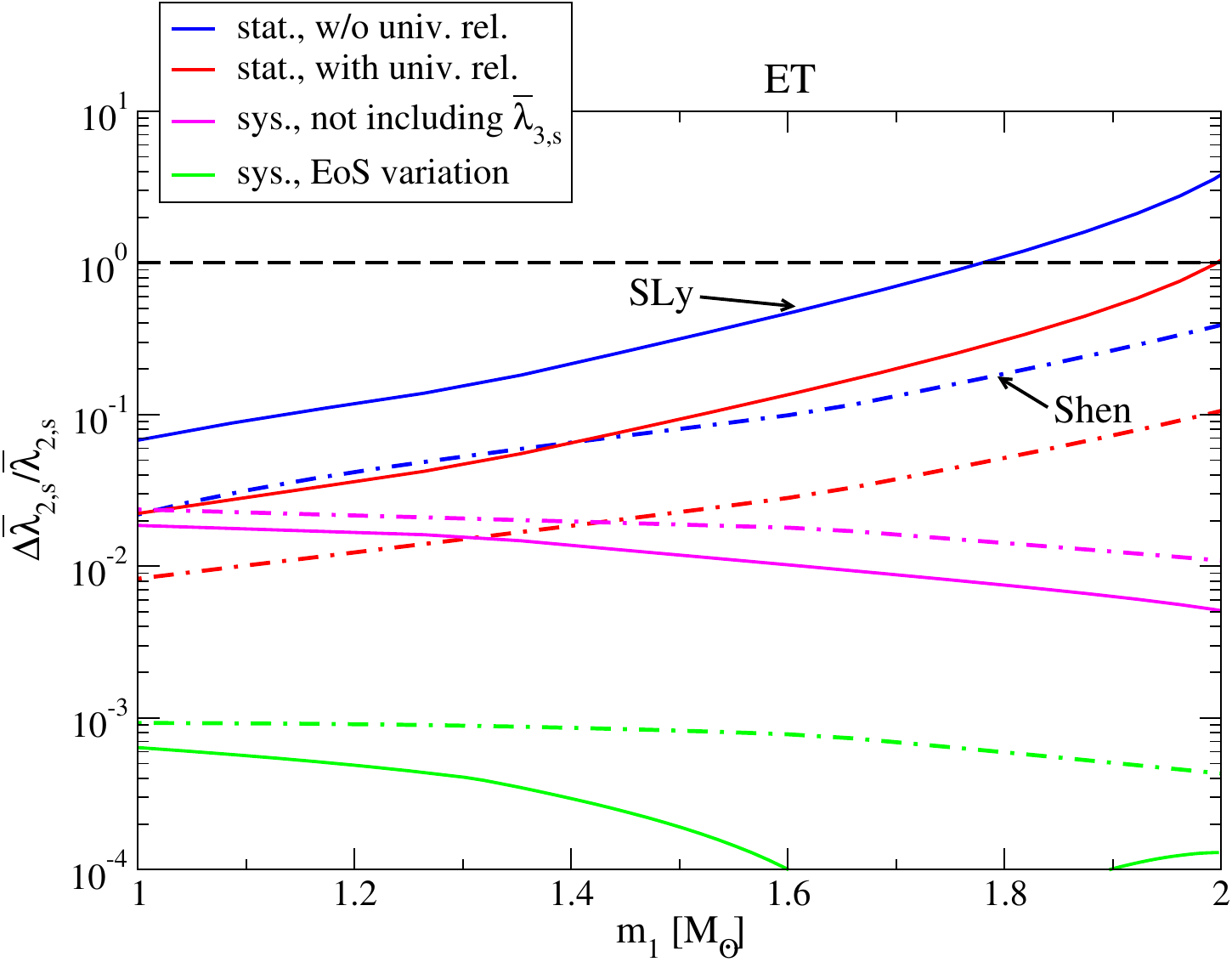}  
\caption{\label{fig:multipole-Love-LIGO3-ET} 
Fractional measurement accuracy of $\bar \lambda_{2,s}$ with Adv.~LIGO (left) and ET (right) as a function of the neutron star mass, assuming that these interferometers detect gravitational wave signals from equal-mass, non-spinning neutron star binaries at 100Mpc. The results are presented for the SLy (solid) and Shen (dotted-dashed) equations of state with (red) and without (blue) using the multipole Love relation. Observe that such a relation reduces the errors by a maximum factor of 3. $\bar \lambda_{2,s}$ is measurable if the curves are below the horizontal dashed unity lines. The magenta curves show the systematic errors due to not including $\bar \lambda_{3,s}$ in the search parameter set, which may dominate the error budget for ET observations of low-mass binaries. Green curves show the fractional systematic errors due to the equation-of-state variation in the multipole Love relation, which are much smaller than fractional statistical errors.
}
\end{center}
\end{figure*}

The  impact of the approximately universal multipole Love relations (see Sec.~\ref{sec:multipole-Love}) in measuring $\bar \lambda_{2,s}$, defined in Eq.~\eqref{eq:lambdas-lambfaa-def}, can be explained as follows. Consider a Fisher analysis as an approximate measure of the accuracy a given parameter $\theta^{i}$ can be measured to. The statistical error on the extraction of parameters $\theta^i$ is approximately given by $\Delta \theta^i = \sqrt{(\Gamma^{-1})_{ii}}$ with the Fisher matrix defined by
\be
\Gamma_{ij} \equiv ( \partial_i h | \partial_j h)\,, \quad (A|B) \equiv 2 \int^{f_\mrm{max}}_{f_\mrm{min}} \frac{\tilde{A}^*(f) \tilde{B}(f) + \tilde{A}(f) \tilde{B}^*(f)}{S_n(f)}  df\,,
\ee
where $*$ represents complex conjugation while the partial derivatives are to be taken with respect to the model parameters, $\partial_i \equiv \partial/\partial \theta^i$. Following Eq.~\eqref{eq:lambdas-lambfaa-def}, let us introduce the symmetric and antisymmetric combination of $\bar \lambda_3$ as $\bar \lambda_{3,s} \equiv (\bar \lambda_{3,1} + \bar \lambda_{3,2})/2$ and $\bar \lambda_{3,a} \equiv (\bar \lambda_{3,1} - \bar \lambda_{3,2})/2$. One can neglect $\bar \lambda_4$ since Fig.~\ref{fig:multipole-love-useful} shows that it is irrelevant for Adv.~LIGO observations. Since an equal-mass binary gives $\bar \lambda_{2,a} = 0 = \bar \lambda_{3,a}$, one can choose the parameter set to be 
\be
\label{eq:Fisher-par}
\theta^i = (\ln \mathcal{M}, \ln \eta, t_c, \phi_c, \ln D_L, \bar \lambda_{2,s}, \bar \lambda_{3,s}), 
\ee
where $\mathcal M \equiv m \eta^{3/5}$ is the chirp mass, $t_c$ and $\phi_c$ are the time and phase of coalescence, and $D_L$ is the luminosity distance to the source. We neglect neutron star spins for simplicity, as they are expected to be small just before coalescence~\cite{bildsten-cutler,Damour:2012yf} (though see Sec.~\ref{sec:GW-astro} on how universal relations may help one to measure neutron star spins). One can further impose Gaussian priors\footnote{Such priors were not considered in Yagi~\cite{Yagi:2013sva}.}~\cite{Cutler:1994ys,Poisson:1995ef,Berti:2004bd} that enforce $\bar \lambda_{2,s} \leq 10^4$ and $\bar \lambda_{3,s} \leq 5 \times 10^4$.
Figure~\ref{fig:multipole-Love-LIGO3-ET} shows the fractional statistical error on the measurement of $\bar \lambda_{2,s}$ as a function of the neutron star mass for an equal-mass, non-spinning neutron star binary with two different equations of state using Adv.~LIGO (left) and ET (right). The blue curves correspond to the case without using the universal $\bar \lambda_3$--$\bar \lambda_2$ relation (which is equivalent to the relation between $\bar \lambda_{3,s}$ and $\bar \lambda_{2,s}$ for an equal-mass binary), while the red curves show the result using this relation. The universal relation allows one to eliminate $\bar \lambda_{3,s}$ from the parameter set, which breaks the partial parameter degeneracy between $\bar \lambda_{3,s}$ and $\bar \lambda_{2,s}$ and improves the measurement accuracy of the latter by a factor of 3.

Although including $\bar \lambda_{3,s}$ and using the universal relations allows for a more accurate extraction of $\bar \lambda_{2,s}$, removing the former from the parameter list does not significantly bias the extraction of other parameters for Adv.~LIGO observations (although this is not quite true for ET). To see this, one can compare the statistical error on $\bar \lambda_{2,s}$ without including $\bar \lambda_{3,s}$ in the search parameter set to the systematic error due to not including $\bar \lambda_{3,s}$. The former is essentially given by the red curves in Fig.~\ref{fig:multipole-Love-LIGO3-ET} as the universal relations allow one to remove $\bar \lambda_{3,s}$ from the search parameter set. Following Cutler and Vallisneri~\cite{Cutler:2007mi}, one can calculate the latter from a Fisher analysis through 
\ba
\label{eq:sys-Cutler-Vallisneri}
\Delta_\sys \bar \lambda_{2,s} &=& \left(  \Gamma^{-1} \right)^{\bar \lambda_{2,s} j} \left( [h_\inj - h_\temp] \bigg| \frac{\partial h_\temp}{\partial \theta^j} \right) \nn \\
&=& \left(  \Gamma^{-1} \right)^{\bar \lambda_{2,s} j} i \left( \delta \Psi \, h_\temp \bigg| \frac{\partial h_\temp}{\partial \theta^j} \right) + \mathcal{O} \left( \delta \Psi^2 \right)\,,
\ea
where $h_\inj$ and $h_\temp$ are the injected waveform and the template waveform respectively and the Fisher matrix $\Gamma_{ij}$ is defined via the template waveform. The only difference between these two waveforms is whether one includes $\Psi_{\bar \lambda_3}$ or not. The Fourier transforms of such waveforms are related by $\tilde h_\inj = \tilde h_\temp \exp\left( i \delta \Psi \right)$, where 
\be
\label{eq:delta-Psi-sys}
\delta \Psi = - \frac{125}{12} \bar \lambda_{3,s} \frac{m_1^5+m_2^5}{m^5} x^{9/2}\,.
\ee
The magenta curves in Fig.~\ref{fig:multipole-Love-LIGO3-ET} present these systematic errors. Observe first that these errors are much smaller than the statistical errors (red curves) for Adv.~LIGO, suggesting that neglecting $\bar \lambda_{3,s}$ does not significantly bias the extraction of other physical parameters with second-generation detectors. On the other hand, observe that such systematics may dominate statistical errors for ET observations of low-mass neutron star binaries. In this case, including $\bar \lambda_{3,s}$ in the parameter set is crucial in decreasing the impact of systematics. This finding is consistent with the Bayes factor analysis in Fig.~\ref{fig:multipole-love-useful}. 

The small equation-of-state variation in the universal relations also produces systematic errors on $\bar \lambda_{2,s}$ when one uses such relations, though they are much smaller than statistical errors, as we now explain in this paragraph. One can estimate such systematics by replacing $\bar \lambda_{3,s}$ with $\bar \lambda_{3,s} - \bar \lambda_{3,s}(\bar \lambda_{2,s})$ in Eq.~\eqref{eq:delta-Psi-sys}\footnote{In~\cite{Yagi:2013sva}, Yagi compared statistical errors directly against the equation-of-state variation in the multipole Love relation and called the latter the systematic error. Here, we estimate systematic errors on $\bar \lambda_{2,s}$ more accurately using a Fisher analysis.}. Green curves in Fig.~\ref{fig:multipole-Love-LIGO3-ET} present such systematic errors. Observe that systematic errors due to the equation-of-state variation in the $\bar \lambda_3$--$\bar \lambda_2$ relation are much smaller than statistical errors. Moreover, the former are much smaller than the 5\% equation-of-state variation in the relation. This is because $\bar \lambda_{2,s}$ enters first at 5PN order while  $\bar \lambda_{3,s}$ enters first at 7PN order. These findings lead to an important conclusion: \emph{the measurability of binary parameters using universal relations is not necessarily limited by the amount of the equation-of-state variation in the universal relations.} The difference in the PN order at which these terms enter suppresses the correlation between $\bar \lambda_{2,s}$ and $\bar \lambda_{3,s}$, which reduces the systematic error. 

\begin{figure}[thb]
\begin{center}
\includegraphics[width=8.0cm,clip=true]{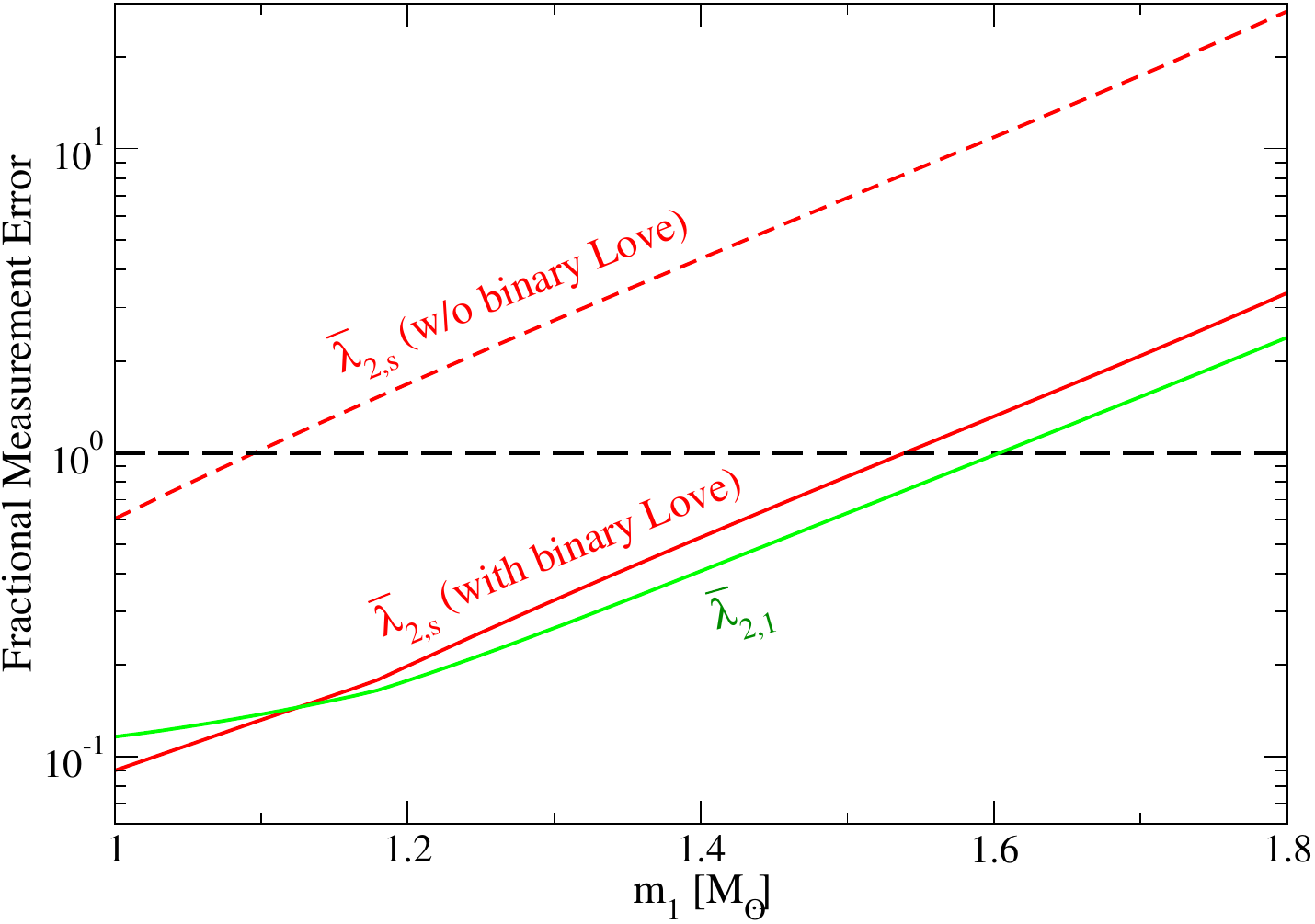}  
\includegraphics[width=8.3cm,clip=true]{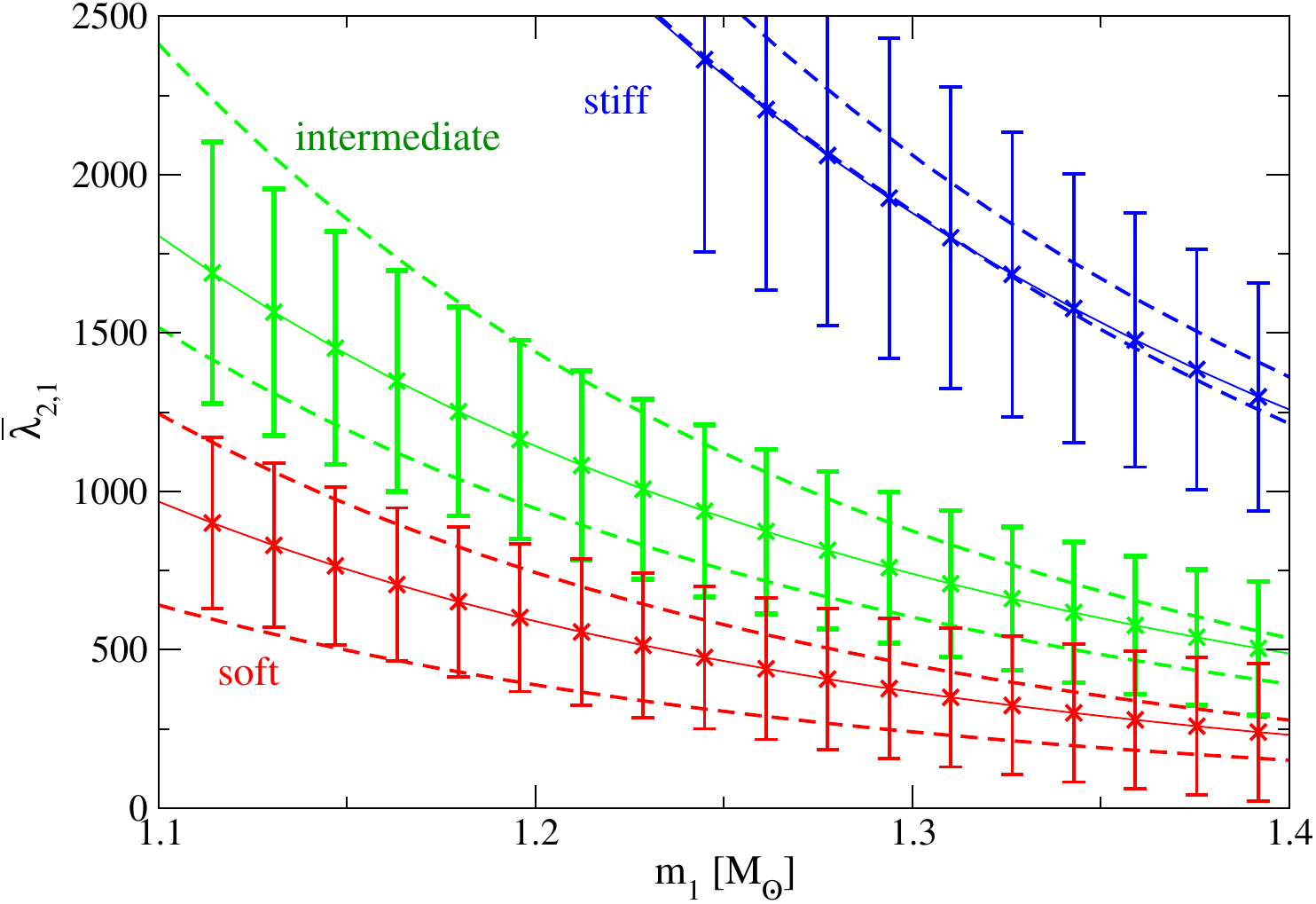}  
\caption{\label{fig:binary-love-nuclear} (Left) Fractional measurement accuracy of $\bar \lambda_{2,s}$ with (red solid) and without (red dashed) using the binary Love relation, and $\bar \lambda_{2,1}$ (green solid) as a function of $m_1$ using Adv.~LIGO. We assume that Adv.~LIGO detects gravitational wave signals from non-spinning neutron star binaries with a mass ratio $q = 0.9$, an signal-to-noise ratio of 30, where the correct equation of state is AP4. Observe that the binary Love relation reduces the measurement error by approximately an order of magnitude. 
(Right) The relation between $\bar \lambda_{2,1}$ and $m_1$ for three different classes of the equations of state, namely soft (red dashed region), intermediate (green dashed region) and stiff (blue dashed region). Within each class, we picked a fiducial equation of state and show 2-$\sigma$ error bars with the same condition as in the left panel (except for the equations of state). Observe that one can distinguish the stiff class easily, while one needs low-mass neutron star observations to distinguish the soft and intermediate classes.
This figure is taken and edited from Yagi and Yunes~\cite{Yagi:2015pkc,Yagi:2016qmr}. 
}
\end{center}
\end{figure}

Let us now focus on gravitational waves emitted by \emph{unequal-mass} neutron star binaries with Adv.~LIGO and see how the universal binary Love relations (Sec.~\ref{sec:binary-Love}) improve the measurability of $\bar \lambda_{2,s}$. In such a case, one needs to include both $\bar \lambda_{2,s}$ and $\bar \lambda_{2,a}$ into the parameter set, while one can neglect $\bar \lambda_3$ and $\bar \lambda_4$ as shown in Fig.~\ref{fig:multipole-love-useful}. The left panel of Fig.~\ref{fig:binary-love-nuclear} presents the fractional measurement accuracy of $\bar \lambda_{2,s}$ as a function of the smaller neutron star mass in a binary, assuming Adv.~LIGO detects a gravitational wave signal from a binary with a mass ratio of $q = 0.9$, a signal-to-noise ratio of 30 and the AP4 equation of state. The red dashed curve shows results without using the universal $\bar \lambda_{2,s}$--$\bar \lambda_{2,a}$ relation, while the red solid curve shows those obtained using the relation. Observe that the universal relation improves the measurement accuracy of $\bar \lambda_{2,s}$ by an order of magnitude. This is because the relation eliminates $\bar \lambda_{2,a}$ from the parameter set, which breaks the partial parameter degeneracy between $\bar \lambda_{2,s}$ and $\bar \lambda_{2,a}$. Combining the measured $\bar \lambda_{2,s}$ and the $\bar \lambda_{2,s}$--$\bar \lambda_{2,a}$ relation, one can then automatically infer the value of $\bar \lambda_{2,a}$. Combining further $\bar \lambda_{2,s}$ and $\bar \lambda_{2,a}$, one can infer the individual tidal deformabilities, $\bar \lambda_{2,1}$ and $\bar \lambda_{2,2}$; the accuracy of inferring the former is shown by the green solid curve in the left panel of Fig.~\ref{fig:binary-love-nuclear}.

How can one use such a measurement of the individual tidal deformabilities to probe nuclear physics? This can be done by combining the inferred values of $\bar \lambda_{2,1}$ and $m_1$, as the relation between these quantities depends strongly on the equation of state. To show how this can be done, let us classify the $\bar \lambda_{2,1}$--$m_1$ relation with respect to three classes of equations of state: stiff, intermediate and soft. The regions inside the dashed curves in the right panel of Fig.~\ref{fig:binary-love-nuclear} show the   $\bar \lambda_{2,1}$--$m_1$ relation within each class. Now, if one selects a representative equation of state and estimates the measurement accuracy of $\bar \lambda_{2,1}$ for different injected masses $m_1$, one can superimpose in these regions the extracted $\bar \lambda_{2,1}$ and its error (the solid lines and error bars in the right panel of Fig.~\ref{fig:binary-love-nuclear}). Thus, if the error bars are small enough, one may be able to separate the different classes of equations of state. Observe that Adv.~LIGO should easily be able to distinguish the stiff class from the other two classes, while it may further be able to distinguish the soft and intermediate classes if the neutron star masses are relatively small $(m_1 \lesssim 1.3 M_\odot)$. 

\begin{figure*}[htb]
\begin{center}
\includegraphics[width=8.cm,clip=true]{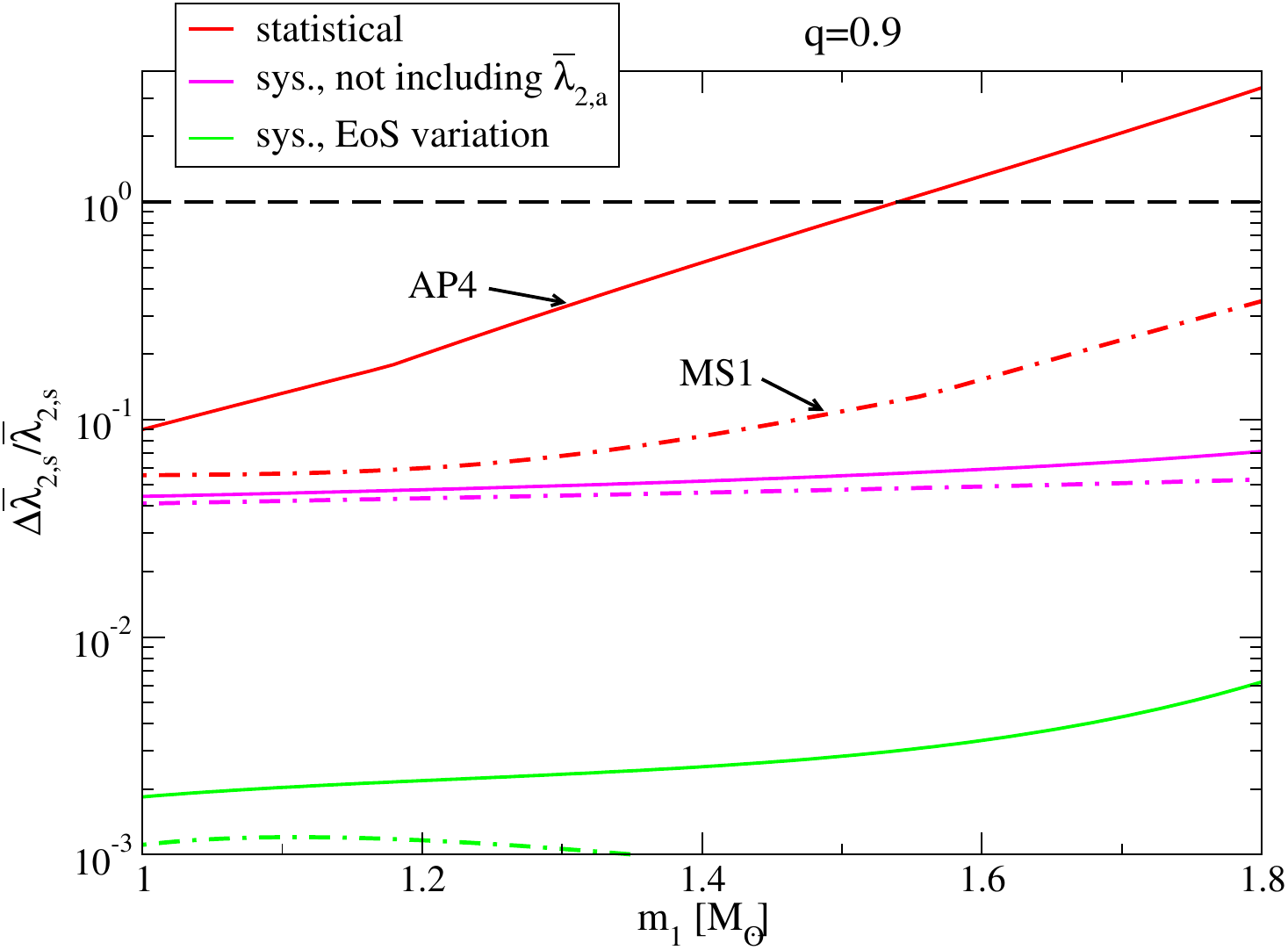}  
\includegraphics[width=8.cm,clip=true]{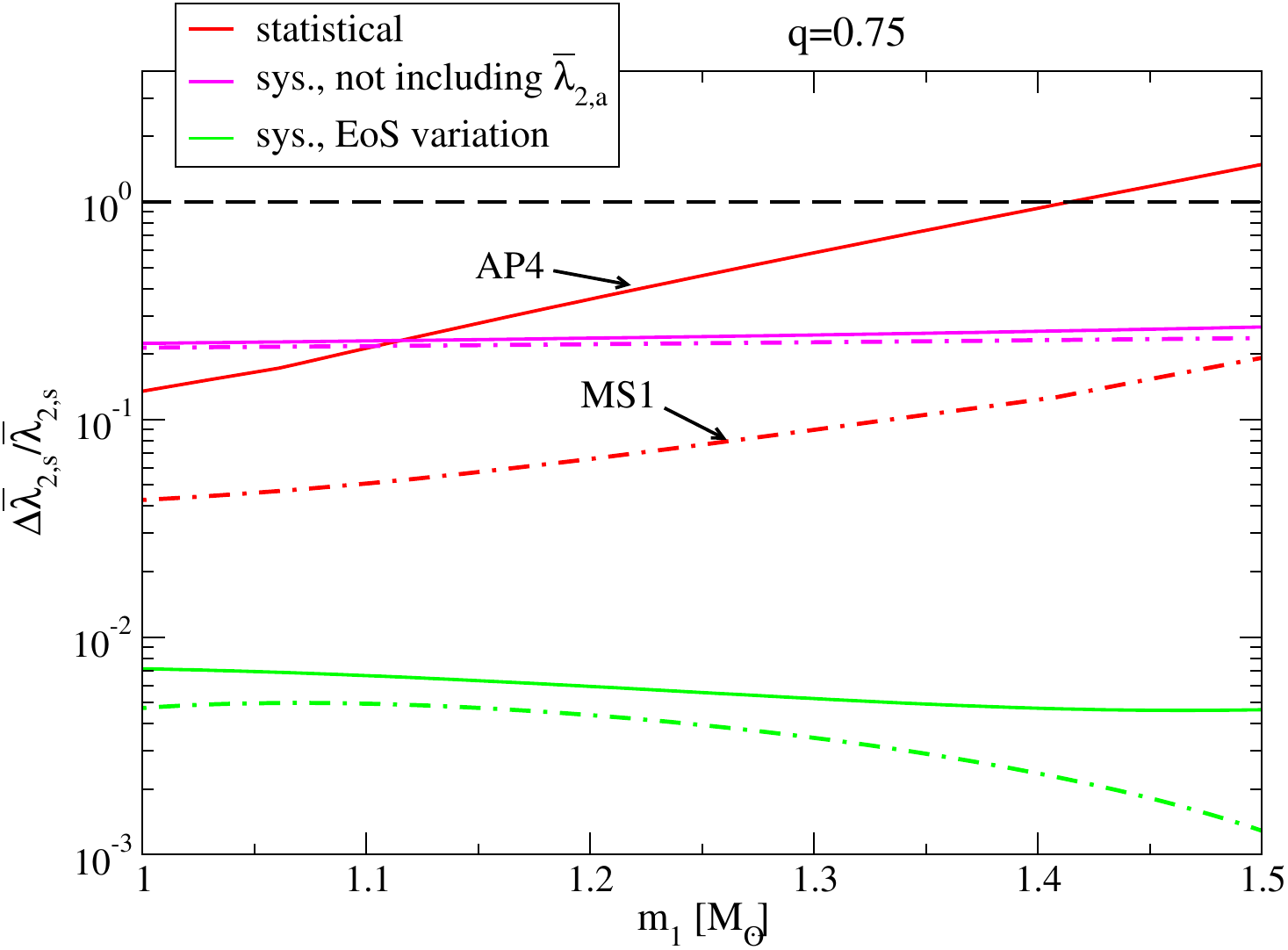}  
\caption{\label{fig:Fisher-sys-q} Fractional statistical (red) and systematic errors on $\bar \lambda_{2,s}$ as a function of $m_1$ with $q=0.9$ (left) and $0.75$ (right). The systematic errors are due to not not including $\bar \lambda_a$ (magenta) and the equation-of-state variation in the $\bar \lambda_{2,s}$--$\bar \lambda_{2,a}$ relation (green). These results are obtained assuming an Adv.~LIGO observation with signal-to-noise ratio of 30. Observe that if one does not take $\bar \lambda_{2,a}$ into account, the systematic errors may dominate the statistical errors for small $q$. Observe also that the systematic errors due to the equation-of-state variation is always smaller than other errors. This figure is taken and edited from Yagi and Yunes~\cite{Yagi:2016qmr}.
}
\end{center}
\end{figure*}

The inclusion of $\bar \lambda_{2,a}$ in the search parameter set is important for binaries with small $q$, as we explain next. The red curves in Fig.~\ref{fig:Fisher-sys-q} show the statistical error on the measurement of $\bar \lambda_{2,s}$ for a $q=0.9$ binary (left) and $q=0.75$ binary (right) without taking $\bar \lambda_{2,a}$ into account. On the other hand, one can calculate the systematic error on the measurement of $\bar \lambda_{2,s}$ due to neglecting $\bar \lambda_{2,a}$ using Eq.~\eqref{eq:sys-Cutler-Vallisneri} with 
\be
\label{eq:sys-inj-temp}
\delta \Psi = - \frac{9}{16} \frac{1 + 9 \eta -11 \eta^2}{\eta} \delta m \, \bar \lambda_{2,a} \,  x^{5/2}\,.
\ee
Here, $\delta m \equiv (m_2-m_1)/(m_1+m_2)$ is the dimensionless mass difference and we only consider the leading tidal effect in the phase for simplicity. Using $i \, h_\temp = \left( \partial \tilde{h}_\temp/\partial \bar \lambda_{2,s} \right) \left( \partial \Psi / \partial \bar \lambda_{2,s} \right)^{-1}$, one can show that the Fisher matrix factor cancels out and the fractional systematic error on $\bar \lambda_{2,s}$ reduces to~\cite{Yagi:2016qmr}
\be
\label{eq:frac-sys-Cutler-Vallisneri}
\frac{\Delta_\sys \bar \lambda_{2,s}}{\bar \lambda_{2,s}} =  \frac{\delta \Psi}{\bar \lambda_{2,s}} \left( \frac{\partial \Psi}{\partial \bar \lambda_{2,s}} \right)^{-1}\,.
\ee
Notice that the partial derivative in the above equation in principle depends on frequency, but this cancels with the frequency dependence of $\delta \Psi$. The magenta curves in Fig.~\ref{fig:Fisher-sys-q} show the systematic error in the extraction of $\bar \lambda_{2,s}$ due to neglecting $\bar \lambda_{2,a}$. Observe that this error becomes more important as one decreases $q$, since $\bar \lambda_{2,a}$ enters the phase multiplied by $\delta m$. In the small $q$ regime, the $\bar \lambda_{2,s}$--$\bar \lambda_{2,a}$ relation is crucial in reducing the systematic error such that it is below the statistical error. Notice also that statistical errors scale inversely with the signal-to-noise ratio, while systematic error is independent of the latter (see Eq.~\eqref{eq:frac-sys-Cutler-Vallisneri}). Thus, even when $q=0.9$, the systematic error may dominate the error budget for 3rd generation gravitational wave interferometers, such as ET. 

\begin{figure}[htb]
\begin{center}
\includegraphics[width=7.cm,clip=true]{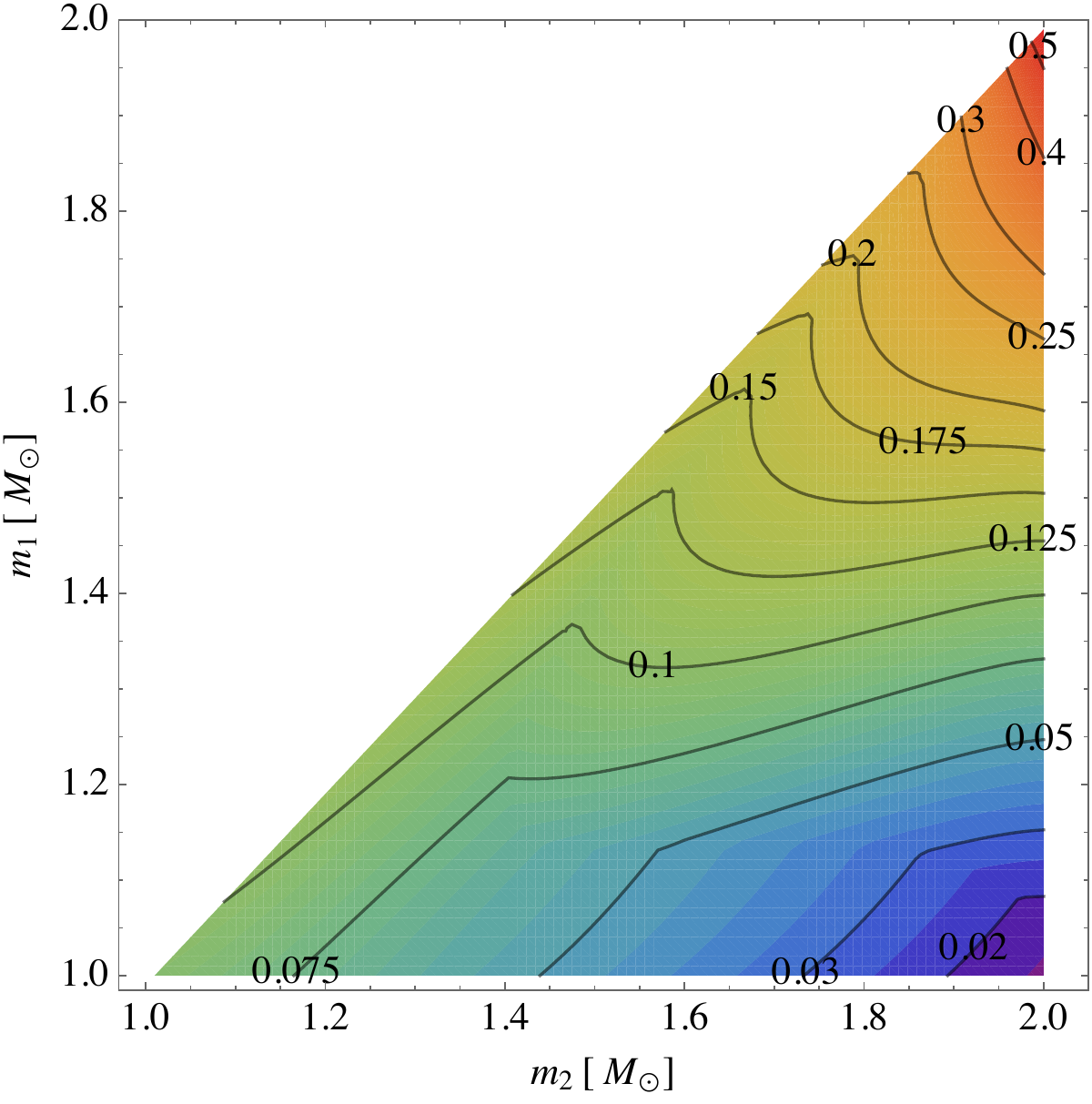}  
\includegraphics[width=7.cm,clip=true]{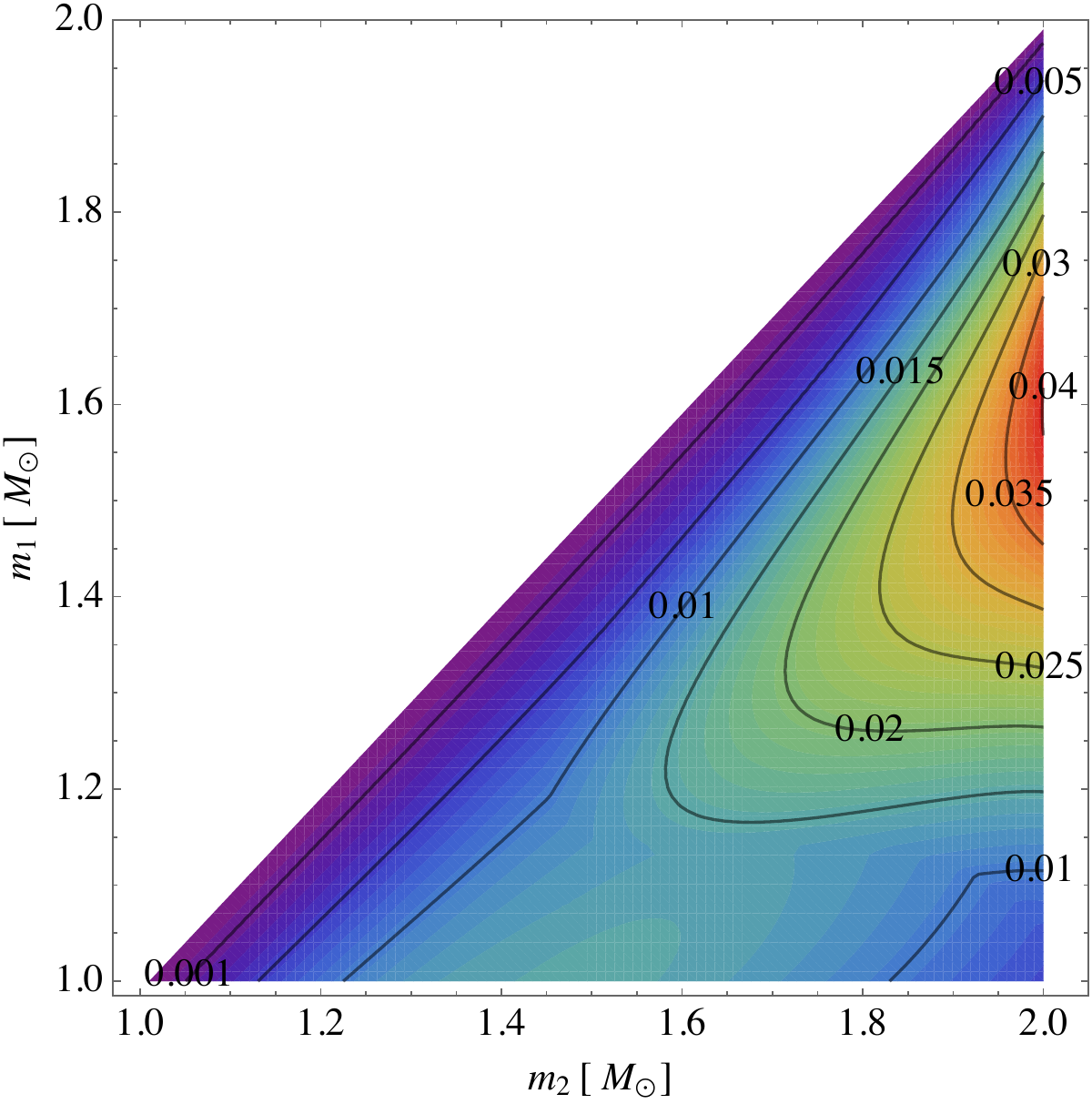}  
\caption{\label{fig:lambdas-lambdaa-frac-diff-max} (Left) Maximum absolute fractional difference between the numerically obtained $\bar \lambda_{2,s}$--$\bar \lambda_{2,a}$ relation and the fit explained in Sec.~\ref{sec:binary-Love} as a function of neutron star masses with $m_1 \leq m_2$. 
(Right) The maximum fractional systematic error on $\bar \lambda_{2,s}$ using the $\bar \lambda_{2,s}$--$\bar \lambda_{2,a}$ relation due to its equation-of-state variation. Observe that such systematic errors are much smaller than the equation-of-state variation in the left panel. This figure is taken from Yagi and Yunes~\cite{Yagi:2016qmr}.
}
\end{center}
\end{figure}

Although the equation-of-state variation in the $\bar \lambda_{2,s}$--$\bar \lambda_{2,a}$ relation can reach 50\%, we now show that the systematic error on $\bar \lambda_{2,s}$ due to this equation-of-state sensitivity is always smaller than 4\%, which is much smaller than the statistical error on the extraction of this parameter. The left panel of Fig.~\ref{fig:lambdas-lambdaa-frac-diff-max} shows the maximum fractional difference in the relation relative to a fit that uses eleven equations of state (AP3, AP4, SLy, WFF1, WFF2, ENG, MPA1, MS1, MS1b, LS220, and Shen) as a function of the component neutron star masses. One can calculate systematic error on the extracted $\bar \lambda_{2,s}$ due to the equation-of-state variation in the $\bar \lambda_{2,s}$--$\bar \lambda_{2,a}$ relation using Eq.~\eqref{eq:frac-sys-Cutler-Vallisneri} and simply replacing $\bar \lambda_{2,a}$ with $ \bar \lambda_{2,a} - \bar \lambda_{2,a} \left( \bar \lambda_{2,s} \right)$ in Eq.~~\eqref{eq:sys-inj-temp}. The right panel of Fig.~\ref{fig:lambdas-lambdaa-frac-diff-max} shows the maximum fractional systematic error on $\bar \lambda_{2,s}$ as a function of the neutron star masses in a binary. Observe that this error is at most $\sim 4\%$, which is much smaller than the maximum fractional equation-of-state variation in the universal relation in the left panel of Fig.~\ref{fig:lambdas-lambdaa-frac-diff-max}. This is because the systematic error is suppressed by $\delta m$ in Eq.~\eqref{eq:sys-inj-temp} for nearly equal-mass systems. The equation-of-state variation becomes maximal when the two masses are large, while $\delta m$ becomes larger for smaller mass ratio systems. This is why the systematic error is maximum when $m_2$ is large while $m_1$ is intermediate. Figure~\ref{fig:Fisher-sys-q} also shows the systematic error on the extracted $\bar \lambda_{2,s}$ due to equation-of-state variation (green curves). Observe that this error is always much smaller than the statistical error and the systematic error due to neglecting $\bar \lambda_{2,a}$ in the search parameter set.

%--------------------------------------
\subsection{Nuclear Physics through X-Ray Observations}
\label{sec:nuclear-Xray}

Another way to probe nuclear physics is to measure the neutron star mass and radius independently from electromagnetic observations~\cite{Miller:2013tca}, as the relation between these two quantities depend strongly on the equation of state. However, most observations of the radius of neutron stars suffer from potential systematic errors~\cite{Miller:2013tca}. A possible exception is the use of energy-dependent models for the interpretation of spectra emitted by hot spots on the surface of rotating neutron stars, a target for the NICER, LOFT and AXTAR missions. Lo \textit{et al.}~\cite{Lo:2013ava} and Miller and Lamb~\cite{Miller:2014mca} showed that one does not find significantly biased results with good fits if the spot shape, spot temperature distribution, beaming pattern, or the spectrum is modeled incorrectly. 

The hot-spot models encode information about the neutron star mass and radius through the stellar shape and the neutron star spacetime on which photons propagate. The waveform, however, depends on other stellar quantities, such as the stellar ellipticity, the moment of inertia and the quadrupole moment~\cite{Psaltis:2013fha}. One needs to take these quantities into account to achieve the desired 5--10\% measurement accuracy on the mass and radius~\cite{Morsink:2007tv,Psaltis:2013zja}. As in the gravitational wave case discussed in Sec.~\ref{sec:nuclear-GW}, universal relations can be used to eliminate some of the parameters from the model, thus analytically breaking degeneracies between the mass, radius and other parameters and improving the measurement accuracy of the former two. For example, the I-Q relation can be used to eliminate the quadrupole moment, while the I-C relation in Fig.~\ref{fig:I-Love-C} can be used to eliminate the moment of inertia from the parameter set. One can further use the approximate universal relation between the stellar ellipticity and compactness~\cite{Morsink:2007tv,Baubock:2013gna} to eliminate the former.   

One can estimate the measurement accuracy of the neutron star radius as follows. After using all of the universal relations described above, the only parameters left in the hot-spot model are the stellar mass, the radius, and a certain combination ($\sin \theta_0 \sin \theta_s$) of the observer's inclination angle $\theta_0$ and the colatitude angle of the hot spot $\theta_s$, assuming that the spot size is negligibly small~\cite{Psaltis:2013fha}. These three quantities can be extracted from three independent observables, such as the amplitudes of the bolometric flux oscillation $C_1$, its second harmonic $C_2$ and the spectral color oscillation; the latter corresponds to the ratio between the number of photons with energies above and below the temperature of blackbody emission. Moreover, the ratio of the first two observables does not depend on the stellar mass. Thus, one can convert the measurement accuracy of the amplitude of the second harmonic to that of the radius to find~\cite{Psaltis:2013fha}
\be
\frac{\Delta R}{R} \sim 0.055 \left( \frac{C_1}{0.3} \right)^{-1} \left( \frac{f}{600\mathrm{Hz}} \right)^{-1} \left( \frac{R}{10\mathrm{km}} \right)^{-1} \left( \frac{\sin \theta_0}{0.5} \right)^{-1} \left( \frac{\sin \theta_s}{0.5} \right)^{-1} \left( \frac{S}{10^6} \right)^{-1}\,.
\ee
Here, $S$ is the total number of source counts, which is assumed to be larger than the number of background counts. This estimate shows that one needs to observe moderately fast spinning neutron star with roughly $10^6$ source counts to achieve a 5\% measurement accuracy for the radius. Such a measurement accuracy, of course, would be impossible without use of the universal relations.

%%%%%%%%%%%%%%%%%%%%%%%%%%%%%%%%%%%%%%%%%%%%%%%%%%%%%%
\subsection{Gravitational Wave Astrophysics}
\label{sec:GW-astro}

We now review how one can apply the universal Q-Love relation to improve the measurement accuracy of neutron star spins with gravitational wave observations~\cite{Yagi:2013bca,Yagi:2013awa}. The effect of spins on gravitational waves emitted by compact binaries enters first at relative 1.5PN order~\cite{Blanchet:2002av}, and it is characterized by the symmetric and antisymmetric combination of the spin angular momentum vectors $\vec{S}_A$ of the binary components: $\vec \chi_s \equiv (\vec \chi_1 + \vec \chi_2)/2$ and $\vec \chi_a \equiv (\vec \chi_1 - \vec \chi_2)/2$, with $\vec \chi_A \equiv \vec{S}_A/m_A^2$. On the other hand, the effect of the quadrupole moments of the compact bodies enters first at 2PN order~\cite{Poisson:1997ha,Mikoczi:2005dn}, and it is characterized by the combinations $\bar Q_s \equiv (\bar Q_1 + \bar Q_2)/2$ and $\bar Q_a \equiv (\bar Q_1 - \bar Q_2)/2$. Without knowledge of the universal relations, the quadrupole moments must be taken as independent search parameters in the waveform models due to their unknown equation-of-state dependance. Degeneracies between the spins and the quadrupole moments, however, greatly deteriorate the measurement accuracy of both sets of quantities. One can break these degeneracies analytically by using the approximately universal Q-Love relation. That is, one can prescribe the quadrupole moments through the tidal deformabilities without a priori knowledge of the correct equation of state, thus eliminating the quadrupole moments from the search parameter set. This procedure is similar to applying the multipole Love and binary Love relations to break degeneracies between tidal parameters, as explained in Sec.~\ref{sec:nuclear-GW}. 

Let us then carry out a Fisher analysis to see how the spin measurement improves when one uses the Q-Love relation in gravitational wave data analysis. Following Yagi and Yunes~\cite{Yagi:2013bca,Yagi:2013awa}, we focus on spin-aligned and nearly-equal mass binaries, and do not consider $\bar Q_a$ and $\bar \lambda_a$ as these parameters contribute negligibly to the waveform for such systems. If one does not use the Q-Love relation, a possible search parameter set is   
\be
\label{eq:Fisher-par2}
\theta^i = (\ln \mathcal{M}, \delta m, \chi_s,\chi_a, t_c, \phi_c, \ln D_L, \bar \lambda_{2,s}, \bar Q_s), 
\ee
where $\chi_s$ ($\chi_a$) is the magnitude of $\vec \chi_s$ ($\vec \chi_a$)\footnote{Yagi and Yunes~\cite{Yagi:2013bca,Yagi:2013awa} used a different parameterization for the spins when one does not apply the Q-Love relation.}.If one uses the Q-Love relation, one removes $\bar Q_s$ from the above expression and replaces $\bar Q_s$ with $\bar Q_s (\bar \lambda_{2,s})$ in the waveform. We will here use the same waveform model as that used in Yagi and Yunes~\cite{Yagi:2013bca,Yagi:2013awa}, with the same Gaussian priors~\cite{Cutler:1994ys,Poisson:1995ef,Berti:2004bd} that physically enforce $\delta m \leq 1/3$, $\chi_s \leq 0.1$ and $\chi_a \leq 0.1$; these priors are crucial in the analysis, especially for determining poorly constrained parameters, such as $\chi_a$. 

\begin{figure}[htb]
\begin{center}
\begin{tabular}{l}
\includegraphics[width=8.5cm,clip=true]{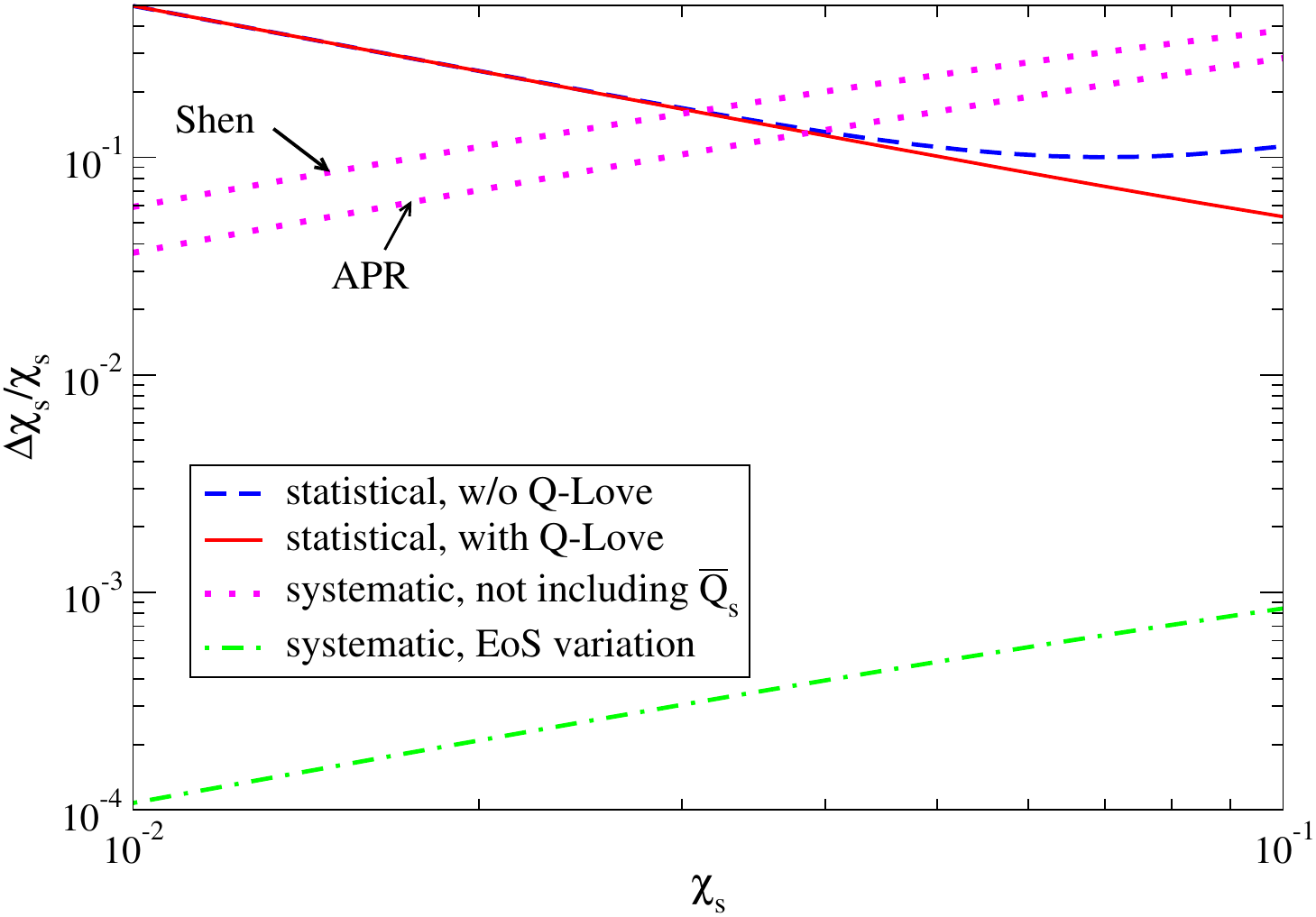} 
\end{tabular}
\caption{\label{fig:chis}
Fractional measurement accuracy (statistical error) of $\chi_s$ as a function of the injected $\chi_s$ with (red solid) and without (blue dashed) using the Q-Love universal relation. We assume that Adv.~LIGO with the zero-detuned configuration detects gravitational wave signals emitted from neutron star binaries with masses $(1.39,1.41)M_\odot$ and equal spins at 100Mpc. Observe that the universal relation improves the measurement accuracy of $\chi_s$ when $\chi_s \gtrsim 0.04$. We also present the fractional systematic error on $\chi_s$ due to not including $\bar Q_s$ in the search parameter set (magenta dotted) and due to the equation-of-state variation in the Q-Love relation (green dotted-dashed). Observe that the former can dominate the error budget for large $\chi_s$, while the latter is negligible compared to statistical errors. These results are insensitive to the equation of state, except for the systematic error due to not including $\bar Q_s$, which we present with two representative equations of state.
}
\end{center}
\end{figure}

Figure~\ref{fig:chis} shows the fractional measurement accuracy of $\chi_s$ as a function of the injected $\chi_s$ with (red solid) and without (blue dashed) using the Q-Love relation. Observe that the relation improves the measurement accuracy of $\chi_s$ when $\chi_s \gtrsim 0.04$. The relation only affects the results for larger spins as the contribution of the quadrupole moment becomes important only in that case. When $\chi_s = 0.1$, the improvement is roughly a factor of two. On the other hand, the measurement accuracy of $\chi_a$ is unaffected by the relation because $\chi_a$ is poorly measured and its measurement accuracy is always dominated by the prior~\cite{Yagi:2013bca,Yagi:2013awa}.

We end this subsection by discussing systematic errors on $\chi_s$. Such errors can be estimated by changing $\bar \lambda_{2,s}$ to $\chi_s$ in Eq.~\eqref{eq:sys-Cutler-Vallisneri} and $\delta \Psi$ is given by
\be
\delta \Psi = - \frac{75}{64} \frac{\delta \bar Q_s}{\eta} \frac{\left( m_1^2 \chi_1^2 + m_2^2 \chi_2^2  \right)}{m^2} x^{-1/2}\,.
\ee
Here either $\delta \bar Q_s = \bar Q_s$ or $\delta \bar Q_s = \bar Q_s - \bar Q_s (\bar \lambda_{2,s})$, depending on whether we wish to estimate the systematic errors due to not including $\bar Q_s$ into the search parameter set, or the errors due to the equation-of-state variation in the Q-Love relation. These systematic errors are shown by the magenta dotted and green dotted-dashed curves in Fig.~\ref{fig:chis}. Observe first that the systematic error due to not including $\bar Q_s$ is larger than the statistical error for large $\chi_s$. In fact, this occurs roughly at the same value ($\chi_{s} \approx 0.04$) at which the Q-Love relation begins to have an effect in reducing the statistical error. Observe also that the systematic error due to the equation-of-state variation is orders of magnitude smaller than statistical errors. The equation-of-state variation in the Q-Love relation with the stellar mass of 1.4$M_\odot$ is $\sim 0.5\%$ at most. The systematic error on $\chi_s$ is much smaller than this equation-of-state variation, as $\chi_s$ and $\bar Q$ enter first at different PN orders (1.5PN and 2PN respectively), which suppresses the amount of correlation between the two. The effect of $\chi_s$ at 1.5PN order is of $\mathcal{O}(\chi_s)$, while that at 2PN order is of $\mathcal{O}(\chi_s^2)$. Thus, the correlation becomes stronger as one increases $\chi_s$, which makes the systematic error larger, as shown in Fig.~\ref{fig:chis}. These results show that taking $\bar Q_s$ into account is important and using the Q-Love relation is crucial in bringing the measurement error of $\chi_s$ down to the red curve in Fig.~\ref{fig:chis} for large $\chi_s$ $(\chi_s \gtrsim 0.04)$.

%%%%%%%%%%%%%%%%%%%%%%%%%%%%%%%%%%%%%%%%%%%%%%%%%%%%%%
\subsection{Experimental Relativity}
\label{sec:application-gravitational}

\begin{figure}[htb]
\begin{center}
\begin{tabular}{l}
\includegraphics[width=9.cm,clip=true]{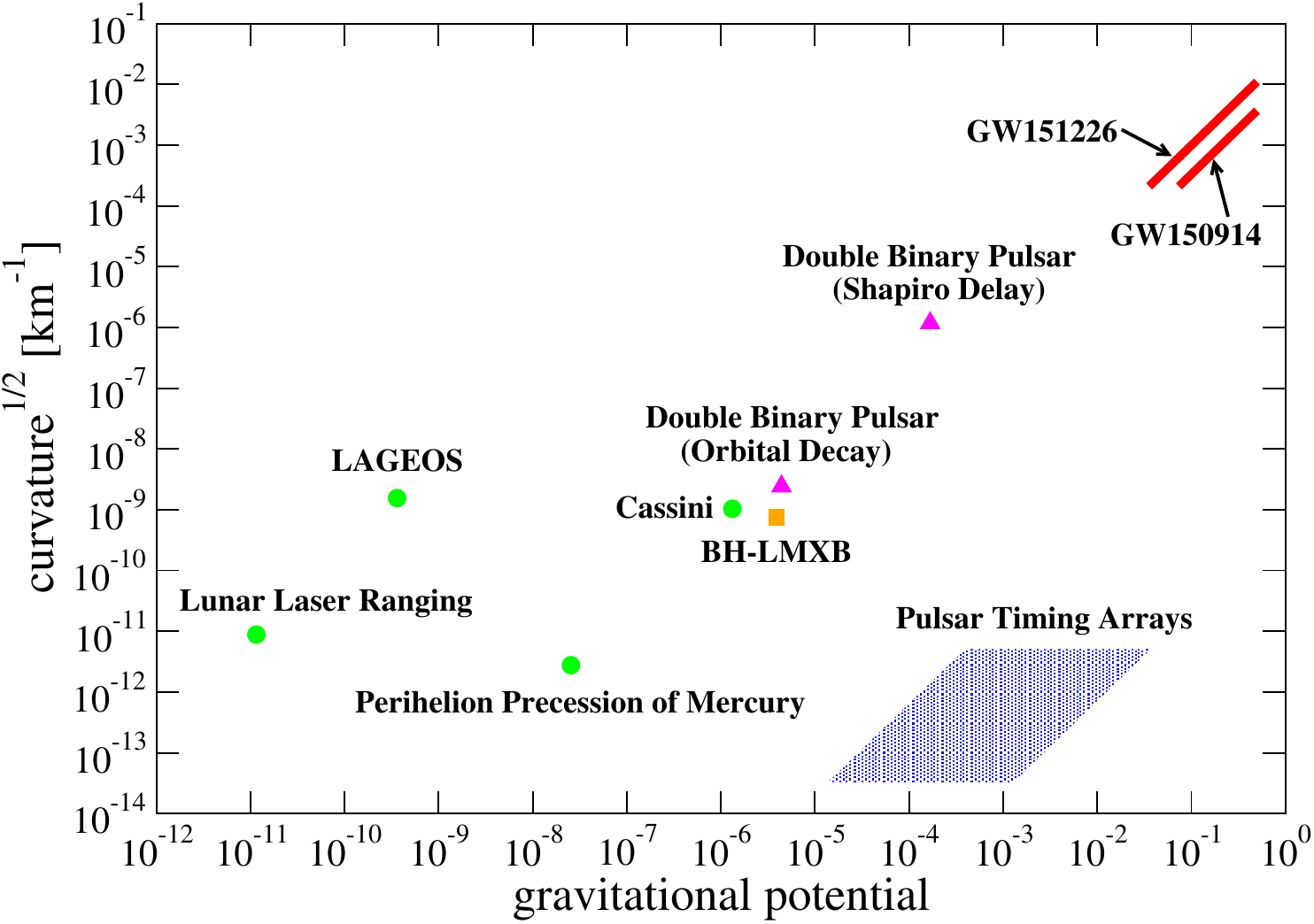} 
\end{tabular}
\caption{\label{fig:phase-diagram}
A phase diagram showing the corresponding strength of the gravitational potential and (the square root of) the curvature of systems probed for tests of General Relativity with Solar System experiments (green circles), binary pulsar observations (magenta triangles), a black hole-LMXB observation (orange square) and gravitational wave observations (red lines). For reference, we also show the region that can be probed with future pulsar time arrays (shaded blue region). Observe that certain binary pulsar observations allow one to probe regions that have a stronger curvature than the GW150914 and GW151226 black hole binaries. Observe also that the gravitational wave sources are shown by lines instead of points, which indicate that such sources are highly dynamical, leading to extreme field tests of gravity. This figure is taken and edited from Yunes \et~\cite{Yunes:2016jcc}. 
}
\end{center}
\end{figure}

We next review applications of the universal I-Love-Q relations aimed at probing extreme gravity~\cite{Yunes:2013dva,Gair:2012nm,Berti:2015itd,Yagi:2016jml}, i.e.~a regime where the gravitational interaction is both strong and highly dynamical. General Relativity has already been constrained rather strongly from Solar System experiments in the weak-field and weakly-dynamical regime~\cite{TEGP,Will:2014kxa}, and from binary pulsar observations in the strong-field but weakly-dynamical regime~\cite{stairs,Wex:2014nva,Berti:2015itd} (see Psaltis~\cite{Psaltis:2008bb} for other tests of General Relativity with neutron star observations). Electromagnetic observations of black holes also allows one to probe the strong-field but weakly-dynamical regime~\cite{Johannsen:2008tm,Johannsen:2008aa,Bambi:2011jq,Horbatsch:2011ye,Yagi:2012gp,Bambi:2012pa,Bambi:2013fea,Bambi:2014nta,Kong:2014wha} (see e.g.~\cite{Bambi:2015ldr,Bambi:2015kza,Johannsen:2016uoh,Yagi:2016jml} for recent reviews), though such observations typically suffer from large systematics due to uncertainties in the astrophysical modeling. The recent GW150914~\cite{Abbott:2016blz} and GW151226~\cite{Abbott:2016nmj} detections offered the first test of General Relativity in the extreme gravity regime~\cite{TheLIGOScientific:2016src,Yunes:2016jcc,TheLIGOScientific:2016pea}. Figure~\ref{fig:phase-diagram} illustrates schematically the regions in the curvature and gravitational potential phase space that each experiment or observation probes~\cite{Psaltis:2008bb,Baker:2014zba,Yunes:2016jcc}. Probing gravity in the extreme gravity regime is important since certain theories (such as dCS gravity described in Sec.~\ref{sec:dCS}) give rise to relatively large deviations from General Relativity \emph{only} in this regime; in such theories, weak-field experiments can only place very weak constraints. 

One of the problems with using neutron star observations to test General Relativity in the strong- or extreme-gravity regime is possible degeneracies between uncertainties in nuclear physics and modified gravity effects. For example, the relation between the neutron star mass and radius depends not only on the underlying gravitational theory, but also on the unknown equation of state. In principle, one could use independent measurements of these quantities to test General Relativity, but in practice this is quite difficult unless the equation of state has been constrained sufficiently strongly \emph{a priori}.

Universal relations are useful precisely because they project out any uncertainties in nuclear physics, thus allowing us to focus on testing General Relativity. Imagine that one makes two independent observations that allow us to determine independently two members of the I-Love-Q relations. Given a single member, the relations themselves allow us to infer the other two. Two members of the trio, therefore, provide redundancy and allow us to verify whether the relation follows its General Relativistic form, or whether there are statistically significant anomalies.   

How can one measure two members of the I-Love-Q trio independently? As already explained at the beginning of this section, the moment of inertia is expected to be measured with future radio binary pulsar observations, while the tidal deformability might be measured with future gravitational wave observations. The I-Love relation then provides an equation-of-state independent and model independent test of General Relativity, where any equation-of-state variation in the universal relations acts as a systematic error. Because of the latter, the power of such an I-Love test will be eventually fundamentally limited by this inherent equation-of-state variation, rather than statistical error. The quadrupole moment, on the other hand, is much more difficult to measure; we will probably have to wait for third- or fourth-generation gravitational wave detectors, such as the Einstein Telescope~\cite{Yagi:2013awa}, or carry out detailed studies of quasi-periodic oscillation of accretion disks around neutron stars~\cite{Pappas:2012nt,Pappas:2015mba} (although the latter suffers from large systematic errors due to modeling uncertainties).

\begin{figure}[htb]
\begin{center}
\begin{tabular}{l}
\includegraphics[width=8.5cm,clip=true]{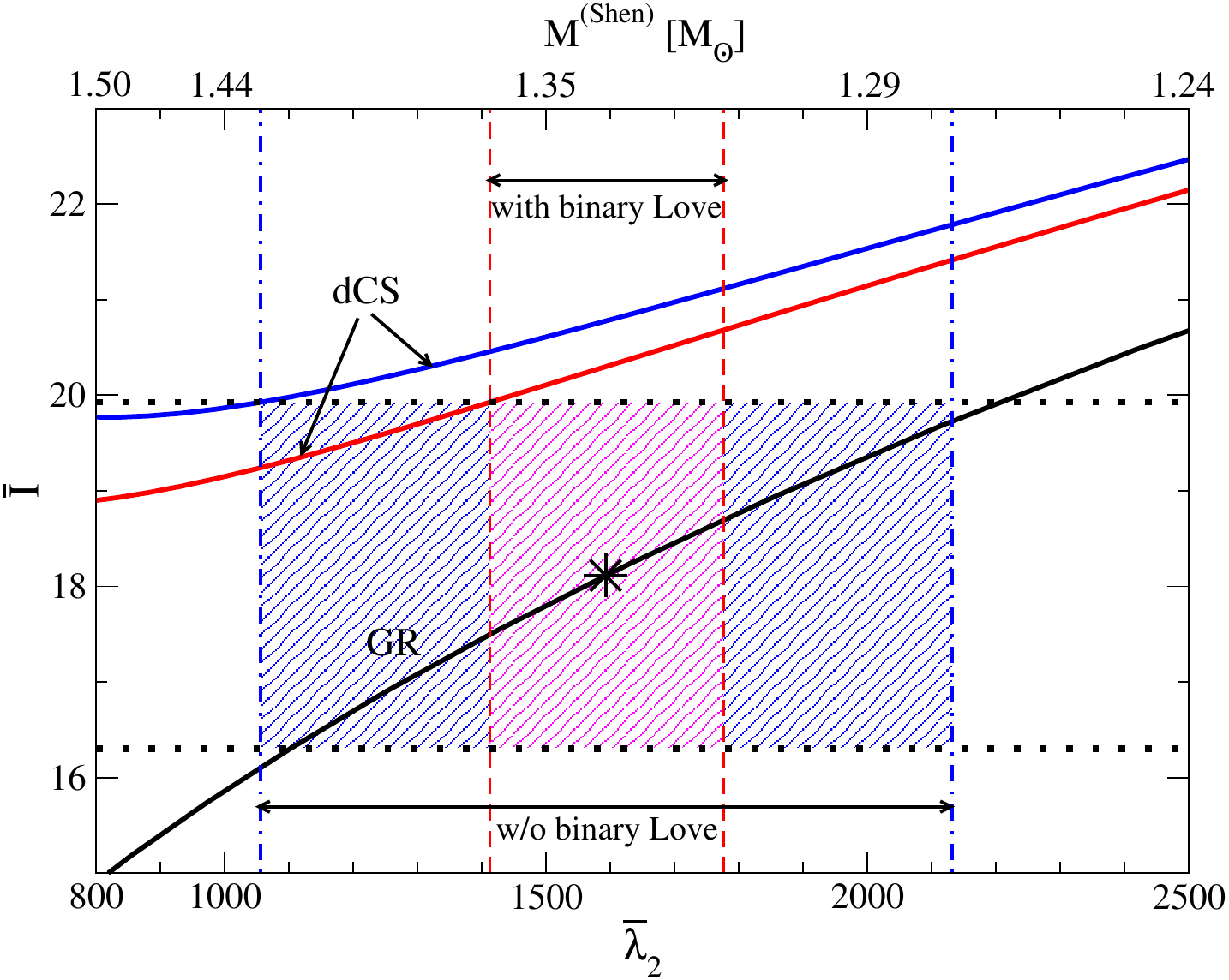} 
\end{tabular}
\caption{\label{fig:I-Love-CS}
The I-Love relation in General Relativity (black solid) and dCS gravity with a fixed dCS coupling constant of $\xi_\CS/M^4 = 1.78 \times 10^4$ (blue solid) and $\xi_\CS/M^4 = 1.46 \times 10^4$ (red solid). For reference, the top axis shows the neutron star mass using the Shen equation of state. The shaded areas show the expected error regions from the future measurements of $\bar I$ with radio double binary pulsar observations (black dotted) and $\bar \lambda_2^{(0)}$ with $m_0 = 1.338M_\odot$ using gravitational wave observations of neutron star binaries with $(1.2,1.4)M_\odot$ and a signal-to-noise ratio of 30, assuming that the measurement is consistent with General Relativity. The blue dotted-dashed (red dashed) vertical lines correspond to measurement accuracy of $\bar \lambda_2^{(0)}$ without (with) the universal $\bar \lambda_2^{(1)}$--$\bar \lambda_2^{(0)}$ relation. 
This figure was taken and edited from Yagi and Yunes~\cite{Yagi:2016qmr}.
}
\end{center}
\end{figure}

Figure~\ref{fig:I-Love-CS} shows the expected error regions in the I-Love plane from future radio and gravitational wave observations. The blue region is constructed without using the $\bar \lambda_2^{(1)}$--$\bar \lambda_2^{(0)}$ relation, while the red region does use this relation. The red region is smaller than the blue one because the universal binary Love relation breaks degeneracies between the tidal parameters, which improves their measurement accuracy (see Sec.~\ref{sec:nuclear-GW}). The figure also shows two I-Love relations in dCS gravity with the coupling parameters $\xi_\CS/M^4 = 1.78 \times 10^4$ $\zeta_\CS = 0.11$) (blue solid) and $\xi_\CS/M^4 = 1.46 \times 10^4$ ($\zeta_\CS = 0.09$) (red solid) that are marginally consistent with the blue and red regions. Notice that the dimensionless coupling constant $\zeta_\CS \equiv \xi_\CS \, M^2/R^6$ is smaller than unity and hence the small coupling approximation remains valid, where we assumed $M=1.338M_\odot$ (the mass of the primary pulsar in J0737-3039~\cite{Kramer:2006nb}) with the Shen equation of state. The binary Love relation allows for slightly stronger constraints, although clearly the strength of the test is essentially dominated by the error in the measurement of the moment of inertia. These projected constraints on $\xi_{\CS}$ from the I-Love relation are six orders of magnitude stronger~\cite{Yagi:2013awa} than current constraints with Solar System~\cite{AliHaimoud:2011fw} and table-top~\cite{Yagi:2012ya} experiments. This is because dCS gravity is only weakly constrained with current experiments, and relatively large deviations may appear in the extreme gravity regime.

The power of these equation-of-state independent and model independent tests depends on how much the universal relations in modified gravity differ from those in General Relativity. This, in turn, depends on how strongly the modified theories have been constrained by existing experiments and observations. For example, with the notable exception of dCS gravity, the I-Love-Q relations in all the theories considered in Sec.~\ref{sec6:ILQ-Mod-Grav} are very similar to the General Relativity ones. This is precisely because these theories have already been tightly constrained and one typically chooses values for the coupling constants that are within experimentally and observationally allowed bounds.

%%%%%%%%%%%%%%%%%%%%%%%%%%%%%%%%%%%%%%%%%%%%%%%%%%%%%%
\subsection{Cosmology}
\label{sec:cosmology}

Let us now review how universal relations allow us to carry out cosmological studies with gravitational waves. Just like Type Ia supernovae can be used as standard candles to probe the cosmological evolution of the universe, one can in principle use neutron star (or black hole) binaries as \emph{standard sirens}~\cite{Schutz:1986gp,Holz:2005df,Dalal:2006qt,Sathyaprakash:2009xt,Nissanke:2009kt,Cutler:2009qv,Zhao:2010sz,Nishizawa:2010xx,Camera:2013xfa,Nissanke:2013fka}. The idea here is that the independent extraction of the luminosity distance and the redshift would suffice to infer the underlying cosmological parameters, such as the Hubble constant and the energy density of dark matter and dark energy~\cite{DelPozzo:2011yh,Nishizawa:2011eq,Yagi:2011bt,Taylor:2011fs,Taylor:2012db,Namikawa:2015prh,Oguri:2016dgk}. 

Accurately extracting both the luminosity distance and the redshift with gravitational waves alone, however, is currently extremely challenging. In principle, gravitational wave observations alone allow us to measure the luminosity distance of the source from the waveform amplitude. In practice, degeneracies with the inclination angle can deteriorate such measurements. Extracting the redshift is even more challenging due to the degeneracies with the intrinsic mass, since the waveform depends on the \emph{redshifted} masses $m_{z,A} \equiv (1+z) m_A$. One way to infer redshift information is to identify the binary's host galaxy and use its electromagnetically-determined redshift, but this is very challenging without a network of detectors due to poor sky localization. Electromagnetic counterparts to gravitational wave observations, such as a short gamma-ray burst after a neutron star merger, would greatly help determine the host galaxy to extract the redshift and break degeneracies between the inclination angle and the luminosity distance~\cite{Sathyaprakash:2009xt}. 

Messenger and Read~\cite{Messenger:2011gi} proposed a novel way of extracting the redshift of a neutron star binary with gravitational waves alone via tidal effects encoded in the late binary inspiral. As mentioned in Sec.~\ref{sec:nuclear-GW}, these tidal effects in the inspiral phase of a neutron star binary are encoded in the tidal deformability parameter $\bar \lambda_{2,A}$, which is a function of the intrinsic mass $m_A$ and \emph{not} the redshifted mass $m_{z,A}$. This means that if one knows the correct equation of state \emph{a priori}, one can in principle extract $m_A$ through the measurement of $\bar \lambda_{2,A}$. Combining this information with the measurement of $m_{z,A}$, one can then break the degeneracy between the intrinsic mass and the redshift, allowing the extraction of the redshift without an electromagnetic counterpart. 

This idea was first tested through a Fisher analysis~\cite{Messenger:2011gi} and then with a Bayesian analysis~\cite{Li:2013via}, by estimating the measurement accuracy of $z$ as a function of the injected redshift with the Einstein Telescope. These studies suggest that the redshift can be measured to $10$--$100\%$, depending sensitively on the equation of state chosen. Messenger \et~\cite{Messenger:2013fya} carried out a similar analysis but used tidal information in the post-merger phase instead of information from the late inspiral. The measurement of the luminosity distance and the redshift were used to infer cosmological parameters in Del Pozzo \et~\cite{DelPozzo:2015bna}. The latter found that the measurement accuracy of cosmological parameters would be comparable to that of Planck~\cite{Ade:2015xua,Ade:2015rim} if the Einstein Telescope detects $\mathcal{O}(10^6$--$10^{7})$ neutron star merger events. 

But assuming we know exactly the equation of state \emph{a priori} is currently too optimistic. A more realistic situation may be the case where only the leading Taylor-expanded coefficient of the tidal deformability $\bar \lambda_2^{(0)}$ in Eq.~\eqref{eq:Taylor-expand-Love} is known from Adv.~LIGO observations (e.g.~by stacking signals from different neutron star binary sources~\cite{DelPozzo:2013ala,Agathos:2015uaa}). In such a case, knowledge of $\bar \lambda_2^{(1)}$ is crucial if one wants to extract the redshift, because the intrinsic mass of a neutron star enters first at $k=1$ in Eq.~\eqref{eq:Taylor-expand-Love}. Yagi and Yunes~\cite{Yagi:2015pkc,Yagi:2016qmr} recently pointed out that one can determine $\bar \lambda_2^{(1)}$ from $\bar \lambda_2^{(0)}$ thanks to the universal binary Love relation (up to an error comparable to the equation-of-state variation in the relation). This study performed a Fisher analysis using the Einstein Telescope, with binary parameters similar to those chosen in Eq.~\eqref{eq:Fisher-par}, a parameter set that included the redshifted chirp mass $(1+z) \mathcal{M}$ and the redshift itself (instead of the non-redshifted chirp mass and the tidal parameters $\bar \lambda_{2,s}$ and $\bar \lambda_{3,s}$), and the universal relation between $\bar \lambda_2^{(0)}$ and $\bar \lambda_2^{(1)}$ in Eq.~\eqref{eq:Taylor-expand-Love} with $m_0 = 1.4M_\odot$.

\begin{figure}[thb]
\begin{center}
\includegraphics[width=8.cm,clip=true]{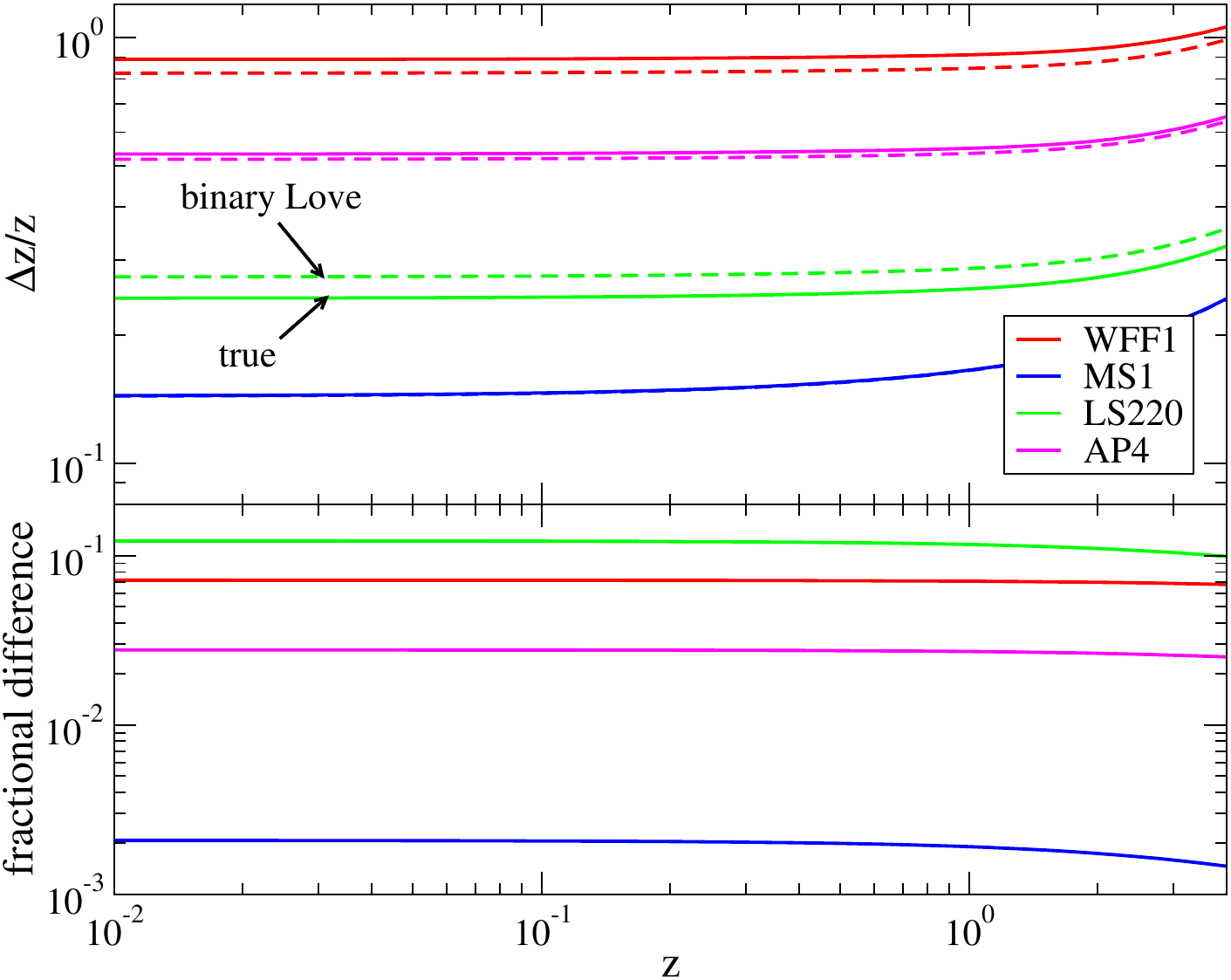}  
\caption{\label{fig:cosmology} (Top) Fractional measurement accuracy of the redshift $z$ as a function of the injected redshift for various equations of state assuming that one knows the true $\bar \lambda_2^{(1)}$ \emph{a priori} (solid), and with $\bar \lambda_2^{(1)}$ obtained from the $\bar \lambda_2^{(0)}$--$\bar \lambda_2^{(1)}$ universal relation (dashed). In both cases, one assumes that ET detects gravitational wave signals emitted from $(1.4,1.4)M_\odot$, non-spinning neutron star binaries and $\bar \lambda_2^{(0)}$ is known \emph{a priori} from e.g.~Adv.~LIGO measurements.  (Bottom) Fractional difference between $\Delta z/z$ with the true $\bar \lambda_2^{(1)}$ and with that obtained from the universal relation. The difference is at most $\sim 10\%$, suggesting that the binary Love relations can be used in place of requiring absolute knowledge of the equation of state \emph{a priori}. 
This figure is taken from Yagi and Yunes~\cite{Yagi:2015pkc}.
}
\end{center}
\end{figure}

The top panel of Fig.~\ref{fig:cosmology} shows the fractional measurement accuracy of $z$ as a function of the signal's injected $z$ for various choices of equation of state, assuming an equal-mass, non-spinning neutron star binary with stellar masses of $1.4M_\odot$. The solid curves correspond to the case where the equation of state is assumed to be known exactly \emph{a priori}~\cite{Messenger:2011gi}. The dashed curves correspond to the case where one knows only $\bar \lambda_2^{(0)}$ from Adv.~LIGO measurements, and $\bar \lambda_2^{(1)}$ is obtained from the universal $\bar \lambda_2^{(1)}$--$\bar \lambda_2^{(0)}$ relation shown in the right panel of Fig.~\ref{fig:lambdas-lambdaa}. The difference between the solid and dashed curves is due to the equation-of-state variability in the $\bar \lambda_2^{(1)}$--$\bar \lambda_2^{(0)}$ relation, which introduces a systematic error. The bottom panel of Fig.~\ref{fig:cosmology} shows the fractional difference between the solid and dashed curves as a function of redshift  for different equations of state. Observe that the fractional difference is $\sim 10\%$ at most, which is smaller than the fractional statistical error on $z$ on the top panel. This suggests that one can still carry out cosmological studies with gravitational wave observations through measurements of the tidal parameters, even if only $\bar \lambda_2^{(0)}$ is known from previous observations and we lack electromagnetic counterparts.

%%%%%%%%%%%%%%%%%%%%%%%%%%%%%%%%%%%%%%%%%%%%%%%%%%%%%%%%
\section{Open Questions}
\label{sec8:OpenQuestions}

This review attempted to summarize the current status of our understanding of the approximate universality present in neutron stars and quark stars, but by no means has this topic been exhausted. Many open questions remain related to essentially every topic covered in this review. Perhaps one of the open questions that would have the largest impact is finding extensions to the universality presented here. For example, universality is clearly destroyed when considering slowly-rotating, strongly-magnetized neutron stars with magnetic fields in excess of roughly $10^{12}$ Gauss~\cite{Haskell:2013vha}. This is simply because the magnetic field and its topology affect the magnetic-induced quadrupolar deformation, thus adding a new, unaccounted for contribution to the standard I-Q relations. Perhaps, however, it would be possible to extend the universality to magnetized neutron stars by adding new parameters to the universal relations that quantify the strength and the topology of the magnetic field. This could be investigated perturbatively by assuming the magnetic-induced moment of inertia and multipole moments are small relative to the rotation-induced ones~\cite{Konno:1999zv,Konno:2001jr}. One can also study how the no-hair like relations among multipole moments for magnetars are different from the unmagnetized case. Alternatively, one can study if the universal relations remain equation-of-state insensitive if one fixes a \emph{dimensionless} combination of the magnetic field strength (other than $B \, P$) or if one changes the normalization of the I-Love-Q and multipole moments. This speculation is motivated from the study of universality in rapidly rotating stars, which seemed to be lost when one fixed the dimensional angular velocity of a star~\cite{Doneva:2013rha}; the universality is, however, preserved if one fixes the dimensionless spin parameters~\cite{Pappas:2013naa,Chakrabarti:2013tca}. The degree to which relations are universal also depends on how one normalizes the stellar quantities~\cite{Majumder:2015kfa}.

An extension of the universal relations to magnetized neutron stars could then directly lead to an application of such relations to proto-neutron stars i.e.~newly-born stars that rotate differentially and are highly magnetized~\cite{Heger:2004qp,Ott:2005wh}. Extensions of the no-hair relations to differentially rotating stars have already been found within the slow- and small differential-rotation approximations in the Newtonian limit~\cite{Bretz:2015rna}. On the other hand, the I-Love-Q relations for non-magnetized proto-neutron stars in a ``quasi-stationary'' phase were studied using non-barotropic equations of state with uniform rotation~\cite{Martinon:2014uua}. Thus, if one could extend these relations to magnetized, rapidly- and differentially-rotating stars in full General Relativity, one could imagine a dynamic investigation of universality during the early stages of neutron star formation. One could then also apply such universal relations for differentially rotating stars to hypermassive neutron stars formed after the merger of neutron star binaries~\cite{Shapiro:2000zh,Paschalidis:2012ff,Hotokezaka:2013iia}.

Another fertile area of research concerns studies of why the universality holds. The only physical explanation of the no-hair relations that has not yet been refuted invokes the emergence of a self-similar symmetry in isodensity profiles due to these being the lowest energy states of the system~\cite{Yagi:2014qua} (see also Chatziioannou \et~\cite{Chatziioannou:2014tha}, Sham \et~\cite{Sham:2014kea} and Chan \et~\cite{Chan:2015iou} for perturbative, mathematical explanations based on incompressible stars). This explanation is based on both numerical calculations in full General Relativity~\cite{Yagi:2014qua} and analytic calculations in the Newtonian limit~\cite{Stein:2014wpa}. The explanation could thus be strengthened by finding a mathematical proof that holds in full General Relativity through the use of coordinate-independent definitions of multipole moments. Even if this were achieved, however, this explanation could still not explain why the universality extends to the tidal deformabilities of neutron stars~\cite{Yagi:2013bca,Yagi:2013awa,Yagi:2013sva,Pani:2015nua,Yagi:2015pkc} or to the neutron star oscillation modes~\cite{Andersson:1997rn,Tsui:2004qd,Lau:2009bu,Chan:2014kua,Chirenti:2015dda}. This explanation also cannot be applied to the universality found in the neutron star to black hole transition~\cite{Yagi:2015upa}. The latter points to perhaps a new type of universality related to a phase transition between black hole and neutron star states. Such a possibility has already been studied in de Boer \et~\cite{deBoer:2009wk} and Arsiwalla \et~\cite{Arsiwalla:2010bt} within the context of the anti-de Sitter/conformal field theory correspondence~\cite{Maldacena:1998re,Aharony:1999ti}. One could investigate whether such universality is truly present in realistic, asymptotically-flat neutron star collapse through numerical simulations with different equations of state. To achieve this, however, one would need to generalize the Geroch-Hansen multipole moments to dynamical spacetimes.

Ultimately, the value of the universal relations rests on their applicability to different observations. One interesting application that has not yet been exhaustively studied is experimental relativity. The I-Love relations, through the measurement of the moment of inertia with binary pulsars and the Love number with gravitational waves, have been shown to be a promising model independent and equation-of-state independent test of General Relativity~\cite{Yagi:2013bca,Yagi:2013awa}. For this test to be effective, however, one must determine whether a large class of modified theories of gravity predict an I-Love relation that is different from the General Relativity one, with the difference proportional to the coupling parameters of the theory. This is crucial to then be able to place concrete constraints on such modified theories once an I-Love observation has been made. As of the writing of this review paper, dCS gravity is the only example that allows for constraints with the I-Love relations that are stronger than all other current bounds. More work is needed to determine how the I-Love-Q relation is modified in other theories, such as in Einstein-\AE ther theory~\cite{Jacobson:2000xp,Eling:2004dk,Jacobson:2008aj} and Ho\v rava-Lifshitz (or khronometric) gravity~\cite{Horava:2009uw,Blas:2009qj,Blas:2010hb}, both of which introduce vector fields. The I-Love relation in massive scalar-tensor theories may show large deviations from GR like in the I-Q relation explained in Sec.~\ref{sec:ST}, as the massive scalar field may allow the evasion of Solar System and binary pulsar observations. Another application in experimental relativity would be to devise new tests that use the universal behavior of the full no-hair relations, instead of just the I-Love universality. This is not difficult in principle, but it seems difficult to implement in practice due to the weak impact that higher multipole moments have on neutron star observables. Another avenue for future work includes studying the I-Love-Q and no-hair relations with the gauge-invariant \emph{extended} Geroch-Hansen multipole moments that are valid in modified theories of gravity~\cite{Pappas:2014gca,Suvorov:2015yfv}.

A final, yet important, application of the universal relations concerns gravitational wave astrophysics. One can apply such relations to break some degeneracies among gravitational waveform parameters, which improve the measurement accuracy of the spins and tidal parameters of individual neutron stars in a binary. Previous work~\cite{Yagi:2013bca,Yagi:2013awa,Yagi:2013sva,Yagi:2015pkc} used a simple Fisher analysis~\cite{Cutler:1994ys} to show the impact of such relations on future gravitational wave observations. However, such an analysis is approximate and only valid for gravitational waves with large signal-to-noise ratios~\cite{Vallisneri:2007ev}. One could thus carry out a more detailed, Bayesian analysis using e.g.~Markov chain Monte Carlo method~\cite{Cornish:2007if,Littenberg:2009bm} to predict how much of an improvement in parameter estimation one obtains when using the universal relations to break parameter degeneracies. 

The long and treacherous road of the universal relations has given us many surprises. The importance of the elliptical isodensity approximation and the possible interpretation of neutron star collapse as a phase transition are only the tip of the iceberg. Only the future will tell what new universal discoveries will be made and how these will be implemented in new observations to further our understanding of the cosmos.

%%%%%%%%%%%%%%%%%%%%%%%%%%%%%%%%%%%%%%%%%%%%%%%%%%%%%%%%
%***********************************************************************************************************************
%***********************************************************************************************************************
%***********************************************************************************************************************
%***********************************************************************************************************************
%***********************************************************************************************************************
\section*{Acknowledgment}   

We would like to thank Haris Apostolatos, Joe Bretz, Katerina Chatziioannou, Koutarou Kyutoku, Barun Majumder, George Pappas and Leo Stein for wonderful and insightful collaborations on the universality of neutron stars. We would also like to thank Neil Cornish, Scott Hughes, Jim Lattimer, Luis Lehner, Ben Owen, Paolo Pani, Madappa Prakash, Frans Pretorius, Luciano Rezzolla and Sanjay Reddy and for many discussions and comments on various manuscripts on universality. We would like to thank Fabrizio Canfora, Vladimir Manko and Kentaro Takami for pointing out important references. We further would like to thank Paolo Pani for sharing with us draft of an erratum to one of his papers. We would also like to conclude by thanking Emanuele Berti, Daniela Doneva, Burkhard Kleihaus, Lap-Ming Lin, Barun Majumder, and Paolo Pani for sharing their data with us, which we used to create some of the figures in this review paper. K.Y. acknowledges support from JSPS Postdoctoral Fellowships for Research Abroad and NSF grant PHY-1305682. N.Y. acknowledges support from NSF CAREER Grant PHY-1250636. 

%%%%%%%%%%%%%%%%%%%%%%%%%%%%%%%%%%%%%%%%%%%%%%%%%%%%%%%
%***********************************************************************************************************************
%***********************************************************************************************************************
%***********************************************************************************************************************
%***********************************************************************************************************************
%***********************************************************************************************************************

%%%%%%%%%%%%%%%%%%%%%%%%%%%%%%%%%%%%%%%%%%%%%%%%%%%%%%%
%***********************************************************************************************************************
%***********************************************************************************************************************
%***********************************************************************************************************************
%***********************************************************************************************************************
%***********************************************************************************************************************

\section*{References}

\bibliographystyle{elsarticle-num}
\bibliography{phyjabb,bibliography}

%%%%%%%%%%%%%%%%%%%%%%%%%%%%%%%%%%%%%%%%%%%%%%%%%%%%%%%
%***********************************************************************************************************************
%***********************************************************************************************************************
%***********************************************************************************************************************
%***********************************************************************************************************************
%***********************************************************************************************************************
\end{document}